\documentclass[12pt]{ruthesis}
\usepackage{amsmath}
\usepackage{amssymb}
\usepackage{amsbsy}
\usepackage{amstext}
\usepackage{amsthm}
\usepackage{amsfonts}
\usepackage{physics}
\usepackage{graphicx}
\usepackage{hyperref}  
\usepackage{pdfpages}
\hypersetup{
	colorlinks = true,
	linkcolor = blue,
	citecolor = blue,
	urlcolor = blue,
	anchorcolor=blue,
	filecolor=blue,
	menucolor=blue,
	linktocpage=true,
	pdfproducer=medialab,
}
\usepackage{bm}
\usepackage{comment}
\usepackage[normalem]{ulem}  
\usepackage{tensor}
\def\t{\tensor}
\usepackage{slashed}
\usepackage{simplewick}
\usepackage{enumitem}
\usepackage{cite}
\usepackage{pdflscape}
\usepackage{setspace}
\let\oldfootnote\footnote
\renewcommand{\footnote}[1]{\oldfootnote{\setstretch{1.0} #1}}

\def\l{\left}
\def\r{\right}
\def\({\l(}
\def\){\r)}
\def\[{\l[}
\def\]{\r]}
\def\la{\langle}
\def\ra{\rangle}

\def\rm{\mathrm}

\def\cal{\mathcal}

\def\pd{\partial}
\def\til{\tilde}
\def\b{\boldsymbol}

\def\MP{M_\mathrm{P}}

\def\bar{\overline}
\def\sl{\slashed}  

\def\j{\varphi}
\def\f{\phi}
\def\ve{\varepsilon}

\def\sep{~,\quad}  

\def\bx{\b x}
\def\by{\b y}
\def\bz{\b z}

\DeclareMathOperator{\sinc}{sinc}

\DeclareMathOperator{\Diag}{Diag}  

\title{Probing ultralight dark fields \\in cosmological and astrophysical systems}
\ctitle{Title}
\author{Hong-Yi Zhang}
\department{Department of Physics and Astronomy}
\school{Rice University}
\degree{Doctor of Philosophy}

\committee {
        Mustafa A. Amin, Chair \\
        Associate Professor of Physics and Astronomy \and
        Andrew J. Long \\
        Assistant Professor of Physics and Astronomy \and
        Milivoje Luki\'{c}\\
        Professor of Mathematics
}

\address{Houston, Texas}
\donemonth{Dec} \doneyear{2023} \makeindex
\begin{document}
\begin{frontmatter}
    \makeatletter
    \providecommand*{\toclevel@frontmatter}{0}
    \makeatother
    
    \pagenumbering{roman}
    \includepdf[width=1.5\textwidth, offset=10mm 0]{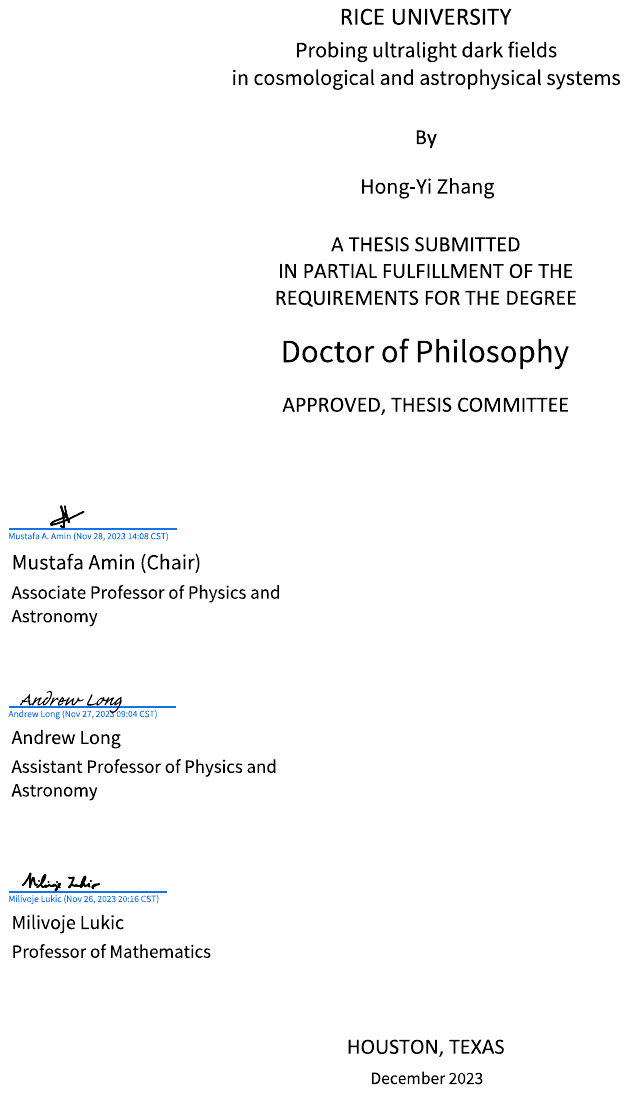}
    \begin{abstract}
Dark matter constitutes $26\%$ of the total energy in our universe, but its nature remains elusive. Among the assortment of viable dark matter candidates, particles and fields with masses lighter than $40 \rm{eV}$, called ultralight dark matter, stand out as particularly promising thanks to their feasible production mechanisms, consistency with current observations, and diverse and testable predictions. In light of ongoing and forthcoming experimental and observational efforts, it is important to advance the understanding of ultralight dark matter from theoretical and phenomenological perspectives: How does it interact with itself, ordinary matter, and gravity? What are some promising ways to detect it?

In this thesis, we aim to explore the dynamics and interaction of ultralight dark matter and other astrophysically accessible hypothetical fields in a relatively model-independent way. Without making specific assumptions about their ultraviolet physics, we first demonstrate a systematic approach for constructing a classical effective field theory for both scalar and vector dark fields and discuss conditions for its validity. Then, we explore the interaction of ultralight dark fields, both gravitational and otherwise, within various contexts such as nontopological solitons, neutron stars, and gravitational waves. 
\end{abstract}
    \thispagestyle{empty}
\begin{acknowledge}
Completing this PhD has not been an easy journey, and it would not have been possible without the guidance, support, and encouragement of many.
	
First and foremost, I would like to express my deepest gratitude to my advisor Prof. Mustafa Amin. His patience, support, and relentless dedication to my growth have been the bedrock of my doctoral journey. His mentorship was not just academic but also profoundly shaped my personal and professional outlook.
	
I am also extremely grateful to Prof. Andrew Long. His insights and vast knowledge has been instrumental in the success of my research. Talking and collaborating with him has been both a privilege and an enriching learning experience.
	
This journey was made even more enlightening by the contributions and collaborations with Dr. Zong-Gang Mou, Dr. Borna Salehian, Ray Hagimoto, Siyang Ling, Dr. Mudit Jain, Dr. Kaloian Lozanov, Prof. David Kaiser, Prof. Edmund Copeland, Prof. Paul Saffin, and Prof. Mohammad Hossein Namjoo. Each one of my collaborators brought unique perspectives and expertise that enriched this work and my personal growth as a researcher.
	
On the personal front, I owe an immense debt of gratitude to my wife, Xia Liu. The challenges of being separated by distance were immense, and yet her unwavering trust, support, and encouragement have been my guiding light. I also want to express my heartfelt thanks to my parents for their constant love and belief in my abilities.
\end{acknowledge}
    \tableofcontents
    \listoffigures
    \listoftables
    \end{frontmatter}
\pagenumbering{arabic}

\linespacing{1.7}

\chapter{Introduction}
\label{sec:intro}

\section{Dark matter}
\label{sec:intro_dm}
Dark matter (DM) remains as one of the biggest mysteries in our universe. The idea of its existence was arguably supported in the 1930s when researchers discovered the high mass-to-light ratio in galaxy clusters \cite{zwicky1933rotshift}. This hypothesis was further validated by the observation of anomalies in the rotation curves of galaxies: the observed flat rotation curves did not align with predictions based on Newtonian mechanics and visible galactic matter alone \cite{Rubin:1970zza}. These discrepancies unveil a universe where Newtonian predictions based solely on visible matter failed. See \cite{Bertone:2016nfn} for a comprehensive introduction to the history of DM.

Modern observations have provided profound insights into DM, presenting evidence across an array of scales -- from small-scale nonlinear structure within galaxies to large-scale structure of the universe. Aligning with theoretical predictions, light element abundances, temperature and polarization anisotropies in the cosmic microwave background (CMB), and multiple probes of galaxy distributions collectively suggest a vivid picture: a universe composed of dark energy, DM and ordinary matter, known as the $\Lambda$CDM model or the concordance model of cosmology \cite{dodelson2020modern, Baumann:2022mni}. These constituents, respectively, account for $69\%$, $26\%$, and $5\%$ of the universe's total energy \cite{Planck:2018vyg}.

\begin{figure}
	\centering
	\includegraphics[width=0.65\linewidth]{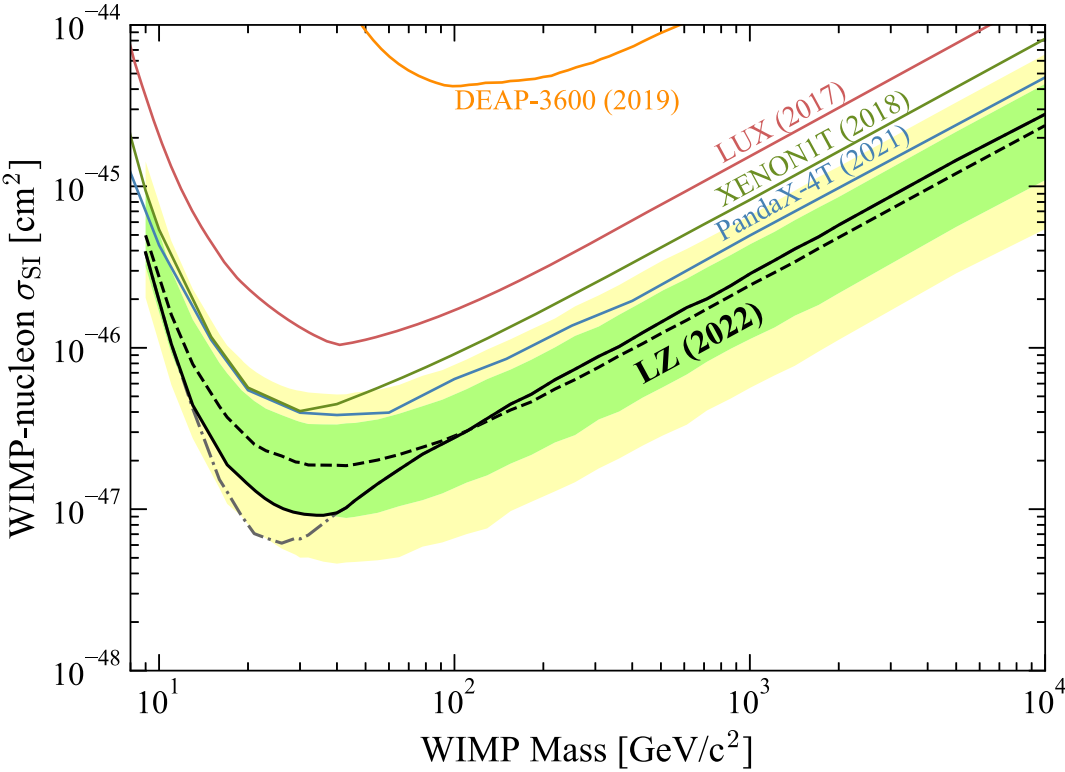}
	\caption{The 90\% confidence limit for the spin-independent WIMP cross section vs WIMP mass. The current best constraint comes from the LUX-ZEPLIN (LZ) experiment (black lines) \cite{LZ:2022lsv}. The dashed line shows the median of the sensitivity projection and the green and yellow bands are the $1\sigma$ and $2\sigma$ sensitivity bands for the LZ experiment. Also shown are the PandaX-4T \cite{PandaX-4T:2021bab}, XENON1T \cite{XENON:2018voc}, LUX \cite{LUX:2016ggv}, and DEAP-3600 \cite{DEAP:2019yzn} limits. Figure taken from \cite{LZ:2022lsv}.}
	\label{fig:wimp}
\end{figure}

DM must be cold, i.e. nonrelativistic (NR) at the time of structure formation, to clump efficiently and account for the formation of galaxies. This simplest scenario of cold dark matter (CDM) is very successful in fitting the angular power spectrum of CMB and the linear power spectrum of galaxy distributions. A notable example is the weakly interacting massive particles (WIMPs), which represent new elementary particles that interact with baryons not only gravitationally but also through the weak force or a new force of comparable strength \cite{Roszkowski:2017nbc, Lin:2019uvt}. WIMPs stand out for a unique reason: if thermally produced in the primordial universe, their current abundance would align with the observed DM abundance. This has led to a fervent experimental hunt for WIMPs and their interactions \cite{Roszkowski:2017nbc, Lin:2019uvt, Arcadi:2017kky, Safdi:2022xkm}, nevertheless conclusive evidence of these particles remains elusive. See figure \ref{fig:wimp} for current best constraints on the interaction cross section of WIMPS with nucleons.

While the gravitational properties of DM on large scales have been extensively mapped, a closer look at DM in sub-galatic scales reveals intriguing complexity. For instance, a discrepancy arises between the observed flat DM density profiles in the centers of galaxies and the predictions of cuspy cores made by numerical simulations of CDM cosmology \cite{Flores:1994gz,Moore:1994yx}. That is, the observed small-scale structure is less prominent than what expected in the CDM scenario. This cusp-core problem poses a challenge, but also reveals an avenue into understanding the microphysics of DM. A particularly interesting candidate motivated by this problem is the \emph{ultralight} DM \cite{Hui:2021tkt, Ferreira:2020fam}, whose de Broglie wavelength is large enough for suppressing the formation of small-scale structure. This DM candidate has rich phenomenology thanks to wavelike effects in macroscopic and even astrophysical scales, and will be the focus of this thesis. Other solutions of the problem include, for example, warm DM \cite{Colin:2000dn} and self-interacting DM \cite{Spergel:1999mh}, for which the small scale structure is suppressed due to a significant thermal velocity dispersion or appropriate nongravitational DM self-interactions.

\begin{figure}
	\centering
	\includegraphics[width=\linewidth]{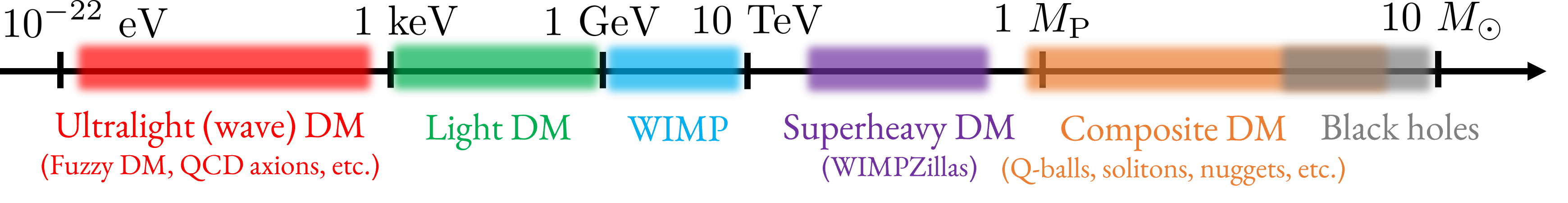}
	\caption{Mass of different DM candidates, ranging from elementary particles, composite objects to primordial black holes. Note that the mass range provided is not to scale and meant to be indicative.}
	\label{fig:dmmass}
\end{figure}

Motivated by various considerations, DM models spanning over 80 orders of magnitude in mass have been postulated, ranging from elementary particles, to composite objects, and up to astrophysical-sized primordial black holes \cite{Jacobs:2014yca, Khlopov:2019qcr, Carr:2020gox, Carr:2020xqk}, as shown in figure \ref{fig:dmmass}. This reflects that the identity and properties of DM remain shrouded in mystery.

\section{Ultralight dark fields}
\label{sec:intro_ultralight}
Dynamical measurements tell us that the DM mass density in the solar neighborhood is about $0.4\rm{GeV~cm^{-3}}$ \cite{deBoer:2010eh, Bovy:2012tw, mckee2015stars, Sivertsson:2017rkp}, based on which we can deduce the average inter-particle separation, given a DM particle mass. It is instructive to compare this scale to the de Broglie wavelength of the particle
\begin{align}
	\lambda_\rm{dB} = \frac{2\pi}{mv} = 6.0 \rm{pc} \( \frac{10^{-20} \rm{eV}}{m} \) \( \frac{200 \rm{km/s}}{v} \) ~,
\end{align}
where $1 \rm{pc}\approx 3\times 10^{16} \rm{m}$. The de Broglie wavelength exceeds the inter-particle separation if the number of particles per de Broglie volume $\lambda_\rm{dB}^3$, given by
\begin{align}
	N_\rm{dB} \sim \( \frac{40 \rm{eV}}{m} \)^4 \( \frac{\rho}{0.4 \rm{GeV/cm^3}} \) \( \frac{200 \rm{km/s}}{v} \) ~,
\end{align}
is larger than unity. Therefore, for $m\ll 40 \rm{eV}$, the small separation between DM particles and large occupation numbers implies that the DM is best described by classical wave mechanics \cite{Guth:2014hsa,Hertzberg:2016tal}. This type of DM is called ultralight/wave DM. Ultralight DM particles must be bosonic for the Pauli exclusion principle precludes such a high occupancy for fermions \cite{Tremaine:1979we}, unless there are many distinct species of DM particles \cite{Davoudiasl:2020uig}.

For the ultralight end, DM particles with a mass $10^{-22}\rm{eV}$--$10^{-20}\rm{eV}$ are sometimes called fuzzy DM \cite{Hu:2000ke}. A mass $m<10^{-22}\rm{eV}$ is possible, but only if the particle constitutes a small fraction of DM, otherwise an excessively large $\lambda_\rm{dB}$ would negate the existence of DM-dominated dwarf galaxies. A characteristic feature of fuzzy DM is the soliton core \cite{Schive:2014hza}, a stable structure at the heart of DM haloes where quantum pressure counteracts gravitational collapse. By hosting a solitonic core, the galaxy can exhibit a flat central density profile, thereby preventing gravitational clustering of matter and resolving the core-cusp problem. The solitonic behavior, when mapped against observational data, can effectively distinguish fuzzy DM from other DM candidates. Moreover, solitons might dominate the bulge dynamics in massive galaxies and could potentially represent the majority of halo mass in smaller dwarf galaxies. Analysis against observed galactic density profiles has provided strong constraints on DM mass \cite{Schive:2014hza, Marsh:2015wka, Chen:2016unw, Gonzalez-Morales:2016yaf, Broadhurst:2019fsl, Bar:2018acw, Veltmaat:2018dfz, DeMartino:2018zkx, Pozo:2023zmx, ParticleDataGroup:2022pth}.

There has been significant effort in recent years to place various astrophysical constraints on the fuzzy DM particle mass \cite{Ferreira:2020fam, Hui:2021tkt}. These constraints generally bifurcate into two groups in moderate disagreement. On one hand, dwarf spheroidal galaxies typically exhibit characteristics consistent with large, low-density DM cores, such as those predicted by fuzzy DM theory with soliton cores at $m\simeq 10^{-22}\rm{eV}$ \cite{Schive:2014dra, Chen:2016unw, Khelashvili:2022ffq}. On the other, the Lyman-$\alpha$ forest observations suggest $m\gtrsim 10^{-21}\rm{eV}$ \cite{Irsic:2017yje, Nori:2018pka, Rogers:2020ltq}, where the wave effects of fuzzy DM become less pronounced. The disparity between these findings has become more pronounced of late. Notably, evaluations of the size and stellar dynamics of ultra-faint dwarf galaxies suggest $m\gtrsim 10^{-19}\rm{eV}$ \cite{Dalal:2022rmp}, while considerations centered on free-streaming effects and white noise in DM isocurvature perturbations lean towards $m\gtrsim 10^{-18}\rm{eV}$ \cite{Amin:2022nlh}. This existing tension hints at a potentially diverse nature of fuzzy DM \cite{Khelashvili:2022ffq}, or that some of the effects should be attributed to baryonic physics \cite{DelPopolo:2016emo, Ferreira:2020fam}.

Another well-known example of ultralight DM is axions \cite{Marsh:2015xka, DiLuzio:2020wdo, Sikivie:2020zpn, ParticleDataGroup:2022pth}, which are pseudo-Goldstone bosons associated with a spontaneously broken global symmetry that is anomalous under the standard model (SM) gauge theory \cite{Kim:1986ax}. Initially proposed as a natural solution to explain the absence of the neutron electric dipole moment \cite{Peccei:1977hh, Weinberg:1977ma, Wilczek:1977pj}, a quantum chromodynamics (QCD) axion is characterized by its decay constant $f_a$ \cite{Kim:1979if, Shifman:1979if, Zhitnitsky:1980tq, Dine:1981rt} and its mass is determined by $m_a \approx 5.7 \rm{\mu eV} (10^{12}\rm{GeV}/f_a)$ \cite{Borsanyi:2016ksw, Gorghetto:2018ocs}. Apart from the QCD axion and DM axions, axionlike particles have also been extensively studied in string theory \cite{Svrcek:2006yi, Arvanitaki:2009fg, Ringwald:2012cu}. For recent reviews, refer to \cite{Marsh:2015xka, DiLuzio:2020wdo, Sikivie:2020zpn, ParticleDataGroup:2022pth}.

\begin{figure}
	\centering
	\includegraphics[width=0.75\linewidth]{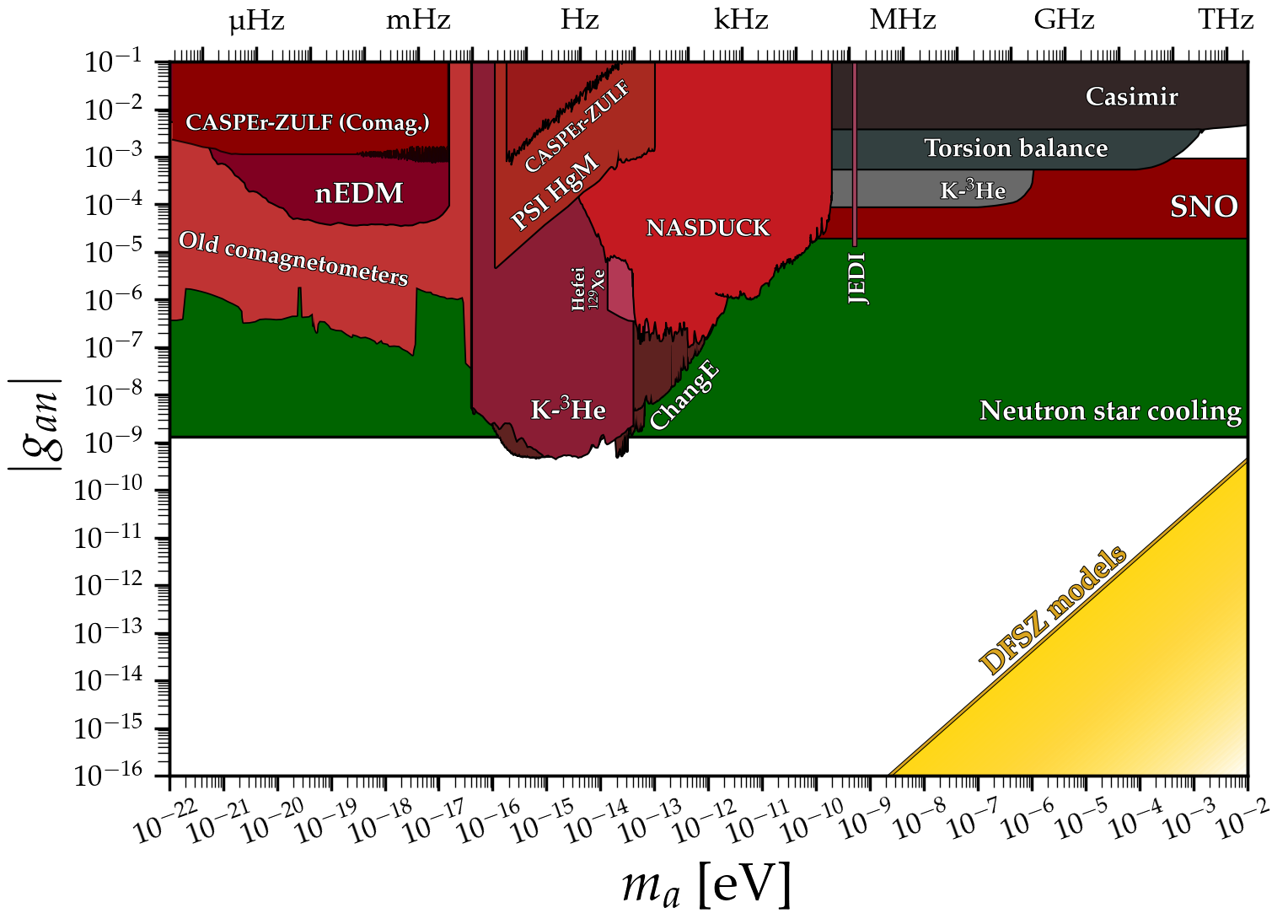}
	\caption{Summary of existing constraints on axions' mass and their coupling to neutrons. Figure taken from \cite{AxionLimits}.}
	\label{fig:anncoupling}
\end{figure}

Due to their weak interactions with SM particles, detecting axions in terrestrial experiments is exceptionally challenging. Therefore, there is a strong motivation to search for evidence of axions in astrophysical systems where their feeble couplings are partially compensated by high temperatures and densities \cite{Raffelt:1990yz}. For instance, the axion's interaction with nucleons is probed by NS cooling \cite{Hamaguchi:2018oqw, Beznogov:2018fda, Buschmann:2021juv} and supernova neutrino emission \cite{Burrows:1988ah, Burrows:1990pk, Keil:1996ju, Hanhart:2000ae, Fischer:2016cyd, Carenza:2019pxu, Carenza:2020cis, Lella:2023bfb}, which imply tight upper limits at the level of $g_{aNN} \lesssim 10^{-10}$, see figure \ref{fig:anncoupling}.

\section{Executive summary}
\label{sec:intro_summary}
In this thesis, we aim to explore the dynamics and interaction of ultralight DM, or hypothetical ultralight fields, in a relatively model-independent way. Many of our insights emerge from the lens of effective field theory (EFT), emphasizing the low-energy behavior of dark fields without making specific assumptions about the ultraviolet physics. In the rest of this section, we will provide a concise overview of the principles underlying the construction of an EFT for DM, and outline ideas to investigate the interaction of ultralight dark fields with itself, ordinary matter, and gravity within cosmological and astrophysical contexts.

\subsection{Exploring the low-energy dynamics of dark fields}
Although physical systems could be nonlinear and too complicated to track in 3+1 spacetime dimensions, their dynamics can often be separated into several hierarchical energy scales. If the energy of involved particles is dominated by their rest mass, then the core dynamics happens on low-energy scales with higher-energy (relativistic) modes treated as fast-varying perturbations. The resulting EFT for NR modes is typically much cleaner and easier to deal with than the original full theory. 

In the use of EFT to describe DM, errors must be quantified to justify adopted assumptions and to compare with observed data. However, in spite of the fact that leading-order NR EFT had been employed for phenomenological studies for decades, the importance of relativistic corrections was not assessed quantitatively in astrophysical/cosmological settings. In section \ref{sec:eft_nreft}, we identify parameters that separate different energy scales and systematically quantify the contribution of fast-varying modes based on perturbative expansions of those parameters.  It turns out that the inclusion of relativistic corrections can significantly improve the accuracy of predictions for certain observables, such as the mass-radius relation of DM solitons.

While the validity of EFT is typically determined by expansion parameters, as illustrated in section \ref{sec:eft_validity}, we discover another type of restriction for spin-1 DM in section \ref{sec:eft_singularity}. This is related to a \emph{singularity} problem in field components at certain amplitudes, whose resolution requires the use of EFT only at lower amplitudes or the use of gauge-invariant interaction with other fields. The latter implication, which was overlooked in the literature, can significantly reduce the number of possible interactions in the construction of a theory involving massive spin-1 fields.

\subsection{Probing the interaction of dark fields with themselves, ordinary matter, and gravity}
The key for observing, detecting, and even using DM is to understand its interactions. Existing observations suggest that the gravitational interaction of DM is described by minimal couplings, following Einstein’s General Relativity, and any direct interaction with ordinary matter must be feeble. However, there are no certain answers for the following questions: How does DM interact with itself? What observable impacts could its feeble interaction with ordinary matter have? Is the gravitational interaction of DM really described by General Relativity? With the established EFT approach, we tackle these problems by investigating the properties of DM oscillons in section \ref{sec:soliton}, emission of ultralight particles from NSs in section \ref{sec:ns}, and phenomenology of nonminimal gravitational interactions in section \ref{sec:nonminimal}.

DM oscillons, solitons supported by a balance between attractive self-interactions and repulsive pressure, are excellent targets to infer DM self-interactions. An observation of their existence and properties could reveal the specific form and coupling strength of the self-interaction. By developing a mathematical framework and carrying out 3+1 dimensional lattice simulations, we conduct thorough studies on their density profiles, stability, radiation, and lifetimes in the presence of self-interactions. In cases where DM particles carry an internal spin, we investigate all possible states of oscillons, which could be polarized and carry a \emph{macroscopic} internal spin. Their stability is confirmed in our fully relativistic lattice simulations.

Despite their feeble interactions with ordinary matter, ultralight particles can be efficiently produced in dense NS matter and yield observable signals. We examine the emission of axions, well-motivated hypothetical particles and DM candidates, from NSs due to novel interactions that violate the lepton flavor. The constraint on the lepton-flavor-violating (LFV) coupling $g_{ae\mu} \lesssim 4\times 10^{-11}$, which is established based on NS cooling and supernova observations, is competitive with the best laboratory limit $g_{ae\mu} < 1.9\times 10^{-11}$.

In the presence of nonminimal gravitational couplings, we study the phenomenology of spin-1 DM, including the mass-radius relation of DM solitons, growth of density perturbations, and propagation of gravitational waves (GWs). Remarkably, we find that the nonminimal coupling would lead to a speed of GWs faster or slower than that of light, depending on the coupling nature. To align with the observations of GWs, gamma ray bursts and large-scale structure, such as GW 170817 and its electromagnetic counterpart GRB 170817A, we place strong constraints on the nonminimal coupling of DM.

\section{Conventions, notations, technical acknowledgments, and frequently used acronyms}
In the rest of this thesis, the upcoming sections are based on my published/submitted papers with my collaborators. Specifically, section \ref{sec:eft} is primarily based on \cite{Salehian:2021khb, Mou:2022hqb}, section \ref{sec:soliton} on \cite{Zhang:2020bec, Zhang:2020ntm, Zhang:2021xxa}, section \ref{sec:ns} on \cite{Zhang:2023vva}, and section \ref{sec:nonminimal} on \cite{Zhang:2023fhs}. I have changed some of the notation and organization compared to the published works, and added additional text, to make this thesis more coherent. Most of the figures are adapted/taken from these works.

We adopt the natural unit $c=\hbar=1$ and mostly plus signature $(-,+,+,+)$ for the spacetime metric.  $\MP \equiv (8\pi G)^{-1/2} = 2.43 \times 10^{18} \rm{GeV}$ is the reduced Planck mass. We use Greek letters and Latin letters standing for indices going within $0-3$ and $1-3$, e.g., $\mu, \nu = 0,\cdots, 3$ and $i,j=1,2,3$. Repeated indices are assumed to be summed over unless otherwise stated. 

In writing this thesis, I acknowledge the use of Mathematica, Python and Anaconda for symbolic and numerical calculations, LaTeX and Youdao Dictionary for thesis writing, ChatGPT and Microsoft Word for grammar checks, the online tool \url{https://feynman.aivazis.com/} and Microsoft PowerPoint for drawing Feynman diagrams and editing figures, the Mathematica package FeynCalc for evaluating matrix elements involving spinor fields, and the Github project BibTeX Tidy \url{https://github.com/FlamingTempura/bibtex-tidy} for LaTeX reference management.

\bigskip

\noindent
\textbf{Frequently used acronyms:}\\
CMB: cosmic microwave background \\
DM: dark matter \\
EFT: effective field theory \\
FRW: Friedmann-Robertson-Walker \\
GW: gravitational wave \\
KGE: Klein-Gordon-Einstein \\
LFP: lepton-flavor-preserving \\
LFV: lepton-flavor-violating \\
NR: nonrelativistic \\
NS: neutron star \\
QCD: quantum chromodynamics \\
SM: standard model of particle physics \\
SP: Schroedinger-Poisson \\
VDM: vector dark matter \\
WIMP: weakly interacting massive particle
\chapter{Classical effective field theory}
\label{sec:eft}

\section{Introduction}
A NR axion field oscillates rapidly on the time-scale of order $m_a^{-1}$, whereas its spatial variations are on length-scales $L\sim (v m_a)^{-1}$, where $v\ll 1$ is the typical velocity of the axion particles. Moreover $m_a/H\gg 1$ (where $H$ is the Hubble parameter) within a few e-folds of expansion after the axion field starts oscillating. Together, these considerations indicate that a NR description of the field, obtained by integrating out the rapid variations in time, might be possible and fruitful for cosmological and astrophysical applications. Such an effective NR theory would be extremely useful both analytically and computationally, since one would no longer need to resolve the rapid oscillations of the field.

Note that the terminology of effective field theory refers to two different approaches. One approach is bottom-up, in which all relevant operators that are consistent with the symmetries are included and then the coefficients are fixed by matching with experiments. This approach is incorporated for example in the EFT of inflation \cite{Cheung:2007st} and large-scale structure formation \cite{Carrasco:2012cv}. In contrast, the approach described above is top-down, in which an EFT is obtained by taking the low-energy limit of a more complete theory. In this case, the coefficients appearing in the EFT are fixed by the parameters given in the more complete theory. This approach has been used for axion DM, for example in \cite{Namjoo:2017nia,Eby:2018ufi,Braaten:2018lmj}. Useful comparisons of the different top-down results are also provided in the same papers.

In section \ref{sec:eft_nreft}, we start from the relativistic Lagrangian of a classical, real-valued scalar field within general relativity. By systematically integrating out relativistic degrees of freedom we obtain an effective NR description for the system. Our specific approach was first used in \cite{Namjoo:2017nia} to obtain an EFT in Minkowski spacetime for a self-interacting scalar field. It was then generalized for curved spacetimes in \cite{Salehian:2020bon}, and more specifically applied to the case of a spatially flat Friedmann-Robertson-Walker (FRW) universe, with the analysis restricted to linearized perturbations. However, one important feature of DM is its ability to form dense, nonlinear structures due to gravitational instability in an expanding universe. The focus of this section is therefore to develop an EFT without any assumptions regarding the amplitude of the density perturbations of DM within an expanding universe. In this sense we obtain an EFT for axion DM in the nonlinear regime. Although metric perturbations are expected to remain small (at least in typical cosmological contexts \cite{Adamek:2013wja,Adamek:2015eda}), we systematically go beyond linear order in the metric perturbations as well.

As we will see in section \ref{sec:eft_nrlimit}, the leading-order result in our EFT is consistent with the Schroedinger-Poisson (SP) system in an expanding universe, which is widely used in the literature \cite{Hu:2000ke}. For example, the SP system has enabled long-time-scale simulations of nonlinear structure formation of axionlike fields \cite{Schive:2014dra,Mocz:2019pyf,Mocz:2019uyd}. It has also been used to understand the cosmological formation, gravitational clustering, and scattering of solitons with strong self-interactions in the early and contemporary universe \cite{Amin:2019ums}. Mirroring the late-universe simulations, purely gravitational growth of structure in the very early universe was pursued in \cite{Musoke:2019ima} with the help of the SP system. The SP system was used for numerically exploring mergers and collisions of solitons with and without self-interactions in axionlike DM \cite{Schwabe:2016rze,Glennon:2020dxs}, along with their non-gravitational consequences \cite{Hertzberg:2020dbk,Amin:2020vja}. The SP system was at the heart of exploring dynamical friction \cite{Lancaster:2019mde}, relaxation \cite{Bar-Or:2018pxz}, turbulence \cite{Mocz:2017wlg}, halo substructure \cite{Du:2016zcv,May:2021wwp}, kinetic nucleation of solitons \cite{Levkov:2018kau,Kirkpatrick:2020fwd}, and the dynamics of transient vortices in fuzzy DM scenarios \cite{Hui:2020hbq}. A number of existing numerical algorithms and codes are being used to explore the nonlinear dynamics of the SP system (see \cite{Veltmaat:2016rxo,Mocz:2017wlg, Edwards:2018ccc} for examples).

\begin{figure}
	\centering
	\includegraphics[width=\linewidth]{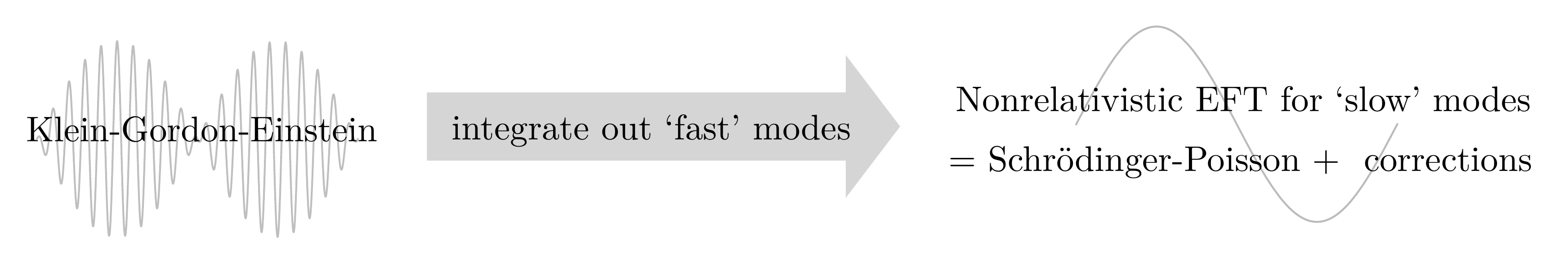} 
	\caption{Schematic approach of our EFT method for identifying systematic corrections to the SP equations.}
	\label{fig:schematic}
\end{figure}

Given its importance and widespread use, it is critical to understand the domain of validity of the SP system as well as expected deviations from it. With our systematic expansion, which relies upon integrating out the dynamics on short time-scales, we go beyond the leading-order SP system of equations and capture quantitative deviations expected due to relativistic corrections. See figure \ref{fig:schematic}. 
These deviations are expected to be small in most cosmological contexts in the late universe, when the fields are essentially NR. Nevertheless, explicit expressions for the relativistic corrections to the SP system can pinpoint which particular physical attribute of the system dominates the corrections. For example, one can hope to understand the relative importance of large gradients in the field, deviations from Newtonian gravity, and self-interactions of the scalar field, and at what order in the relativistic corrections vector and tensor perturbations of the metric become relevant as one moves beyond the SP system. This understanding, in turn, can clarify the domain of validity of the SP system, and provide insights into the most efficient way of including relativistic corrections in different physical contexts (such as those discussed in the previous paragraph). The corrections can also point the way towards exploring deviations from general relativity or characterizing the type of field content making up the DM. Furthermore, they might point to symmetries in the problem that are lost or restored as we go beyond the SP system.

As an explicit application of our NR EFT equations, we explore the mass-radius relation for solitons in the axion field in section \ref{sec:eft_soliton}. We demonstrate that our EFT better approximates the fully relativistic solution within the mildly relativistic regime compared to the SP equations alone.  Although our EFT with leading-order relativistic corrections is more complicated than the SP system, it is still easier to use numerically and analytically than the fully relativistic Klein-Gordon-Einstein (KGE) equations.

In section \ref{sec:eft_validity}, we discuss the conditions necessary for the validity of the NR EFT. Specifically, our approach hinges on the requirement that certain dimensionless quantities (e.g., the gravitational potential) are much less than one, a characteristic feature of DM being NR. We confirm the validity of using the NR EFT in our soliton studies and identify the circumstances under which the EFT fails.

By considering a similar procedure, the NR EFT for a massive vector (Proca) field $A_\mu$ can be derived in terms of a corresponding NR field $\psi_i$. While it is tempting to conclude that the EFT for Proca fields remains valid in cases where all dimensionless expansion parameters are less than unity, the situation is more complicated than the scalar case as the theory is a nontrivially \emph{constrained} system, where the auxiliary component $A_0$ is supposed to be uniquely solved in terms of the canonical fields $A_i$. The problem is that the interaction of Proca fields may lead to a singular problem for $A_0$, signaling a breakdown of the theory. We discuss it in section \ref{sec:eft_singularity}.

In general, Proca fields could have a variety of interactions and thus very rich dynamics. For example, the coupling to an axion field may allow for a significant energy transfer from axions to dark photons, and make the latter the dominant component of DM in the present-day universe \cite{Co:2018lka, Agrawal:2018vin}. If the Proca field possesses a nonlinear self-interacting potential or a nonminimal coupling to gravity, it may drive the cosmic inflation in the early universe \cite{Ford:1989me, Golovnev:2008cf}, and support coherently oscillating localized solitonic field configurations called vector oscillons \cite{Zhang:2021xxa}. The existence of strong self-interactions would also weaken the superradiance bounds on ultralight vectors \cite{Baryakhtar:2017ngi, Fukuda:2019ewf, Wang:2022hra, Clough:2022ygm, East:2022ppo}. Moreover, the theory of vector Galileons, whose effective action contains self-interactions with higher-order derivatives, has been constructed by requiring that the equation of motion has 2nd-order time derivatives and yields three healthy propagating degrees of freedom \cite{Tasinato:2014eka, Heisenberg:2014rta}. As an IR modification of gravity, it has been shown that these self-interacting Proca fields can lead to a viable cosmic expansion history and even alleviate the Hubble tension without sabotaging the success of General Relativity on scales of the solar system \cite{DeFelice:2016yws, deFelice:2017paw, Heisenberg:2020xak}.

In section \ref{sec:eft_singularity}, we illustrate the singularity problem by taking the quartic self-interaction of a real Proca field as an example and then generalize the discussion to complex fields and more generic interactions. We point out that either some constraints on field values must be respected to use the vector field theory in a self-consistent way or the specific interaction is forbidden in cases where the theory is meant to be fundamental.

{
\newcommand{\nc}{\newcommand}
\nc{\ba}{\begin{eqnarray}}
	\nc{\ea}{\end{eqnarray}}

\newcommand{\p}{{\partial}}
\newcommand{\Mpl}{M_{\rm P}}

\newcommand{\eq}[1]{\begin{equation}#1\end{equation}}
\newcommand{\eqa}[1]{\begin{align}#1\end{align}}
\newcommand{\spl}[1]{\begin{split} #1 \end{split}}
\newcommand{\fg}[1]{\begin{figure}[tbp]\centering #1 \end{figure}}
\newcommand{\bxeq}[1]{\begin{tcolorbox}[colback=gray!12,colframe=black!40,boxrule=0.15mm]\begin{equation}#1\end{equation}\end{tcolorbox}}

\newcommand{\vep}{\epsilon}
\newcommand{\vp}{\phi}
\def\blue{\textcolor{blue}}
\newcommand{\phiw}{\xi}
\renewcommand{\vb}{\b}

\section{Nonrelativistic effective field theory}
\label{sec:eft_nreft}
To illustrate the procedure to derive a NR EFT, we consider the nonlinear and inhomogeneous dynamics of a scalar field. Having in mind cosmological applications, we consider an expanding universe which contains a perfect fluid in addition to the scalar field; the additional fluid contributes to the homogeneous and isotropic background. However, we neglect perturbations of the perfect fluid and assume that gravitational collapse is only sourced by the scalar field. Our theory takes the following form:
\eq{
	\label{action}
	S=\int\dd[4]{x}\sqrt{-g}\left[\frac{1}{2}\Mpl^2R+\mathcal{L}_\vp\right]+\text{Background fluid}\,,
}
The scalar field and the background fluid are described by a Lagrangian density (${\cal L}_\varphi$) and an energy-momentum tensor ($T_f$), respectively,
\ba 
\label{lag}
\mathcal{L}_\vp&=&-\frac{1}{2} \p_\mu\vp\p^\mu\vp-\frac{1}{2}m^2\vp^2-V_\rm{int} \sep
V_\rm{int}=\frac{1}{4!}\lambda\vp^4+\frac{1}{6!}\frac{\kappa}{\Lambda^2}\vp^6+\dots ~,
\\
(T_f)^\mu{}_\nu&=&p\,\delta^\mu{}_\nu+(p+\rho)u^\mu u_\nu,
\ea 
where $m$ is the mass of the scalar field, $\lambda$ and $\kappa$ are dimensionless coupling constants, and $\Lambda$ is a large mass scale (compared to $m$). The $\kappa$ term is expected to be suppressed compared to the $\lambda$ term for a sufficiently large cutoff $\Lambda$. In what follows, we will assume such a hierarchy but our approach for obtaining the NR EFT can easily be extended to a situation with no hierarchy.

In the NR regime the dynamics of the scalar field is dominated by oscillations with frequency almost equal to its mass $m$. Thus it is reasonable to rewrite the scalar field in terms of a new, complex variable $\psi$ by
\eq{
	\label{redef1}
	\vp(t,\vb{x})= \frac{1}{\sqrt{2m}}\left[ e^{-imt}\psi(t,\vb{x})+e^{imt}\psi^*(t,\vb{x})\right]\,.
}
The remaining time or space dependence, encoded in $\psi$, is expected to be dominated by low-energy physics (i.e., lower than the mass scale), so that $\psi$ is a slowly varying function of time and space (compared to the dominant frequency of the system given by $m$). However, due to the nonlinearities involved in the system, high-frequency oscillations appear in $\psi$ with small amplitudes. The task is to integrate out such high-frequency modes and obtain an effective theory for the slowly varying mode.

It should be noted that the field redefinition of \eqref{redef1} is not a one-to-one correspondence between the real field $\vp$ and the complex NR field $\psi$. In \cite{Namjoo:2017nia}, the authors assumed a relation similar to \eqref{redef1} as a transformation in phase space with an accompanying redefinition for the conjugate momentum, which together make the transformation canonical and invertible. A nonlocal operator was also introduced in \cite{Namjoo:2017nia}, which simplifies the derivation of the EFT in Minkowski spacetime. However, as discussed in \cite{Salehian:2020bon}, this strategy is not very helpful for the case of curved spacetimes. An alternative approach is to remove the redundancy in \eqref{redef1} by adding a constraint on the NR field $\psi(t,\vb{x})$. One convenient choice for the constraint is \cite{Salehian:2020bon}
\eq{
	\label{cons}
	e^{-imt}\dot{\psi}+e^{imt}\dot{\psi}^*=0\,,
}    
where the overdot denotes a time derivative. This constraint implies an equation of motion for $\psi$ that is first order in time derivatives. By using \eqref{cons} one can show that
\eq{
	\label{redef2}
	\dot{\vp}(t,\vb{x})=-i\sqrt{\frac{m}{2}}\left[e^{-imt}\psi(t,\vb{x})-e^{imt}\psi^*(t,\vb{x})\right]\,.
} 

Applying \eqref{redef1} and \eqref{redef2} to the Klein-Gordon equation yields
\eq{
	\label{eq:psidotD}
	ig^{00}\dot{\psi}+\mathcal{D}\psi+e^{2imt}\mathcal{D}^*\psi^*=-\frac{e^{imt}}{\sqrt{2m}}\dv{V_\rm{int}}{\vp}{}(\psi,\psi^*)\,,
}
where $\mathcal{D}$ is a differential operator defined by
\eq{
	\label{Dorg}
	\mathcal{D}=\frac{m}{2}\left(g^{00}+1\right)+\frac{i\,\p_\mu(\sqrt{-g}g^{0\mu})}{2\sqrt{-g}}+\left(ig^{0i}-\frac{\p_\mu(\sqrt{-g}g^{\mu i})}{2m\sqrt{-g}}\right)\p_i-\frac{1}{2m}g^{ij}\p_i\p_j\,,
}
and $V_\rm{int}$ is given in \eqref{lag}, which here is written in terms of $\psi$ and $\psi^*$ using \eqref{redef1}. This is a generalized Schroedinger equation in an arbitrary spacetime. Notice that \eqref{eq:psidotD} is exact and there exists a one-to-one map from the complex field $\psi$ to the real field $\varphi$ and its conjugate momentum. Also note the appearance of rapidly oscillating factors. A common approximation is to neglect such terms, under the assumption that they average to zero. Whereas this is true at leading order, these terms will play a crucial role in the derivation of our EFT, as we will see shortly. Note that the oscillatory terms are present even in a free theory with $V_\rm{int}=0$, which then leads to a tower of higher spatial-derivative terms in the free EFT; these terms can be thought of as the expansion of the relativistic energy in the large-mass limit \cite{Namjoo:2017nia}. 

\subsection{Metric perturbations in an expanding universe}
To fully describe the system, the Schroedinger equation \eqref{eq:psidotD} must be accompanied by the Einstein field equations as well as the energy-momentum conservation for the fluid,
\eq{
	G^\mu{}_\nu=\frac{1}{\Mpl^2}\big[T^\mu{}_\nu+(T_f)^\mu{}_\nu\big]\,,\qquad (T_f)^\mu{}_{\nu;\mu}=0\,.
}
Let us emphasize again that in what follows, for simplicity, we will ignore perturbations of the background fluid. For application of our EFT to cosmology, we consider a perturbed expanding universe. As noted above, we intend to study the inhomogeneities in the scalar field $\psi (t, {\bf x})$ nonlinearly. This implies that, contrary to the case for linear perturbation theory, the vector and tensor modes of the spacetime metric may play a nontrivial role in the dynamics of the scalar degrees of freedom. As a result, here we start from a general metric, including all forms of the metric perturbations, and then estimate the contribution of each type of modes to the dynamics of the scalar field. It is convenient to work with the ADM metric decomposition, which is given by
\eq{
	\label{metric1}
	\dd{s}^2=-N^2\dd{t}^2+\gamma_{ij}(\dd{x^i}+N^i\dd{t})(\dd{x^j}+N^j\dd{t})\,,
}
where $N$, $N^i$ and $\gamma_{ij}$ are the lapse function, shift vector, and the first fundamental form, respectively. We remove the gauge redundancy by the following choice of the metric components:
\eq{
	N=e^\Phi\,,\qquad N^i=\frac{1}{a}\sigma^i\,,\qquad\gamma_{ij}=a^2e^{-2\Psi}(e^h)_{ij}\,,
} 
where we have
\eq{
	\label{eqTT}
	\p_i\sigma^i=\delta^{ij}h_{ij}=\p_ih^i{}_j=0\,,
}
and we lower and raise the Latin indices with $\delta_{ij}$ and $\delta^{ij}$. The background geometry is FRW spacetime and $a(t)$ is the scale factor. We have fixed the gauge by requiring the scalar mode of $g_{0i}$ and the vector and some of the scalar modes of $g_{ij}$ to vanish. This  choice of gauge can be retained to all orders of perturbations and is a natural generalization of the Newtonian gauge, which is particularly convenient for the system in the NR regime. Note that we think of the above metric as perturbative in $\Phi$, $\Psi$, $\sigma^i$ and $h_{ij}$ (while we treat the scalar field $\psi$ nonperturbatively), which we will justify shortly. From \eqref{metric1}-\eqref{eqTT}, one has 
\eq{
	\label{metric2}
	\spl{
		\sqrt{-g}=N\sqrt{\gamma}=a^3e^{\Phi-3\Psi}\,,\qquad g^{00}&=\frac{-1}{N^2}=-e^{-2\Phi}\,,\qquad g^{0i}=\frac{N^i}{N^2}=\frac{1}{a}e^{-2\Phi}\sigma^i\,,\\ g^{ij}=\gamma^{ij}-\frac{1}{N^2}N^iN^j&=\frac{1}{a^2}\Big[e^{2\Psi}(e^{-h})^{ij}-e^{-2\Phi}\sigma^i\sigma^j\Big]\,,\qquad 
}}    
which may be used in \eqref{eq:psidotD} as well as for the Einstein field equations.

\subsection{Power counting}
As we take the NR limit of a relativistic theory, several small parameters/operators appear, which allows us to organize different terms that arise in the EFT. In this subsection we shall identify these small parameters. Furthermore, in the NR limit and as a result of the source of gravitational perturbations being a scalar field, there exists a hierarchy among the amplitudes of the scalar, vector, and tensor modes of the perturbed spacetime metric. As we will see below, the scalar modes dominate and the amplitude of the vector mode is larger than that of the tensor modes. 

Ignoring metric perturbations, the Klein-Gordon equation is
\eq{
	\label{eqphi}
	\spl{
		&\ddot{\vp}+m^2\vp\\
		&+3H\dot{\vp}-\frac{\nabla^2\vp}{a^2}+\frac{1}{3!}\lambda\vp^3+\frac{1}{5!}\frac{\kappa}{\Lambda^2}\vp^5+\dots=0\,.
	}
}
As stated above, in the NR regime the mass term is the dominant contribution to the time evolution of the scalar field, and all other terms are suppressed. We have written the equation of motion in \eqref{eqphi} in two different lines to emphasize this hierarchy. Demanding that the terms on the second line are smaller than those on the first line, we identify the following small parameters in the NR limit:
\eq{
	\epsilon_{\scriptscriptstyle H}\sim\frac{H}{m}\ll1 \sep
	\epsilon_x\sim\Bigg|\frac{\nabla^2}{m^2a^2}\Bigg|\ll1 \sep
	\epsilon_\lambda\sim |\lambda|\frac{\vp^2}{m^2}\ll1\sep
	\epsilon_\kappa\sim|\kappa|\frac{\vp^4}{m^2\Lambda^2}\ll1\,.
} 
The first parameter quantifies the smallness of the expansion rate compared to the mass scale. In the opposite regime, when the Hubble scale is larger than $m$, the scalar field does not oscillate and cannot mimic DM behavior. The second parameter quantifies the smallness of the typical momentum of the DM ``particles" compared to $m$, while the last two parameters specify the smallness of self-interactions. Note that if we assume that $\lambda$ and $\kappa$ are of the same order and $\Lambda$ is a very large mass scale, then one can see that $\epsilon_\kappa$ is much smaller than $\epsilon_\lambda$. In fact for the special case of the axionlike field we have $\epsilon_\kappa=\epsilon_\lambda^2$. Although we do not restrict ourselves to the axion, we assume a similar hierarchy between these two parameters, with $\epsilon_\kappa\sim {\cal O} (\epsilon_\lambda^2)$. 

Next we study the hierarchy among the dynamical variables, which follows as a consequence of the system being NR. For these estimates, one can use the Einstein field equations. However, most of the approximate relations can also be estimated by considering symmetries and other simple relationships. First we note that in order for the system to remain NR even amid the gravitational dynamics, the gravitational potentials must remain small,  
\eq{
	\epsilon_g\equiv |\Phi|\,\sim|\Psi|\ll1\,.
} 
The fact that $\Phi$ and $\Psi$ are expected to be of the same order in $\epsilon_g$ is discussed below. From the 00 component of the Einstein field equations one can obtain the Poisson-like equation for $\Psi$,  leading to
\eq{
	\label{phi}
	\frac{\nabla^2\Psi}{a^2}\sim\frac{m^2\vp^2}{\Mpl^2}\,.
}
Further, the Poisson equation implies that there is another small parameter related to the amplitude of the scalar field, 
\eq{
	\label{ephi}
	\epsilon_\vp\equiv \frac{|\vp|}{\Mpl}\sim\frac{|\psi|}{\Mpl\sqrt{m}}\ll1\,,
} 
from which we find $\epsilon_\vp^2\sim\epsilon_x\epsilon_g$. In addition, by using the trace-free part of the $ij$-component of the Einstein field equations we can see that
\eq{
	\label{phi-psi-diff}
	\nabla^2(\Phi-\Psi)\sim\frac{(\nabla\vp)^2}{\Mpl^2}\,,\quad\implies\quad\Phi-\Psi\sim\epsilon_\vp^2\, , 
}
that is, the difference between the two gravitational potentials is typically one order smaller than the gravitational potentials themselves. 
Further, from the $0i$-component of the Einstein field equations we find
\eq{
	\label{sigma1}
	\frac{1}{a}\nabla^2\sigma_i\sim\frac{1}{\Mpl^2}\dot{\vp}\p_i\vp\,,\quad\implies\quad	\sigma_j\sim\epsilon_x^{1/2}\epsilon_g\,,
} 
where we have used $\dot{\vp}\sim m\vp$. Note that the relation between the vector mode and the scalar field (which acts as its source) could also be identified from symmetries and dimensional analysis. Finally, by using the $ij$-component of the Einstein field equations (or, again, by  symmetries), we find 
\eq{
	\label{hij1}
	\nabla^2h_{ij}\sim\frac{\p_i\vp\p_j\vp}{\Mpl^2}\,,\quad\implies\quad h_{ij}\sim\epsilon_\vp^2\,.
}
In the following, instead of keep tracking of these small parameters individually, we collectively denote all of them by $\epsilon=\{\epsilon_{\scriptscriptstyle H},\epsilon_x,\epsilon_\lambda,\epsilon_g,\epsilon_\vp\}$ and work up to appropriate order in $\epsilon$. This effectively means that we assume all small parameters are of the same order (except for $\epsilon_\kappa$, which is one order smaller). Depending on the application, it is expected that a hierarchy among the small parameters exists, in which case our EFT would be simplified. By using our approach, any higher-order term in the EFT can be derived systematically.

\subsection{Scalar and vector equations}
By using our power-counting arguments, one can obtain a set of equations for the gravitating scalar DM in an expanding background at the requisite order in $\epsilon$.
Because our primary interest is the evolution of the scalar modes, we will be dealing with scalar equations of motion; hence the tensor modes $h_{ij}$ cannot appear by themselves, but will always enter the equations of motion accompanied by at least two spatial derivatives. This implies that the tensor modes would only appear at $\order{\epsilon^3}$, according to \eqref{hij1}. Similarly, the vector mode $\sigma^i$ appears with at least one spatial derivative which, upon using \eqref{sigma1}, implies that it appears at $\order{\epsilon^2}$ in the scalar equations of motion.

Based on the above considerations, the generalized Schroedinger equation \eqref{eq:psidotD}, the Einstein field equations that reduce to the Poisson equation in the NR limit, and the equation for the vector mode take the following form
\small
\begin{align}
	\label{eq:psidotD3}
	&i\dot{\psi}+\tilde{\mathcal{D}}\psi+e^{2imt}\tilde{\mathcal{D}}^*\psi^* =e^{imt}e^{2\Phi}\mathcal{J}+\order{\epsilon^{4}}\,,\\
	\label{eqPhi-simp}
	&\frac{\nabla^2}{a^2}\Phi+3e^{-2(\Phi+\Psi)}\left(H\dot{\Phi}+2H\dot{\Psi}-\dot{\Phi}\dot{\Psi}-\dot{\Psi}^2+\ddot{\Psi}-\frac{\ddot{a}}{a}\right)
	= \frac{e^{-2\Psi}}{2\Mpl^2}\Big[(\rho+3p)+\mathcal{S}_\Phi\Big]+\order{\epsilon^{4}}\,,\\
	\label{eqPsi-simp}
	&\frac{\nabla^2\Psi}{a^2}-\frac{(\nabla\Phi)^2}{2a^2}+\frac{3}{2}e^{-2(\Phi+\Psi)}\left(H^2+\dot{\Psi}^2-2H\dot{\Psi}\right)= \frac{	e^{-2\Psi}}{2\Mpl^2}\Big[\rho+\mathcal{S}_\Psi\Big]+\order{\epsilon^{4}}\,,\\
	\label{eqsigma-simp}
	&\frac{\nabla^4}{a}\sigma_i =\frac{2i}{\Mpl^2}\Big[ \left( \nabla^2\psi+(\nabla\psi \cdot \nabla) \right) \p_j\psi^*- \left( \nabla^2\psi^*+(\nabla\psi^* \cdot \nabla) \right) \p_j\psi\Big]+\order{\epsilon^{9/2}}\, ,
\end{align}\normalsize
where we have defined
\eqa{
	\label{D-simp}
	\tilde{\mathcal{D}}&=\frac{m}{2}\left(1-e^{2\Phi} \right)+\frac{e^{4\Phi}}{2ma^2}\nabla^2+\frac{i}{2}\big(3H-\dot{\Phi}-3\dot{\Psi}\big)-\frac{i}{a}\vec{\sigma} \cdot \nabla\,,\\
	\label{J-simp}
	\mathcal{J}&=\frac{1}{3!}\lambda\vp^3+\frac{1}{5!}\frac{\kappa}{\Lambda^2}\vp^5\, , \qquad 
	\mathcal{S}_\Phi = 2e^{-2\Phi}\dot \vp^2 -m^2\vp^2 -\dfrac{2\lambda}{4!}\vp^4,
	\\
	\label{Spsi-simp}
	\mathcal{S}_\Psi   &= \dfrac12 e^{-2\Phi}\dot \vp^2 +\dfrac{e^{2\Phi}}{2a^2}(\nabla \vp)^2+\dfrac12 m^2\vp^2 +\dfrac{\lambda}{4!}\vp^4 \, .
}
Within the expressions (\ref{D-simp})-(\ref{Spsi-simp}), the fields $\vp$ and $\dot \vp$ need to be replaced by $\psi$ and $\psi^*$ using \eqref{redef1} and \eqref{redef2}. Note that we have replaced $\Psi$ with $\Phi$ in several terms, because the difference between the two gravitational potentials is one order smaller than $\Phi$ and $\Psi$ themselves. To avoid clutter we did not expand the exponential factors, but one needs to keep in mind that they are only relevant to appropriate order in their Taylor expansion. Notice that since, at this stage, different variables may contain highly oscillating contributions, it is not necessarily the case that the time derivative operator is small. Moreover, as a result of the assumed hierarchy between the self-interaction terms (i.e., $\epsilon_\kappa \sim \epsilon_\lambda^2$), the $\kappa$ term only appears in the Schroedinger equation at the current working order. 

In general, the order of terms that are neglected must be compared with the leading-order terms. For example, in the Schroedinger equation \eqref{eq:psidotD3}, the leading-order terms (such as $\nabla^2 \psi/m^2a^2$) are already of $\order{\epsilon^2}$, according to our power counting. This implies that we are neglecting some terms that are at least two orders higher in $\epsilon$. It is thus evident that we have only kept the leading-order nontrivial corrections, which is indeed the case for all other equations. To go beyond that, one needs a more accurate set of equations.

One can obtain the SP equations from \eqref{eq:psidotD3} and \eqref{eqPhi-simp} to the leading order in the EFT, while corrections appear at the next order. As a final remark, note that at the background level the above set of equations reduce to
\small
\eqa{
	&i\dot{\bar{\psi}}+\frac{3i}{2}H \left( \bar{\psi}-e^{2imt}\bar{\psi}^* \right)=e^{imt}\mathcal{\bar{J}} \sep
	3\Mpl^2H^2=\rho+\bar{\mathcal{S}}_\Psi \sep 
	\frac{\ddot{a}}{a}=-\frac{1}{6\Mpl^2}\left[(\rho+3p)+\bar{\mathcal{S}}_\Phi\right] ~,
	\nonumber 	\\ 
	\label{bg}
	&\dot{\rho}+3H(\rho+p)=0\,,
}\normalsize
where an overbar indicates that the quantity is evaluated at the spatially homogeneous background level. In (\ref{bg}) we have also included the continuity equation for the additional perfect fluid, which is assumed to be spatially homogeneous.

So far, by removing unnecessary terms, we have already taken the first step toward identifying the leading-order corrections in the NR EFT. In principle, one can solve (\ref{eq:psidotD3})-(\ref{eqsigma-simp}) numerically to obtain the dynamics of the scalar field. However, such numerical computation is, in general, a difficult task due to the rapidly oscillating factors appearing in those equations. Next, we will remove such factors in a systematic way (instead of naively neglecting them) and obtain their corresponding corrections to the slowly varying variables. We will see that this procedure leads to nontrivial corrections at a given working order in $\epsilon$, and hence cannot be neglected.

\subsection{Nonrelativistic limit and relativistic corrections}
\label{sec:eft_nrlimit}
As stated in the previous section, we are interested in the slowly varying modes of dynamical variables. However, due to the appearance of oscillatory factors in the equations of motion, the dynamics of slowly varying quantities will be affected by rapidly oscillating terms. The situation is illustrated in figure \ref{fig:modes}, which shows the typical behavior of the variables in the frequency domain (i.e., the Fourier transform of time-dependent variables). We see that the zero mode (which translates to the slowly varying mode in the time domain) dominates,   although modes with nonzero frequencies, close to integer multiples of the mass $m$, also exist in the spectrum, albeit with subdominant amplitudes.

To obtain a theory entirely in terms of slowly varying quantities, we must integrate out modes associated with these rapid oscillations. This is a nontrivial task because the rapid oscillations are sourced by the slowly varying mode, and they in turn backreact on the evolution of the slowly varying mode.

Working in the time domain, we may apply a smearing operator to each variable in order to extract the slowly varying part of the variables. That is, we may take a time average of each variable with a suitable choice of window function \cite{Salehian:2020bon}: 
\fg{
	\centering
	\includegraphics[width=0.9\textwidth]{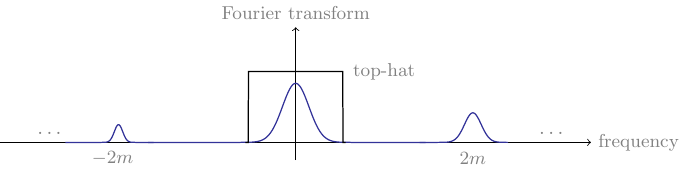} 
	\caption{Typical frequency  spectrum of the variables in the problem.  The system is dominated by the slowly varying mode, though modes associated with rapid oscillations also arise in the spectrum. By applying the smearing operator of (\ref{eq:smearing}), we may extract the slowly varying part.}\label{fig:modes}
}
\eq{
	\label{eq:smearing}
	X_s=\ev{X}\equiv\int\dd{t'}W(t-t')X(t')\,,
}
in which $W(t)$ is the window function and $X_s$ is the ``slow mode" of the variable $X$. In \cite{Salehian:2020bon} it has been shown that the top-hat window function in the frequency domain, which becomes $\rm{sinc}(t)$ in  the time domain, is a suitable choice. Besides the slow mode $X_s$, each variable also contains a tower of modes  associated with rapid oscillations. Quite generally, one has
\eq{
	\label{eq:modexp}
	X=\sum_{\nu=-\infty}^{\infty}X_\nu\, e^{i\nu m t}\,,\qquad X_\nu=\ev{X\,e^{-i\nu m t}}\,,
} 
where the coefficients $X_\nu$ depend on both time and space.  We define the ``slow mode" as $X_s =X_{\nu = 0}$, and refer to the modes associated with rapid oscillations as ``nonzero modes," that is, modes $X_\nu$ with $\nu \neq 0$. According to the definition in \eqref{eq:modexp}, we have $(X^*)_\nu=X_{-\nu}^*$, and therefore if $X$ is real-valued then the modes $X_\nu$ obey the constraint $X_\nu^*=X_{-\nu}$. 

Note first that the expansion in (\ref{eq:modexp}) is exact, as a result of the appropriate  choice of the window function (see \cite{Salehian:2020bon} for an outline of the proof.) Second, we emphasize that the coefficients $X_\nu$ (even with $\nu \neq 0$) are themselves slowly varying functions of time (compared to the frequency of the oscillations) since, as noted in the second expression in (\ref{eq:modexp}), the $X_\nu$ may be represented by the smearing operator acting on $X$ (weighted by an appropriate phase). In fact, as shall be made explicit below, the $\nu\ne 0$ modes can be written in terms of the slow mode $X_s$ (since, to reiterate, they are sourced by the slow mode). We may therefore identify a new small operator, namely, the time-derivative operator acting on the slow modes,
\eq{
	\epsilon_t \sim \left\vert \frac{\dot{X_s}}{mX_s} \right\vert \ll 1\, ,
}
where $X$ can be any of our variables after the field redefinition, including $\psi$, $\Phi$, $\Psi$, $\sigma_i$, $h_{ij}$, and $a$; the subscript indicates that the slow mode (with $\nu = 0$) is considered.\footnote{Because the functions $X_\nu$ with $\nu \neq 0$ are also slowly varying and can be expressed in terms of $X_s$, we also have $\epsilon_t \sim \vert \dot X_\nu/mX_\nu \vert$ -- at least for $\nu$ not too large.} Thus, we can include the time-derivative operator within the set of small parameters/operators identified below (\ref{hij1}), and our EFT will be an expansion in powers of $\epsilon=\{\epsilon_t,\epsilon_{\scriptscriptstyle H},\epsilon_x,\epsilon_\lambda,\epsilon_g,\epsilon_\vp\}$. Note that these small parameters are not all independent; one may derive relations among them by using the equations of motion. Let us emphasize that $\epsilon_t$ is defined as the operator that acts on modes $X_\nu$, rather than on the full fields. In the latter case, according to the definition in \eqref{eq:modexp}, the time derivative would act on the oscillatory factors, which are not slowly varying, which is why, for example, $\dot{\Psi}$ in \eqref{eqPhi-simp} should not be considered as $\order{\epsilon^2}$ in the power counting.

We will be interested in the effective equations for the slow modes $X_s = X_{\nu = 0}$; therefore we will systematically remove nonzero modes $X_\nu$ with $\nu\neq 0$ from the theory. For a NR system, all $\nu\ne0$ modes are suppressed by factors of the small parameters/operators denoted collectively by $\epsilon$. Using power counting to estimate the size of each term that appears in the equations of motion for the nonzero modes, we may solve for them perturbatively in $\epsilon$. To achieve this, we expand the nonzero modes as a power series in $\epsilon$: 
\eq{
	\label{pertexp}
	X_\nu=\sum_{n=1}^{\infty}X_\nu^{(n)}=X_\nu^{(1)}+X_\nu^{(2)}+\dots\,,\qquad(\nu\neq0)\,.
} 
The superscript $(n)$ denotes the order of magnitude relative to the slowly varying mode $X^{(n)}_\nu/X_s\sim\order{\epsilon^n}$.\footnote{Note that the power series we use here is slightly different from the one introduced in \cite{Salehian:2020bon}. $X^{(n)}_\nu$ here corresponds to what was denoted $X^{(n+1)}_\nu$ in \cite{Salehian:2020bon}. } This expansion allows us to solve for the $\nu\ne 0$ modes perturbatively and substitute the solutions back into the equations for the slow mode, resulting in an EFT for the slow modes. This procedure has been done explicitly for an interacting theory in Minkowski spacetime in \cite{Namjoo:2017nia} and extended to the case of a linearly perturbed FRW universe in \cite{Salehian:2020bon}. Here we outline essential steps in the derivation, focusing mainly on the Schroedinger equation, and omit additional details.

Applying the mode expansion \eqref{eq:modexp} to \eqref{eq:psidotD3} yields\footnote{The mode decomposition of equations of motion can be understood as the result of multiplying each equation by $e^{-i \nu m t}$ and then applying the smearing operator. }
\eq{
	\label{psinu}
	i\dot{\psi}_\nu-\nu m\psi_\nu+\tilde{\mathcal{D}}_\alpha\psi_{\nu-\alpha}+\tilde{\mathcal{D}}^*_{-\alpha}\psi^*_{2+\alpha-\nu}=\left(e^{2\Phi}\right)_\alpha\mathcal{J}_{\nu-\alpha-1}\,.
}    
This equation makes it evident that modes of different $\nu$ couple to each other; in particular, nonzero modes ($\nu\ne 0$) affect the dynamics of the slow mode ($\nu=0$) and the nonzero modes are sourced by the slow mode. Similar results follow from \eqref{eqPhi-simp} for $\Phi$ and \eqref{eqPsi-simp} for $\Psi$. We can then solve for the nonzero modes perturbatively. At leading order we find
\eqa{
	\label{nonzero}
	\psi^{(1)}_\nu&=\left(\frac{\nabla^2}{4m^2a_s^2}-\frac{3iH_s}{4m}-\frac{\lambda|\psi_s|^2}{16m^3}-\frac{1}{2}\Phi_s\right)\psi_s^*\delta_{\nu,2}+\frac{\lambda\psi_s^3}{48m^3}\delta_{\nu,-2}-\frac{\lambda\psi_s^*{}^3}{96m^3}\delta_{\nu,4}\\
	\Psi^{(1)}_\nu&=\frac{\psi_s^*{}^2-\bar{\psi}_s^*{}^2}{16m\Mpl^2}\delta_{\nu,2}+\frac{\psi_s^2-\bar{\psi}_s^2}{16m\Mpl^2}\delta_{\nu,-2},\,
	\quad 
	H^{(1)}_\nu=-\frac{i\bar{\psi}_s^*{}^2}{8\Mpl^2}\delta_{\nu,2}+\frac{i\bar{\psi}_s^2}{8\Mpl^2}\delta_{\nu,-2},
}
where $\psi^{(1)}_\nu$ is derived from \eqref{psinu} while $\Psi^{(1)}_\nu$ and $H^{(1)}_\nu$ can be obtained by solving equations after the mode decomposition \eqref{eqPhi-simp} and \eqref{bg}, respectively. Note that $\delta_{\nu,i}$ is the Kronecker delta function, and the superscript $(1)$ denotes that each term on the right hand side is suppressed by $\order{\epsilon}$ compared to $\psi_s$, $\Psi_s$ or $H_s$ . An expression can also be derived for the $\nu\ne 0$ modes of $\Phi$, with the additional complication that the solution would be nonlocal. Fortunately, among the leading-order corrections, nonzero modes of $\Phi$ do not contribute. Note also that the leading-order nonzero modes of the scale factor vanish. 

We can now use these solutions to replace nonzero modes that appear in the equations for the slow modes. Furthermore, based on the power counting, we can neglect terms that are at higher order compared to the leading-order corrections. After significant algebraic simplification, for the Schroedinger equation we find
\begin{align}
	\label{eqftkg}
	\nonumber
	&i\dot{\psi}_s+\frac{3i}{2}H_s\psi_s+\frac{1}{2ma_s^2}\nabla^2\psi_s-m\Phi_s\psi_s-\frac{\lambda}{8m^2}|\psi_s|^2\psi_s\\
	\nonumber
	&+\left(\frac{3\rho_s}{8m\Mpl^2}+\frac{|\bar{\psi}_s|^2}{2\Mpl^2}+\frac{|\psi_s|^2}{16\Mpl^2}-\frac{m}{2}\Phi_s^2\right)\psi_s-2i\dot{\Phi}_s\psi_s+\frac{\nabla^4\psi_s}{8m^3a_s^4}\\
	\nonumber
	&+3\Phi_s\frac{\nabla^2\psi_s}{2ma_s^2}-\frac{\nabla\Phi_s \cdot \nabla\psi_s}{2ma_s^2}-i\frac{\vec{\sigma}_s\cdot\nabla\psi_s}{a_s}+\left(\frac{17\lambda^2}{8m^2}-\frac{\kappa}{\Lambda^2}\right)\frac{|\psi_s|^4\psi_s}{96m^3}\\
	&-\frac{\lambda}{16m^4a_s^2}\left(2|\nabla\psi_s|^2\psi_s+\psi_s^2\nabla^2\psi_s^*+2|\psi_s|^2\nabla^2\psi_s+\psi_s^*(\nabla\psi_s)^2\right)=0+\order{\epsilon^4}\,,
\end{align}
where the background equations as well as the leading-order Poisson equation are used to simplify the subleading terms. Because of $\Psi_s-\Phi_s\sim \mathcal{O}(\epsilon^2)$, we have replaced $\Psi_s$ by $\Phi_s$ if it appears anywhere but the first line. Note that since $\rho_s\leq 3\Mpl^2 H^2$ we have that $\rho_s/m^2\Mpl^2 \sim \order{\epsilon^2}$ or smaller.

In a similar way, we obtain the effective equation for the gravitational potential $\Phi_s$, starting from \eqref{eqPhi-simp}:
\begin{align}
	\label{eftPhi}
	\nonumber
	&\frac{\nabla^2\Phi_s}{a_s^2}-\frac{m}{2\Mpl^2}(|\psi_s|^2-|\bar{\psi}_s|^2)\\
	\nonumber
	&+3(3H_s\dot{\Phi}_s+\ddot{\Phi}_s)-\dfrac{1}{\Mpl^2}\left(\rho_s+3p_s+2m|\bar{\psi}_s|^2-(3/2)m|\psi_s|^2\right)\Phi_s\\
	&+\frac{3}{8m\Mpl^2a_s^2} \left( \psi_s\nabla^2\psi_s^*+\psi_s^*\nabla^2\psi_s \right) - \frac{\lambda}{8m^2\Mpl^2} \left( |\psi_s|^4-|\bar{\psi}_s|^4 \right)=0+\order{\epsilon^4}\, .
\end{align}
Interestingly, notice that the other gravitational potential, $\Psi_s$, decouples from $\Phi_s$ and $\psi_s$ to this order. However,
to close the system of equations, we must add one for the vector modes, which at this order is simply given by \eqref{eqsigma-simp} with all variables replaced by their corresponding slow modes:
\begin{align}
	\label{eftsigma}
	\frac{\nabla^4\vec{\sigma}_s}{a_s}=\frac{2i}{\Mpl^2}\Big[\left( \nabla^2\psi_s+(\nabla\psi_s \cdot \nabla) \right) \nabla\psi_s^* - \left( \nabla^2\psi_s^*+(\nabla\psi_s^* \cdot \nabla) \right)\nabla\psi_s\Big]+\order{\epsilon^{9/2}}\,.
\end{align}
Equations (\ref{eqftkg})-(\ref{eftsigma}) are sufficient for obtaining the leading-order corrections to the SP equations. However, the gravitational potential $\Psi_s$ might also be of  interest for some purposes, such as lensing effects of compact objects. Based on \eqref{eqPsi-simp}, its effective equation is
\begin{align}
	\label{eftPsi}
	\nonumber
	&\frac{\nabla^2\Psi_s}{a_s^2}-\frac{m}{2\Mpl^2}(|\psi_s|^2-|\bar{\psi}_s|^2)\\
	\nonumber
	&-\frac{(\nabla\Phi_s)^2}{2a_s^2}-3H_s\dot{\Phi}_s-\dfrac{1}{\Mpl^2}\left(\rho_s+2m|\bar{\psi}_s|^2- \frac{3}{2} m |\psi_s|^2\right)\Phi_s
	\\
	&-\frac{|\nabla\psi_s|^2}{4m\Mpl^2a_s^2}-\frac{\lambda}{32m^2\Mpl^2} \left( |\psi_s|^4-|\bar{\psi}_s|^4 \right) =0+\order{\epsilon^4}.
\end{align}

Equations~\eqref{eqftkg}-\eqref{eftPsi} are the main results of this section. It is worth noting that the first lines of \eqref{eqftkg} and \eqref{eftPhi} yield the familiar SP equations
\begin{equation}
	\begin{aligned}
		\label{eq:SP}
		&i\dot{\psi}_s+\frac{3i}{2}H_s\psi_s+\frac{1}{2ma_s^2}\nabla^2\psi_s-m\Phi_s\psi_s-\frac{\lambda}{8m^2}|\psi_s|^2\psi_s=0+\order{\epsilon^3}\,,\\
		&\frac{\nabla^2\Phi_s}{a_s^2}-\frac{m}{2\Mpl^2} \left( |\psi_s|^2-|\bar{\psi}_s|^2 \right)=0+\order{\epsilon^3}\,.
	\end{aligned}
\end{equation}
We do not need \eqref{eftsigma} and \eqref{eftPsi} for the evolution of $\psi_s$ at this order. 

At the FRW background level, there are no corrections to the Friedmann equation or the continuity equation at leading order. We have 
\eqa{
	\label{eq:SPe}
	3\Mpl^2H_s^2=\rho_s+m|\bar{\psi}_s|^2+\textcolor{black}{\frac{\lambda|\bar{\psi}_s|^4}{16m^2}}+\order{\epsilon^4}, \quad \dot{\rho}_s+3H_s(\rho_s+p_s)=0+\order{\epsilon^4}.
}
The background Schroedinger equation receives corrections, which can be obtained by the replacement $\psi_s \to \bar \psi_s$ in \eqref{eqftkg} and setting metric perturbations to zero.

Note that from the bottom-up EFT point of view one can expect the appearance of all correction terms in \eqref{eqftkg} but with unknown coefficients. However, it is not the case that all terms consistent with the symmetries appear: for example, terms like $\Phi_s\lambda|\psi_s|^2\psi_s$ or $H_s\nabla^2\psi_s$ did not appear. This can be thought as the consequence of the original theory with which we started, namely general relativity with a scalar field minimally coupled to gravity. One may expect new terms to appear if one considers a modified theory of gravity, which may also change the coefficients of terms already identified in (\ref{eqftkg})-(\ref{eftPsi}). It would be interesting to explore such a possibility in the future. Furthermore, note that a term proportional to $\lambda^2$ has appeared in \eqref{eqftkg} with the same structure as the $\kappa$ term. Therefore, we see that a single term in the original theory gives rise a tower of terms in the low-energy EFT as a result of integrating out high-energy modes.

\subsection{Case study: Mass-radius relation of solitons}
\label{sec:eft_soliton}
Equations \eqref{eqftkg}-\eqref{eftPsi} of our EFT, which include relativistic corrections to the SP system, can be incorporated in many different contexts and the solutions will take different forms. In this section we study one of the simplest solutions: spherically symmetric, stationary solitonic solutions of the form
\eq{
	\label{psistat}
	\psi_s(t,r) = f(r) e^{i\mu t} ~,
}
where $\mu/m\sim\epsilon_t\ll1$. Under spherical symmetry, the vector and tensor modes vanish identically. The expansion of the universe is not relevant due to the small size of solitons, so we set $a(t) = 1$ and ignore contributions from the background fluid.   

The specific form of the field in \eqref{psistat} resembles the wavefunction of stationary states in quantum mechanics. Although strictly speaking the field $\psi_s$ is not a wavefunction, its time evolution is governed by \eqref{eqftkg} which, at leading order, resembles the conventional Schroedinger equation. The ansatz \eqref{psistat} corresponds to a time-independent energy density.\footnote{One crucial difference between our system and conventional quantum mechanics is that our system is nonlinear (even without relativistic corrections), so that a superposition of solutions fails to be a solution. As a result, unlike in quantum mechanics, one cannot express the general time evolution in terms of a superposition of various stationary states.} 

Using (\ref{psistat}) and the time independence of $\Phi_s$ in \eqref{eqftkg} and \eqref{eftPhi}, we find
\begin{align}
	\label{eftkgsss}
	\nonumber
	&\frac{\nabla^2f}{2m}-\Big(\Phi_s+\frac{\mu}{m}\Big)mf-\frac{\lambda f^3}{8m^2}\\
	&+\Big(3\Phi_s^2+\frac{4\mu}{m}\Phi_s+\frac{\mu^2}{2m^2}\Big)mf+\frac{\lambda f^3}{8m^2}\Big(2\Phi_s-\frac{\mu}{m}\Big)+\frac{3f^3}{16\Mpl^2}-\frac{\lambda^2f^5}{768m^5}=0+\order{\epsilon^4} ~,
\end{align}
and
\begin{align}
	\label{eftPhisss}
	\nonumber
	&\nabla^2\Phi_s-\frac{mf^2}{2\Mpl^2} \\
	&-\frac{mf^2}{2\Mpl^2}\Big(-6\Phi_s-3\frac{\mu}{m}\Big)+\frac{\lambda f^4}{16m^2\Mpl^2}=0+\order{\epsilon^4}\,,
\end{align}                             
where for simplicity we have set $\kappa=0$. The above equations are the time-independent NR EFT system of equations; in both (\ref{eftkgsss}) and (\ref{eftPhisss}), terms on the second line are smaller than those on the first by $\order{\epsilon}$ (while the first lines are already $\order{\epsilon^2}$ according to our power counting). Note the explicit appearance of the parameter $\mu$ in these equations. Since the stationary ansatz \eqref{psistat} removes all time derivatives from \eqref{eqftkg} and \eqref{eftPhi}, we have used the leading-order equations to remove all spatial derivatives in subleading terms, which yields multiple terms proportional to $\mu$ in the final result.\footnote{Some appropriate field redefinitions can remove $\mu$ completely from the equations, but change the asymptotic behavior of $\Phi$ to a nonzero constant. In this case $\Psi$ can no longer be replaced by $\Phi$ to the working order, so that both gravitational potentials must be solved simultaneously. Such a system is easier to solve numerically, and the plots in this section take advantage of this procedure.} Moreover, the $\nabla^4$ term in \eqref{eqftkg}, which appears due to integrating out nonzero modes $X_\nu$ with $\nu \neq 0$, can also be removed by a similar manipulation. This removal of higher spatial derivatives makes the system more suitable for numerical calculations. We look for spatially localized, nodeless and regular solutions. That is, we demand that $f(r)$ and $\Phi_s(r)$ vanish fast enough at infinity; that $f'(0)=0$; and that the solutions are monotonic. Such solutions are expected to describe long-lived solitonic solutions. 

We wish to compare solutions of our EFT, \eqref{eftkgsss} and \eqref{eftPhisss}, with corresponding solutions of the SP equations, as well as  solutions of a fully relativistic theory. This will allow us to see whether the EFT equations provide an improvement over the SP equations. Before doing so, however, we must address two questions: (1) What solution of the Klein-Gordon equation corresponds to the solution of \eqref{eftkgsss} and \eqref{eftPhisss}? (2) What observable should we choose in order to compare the solutions?

To answer the first question we try to reconstruct the scalar field $\vp$, or equivalently $\psi$, from the knowledge of $\psi_s$ and nonzero modes $\psi_\nu$. From \eqref{nonzero} for the stationary solution we have
\eq{
	\psi^{(1)}_\nu=\left[\frac{\mu}{2m}e^{-i\mu t}\delta_{\nu,2}+\frac{\lambda f^2}{48m^3}e^{3i\mu t}\delta_{\nu,-2}-\frac{\lambda f^2}{96m^3}e^{-3i\mu t}\delta_{\nu,4}\right]f\,,
}        
where we have used the leading-order Schroedinger equation to simplify terms. Using $\psi=\psi_s+\sum_{\nu\ne0}\psi_\nu^{(1)}e^{i\nu m t}+\order{\epsilon^3}$ and \eqref{redef1}, we see that the relativistic scalar field will take the form
\eq{
	\vp=\sqrt{\frac{2}{m}}\left[\vp_1\cos(\omega t)+\vp_3\cos(3\omega t)+\order{\epsilon^3}\right]\,,
} 
where the higher-order terms also include higher multiples of the frequency $\omega \equiv m-\mu$. We have defined time-independent coefficients $\vp_1=(1+\mu/2m)f$ and $\vp_3=\lambda f^3/96m^3$ at this working order. As a result, the specific form of \eqref{psistat} implies a periodic solution for the scalar field with a period ${2\pi}/{\omega}$, and we must look for this type of solution in the relativistic theory; keeping in mind that the true relativistic solutions also include radiating modes (leading to deviations from periodicity) \cite{Fodor:2008du,Fodor:2019ftc,Zhang:2020bec}, which are not captured here.

Solutions in which the field configuration is spatially localized, coherently oscillating in the core, and the configuration is exceptionally long lived, are well known and are called oscillons \cite{Bogolyubsky:1976yu,Gleiser:1993pt,Copeland:1995fq,Kasuya:2002zs,Amin:2010jq,Amin:2013ika,Fodor:2019ftc}, axion stars, scalar stars \cite{Seidel:1991zh,Visinelli:2017ooc,Chavanis:2017loo, Eby:2019ntd} depending on the context. They are approximate, time-dependent solitons of the relativistic theory. Such solitons are relevant in many cosmological contexts, both in the early and contemporary universe (for example, see \cite{Copeland:1995fq,  Amin:2010xe,Amin:2011hj,Gleiser:2011xj, Grandclement:2011wz, Lozanov:2017hjm,Hong:2017ooe,Ikeda:2017qev,Levkov:2018kau, Bond:2015zfa,Antusch:2017flz, Arvanitaki:2019rax, Brax:2020oye,Kawasaki:2020jnw,Amin:2020vja}). They owe their localization to gravitational interactions\cite{Seidel:1991zh, Seidel:1993zk, UrenaLopez:2001tw, Alcubierre:2003sx}, or self-interactions \cite{Amin:2019ums, Zhang:2020bec, Zhang:2020ntm} or a combination of both.

As for the second question, one important and reliable observable we have for solitonic solutions is their mass. We use the ADM definition of mass \cite{Arnowitt:1962hi} (see also \cite{Gourgoulhon:2007ue}), which is the Schwarzschild mass for an observer at infinity,
\eq{
	\label{mass}
	M\equiv\lim_{r\to\infty}\frac{r^2\p_r\Psi}{G}=-\int_{0}^{\infty}\dd{r}4\pi r^2\left[T^0{}_0+3\Mpl^2e^{-2\Phi}\dot{\Psi}^2\right]e^{-5\Psi/2}\,.
}   
In the second equality we have used the Einstein field equations and the expression for the $G^0{}_0$ component of the Einstein tensor. Since the ADM mass is time independent, in the language of the mode expansion of \eqref{eq:modexp}, it only depends on the slow mode: $M=\ev{M}=M_s$. As a result, it is also possible to compute $M$ with the help of EFT variables $f$ and $\Phi_s$ as
\eq{
	\label{masss}
	M_s=\int_{0}^{\infty}\dd{r}4\pi r^2\left[m f^2\Big(1-\frac{7\Phi_s}{2}\Big)+\frac{(\nabla f)^2}{2m}+\frac{\lambda f^4}{16m^2}\right]~,
} 
to working order. This means that, although the KGE and SP (plus suitable corrections) systems are two different theories, we can compare their solutions by demanding that they both yield the same solitonic mass. We define the radius $R_{95}$ as the distance enclosing 95 percent of the mass after taking the time average of the mass density:
\eq{0.95 M=\int_0^{R_{95}}\langle [\hdots ]\rangle 4\pi r^2 dr \, ,}
where the integrand is given by the corresponding term in (\ref{masss}). 

\fg{
	\centering
	\includegraphics[width=0.5\textwidth]{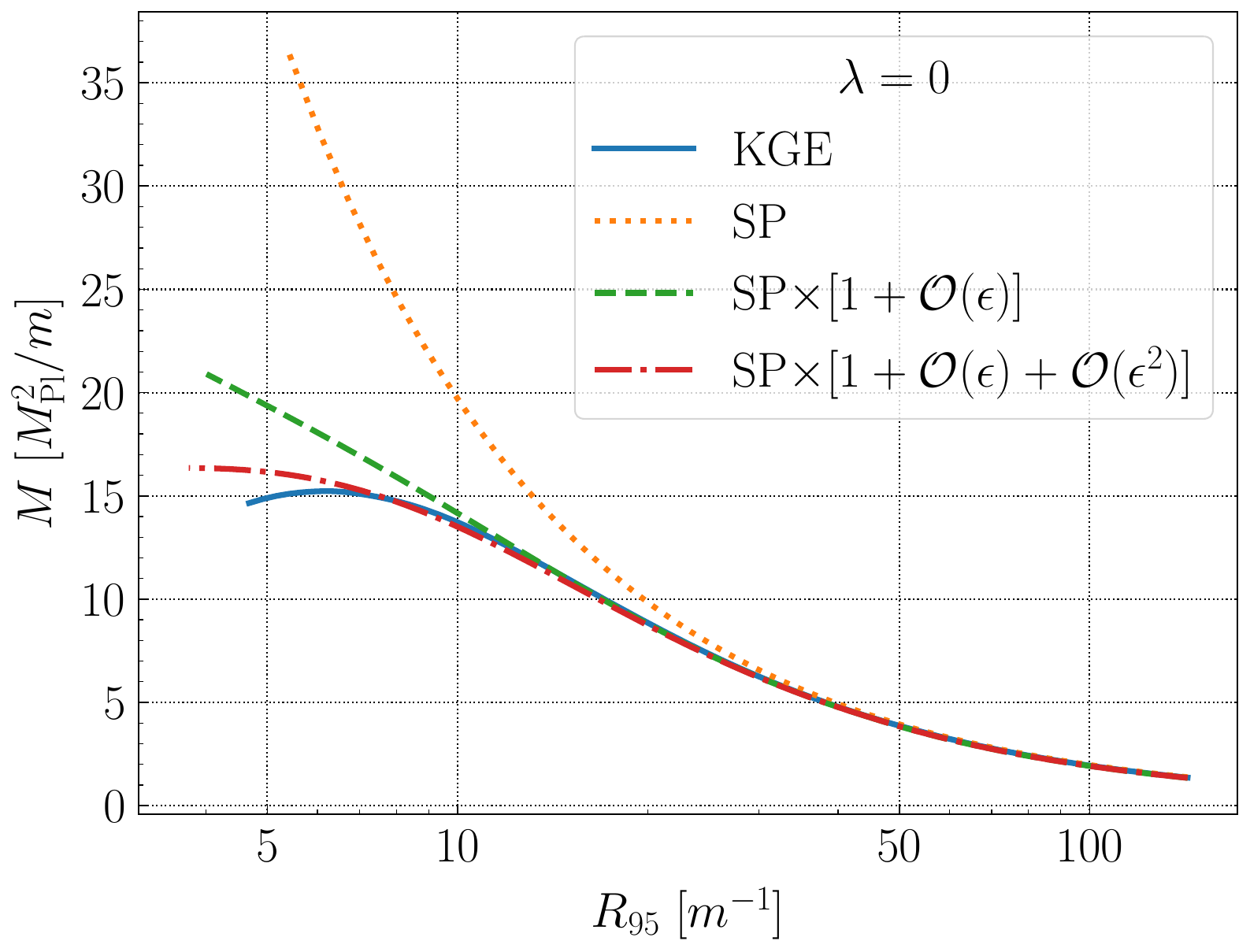} 
	\caption{A comparison of the mass-radius relation for the free theory ($\lambda = 0$) obtained from the Klein-Gordon-Einstein (KGE) equations, the Schroedinger-Poisson (SP) equations, and our NR EFT that includes $\mathcal O(\epsilon)$ and $\mathcal O(\epsilon^2)$ corrections beyond the SP equations. When the system becomes mildly relativistic, $R_{95}\sim10\,m^{-1}$, the SP equations show increasing disagreement with the fully relativistic results obtained from the KGE equations. On the other hand, our effective equations with just the leading relativistic corrections improve the results significantly. Furthermore, the $\mathcal O(\epsilon^2)$ corrections also capture the qualitative behavior when the system becomes highly relativistic.}
	\label{fig:free}
}
In practice, it is easier to fix the value of the field at the origin, rather than the mass, and find the corresponding solution in both theories. The mass and radius of the solution can then be obtained from \eqref{masss}. In figure \ref{fig:free} we compare the mass-radius relationship for solitons in the free theory ($\lambda = 0$) obtained with the KGE and with SP equations (with and without corrections). In the figure, ``SP" refers to results from the lowest-order SP equations, neglecting all corrections, as given by the first lines of \eqref{eftkgsss} and \eqref{eftPhisss}. The EFT including $\order{\epsilon}$ corrections to the SP equations is given by \eqref{eftkgsss} and \eqref{eftPhisss} (first and second lines). The $\order{\epsilon^2}$ corrections are derived using the same procedure, and the details are omitted here. Compared with the SP equations, our effective equations with just the $\order{\epsilon}$ corrections improve the mass-radius relation significantly in the mildly relativistic regime, for $R_{95}\simeq 10 ~m^{-1}$. Note that at this radius, the mass calculated using the SP equations differs from that obtained from the relativistic KGE calculation by $>50\%$, whereas including $\order{\epsilon}$ corrections in our EFT leads to a discrepancy of $<10\%$ from the relativistic KGE solutions. We can improve the results further by including $\order{\epsilon^2}$ corrections, which match the relativistic KGE calculation to within $\sim 1\%$ around $R_{95} \simeq 10~m^{-1}$.

\fg{
	\centering
	\includegraphics[width=0.49\textwidth]{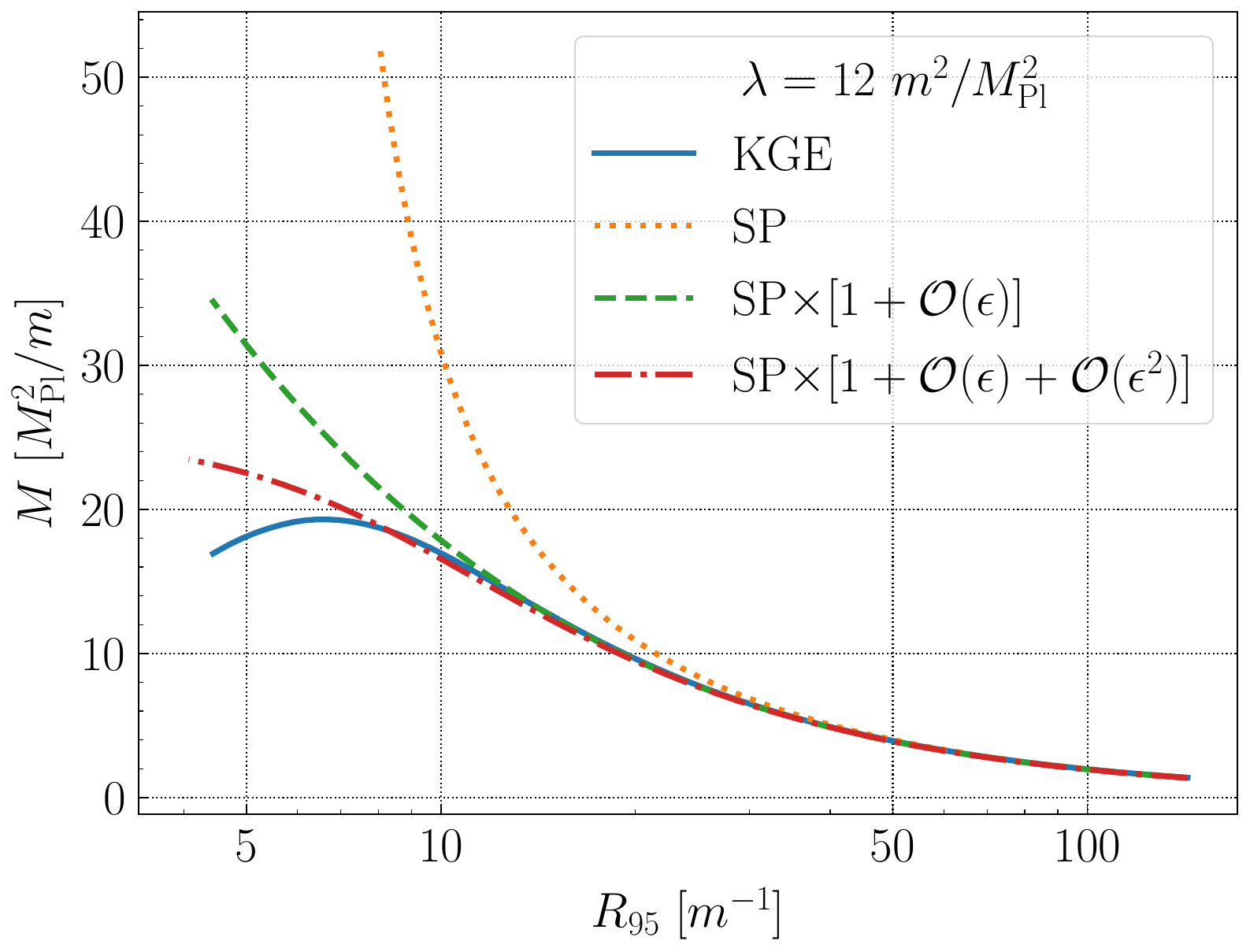} 
	\includegraphics[width=0.49\textwidth]{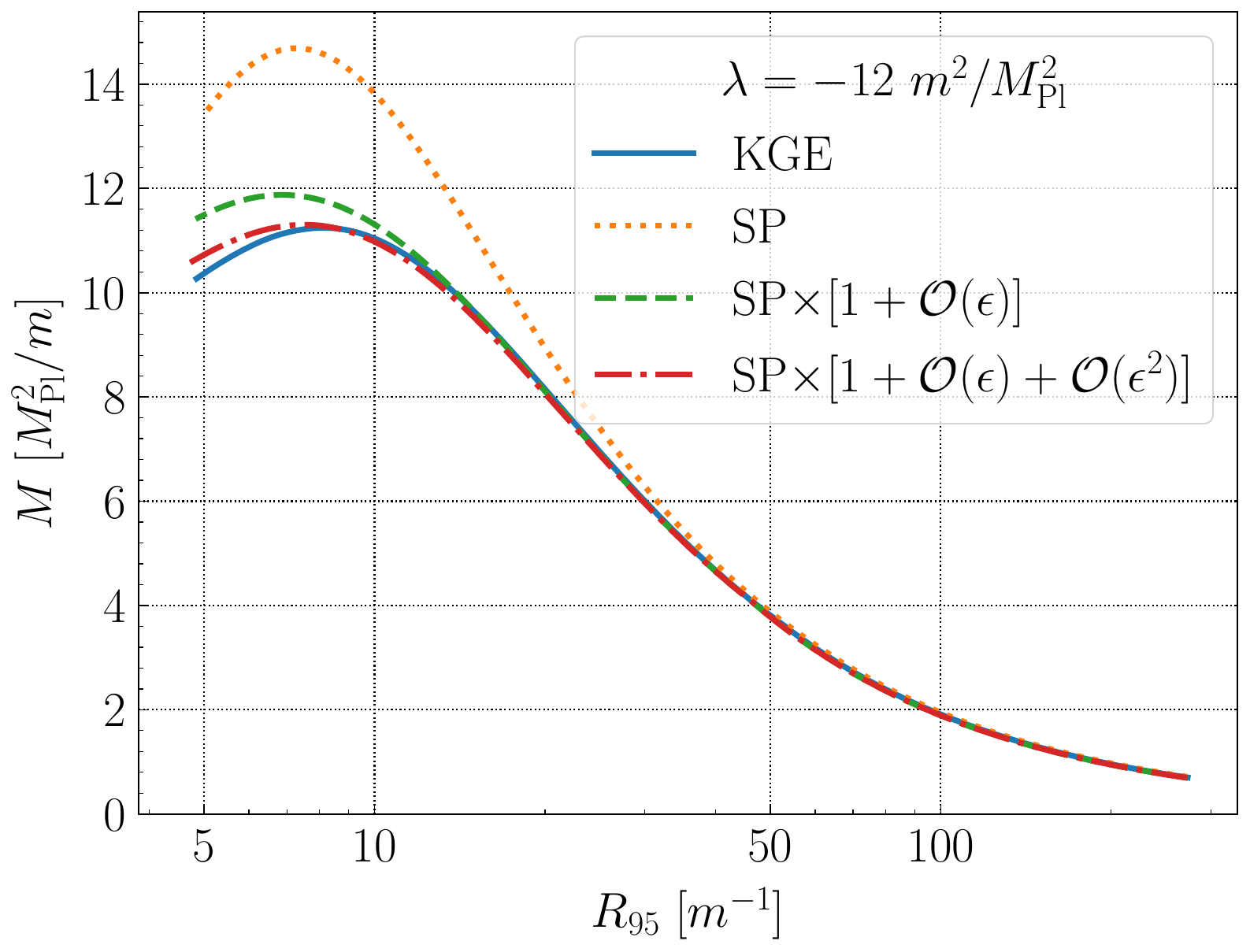}
	\caption{A comparison of the mass-radius relation for theories with repulsive (left) and attractive (right) self-interactions. The results for much larger $|\lambda\Mpl^2/m^2|$ (as would be the case for QCD and ultra-light axions) can also be obtained within our EFT. In this case the deviations from the SP system with attractive self-interactions still appear at $mR_{95}\lesssim 50$. For the repulsive case, the deviation from the SP system becomes significant at larger and larger $mR_{95}$ as $|\lambda\Mpl^2/m^2|$ increases.}
	\label{fig:int}
}
We have also confirmed the improvement in the mass-radius relationship obtained from our equations compared to the SP system in theories with repulsive ($\lambda>0$) and attractive ($\lambda<0$) self-interactions in figure \ref{fig:int}. The reasons for choosing $\lambda=\pm 12 m^2/\Mpl^2$  in figure \ref{fig:int} are that (i) by this choice all small parameters become the same order of magnitude in the mildly relativistic regime and (ii) it can make the comparison with the $\lambda=0$ easier. With more canonical parameter choices, the value of $|\lambda\Mpl^2/m^2|$ can be very large, for example, for QCD axions $\lambda\Mpl^2/m^2\sim -\Mpl^2/f_a^2$ where $f_a\sim 10^{11} \mathrm{GeV}$.  A natural question is at what $mR_{95}$ does the mass-radius relation of SP equations start to deviate significantly from that of KGE equations when $|\lambda\Mpl^2/m^2|\gg 10$? We have confirmed that few percent level differences from the SP system always start appearing at $mR_{95}\lesssim 50$ as long as $\lambda\Mpl^2/m^2\ll -10$.\footnote{In practice, as a proxy for the detailed mass radius curve, we simply construct $mR_{95}$ vs. $|\lambda|$ for $\epsilon_\lambda=0.1$, and see that this $mR_{95}$ initially grows slowly with $|\lambda|$ and then approaches a constant at sufficiently large $|\lambda|$.}

For the repulsive case, and with $\lambda\Mpl^2/m^2\gg 10$, there is a minimum radius for the soliton (in the SP system). Around the minimum, the soliton is formed by a balance between gravity and self-interactions which implies that $mR_\mathrm{min}\propto \sqrt{\lambda\Mpl^2/m^2}$. The deviation between results from SP and KGE equations is large at $mR_\mathrm{min}$. As a result, relativistic corrections become important at larger $mR_{95}$ as we increase the value of $\lambda\Mpl^2/m^2$ -- which shifts the minimum (and therefore the whole curve) to the right in the mass-radius plot. The difference between attractive and repulsive cases is because the former has a balance between the gradient term and attractive self-interactions. For the repulsive case, also see \cite{Croon:2018ybs}.

\section{Validity for the nonrelativistic expansion}
\label{sec:eft_validity}
Regarding the consistency of an EFT, one guiding principle that often comes into play is the validity of an effective description of the interaction. Another standard lore is that theories with ghosts or energies unbounded from below are usually unstable and problematic \cite{Woodard:2015zca, Cline:2003gs, Sawicki:2012pz} and the initial conditions must be restricted in ``islands of stability'' if possible \cite{Pagani:1987ue, Smilga:2004cy}, although there may be some exceptions \cite{Deffayet:2021nnt}. For example, it is pointed out that if massive vectors are nonminimally coupled to gravity, the longitudinal mode may exhibit ghost instabilities and one can not discuss the vector field dynamics in a healthy way \cite{Nakayama:2019rhg, Kolb:2020fwh}. In practice, one performs as many sanity checks as possible to determine the scope of application for the theory in hand.

To derive the NR EFT, we have assumed in section \ref{sec:eft_nreft} that all dimensionless $\epsilon$ parameters are less than unity. To check the validity for our use of the EFT in soliton investigations in section \ref{sec:eft_soliton}, we plot various small parameters $\epsilon$ for the solitons with repulsive self-interactions in figure \ref{fig:epsilon}. The left panel shows the profile of $\epsilon$ in terms of the radius, while the right panel shows the maximum value of $\epsilon$ in terms of the 95\% radius. Our perturbative scheme fails when $\epsilon \sim 1$. Also note that $\epsilon_x$ is a measure of momentum, and we see that particles are indeed mildly relativistic when $R_{95}\simeq 10 ~m^{-1}$.
\fg{
	\centering
	\includegraphics[width=0.49\textwidth]{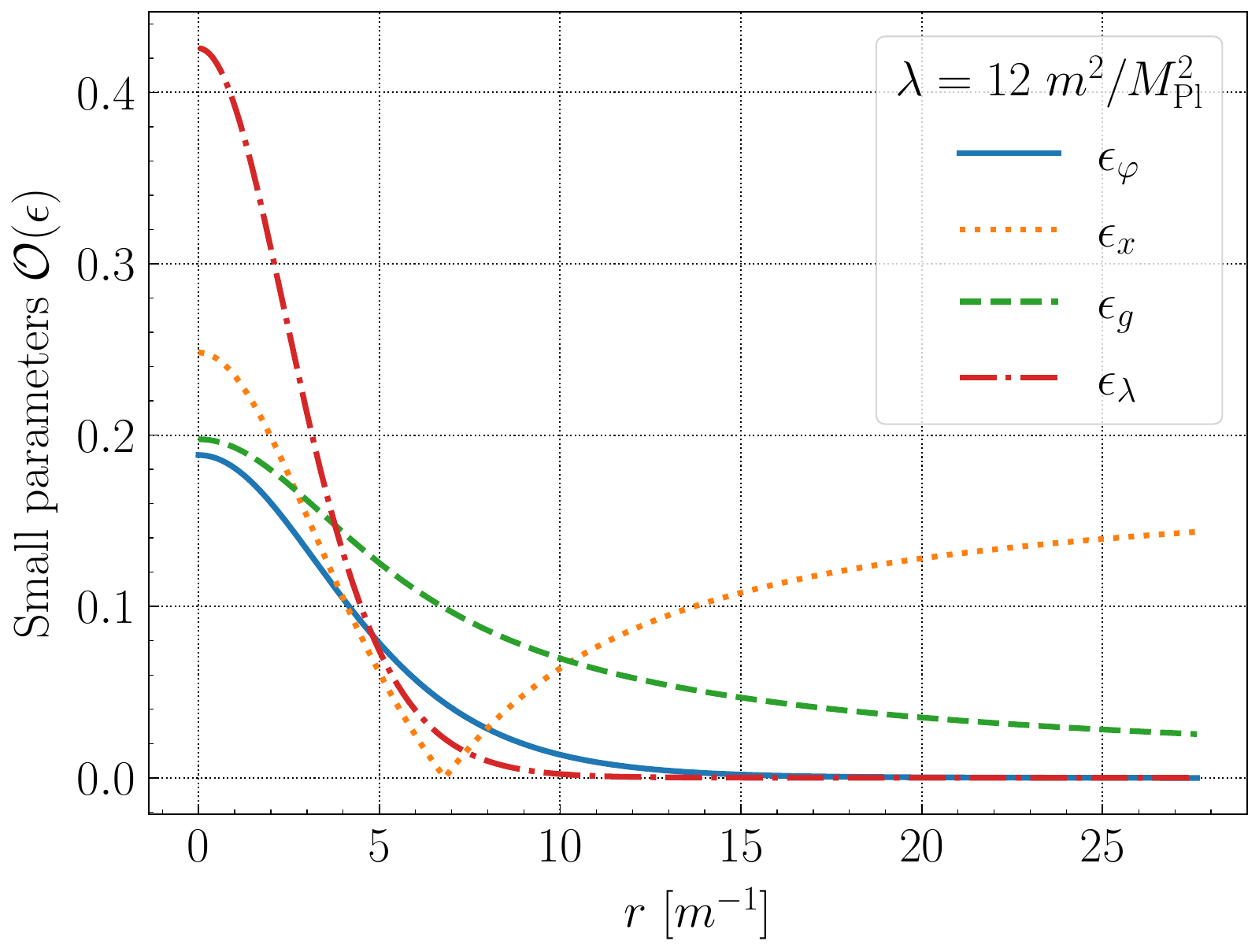} 
	\includegraphics[width=0.49\textwidth]{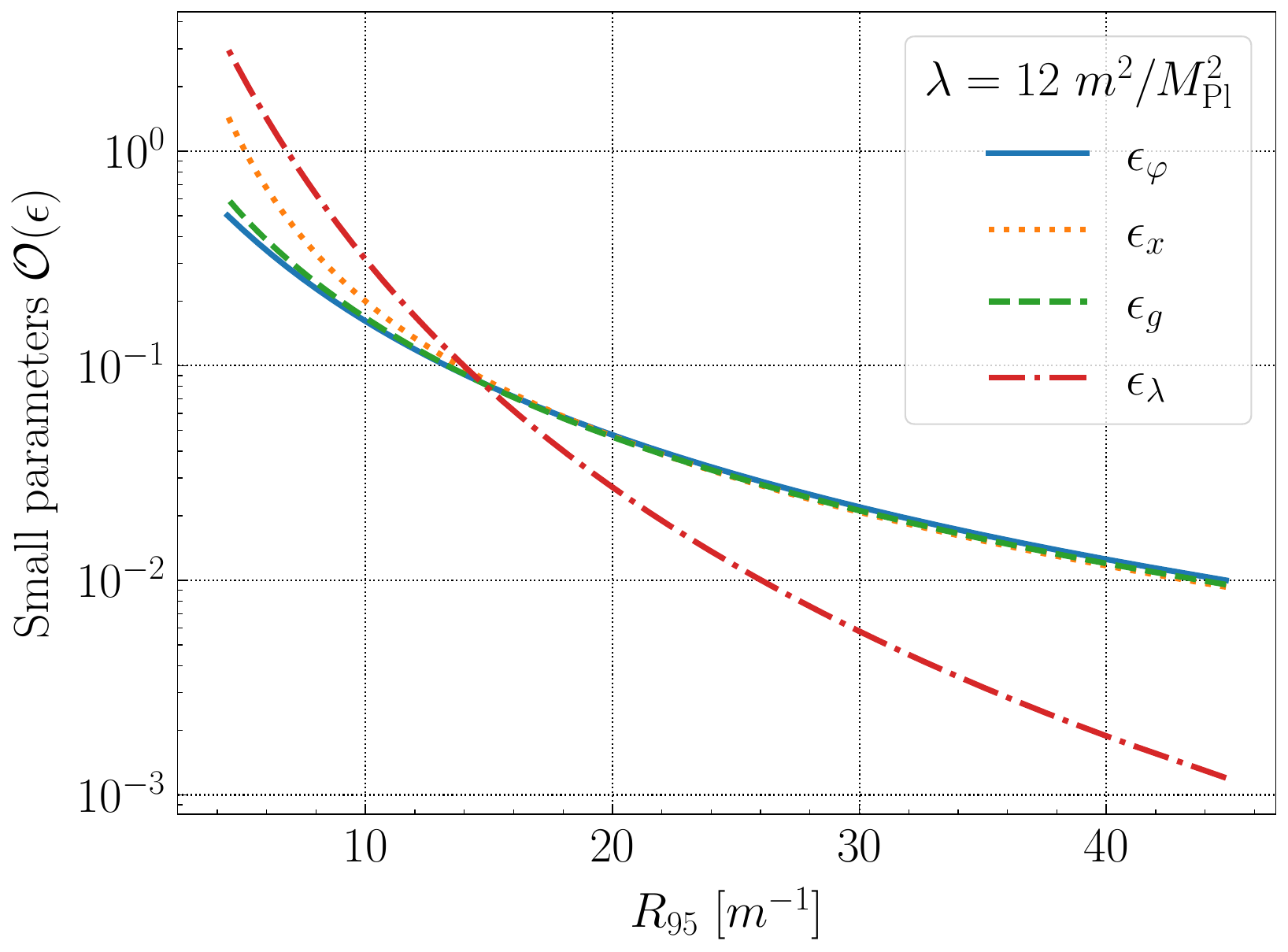}
	\caption{Small parameters in the EFT for the soliton solutions. In the left panel, we plot the spatial dependence of the small parameters $\epsilon$ for a soliton with $R_{95}=9.08 ~m^{-1}$. In the right panel, we plot the maximum value of $\epsilon$ as a function of the 95\% radius. The data is obtained based on our effective equations with $\mathcal O(\epsilon^2)$ corrections.}
	\label{fig:epsilon}
}
}

\section{Validity for vector field theory}
\label{sec:eft_singularity}
Unlike scalar field theory, an interacting Proca theory is a nontrivially constrained system, where a singularity problem for the auxiliary component $A_0$ could arise if it cannot be uniquely solved in terms of the canonical fields, signaling a breakdown of the theory. Note that this issue is not exclusive to the NR EFT described above but is more broadly applicable and could arise in any limits of a classical EFT. Therefore, instead of expressing the original field in terms of an NR field as done previously, we will retain the original field to describe the problem as generally as possible.

In what follows, we will first clarify three consistency conditions and introduce the singularity problem by taking real-valued self-interacting vectors as an example, then a specific model is carried out in detail both analytically and numerically. Finally we discuss the implication of our results and illustrate that the singularity problem can also exist for complex fields and general types of interactions.

\subsection{Singularity problem}
For definiteness, let us consider a real-valued massive vector field $A_\mu=(A_0, \b A)$ with the Lagrangian
\begin{align}
	\label{Lagrangian}
	\cal L = -\frac{1}{4}F_{\mu\nu} F^{\mu\nu} - V(A_\mu A^\mu) ~,
\end{align}
where $F_{\mu\nu} = \pd_\mu A_\nu - \pd_\nu A_\mu$ and there is no gauge invariance thanks to the potential $V$, which includes a mass term along with self-interactions. A concrete example is the Abelian-Higgs model, where a quartic self-interaction is induced by Higgs exchange in the low-energy limit. Later on, we will generalize the discussion to complex fields and more generic interactions. By varying the action $S = \int d^4x \cal L$ with respect to the field $A_\nu$, we find the Euler-Lagrange equation $\pd_\mu F^{\mu\nu} - 2V'(A_\mu A^\mu) A^\nu = 0$. In vector notation, it becomes
\begin{align}
	\label{EOM_A0}
	\nabla\cdot \b \Pi + 2V' A_0 = 0 ~,\\
	\label{EOM_Ai}
	\dot{\b \Pi} + \nabla\cp\nabla\cp\b A + 2V'\b A = 0 ~,
\end{align}
where the prime denotes the derivative of the potential in terms of $A_\mu A^\mu$ and we have defined the conjugate field $\Pi_\mu \equiv \pd\cal L/\pd\dot A^\mu = F_{0\mu}$, so
\begin{align}
	\label{EOM_Pi}
	\dot{\b A} = \b\Pi + \nabla A_0 ~.
\end{align}
One more useful equation can be obtained by noting that $F^{\mu\nu}$ is antisymmetric, so that $\pd_\mu (V'A^\mu)=0$. That is
\begin{align}
	\label{EOM_extra}
	-(V' - 2V'' A_0^2)\dot A_0 - 2 A_0 V'' (\b A\cdot \dot{\b A}) + \nabla\cdot ( V' \b A) = 0 ~.
\end{align}
In the language of Hamiltonian mechanics, $\Pi_0=0$ is a primary constraint, equation \eqref{EOM_A0} is a secondary constraint obtained by requiring $\dot \Pi_0 = \delta H/\delta A^0=0$, and equation \eqref{EOM_extra} is a tertiary constraint obtained by requiring the secondary constraint to be preserved in time \cite{Weinberg:1995mt}. The foundation for these derivations is the stationary action principle, in which we have implicitly assumed that the field $A_\mu$ is continuous otherwise the infinitesimal variation $\delta A_\mu$ is ill-defined. By applying the above formalism, therefore, we require a consistent classical system to satisfy at least three conditions everywhere:
\begin{enumerate}[label=(\roman*)]
	\item \label{consistency1} The field $A_\mu(t,\b x)$ is real-valued;
	\item \label{consistency2} The field $A_\mu(t,\b x)$ is continuous;
	\item \label{consistency3} The second-class constraints, e.g. \eqref{EOM_A0} and \eqref{EOM_extra}, are respected.
\end{enumerate}

These conditions are not trivial, and we may gain some insights about them by using equations \eqref{EOM_Ai}-\eqref{EOM_extra} and numerically evolving $\b\Pi$, $\b A$ and $A_0$. Given appropriate initial conditions, suppose that $V'-2V''A_0^2$ never becomes 0, then the infinitesimal variations $\delta\b\Pi$, $\delta\b A$ and $\delta A_0$ are always well defined in a infinitesimal time interval $\delta t$, and the field $A_\mu$ will remain smooth and unique all the time. This is indeed the case for free massive fields, where $V(A_\mu A^\mu) = m^2 A_\mu A^\mu/2$. For theories with self-interactions, however, a singularity is encountered if $V'-2V''A_0^2$ becomes 0 at some spacetime point unless $- 2 A_0 V'' (\b A\cdot \dot{\b A}) + \nabla\cdot ( V' \b A)$ also vanishes in an appropriate way to ensure a finite $\dot A_0$, which otherwise causes a discontinuity in $A_0$ and violate at least one of the consistency conditions. Thus maintaining the continuity of $A_0$ at this point needs an over-constraint and requires fine tuning of initial conditions. It is seen that any plausible interacting Proca theories should ensure that such a problem is avoided in its validity, to wit, the field value should never cross the \emph{boundary} in field space $\{\abs{A_0}, \abs{\b A}\}$ specified by 
\begin{align}
	\label{boundary}
	V'-2V'' A_0^2=0 ~.
\end{align}
One may think that the problem identified here can be easily avoided if we use equation \eqref{EOM_A0} instead of \eqref{EOM_extra} to obtain $A_0$. As will be shown shortly, this difficulty is actually independent of whether and how we evolve the system numerically.

\subsection{A concrete model}
In order to understand the significance of this singularity bound, it is illuminating to consider the simplest possibility of a self-interaction
\begin{align}
	\label{potential_A4}
	V(A_\mu A^\mu) = \frac{m^2}{2} A_\mu A^\mu + \frac{\lambda}{4}(A_\mu A^\mu)^2 ~,
\end{align}
where $A_0$ can be solved in closed form. We are going to show that if we stick with condition \ref{consistency1} and \ref{consistency3}, then condition \ref{consistency2} will necessarily be violated if the field system hits the boundary \eqref{boundary} during its evolution.

\begin{figure}
	\centering
	\includegraphics[width=0.35\linewidth]{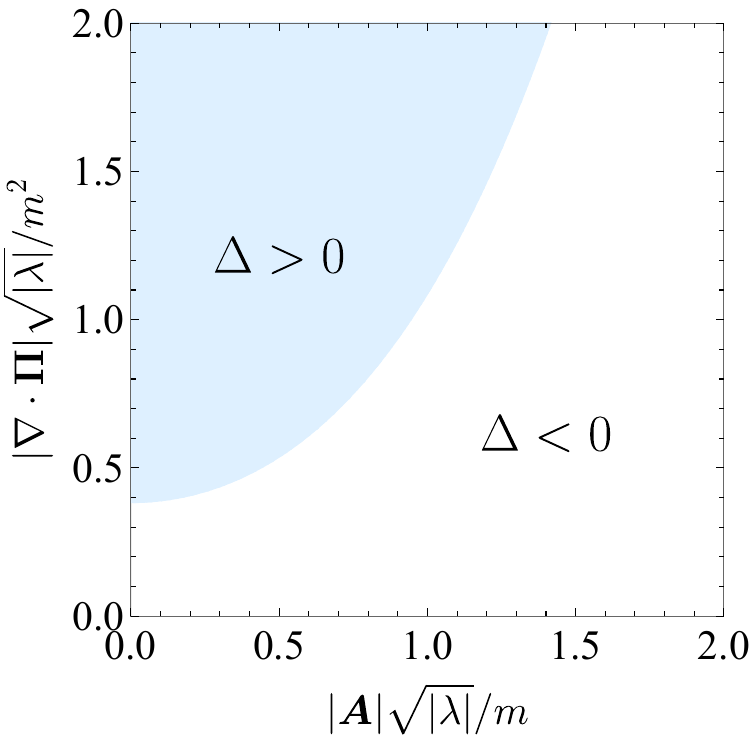}
	\quad\quad\quad
	\includegraphics[width=0.35\linewidth]{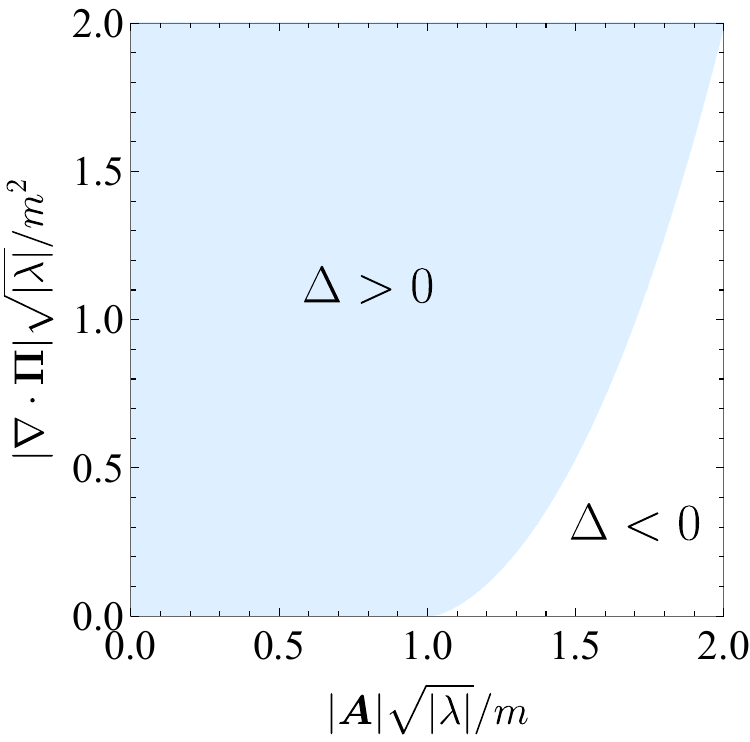}
	\caption{The value of the discriminant $\Delta$ in terms of $\nabla\cdot\b\Pi$ and $\b A$, defined by equation \eqref{Delta}, for repulsive ($\lambda>0$, left) and attractive ($\lambda<0$, right) self-interactions. There are one, two or three different real roots of $A_0$ in equation \eqref{cubic_eq} when $\Delta$ is $>,=$ or $<0$.
	}
	\label{fig:boundary_divPi_Ai}
\end{figure}

The secondary constraint \eqref{EOM_A0} in this case becomes 
\begin{align}
	\label{cubic_eq}
	A_0^3 + c_1 A_0 + c_2=0 ~,
\end{align}
where $c_1 = - m^2/\lambda - \b A^2$ and $c_2 = -\nabla\cdot\b\Pi/\lambda$. The general solution of this cubic equation can be given by the Cardano's formula,
\begin{align}
	\label{root1}
	A_0^{(1)} &= u+v ~,\\
	\label{root2}
	A_0^{(2)} &= \frac{-1+i\sqrt{3}}{2}u + \frac{-1-i\sqrt{3}}{2}v ~,\\
	\label{root3}
	A_0^{(3)} &= \frac{-1-i\sqrt{3}}{2}u + \frac{-1+i\sqrt{3}}{2}v ~,
\end{align}
where $u = \sqrt[3]{- c_2/2 + \sqrt{\Delta}}$ and $v = \sqrt[3]{- c_2/2 - \sqrt{\Delta}}$. The $A_0^{(1)}$ is a real root and the other two are complex conjugate if $\Delta >0$. All three are real roots with $A_0^{(2)}$ and $A_0^{(3)}$ being the same if $\Delta=0$. And all three are different real roots if $\Delta <0$.\footnote{Depending on the sign of $\lambda$, there are three or one real roots when $\lambda\rightarrow 0^+$ or $0^-$. In either case there is only one finite root as expected since the free theory should be recovered in this limit.} Here the discriminant is defined as
\begin{align}
	\label{Delta}
	\Delta\equiv \(\frac{c_2}{2}\)^2 + \(\frac{c_1}{3}\)^3 = \( \frac{\nabla\cdot\b\Pi}{2\lambda} \)^2 - \( \frac{m^2}{3\lambda} + \frac{\b A^2}{3} \)^3 ~.
\end{align}
The value of $\Delta$ in terms of $\nabla\cdot\b\Pi$ and $\b A$ is shown in figure \ref{fig:boundary_divPi_Ai}. 

A few subtleties need to be clarified when we apply the Cardano's formula \eqref{root1}-\eqref{root3}. First, we have defined the square root of any number by its principal value. Second, we have defined the cube root of any number by its \emph{principal} value when $c_2 \le 0$ and its \emph{anti-principal} value when $c_2 >0$. The (anti-)principal cube root returns the real cube root for a real number, and the root with the (smallest) greatest real part for a complex number. These conventions are adopted such that $A_0^{(1)}$ is always real and all three roots are continuous everywhere except at $\nabla\cdot\b\Pi=0$, where $A_0$ can actually remain continuous by switching roots.

If a field system crosses the boundary $\Delta=0$ (and $\nabla\cdot\b\Pi\neq 0$) during its evolution, then the real roots $A_0^{(2,3)}$ will be annihilated or created depending on which region in figure \ref{fig:boundary_divPi_Ai} the system is in before the crossing. It is easy to see that the discontinuity of $A_0$ is an inevitable consequence if $A_0$ follows either $A_0^{(2)}$ or $A_0^{(3)}$, and if the system hits $\Delta=0$ from the white region where $\Delta<0$. 

\begin{figure}
	\centering
	\includegraphics[width=0.35\linewidth]{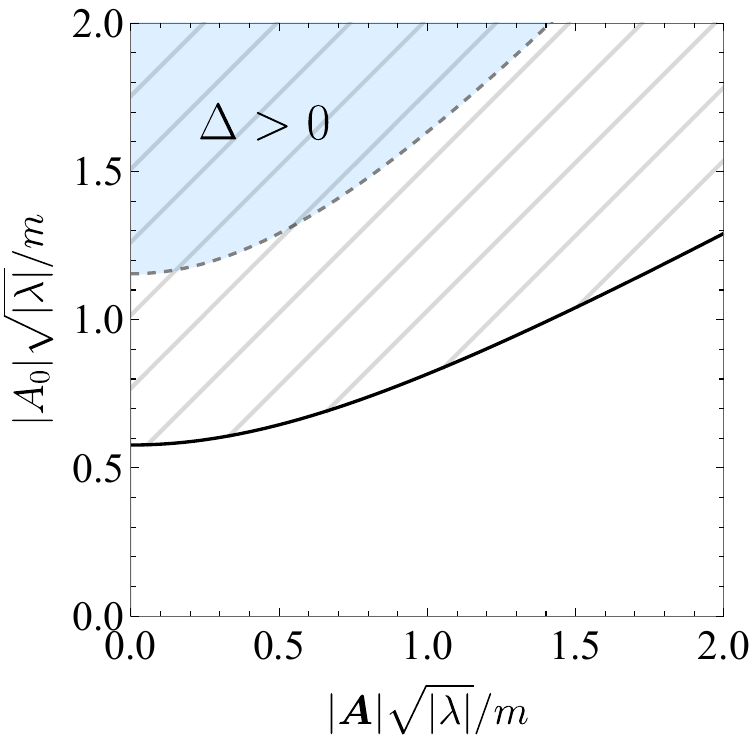}
	\quad\quad\quad
	\includegraphics[width=0.35\linewidth]{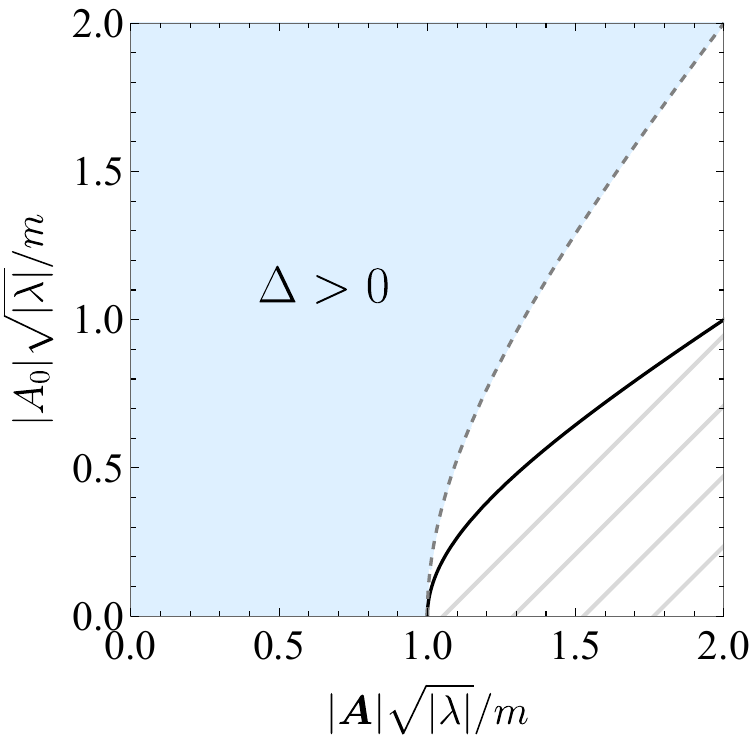}
	\caption{The value of the discriminant $\Delta$ in terms of $A_0$ and $\b A$ for repulsive ($\lambda>0$, left) and attractive ($\lambda<0$, right) self-interactions. The colored and white regions correspond to $\Delta>0$ and $\Delta<0$ as in figure \ref{fig:boundary_divPi_Ai}, and the gray dashed and black solid curves represent $A_0=2 A_{0,\rm{crit}}$ and $A_0=A_{0,\rm{crit}}$ at which $\Delta=0$. A consistent classical system should never cross the black solid curve, which is exactly the one specified by equation \eqref{boundary} (see the texts for proof). Allowing field values to be small, the system during the evolution should never enter into the meshed region.
	}
	\label{fig:boundary_A0_Ai}
\end{figure}

Now we will show that $A_0$ can not remain continuous if the system hits the boundary specified by equation \eqref{boundary}. To do this, we can judiciously rewrite the discriminant $\Delta$ in terms of $|A_0|$ and $|\b A|$ by using the secondary constraint \eqref{EOM_A0}. The value of $\Delta$ in terms of $|A_0|$ and $|\b A|$ is shown in figure \ref{fig:boundary_A0_Ai}. At $\Delta=0$, the three roots \eqref{root1}-\eqref{root3} become $|A_0^{(1)}| = 2A_{0,\rm{crit}}$, $|A_0^{(2,3)}|=A_{0,\rm{crit}}$, where\footnote{We are only interested in the case where $A_{0,\rm{crit}}$ is real.}
\begin{align}
	A_{0,\rm{crit}}(\b A) = \sqrt{\frac{m^2 + \lambda \b A^2}{3\lambda}} ~,
\end{align}
and $A_0=2A_{0,\rm{crit}}$ and $A_0=A_{0,\rm{crit}}$ are visualized as the gray dashed and solid black curves respectively in figure \ref{fig:boundary_A0_Ai}. Note that only $A_0 = A_{0,\rm{crit}}$ (solid black line) corresponds to the boundary $V'-2V''A_0^2=0$, and the adjacent regions separated by this line both have $\Delta<0$, which justifies the foregoing claim.

On the other hand, it is always safe to cross the gray dashed line, since in this case $A_0$ follows the root $A_0^{(1)}$ and $A_0^{(1)}$ is real and continuous. But there is no guarantee that the evolution will be healthy if the root $A_0^{(1)}$ is chosen for $A_0$ initially, because $A_0$ switches roots at $\nabla\cdot\b \Pi = 0$. In order to avoid the singularity problem and also allowing field values to be small, we conclude that the field evolution should be restricted in the non-meshed region in figure \ref{fig:boundary_A0_Ai}.

A minimal model of \eqref{potential_A4} is carried out numerically in $1+1$-dimensional spacetime to support the above analysis. We present field-space trajectories for repulsive self-interactions in figure \ref{fig:numerical1}. The case of attractive self-interactions is similar, and thus only shown in the Supplemental Material I, where numerical details are also provided.

\begin{figure}
	\centering
	\includegraphics[width=0.35\linewidth]{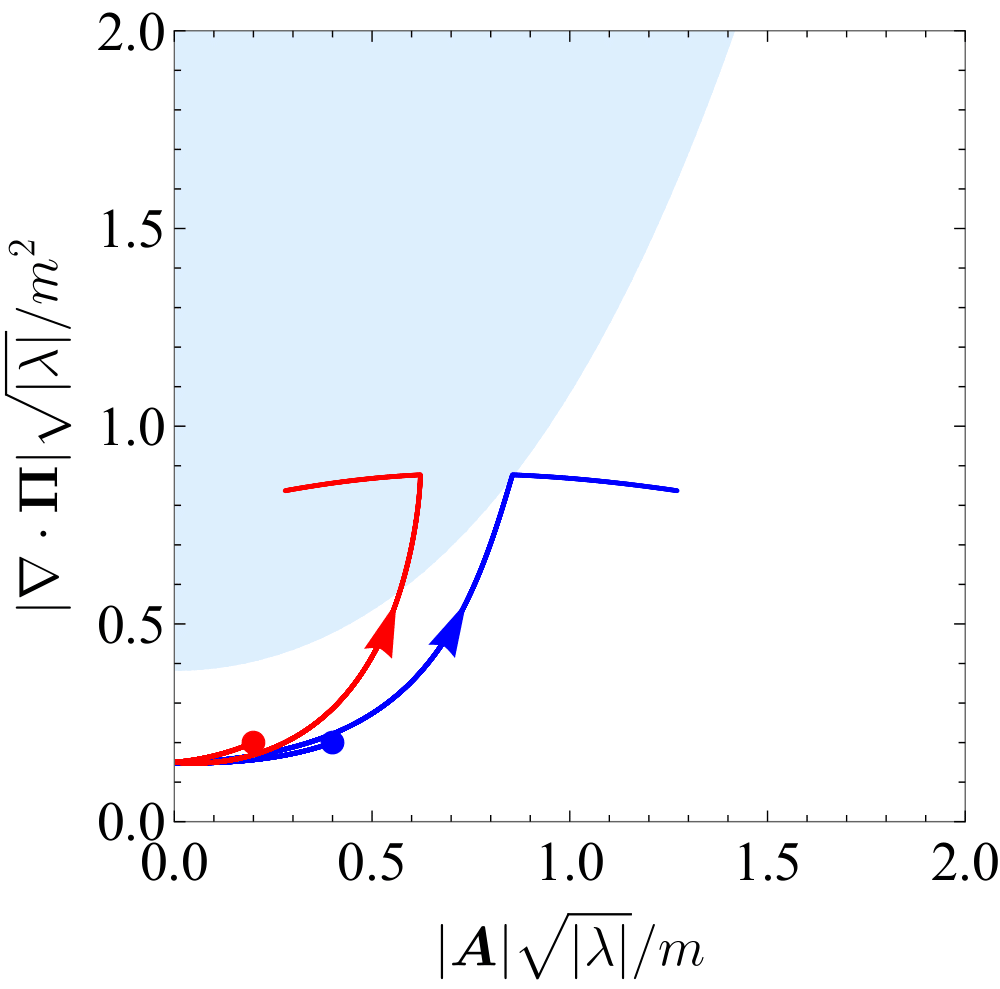}
	\quad\quad\quad
	\includegraphics[width=0.35\linewidth]{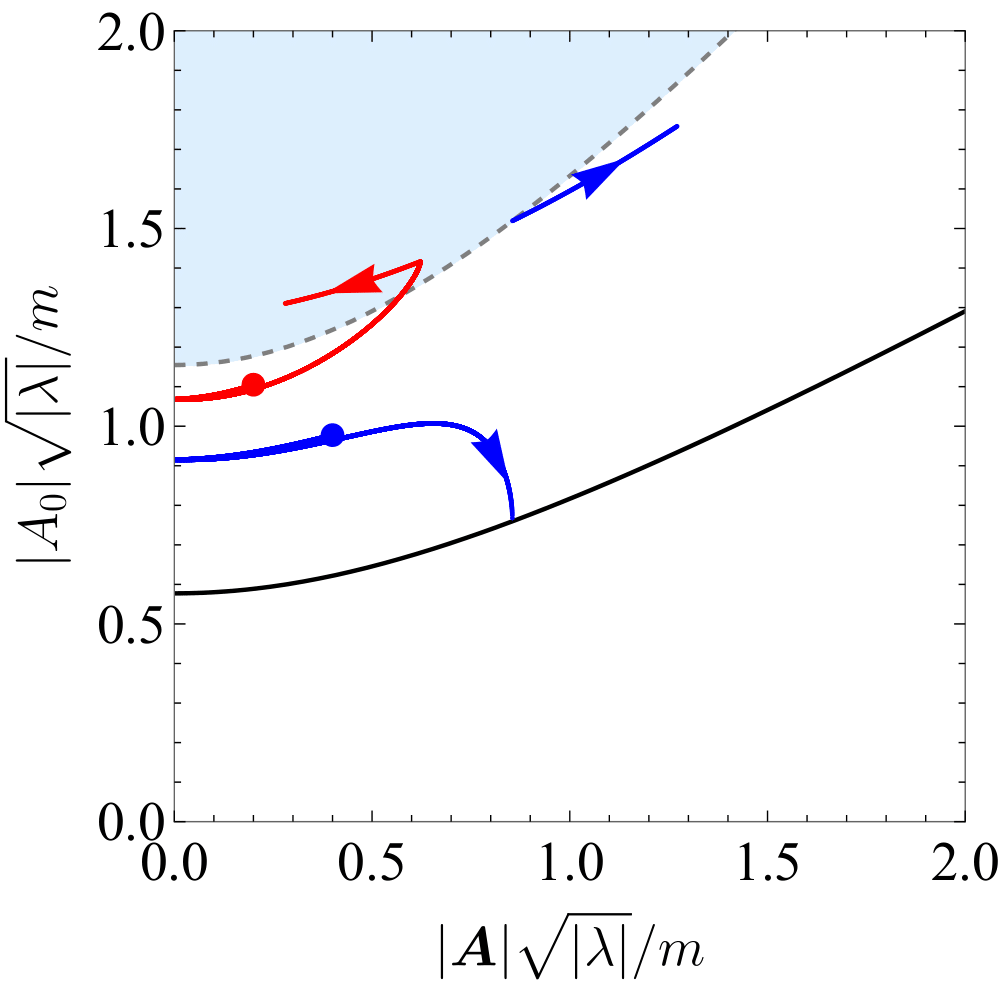}
	\caption{Field-space trajectories of a numerical example for repulsive self-interactions, where the system crosses the black solid boundary specified by equation \eqref{boundary}. The colored and white regions, and the gray dashed and black solid curves have the same meaning as in figure \ref{fig:boundary_divPi_Ai} and \ref{fig:boundary_A0_Ai}. The blue and red trajectories represent the time evolution of fields at two adjacent spatial locations starting from the solid point. As shown by the blue trajectory, when the system meets the black solid boundary, the value of $A_0$ can no longer remain continuous and suddenly jumps to the gray dashed line, which violates the consistency conditions. 
	}
	\label{fig:numerical1}
\end{figure}

Up to this point, it looks like that the ``discontinuity problem'' is a more appropriate name inasmuch as the temporal component $A_0$ can not be continuous when the field system hits the boundary specified by \eqref{boundary}. In fact, the discontinuity is just an artificial phenomenon because in principle we can stick with the condition \ref{consistency1} and \ref{consistency2} instead, and then we will reach a conclusion that the second-class constraints can not be obeyed. The real problem is that at least one of the consistency conditions would be violated if the boundary \eqref{boundary} is hit, which is closely related to a singularity in $\dot A_0$. In \cite{Clough:2022ygm, Coates:2022qia}, the authors also consider self-interactions that are described by equation \eqref{Lagrangian} and interpret the singularity problem discussed here as a ghost instability or a loss of hyperbolicity, by rewriting the field equations for $A_\mu$ in the form of wave equations (up to some terms with derivatives) and by identifying an effective metric $\hat g_{\mu\nu}$. The condition $\hat g_{00}=0$ turns out to be the same as \eqref{boundary}.\footnote{There are some caveats, though: (i) The vector field would become ghosts by diverging, at which point the entire theory actually breaks down (so the field would not get a chance to acquire kinetic terms with a wrong sign), and (ii) The loss of hyperbolicity is often dangerous, but not always fatal for physical systems (see, for example \cite{castro2011numerical, Frolov:2002rr, Kroger:2003qh}).}

\subsection{Potential applications}
Loosely speaking, trajectories in phase space (if we could ever visualize them for PDEs) would intersect at the singularity bound \eqref{boundary}, indicating that $A_0$ can no longer be solved uniquely. This situation is usually avoided in physics equations because of the Picard-Lindelof theorem (also called the existence and uniqueness theorem), which states that the existence and uniqueness of solutions are guaranteed if the derivative of the variable is continuously differentiable. We also note in there the existence of a basin of attraction towards the singularity bound. If the same goes for interacting massive vectors, the allowed field space is further restricted.

Our procedure to obtain the singularity bound is systematic, and also works for more general interactions (such as derivative interactions and interactions with external fields) -- finding the tertiary constraint in the theory and then picking out the coefficient of the time derivative of the auxiliary component.\footnote{This procedure may even work for spin-2 fields, and we leave this investigation for future work.} Following the procedure, we can also find that the singularity bound equally exists for complex fields. To see this, we may separate the real and imaginary part of a complex field $A_\mu = R_\mu + i I_\mu$, then the theory of $A_\mu$ becomes a theory of two interacting real fields $R_\mu$ and $I_\mu$, which are both constrained by demanding the absence of the singularity for $\dot R_0$ and $\dot I_0$.

We note that the singularity problem can be avoided by \emph{gauge-invariant} interactions, e.g. those only involving $F_{\mu\nu}$, although this may not be the only solution. This is because the gauge-invariant part in action must satisfy $\pd_\mu (\delta S_\rm{GI}/\delta A_{\mu})=0$ while we have $\pd_\mu (\delta S /\delta A_{\mu})=0$ if the equation of motion is satisfied.\footnote{By gauge invariance here we mean that the action is invariant under the gauge transformation $A_\mu\rightarrow A_\mu + \pd_\mu \Lambda$. More generally, a combined gauge transformation with another scalar field can be invented for massive vectors by using the Stueckelberg trick \cite{Hinterbichler:2011tt}. In this case one must generalize $\pd_\mu (\delta S_\rm{GI}/\delta A_{\mu})=0$ to include the scalar field.} Thus the tertiary constraint like \eqref{EOM_extra} can be obtained solely from gauge-symmetry-breaking terms. This is an example where gauge invariance plays a role even in theories without gauge invariance. The existence of the singularity problem in a theory indicates that the theory can not be the complete story. Learning from perturbative unitarity \cite{Lee:1977eg, Schwartz:2014sze}, the standard solution would be to introduce new particles or to look for a ultraviolet completion above the scale where the singularity bound is met. For example, we can introduce a Higgs boson to rescue the quartic theory \eqref{potential_A4}.

\chapter{Oscillons as a probe of dark matter self-interactions}
\label{sec:soliton}

\section{Introduction}
Exceptionally long-lived, spatially-localized and oscillatory field configurations, called {\it{oscillons}},  exist in real-valued scalar field theories with attractive self-interactions \cite{Bogolyubsky:1976yu,Gleiser:1993pt,Copeland:1995fq,Kasuya:2002zs,Amin:2010jq}. Oscillons emerge naturally from rather generic initial conditions making them relevant for wide ranging physical contexts including reheating after inflation \cite{Amin:2010xe,Amin:2010dc,Amin:2011hj,Gleiser:2011xj,Lozanov:2017hjm,Hong:2017ooe} and other phase transitions \cite{Farhi:2007wj,Gleiser:2010qt,Bond:2015zfa}, moduli field dynamics in the early universe \cite{Antusch:2017flz}, and  structure formation in scalar field DM \cite{Kolb:1993hw,Olle:2019kbo,Arvanitaki:2019rax,Kawasaki:2019czd}. Oscillons can have gravitational  implications in the form of clustering \cite{Amin:2019ums}, gravitational waves \cite{Zhou:2013tsa,Antusch:2016con, Liu:2017hua, Lozanov:2019ylm,Amin:2018xfe} and even formation of primordial black holes \cite{Cotner:2019ykd,Kou:2019bbc}. They can also have non-gravitational  implications, for example in the generation of matter-antimatter asymmetry \cite{Lozanov:2014zfa}. Besides single scalar fields with canonical kinetic terms, oscillons can be found in theories with non-canonical kinetic terms \cite{Amin:2013ika,Sakstein:2018pfd} as well as multi-field systems beyond scalar fields \cite{Graham:2006vy,Gleiser:2008dt,Sfakianakis:2012bq}.\footnote{When gravity is more important than scalar-field self-interactions, oscillons are called ``oscillatons" \cite{UrenaLopez:2001tw,Alcubierre:2003sx,Ikeda:2017qev}. Oscillons are also intimately connected with Q-balls \cite{Coleman:1985ki,Nugaev:2019vru}, and boson stars \cite{Kaup:1968, Liebling:2012fv} which are related configurations in complex valued fields (without and with gravity respectively). Oscillons also have NR analogs in Bose-Einstein condensates \cite{2017Sci...356..422N}, as well as in the NR, and weak field gravity regime in astrophysical contexts \cite{Schive:2014dra, Niemeyer:2019aqm}.}

The longevity, and decay rates of oscillons has long been a subject of interest. A decade after the discovery of oscillons (initially called ``pulsons" \cite{Bogolyubsky:1976yu}), Kruskal and Segur provided an estimate of their exceptionally suppressed decay rates in the small amplitude limit \cite{Segur:1987mg} (also see,  \cite{Fodor:2008du,Fodor:2009kf,Fodor:2019ftc}). However, oscillons of interest in cosmology do not have small amplitudes because there exists a long-wavelength instability in small amplitude oscillons in 3+1 dimensions, whereas larger amplitude ones are safe from such long-wavelength instabilities (see, for example \cite{Amin:2010jq}).\footnote{For shorter wavelength instabilities in the small amplitude limit, which are related to quantum instabilities, see for example, \cite{Hertzberg:2010yz}.} Moreover, for many potentials relevant for cosmology, polynomial approximations to the potential are not sufficient (for example, in the context of inflationary physics \cite{Amin:2011hj}). The characteristics of the radiation from oscillons in non-polynomial, flattened potentials was explored numerically in \cite{Salmi:2012ta}. 

Motivated by, and building upon earlier works, in section \ref{sec:soliton_oscillon} and \ref{sec:soliton_case} we provide a semi-analytical calculation of the decay rate of oscillons in 3+1 dimensions. Our technique is applicable to large amplitude oscillons in polynomial and non-polynomial potentials. Crucially, by including the effects of a spacetime dependent effective mass, we are able to capture the decay rates accurately -- an improvement by many orders of magnitude in certain cases compared to earlier techniques \cite{Mukaida:2016hwd,Ibe:2019vyo,Eby:2018ufi}. Our results match well with detailed numerical simulations.

The gravitational effects on oscillon decay rates are expected to depend on the relative magnitude between gravitational and self- interactions. In 3+1 dimensions, the lifetime of small-amplitude \emph{dilute} oscillons (whose self-interactions are negligible) was shown to exceed the present age of the universe \cite{Grandclement:2011wz,Eby:2015hyx, Eby:2020ply}.\footnote{The stability of dilute oscillons is ensured by gravitational attraction, for example, oscillatons \cite{Seidel:1991zh,UrenaLopez:2002gx,Alcubierre:2003sx} and dilute axion stars \cite{Visinelli:2017ooc, Eby:2019ntd}. Their size can be cosmological scales, e.g. $\sim \mathrm{kpc}$ for fuzzy DM \cite{Hui:2016ltb}.} What was not clear to us is whether the existence of gravity stablizes large-amplitude \emph{dense} ones, whose self-interactions are more or at least equally important. We investigate the gravitational impacts on oscillon lifetimes in section \ref{sec:soliton_gravity}.

Apart from scalar fields, nature provides us with many examples of higher spin fields. For instance, $W$ and $Z$ bosons in the SM, or  speculatively, as (some or all of) DM \cite{Graham:2015rva, Co:2018lka, Bastero-Gil:2018uel, Agrawal:2018vin, Kolb:2021xfn, Kolb:2021nob, Babichev:2016bxi, Alexander:2020gmv, ParticleDataGroup:2020ssz}. In section \ref{sec:soliton_vector}, we study oscillons in real-valued massive vector fields with attractive self-interactions. These spatially localized objects could be maximally polarized (with respect to a particular direction), i.e. either the vector field configuration is primarily linearly polarized which we call a directional oscillon, or it is mostly circularly polarized that we refer to as a spinning oscillon (see figure \ref{fig:vectoroscillon} for a quick description). 

The spin nature of the vector field, manifest in these oscillons, can lead to novel phenomenological implications. Collisions and mergers of dense vector oscillons can lead to gravitational wave production, which might be distinct from the scalar case ~\cite{Zhou:2013tsa, Palenzuela:2017kcg, Liu:2017hua, Helfer:2018vtq, Amin:2018xfe, Dietrich:2018jov, Lozanov:2019ylm}. If the massive (dark) vector field kinetically mixes with the visible photon, namely $\cal L \supset (\sin\alpha/2) X^{\mu\nu} F_{\mu\nu}$ where $\sin\alpha$ is the mixing parameter and $X_{\mu\nu},F_{\mu\nu}$ are the field strength of dark photons and regular photons \cite{Caputo:2021eaa}, collisions between polarized vector oscillons, or interaction with strong magnetic fields can also lead to specific outgoing radiation patterns based on oscillon polarization (see~\cite{Hertzberg:2020dbk,Levkov:2020txo,Amin:2020vja} for scalar case). If such vector oscillons exist today, and interact with terrestrial experiments \cite{Caldwell:2016dcw, Baryakhtar:2018doz, Chiles:2021gxk, Chen:2021bdr}, detectable signatures that depend on the polarization state of the vector field might be possible. Therefore, oscillons are excellent targets for probing the self-interaction, and potentially the spin nature, of DM.

\section{Scalar oscillons}
\label{sec:soliton_oscillon}
\begin{figure}
	\centering
	\includegraphics[width=0.55\linewidth]{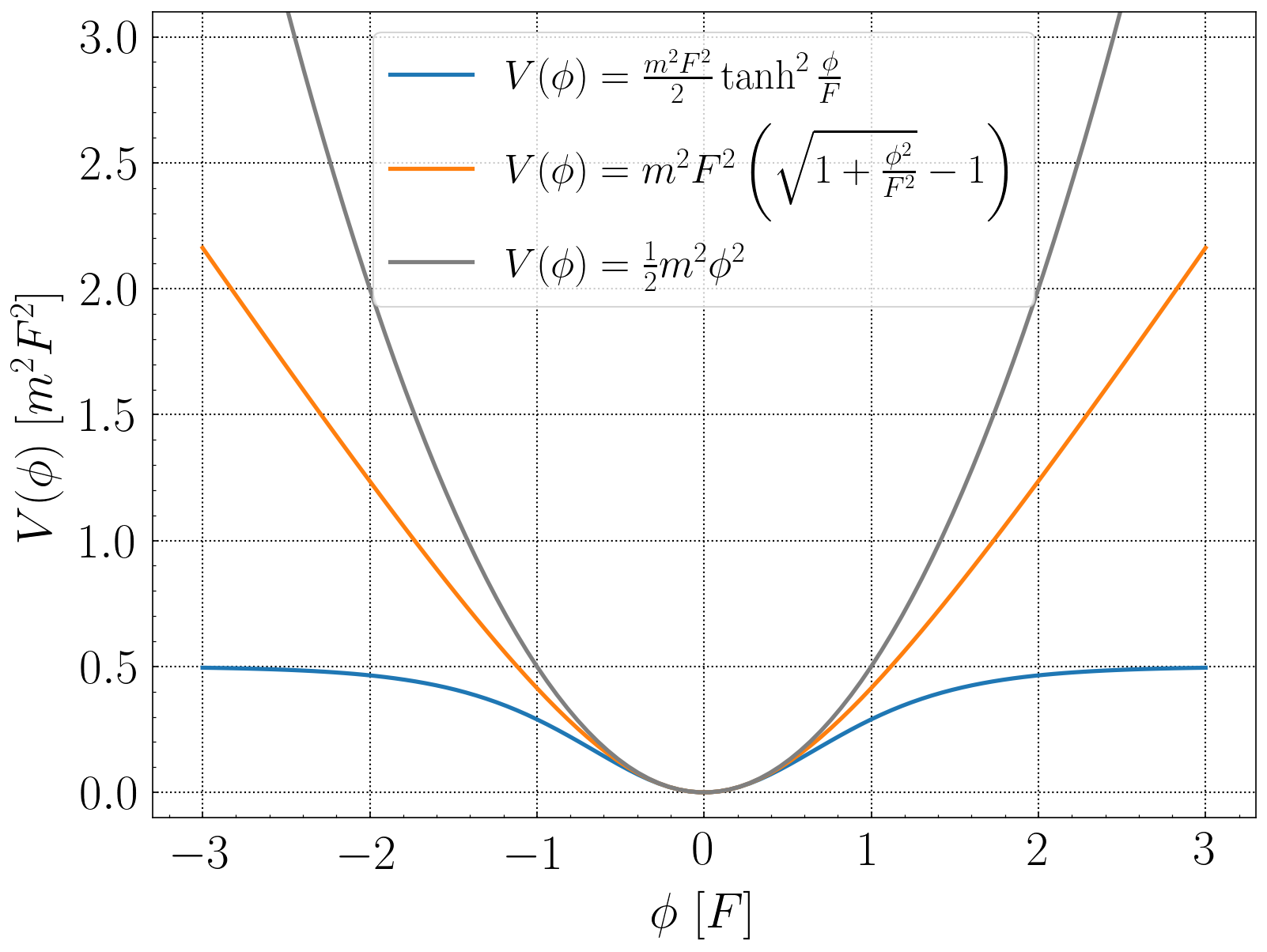}
	\caption{Potential of scalar fields. For the $\alpha$-attractor T-model of inflation and the axion monodromy model, the deviation from a quadratic minimum is significant for $\phi\gtrsim F$.}
	\label{fig:potentials}
\end{figure}

To understand the basic properties of oscillons, consider a real-valued scalar field with the lagrangian given by
\begin{align}
	\cal L = -\frac{1}{2} g^{\mu\nu} \pd_\mu\f \pd_\nu\f - \frac{1}{2}m^2\f^2 - V_\mathrm{nl}(\f) ~,
\end{align}
where $g_{\mu\nu}=\Diag(-1,1,1,1)$ is the Minkowski metric and $V_\mathrm{nl}(\f)$ is the nonlinear part of the potential $V(\f)$. In this section, we will study two well-motivated examples, the $\alpha$-attractor T-model of inflation \cite{Kallosh:2013hoa,Lozanov:2017hjm} and the axion monodromy model \cite{Silverstein:2008sg,Amin:2011hj,McAllister:2014mpa} 
\begin{align}
	\label{potentials}
	V(\f) = \frac{m^2F^2}{2}\tanh^2 \frac{\f}{F} \sep V(\f) = m^2F^2\( \sqrt{1+ \frac{\phi^2}{F^2} } - 1 \) ~,
\end{align}
where $F$ is the amplitude scale that indicates a significant deviation from a quadratic minimum, see figure \ref{fig:potentials}.

\begin{figure}
	\centering
	\includegraphics[width=0.55\linewidth]{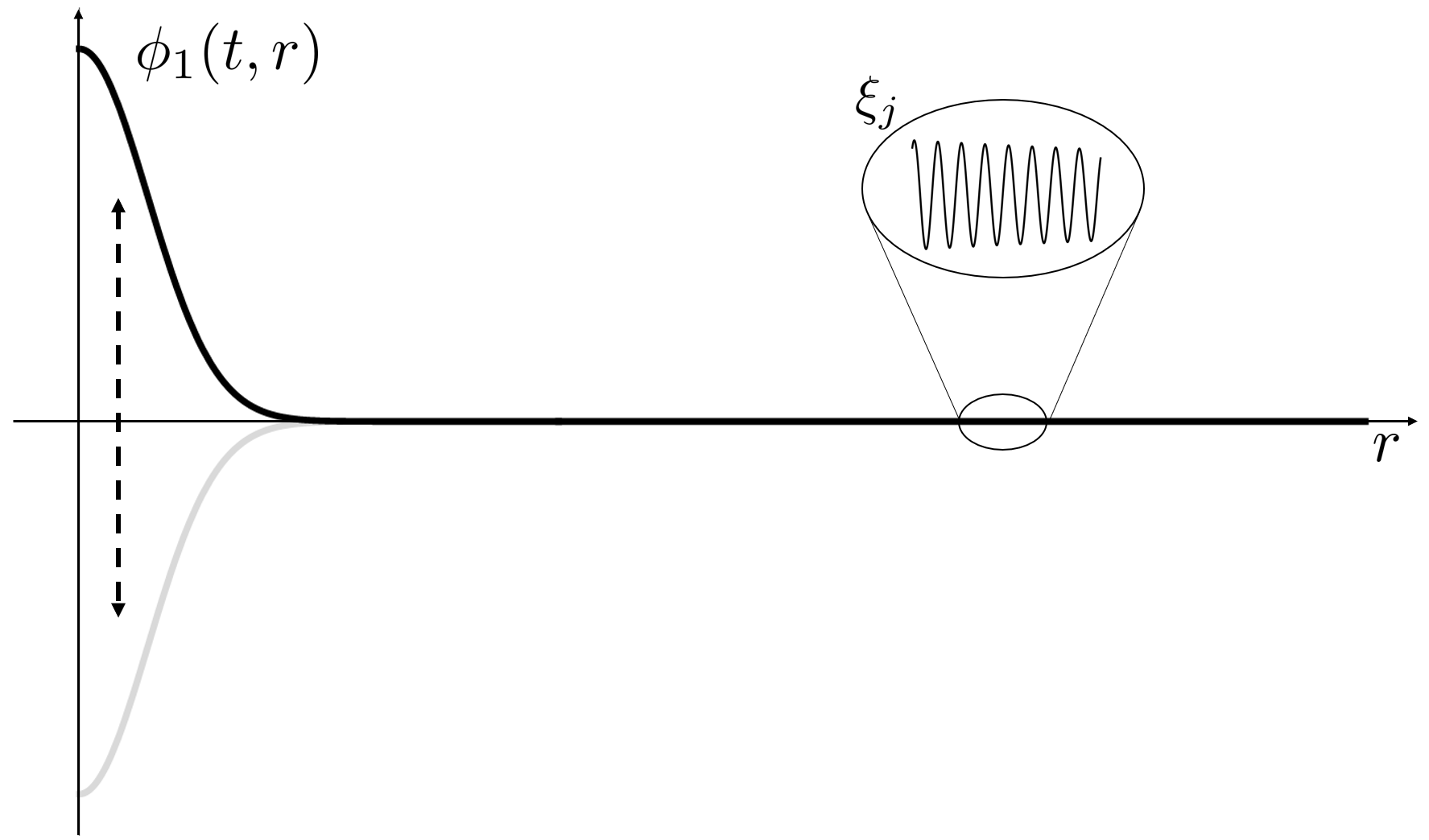}
	\caption{A schematic plot of an oscillon: a spatially-localized, oscillating field configuration and its small radiation tail.}
	\label{fig:oscillon}
\end{figure}

The equation of motion is the Klein-Gordon equation
\begin{align}
	\label{eq:KG}
	\[ -\pd_t^2 + \nabla^2 - m^2 \] \f - V_\mathrm{nl}'(\f) = 0 ~,
\end{align}
where $\nabla^2\equiv \pd_r^2+(2/r)\pd_r$. For simplicity we will only consider symmetric potentials, but the method developed in this paper should also be applicable to asymmetric ones. As suggested in \cite{Seidel:1991zh}, we approximate oscillons by a cosine series
\begin{align}\label{profile_single_freq}
	\f(t,r) = \f_\mathrm{osc}(t,r) + \xi(t,r) = \phi_1(r)\cos(\omega t) + \sum_{j=3}^{\infty}\xi_j(r) \cos(j\omega t) ~,
\end{align}
where $j$ is odd, $\f_\mathrm{osc}$ is a single-frequency profile and $\xi(t,r)$ includes all the radiating modes. See figure \ref{fig:oscillon} for a schematic plot of oscillons. The $\xi_j(r)\cos(j\omega t)$ mode is a radiating mode if $j\omega>m$. Notice that this expansion is actually a balance between ingoing and outgoing waves, and we must manually ignore the ingoing contributions in the end. Typically $|\xi|\ll|\f_\mathrm{osc}|$ inside oscillons.

As a result, the potential and its derivatives can also be written in terms of the Fourier cosine series
\begin{align}
    \label{V_expansion}
	U&\equiv V_\mathrm{nl}(\f_\mathrm{osc}) = \frac{1}{2}U_0(r) + \sum_{j=2}^{\infty} U_j(r)\cos(j\omega t) \sep U_j = \frac{\omega}{\pi} \int_{-\frac{\pi}{\omega}}^{\frac{\pi}{\omega}} V_\mathrm{nl}(\f_\mathrm{osc}) \cos(j\omega t) dt ~,\\
	\label{M_expansion}
	M &\equiv V_\mathrm{nl}''(\f_\mathrm{osc}) = \frac{1}{2}M_0(r) + \sum_{j=2}^{\infty} M_j(r)\cos(j\omega t) \sep M_j = \frac{\omega}{\pi} \int_{\frac{\pi}{\omega}}^{\frac{\pi}{\omega}} V_\mathrm{nl}''(\f_\mathrm{osc}) \cos(j\omega t) dt ~.
\end{align}
where $j$ is even, and 
\begin{align}
	\label{J_expansion}
	J\equiv V_\mathrm{nl}'(\f_\mathrm{osc}) = \sum_{j=1}^{\infty} J_j(r)\cos(j\omega t) &\quad\text{and}\quad J_j = \frac{\omega}{\pi} \int_{-\frac{\pi}{\omega}}^{\frac{\pi}{\omega}} V_\mathrm{nl}'(\f_\mathrm{osc}) \cos(j\omega t) dt ~,
\end{align}
where $j$ is odd. The $U_j$ is a functional of $\f_1$ hence a function of $r$, namely $U_j(\f_1)\equiv U_j(r)$. We will mix the notation $U_j(\f_1)$ and $U_j(r)$, and similarly for $J_j$ and $M_j$. For polynomial potentials, it is possible to find analytical expressions for $U_j,M_j$ and $J_j$ \cite{Zhang:2020ntm}. 

\subsection{Spatial profiles}
\label{sec:soliton_oscillon_profile}
Plugging the single-frequency profile $\f_\mathrm{osc}$ into the Klein-Gordon equation and collecting the coefficient of $\cos(\omega t)$, we obtain the radial profile equation
\begin{align}\label{Minkowski_radial_eq}
	(\nabla^2+\kappa_1^2) \phi_1(r) = J_1(r) ~,
\end{align}
where $\kappa_j^2\equiv (j\omega)^2-m^2$ is the square of the momentum for a particular mode and we have used $J_1(\f_1) = U_0'(\f_1)$. Oscillon profiles can be found by using the numerical shooting method and by demanding a localized, smooth and no-node solution for each $\omega$. Without loss of generality, we assume that the minimum of $V(\f)$ is located at $\f=0$ thus the boundary condition is $\phi_{1}(\infty)= 0$.

A few words regarding the single frequency assumption are in order. We assumed that a good approximation to $\phi_\rm{osc}$ is provided by the single frequency solution as shown in \eqref{profile_single_freq}. More generally, $\phi_\rm{osc}(t,r)= \sum_{j=0}^{\infty}\f_j(r)\cos(j\omega t+\theta_j)$ can be used. For small amplitude oscillons, the profiles can be solved order by order (see chapter \ref{sec:eft} and \cite{Fodor:2009kf}). For the models (and large amplitudes) considered here, we have checked numerically that the Fourier transform of oscillons in the temporal domain show a rich structure in other frequencies (also see \cite{Salmi:2012ta}), including frequencies other than multiples of $\omega$. Nevertheless, there is typically a single dominant frequency, and the single frequency solution remains a good approximation up to moderately large field amplitudes. We use this check to justify our single frequency approximation. The dropping of higher harmonics, however, might have consequences in the form of a somewhat larger than expected amplitude for the radiation modes $\xi$.\footnote{Heuristically, this could be because we have transferred these higher multipoles to the radiation sector in our calculation.} Also recall that even with the single-frequency assumption, we have ignored contributions from $\xi$ in the profile equation \eqref{Minkowski_radial_eq}. The assumption of a single-frequency profile becomes invalid when $\omega\ll m$, however we never consider the $\omega\ll m$ limit because: (1) The particles that make up the oscillons in this regime are relativistic as we will see in section \ref{sec:soliton_oscillon_virial}. In this case, a large number of particles can easily pop in and out of the condensate and we are unlikely to have a stable long-lived condensate. (2) The radiating modes, and hence decay rates, are typically too large for the oscillon to maintain a stable configuration and (3) The size of the oscillon approaches the Schwarzschild radius and the nonlinearity of gravity becomes important, which is beyond the scope of this section.

\begin{figure}
	\centering
	\includegraphics[width=0.55\textwidth]{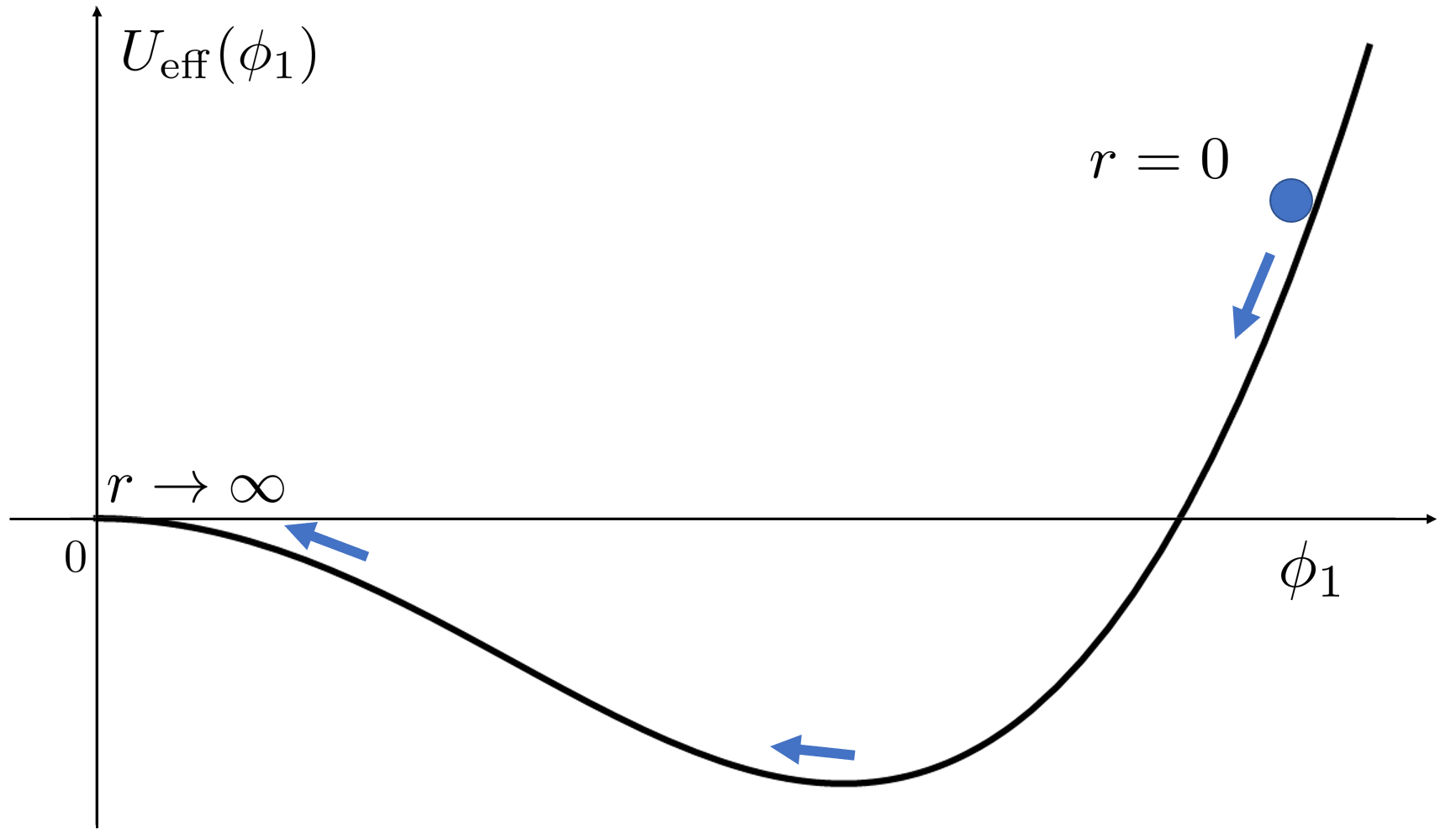} 
	\caption{The shape of the effective potential $U_{\textrm{eff}}(\f_1)$ for $V(\phi)$ potentials that open up away from the minimum. The profile solution can be obtained by thinking about $\f_1(r)$ as the spatial coordinate of a ball rolling down the $U_{\textrm{eff}}(\f_1)$ in the presence of friction $(2/r)\partial_r\f_1(r)$. Note that $r$ plays the role of the time coordinate.}
	\label{fig:Ueff}
\end{figure}

Not all forms of $U_0(\f_1)$ and $\omega$ allow spatially localized, nodeless solutions. The necessary condition for such solutions to exist are:
\begin{align}
	\frac{2 U_0(\f_1)}{\f_1^2}>m^2-\omega^2>0\,.
\end{align}
To understand why $U_0(\f_1)/\f_1 >(m^2-\omega^2)/2$ is necessary, note that \eqref{Minkowski_radial_eq} can be regarded as an equation of motion for a rolling ball with $r$ playing the role of a time variable, see figure \ref{fig:Ueff}. Explicitly,
\begin{align}
	\label{eq:Ueff}
	\partial_r^2\f_1+\frac{2}{r}\partial_r\f_1=-U_{\textrm{eff}}'(\f_1)
\end{align}
where 
\begin{align}
	U_\textrm{eff}(\f_1)\equiv -\frac{1}{2}(m^2-\omega^2)\f_1^2 + U_0(\f_1)
\end{align}
We think of $(2/r) \partial_r\f_1$ as a friction term. We would like a monotonic solution with $\f_1|_{r= 0}\ne 0$, $\partial_r\f_1|_{r=0}=0$ and $\f_1,\partial_r\f_1|_{r\rightarrow\infty} \rightarrow 0$. If $U_0(0)=0$, then the final effective energy at $r\rightarrow \infty$ is $0$.  Now, since there is friction in the system, we must have the ``initial energy" at $r=0$ satisfy $U_{\textrm{eff}}(\f_1(r=0))>0$. That is, for some $\f_1\ne 0$, we must have ${U_0(\f_1)}>(m^2-\omega^2)\f_1^2/2$. To understand the second inequality, $m^2-\omega^2>0$, note that for large $r$, \eqref{eq:Ueff} has solutions of the form $\f_1|_{r\rightarrow\infty}\sim  e^{\pm i\sqrt{\omega^2-m^2}r}/r$ (assuming we can ignore $U_0$ at large $r$ since $\f_1$ will be small). Hence, if we want localized solutions, we need $m^2>\omega^2$.

Once the profile is found, the energy of oscillons can be obtained by time averaging the energy density over a period, and is given by
\begin{align} \label{Minkowski_energy}
	E_{\mathrm{osc}} = \int_0^\infty \[ \frac{1}{4}\(\pd_r\phi_1 \)^2 + \frac{1}{4}(\omega^2 + m^2)\phi_1^2 + \frac{1}{2}U_0  \] 4\pi r^2dr~.
\end{align}
For these single frequency objects, we can define the particle number \cite{Mukaida:2016hwd}
\begin{align}
	N_\mathrm{osc} = \frac{\omega}{2}\int_0^\infty \f_1^2 ~4\pi r^2 dr ~.
\end{align}
The stability condition of oscillons against small perturbations is given by \cite{Friedberg:1976me}
\begin{align}\label{stability_condition2}
	\frac{dN_\mathrm{osc}}{d\omega} < 0 \quad\text{or}\quad \frac{dE_\mathrm{osc}}{d\omega} < 0~,
\end{align}
and the critical frequency $\omega_\mathrm{crit}$ can be obtained by setting $dE_\mathrm{osc}/d\omega = 0$.

The profile of oscillons is unique for each $\omega$. For $V(\phi)$ that have quadratic minima, and  flatten to shallower than quadratic power laws at larger field values, $U_{\textrm{eff}}(\f_1)$ will be negative for small $\f_1$ and positive and monotonic for large $\f_1$ (for any $\omega<m$). Following our ball-rolling on a hill analogy with $r$ as the time coordinate, it is clear that there will be a unique initial condition $\f_1|_{r= 0}\ne 0$, $\partial_r\f_1|_{r=0}=0$, where $U_{\textrm{eff}}(\f_1|_{r=0})>0$ for which $\f_1(r)$ will be localized. We note that for polynomial potentials, $U_{\textrm{eff}}(\f_1)$ can have a positive local maximum at some $\f_1\ne 0$. For the case of the $\phi^6$ potential, such local maxima indicate the existence of ``flat-top" oscillons \cite{Amin:2010jq}.

Multiple studies have shown that oscillons tend to be attractors in the space of solutions (see for example \cite{Andersen:2012wg}). The existence of a unique profile for each $\omega$ allows for some freedom in setting up initial conditions for the profiles numerically. Once an approximate initial profile is set up at some sufficiently small $\omega<m$, the oscillons radiate energy quickly, and latch on to an oscillon configuration. This oscillon configuration then adiabatically passes through a unique set of subsequent oscillon configurations with slowly increasing $\omega$. The configurations continue to evolve adiabatically, emitting a small amount of radiation, until they collapse at $\omega_{\textrm{crit}}$. 

\subsection{Decay rates}
\label{sec:soliton_oscillon_decayrate}
Radial equations for radiating modes can be obtained by plugging the full expansion \eqref{profile_single_freq} into the Klein-Gordon equation and collecting the coefficient of $\cos(j\omega t)$, i.e.
\begin{align}\label{radial_radiation}
	\[ \nabla^2 + \kappa_j^2 \]\xi_j(r) = S_j(r) \equiv J_j + \frac{1}{2}\sum_{k=3}^{\infty}\xi_k\( M_{|k+j|} + M_{|k-j|} \) ~,
\end{align}
where $j$ and $k$ are both odd, $S_j$ is the effective source and $M_j$ is the effective mass. For $\kappa_{j}^2>0$, the central amplitude of a radiating mode is
\begin{align}\label{xi_initial}
	\xi_j(0) = - \int_{0}^{\infty} dr' ~S_j(r')~r'\cos(\kappa_j r') ~.
\end{align}
This is not in closed form, and we can find the solution of $\xi_j(r)$ by using the \emph{iterative method} developed in \cite{Zhang:2020bec}.\footnote{We will no longer call this method ``shooting'' as we did in \cite{Zhang:2020bec}, because technically we are not solving a boundary value problem.} The energy loss rate of oscillons is determined by radiation at large radius
\begin{align}
	\frac{dE_\mathrm{osc}}{dt} = -4\pi r^2 \t{T}{^1_0} |_{r\rightarrow\infty} = \l. -4\pi r^2 \pd_t\xi\pd_r\xi \r|_{r\rightarrow\infty} ~,
\end{align}
where $\t{T}{^\mu_\nu} = \pd^\mu\f \pd_\nu\f + \delta^\mu_\nu \cal L$ is the energy-momentum tensor and the radiation is given by
\begin{align}\label{xi_inf}
	\l. \xi(t,r) \r|_{r\rightarrow\infty} = -\frac{1}{4\pi r} \sum_{j=3}^\infty \til S_j(\kappa_j) \cos(\kappa_jr - j\omega t) ~.
\end{align}
Here $\til{S}_j(p)$ is the Fourier transform of the effective source
\begin{align}
	\til S_j(p) = \int_0^\infty dr~ 4\pi r^2 \sinc(p r) S(r) ~.
\end{align}
The (absolute value of) decay rate is defined
\begin{align}\label{Gamma_Minkowski}
	\Gamma \equiv \l| \la\dot E_\mathrm{osc}/E_\mathrm{osc} \ra_T \r|  = \frac{1}{8 \pi E_\mathrm{osc}} \sum_{j=3}^{\infty}\left[\tilde{S}_{j}(\kappa_{j})\right]^{2} j\omega~ \kappa_{j} \equiv \sum_{j=3}^{\infty} \Gamma_{j} ~,
\end{align}
where $\Gamma_j$ is the contribution due to $S_j$. Typically the radiating mode safisfies $|\xi_j|\gg|\xi_{j+2}|$, which means only finite terms are needed in the radial equation \eqref{radial_radiation}. However, the leading channel of decay rates $\Gamma_3$ might vanish for some $\omega$, causing a dip structure in $\Gamma-\omega$ plots.

If we start with an oscillon with $\omega<\omega_\rm{dip}$, the frequency of the oscillons evolves to larger values by slowly emitting scalar radiation (and the profile changes correspondingly). Since the leading decay channel $\Gamma_3$ is vanishing in the dip, the oscillon configuration with $\omega\approx \omega_\rm{dip}$ is expected to have a long lifetime. When considering the total lifetime of oscillons that start out at $\omega< \omega_\rm{dip}$, the oscillon will spend most of its lifetime in such a dip. Generally speaking, we will use $\Gamma_3 + \Gamma_{5}$ to estimate the lifetime of oscillons, which is just the area enclosed by the evolution curve in $dt/dE_\mathrm{osc}$ versus $E_\mathrm{osc}$ plot, where $dt/dE_\mathrm{osc}=1/(\Gamma E_\mathrm{osc})$. For future reference, if we keep only $\xi_3$ and $\xi_5$, the effective source becomes
\begin{align}
	\label{S3_Minkowski}
	S_3 &= J_3 + \frac{1}{2}\xi_3 (M_0+M_6) + \frac{1}{2}\xi_5(M_2 + M_8) ~,\\
	\label{S5_Minkowski}
	S_5 &= J_5  + \frac{1}{2}\xi_3(M_2 + M_8) + \frac{1}{2}\xi_5 (M_0+M_{10}) ~.
\end{align}
The generalization to including more $\xi_j$ terms is straightforward. Note that the effective source $S_j$ receives a contribution from the oscillon background $J_j$ as well as corrections due to radiation $\xi_j$.

\subsection{Lattice simulations}
\label{sec:soliton_oscillon_numerics}
We apply the following numerical strategy to verify our analytical results. The main goal is to solve the nonlinear Klein-Gordon equation \eqref{eq:KG} and obtain a decay rate as a function of time. Since at each instant in time we are passing adiabatically through different oscillon configurations (specified by an $\omega (t)$), these results can be directly compared to the analytically obtained decay rates from the previous sections. 

We solve the nonlinear Klein-Gordon equation \eqref{eq:KG} (assuming spherical symmetry) using a Verlet method (a 2nd-order symplectic method) while the spatial derivative is characterized by centered difference. The simulations are performed on a box of size $r_\rm{max}=60m^{-1}$ with $dt=dr/5=0.005 m^{-1}$. We have checked that changing the box size or the spatial/temporal step size does not change our results qualitatively. The size of the box is much larger than the typical width of the oscillon profile which is $<\mathcal{O}[10]\,m^{-1}$. 

At the boundary $r\rightarrow r_\rm{max}$ we impose the absorbing boundary condition \cite{Salmi:2012ta}, i.e. 
\begin{align}
	\partial_{t}^{2} \phi+\partial_{t} \partial_{r} \phi+\frac{1}{r} \partial_{t} \phi+\frac{1}{2} m^{2} \phi=0 ~,
\end{align}
which uses a backward-in-time, and centered-in-space discretization. This boundary condition provides an alternative approach to remove the dispersive waves from the lattice, requiring no extra lattice sites for its operation.

We begin with spatial profile $\phi(t,r)|_{t=0}$ and $\partial_t\phi(t,r)|_{t=0}$ which is smooth at the origin $r=0$. After picking an $\omega$, we find the profile using the shooting algorithm discussed in section \ref{sec:soliton_oscillon_profile}. Once set up in this way, the system evolves primarily via radiation of scalar modes which are approximately removed at $r=r_\rm{max}$. Typically the characteristic $\omega$ of the solution (near $r=0$) increases with time, and the oscillon undergoes an adiabatic evolution, passing through many oscillon configurations with increasing $\omega$. The frequency $\omega$ is measured using the interval between the field maxima at $r=0$. 

The decay rate of the oscillons is numerically calculated using
\begin{align}
	\Gamma(t) = \frac{1}{T_\rm{ave}} \int_{t-T_\rm{ave}/2}^{t+T_\rm{ave}/2} \frac{1}{E_\rm{osc}(t')}\frac{dE_\rm{osc}(t')}{dt'} dt' ~,
\end{align}
where we use $T_\rm{ave}=200m^{-1}$ for convenience. The time-dependent, but slowly decreasing energy $E_\rm{osc}(t)$ of the oscillon is calculated using
\begin{align}
	E_\rm{osc}(t)=\int_0^{r_{\max}/2} dr\, 4\pi r^2 \left[\frac{1}{2} (\partial_t\phi)^2 + \frac{1}{2}(\pd_r\phi)^2 + \frac{1}{2}m^2\phi^2 + V_\rm{nl}(\phi_\rm{osc})\right]\,.
\end{align}
Our choice of bounding radius $r_\rm{max}/2$ is arbitrary. However, as long as the bounding radius is $
\gtrsim\mathcal{O}[10]m^{-1}$, the decay rate is approximately independent of this choice.

We also keep track of the time averaged frequency, central amplitude, energy and decay rates of oscillons obtained by an average over a time period $T_\rm{ave}=200m^{-1}$ unless otherwise stated.

To get a more refined picture of the frequency content of the oscillons and the radiation, we calculate Fourier Transform of the time dependence of the field at $r=0$ and $r=r_\rm{rad}=50 m^{-1}$ respectively. Such Fourier Transforms are calculated over a time interval of $T_\rm{fourier}=5000m^{-1}$. 

We have confirmed that a slight change of parameters ($r_\rm{max},dt,T_\rm{ave},T_\rm{fourier}$) will not affect the results significantly.

\subsection{Virial theorem}
\label{sec:soliton_oscillon_virial}
Assume that oscillons are single-frequency objects like Q-balls and Boson stars \cite{Lee:1991ax}, then one way to derive a virial theorem is to use the variational principle. The Legendre transformation
\begin{align}
	F_\mathrm{osc} = \omega N_\mathrm{osc} - E_\mathrm{osc}
\end{align}
defines a functional of $\f_1$ and a function of $\omega$ (one may recognize that $F_\mathrm{osc}$ is just the lagrangian). The variation of $F_\mathrm{osc}$ in terms of $\f_1$ by keeping $\omega$ fixed gives the profile equation of oscillons \eqref{Minkowski_radial_eq}. A virial theorem can be obtained by considering the variation $\f_1(r)\rightarrow \f_1(\lambda r)$ for an oscillons solution. By setting $\( \partial F_\mathrm{osc}/\pd \lambda \)_\omega = 0$
at $\lambda=1$, we find 
\begin{align}\label{virial_theorem}
	E_\mathrm{S}/3 + E_\mathrm{V} = E_\mathrm{K} ~,
\end{align}
where the surface energy, potential energy and kinetic energy are defined
\begin{align}
	E_\mathrm{S} = \int \frac{1}{4}(\pd_r\f_1)^2 d^3r \sep
	E_\mathrm{V} = \int \( \frac{1}{4}m^2\f_1^2 + \frac{1}{2}U_0\) d^3r \sep
	E_\mathrm{K} = \int \frac{1}{4}\omega^2\f_1^2 ~d^3r ~.
\end{align}

For oscillons with $\omega\lesssim m$, we can identify three small parameters immediately
\begin{align}\label{small_quantity}
	\epsilon_r\equiv 1-\omega^2/m^2 \sep
	\epsilon_V\sim \frac{U_0}{m^2\f_1^2/2} \sep
	\epsilon_\xi \sim \frac{\xi_j}{\f_1} ~.
\end{align}
The parameter $\epsilon_r$ is a measure of how relativistic the particles inside the oscillon are. Equation \eqref{Minkowski_radial_eq} implies that at large radius $\f_1(r) \propto r^{-1} \exp\[ -(m^2-\omega^2)^{1/2}r \]$, hence a typical spatial derivative brings a factor $\sqrt{\epsilon_r} m$, that is, $\pd_r\f_1 \sim -\sqrt{\epsilon_r} m \f_1$, and
\begin{align}
	\nabla^2\f_1\ll \omega^2\f_1 ~.
\end{align}
This means that the particles that make up the oscillon are NR. And from equations \eqref{radial_radiation} and \eqref{virial_theorem}, we see $\epsilon_\xi\sim - \epsilon_V$ and $\epsilon_V\sim -\epsilon_r$.

For oscillons with $\omega \ll m$, we may not regard surface energy as a small quantity anymore. Take the tanh potential in \eqref{potentials} for example and assume a Gaussian profile
\begin{align}
	\f_1(r) = C ~e^{-r^2/R^2} ~,
\end{align}
where $C\gg F$. Then each energy component becomes
\begin{align}
	E_\mathrm{S} \sim C^2 R \sep
	E_\mathrm{K} \sim C^2 R^3 \omega^2 \sim \frac{N_\mathrm{osc}^2}{C^2R^3}\sep
	E_\mathrm{V} \sim R_\mathrm{V}^3 m^2 F^2 ~,
\end{align}
where we have taken advantage of the flatness of the potential at large $\f$ and $R_\mathrm{V}$ is the length scale satisfying $\f_1(R_\mathrm{V})\sim \cal O(1)$, i.e. 
\begin{align}
	R_\mathrm{V}\sim R \log^{1/2} \( \frac{C}{F} \) \sim R ~.
\end{align}
By setting $\pd E_\mathrm{osc}/\pd R=\pd E_\mathrm{osc}/\pd C=0$ and keeping $N_\mathrm{osc}$ fixed, we obtain
\begin{align}
	R\sim \omega^{-1}\sep
	C\sim \omega^{-1} ~.
\end{align}
We see that three components of energy now are all comparable, and thus oscillon particles are relativistic. This phenomenon has been witnessed numerically in the context of dense axion stars \cite{Visinelli:2017ooc,Eby:2019ntd}.

\section{Case study}
\label{sec:soliton_case}
In this subsection, we study the decay rate of oscillons with the field potential \eqref{potentials}. We will compare the prediction of the foregoing analytical framework with results of lattice simulations. 

Analytically, we can calculate the leading decay rate mode $\Gamma_3$ as a good approximation of the total decay rate. However, occasionally $\Gamma_3$ vanishes at some frequency $\omega_\rm{dip}$, and the leading decay rate mode becomes $\Gamma_5$. It is convenient to define
\begin{align}
	\Gamma_{(N)} = \sum_{j=3}^{N} \Gamma_j ~,
\end{align}
where $j$ is an odd integer. The total decay rate is the sum of all components $\Gamma_{(\infty)}$.

On the other hand, we can numerically evolve the nonlinear Klein-Gordon equation \eqref{eq:KG} assuming spherical symmetry and calculate the decay rate as a function of time. This time dependence of the decay rate is translated to an $\omega$ dependence since the solution evolves slowly, and continuously through different oscillon configurations characterized by an adiabatically changing $\omega(t)$. We typically start the calculation by evolving the nonlinear Klein-Gordon equation \eqref{eq:KG} with field configurations corresponding to $\omega$ that are smaller than the ones shown in the upcoming plots. Regardless of the starting points, we always end up on the same $\Gamma-\omega$ trajectory numerically. This is a consequence of oscillons being attractors in the space of solutions, and the fact that there is a unique oscillon profile for each $\omega$.

\subsection{\texorpdfstring{$\alpha$}{alpha}-attractor T-model}
\begin{figure}
	\centering
	\begin{minipage}{0.45\linewidth}
		\includegraphics[width=\linewidth]{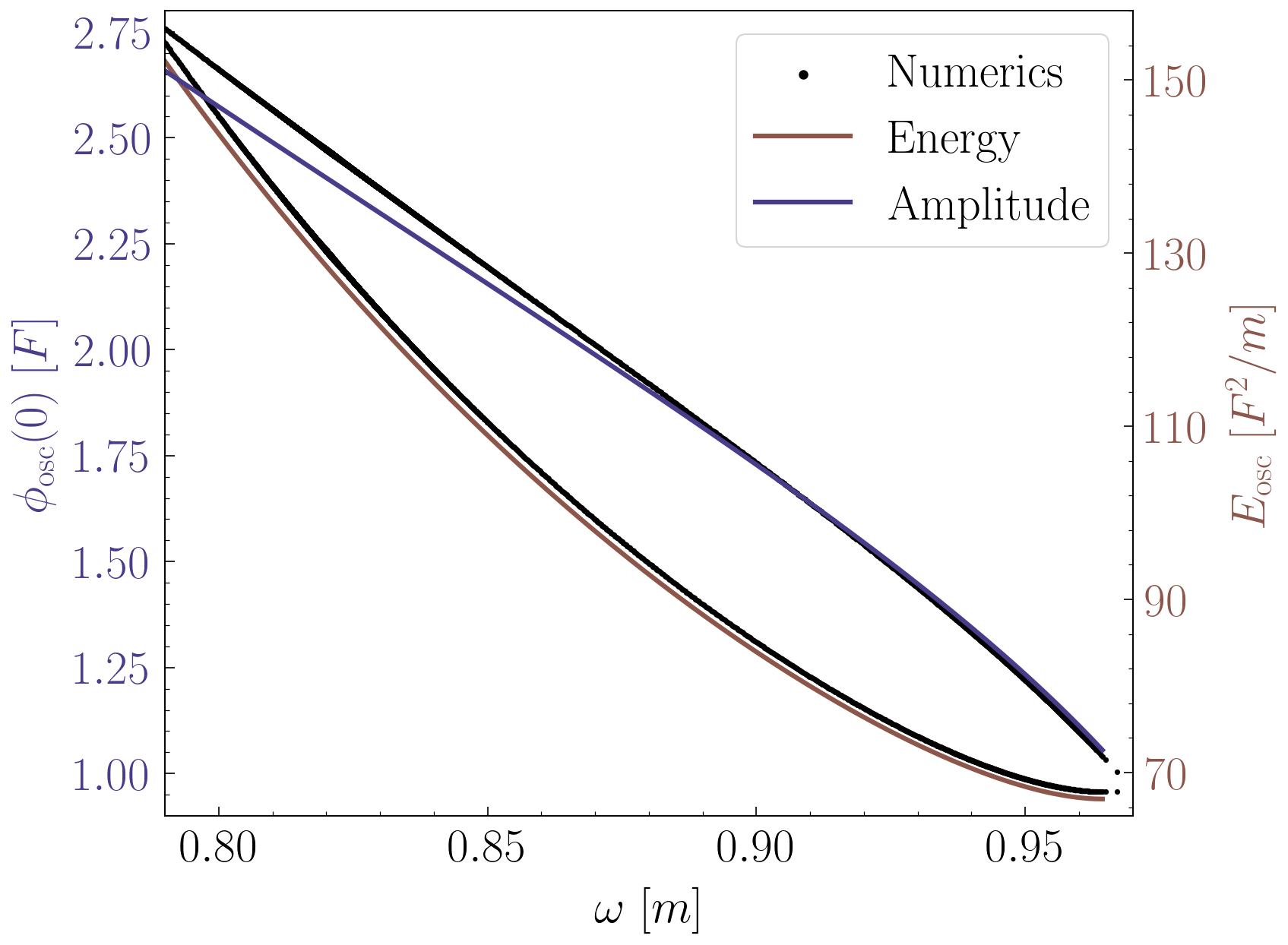}
	\end{minipage}
	\qquad
	\begin{minipage}{0.45\linewidth}
		\includegraphics[width=\linewidth]{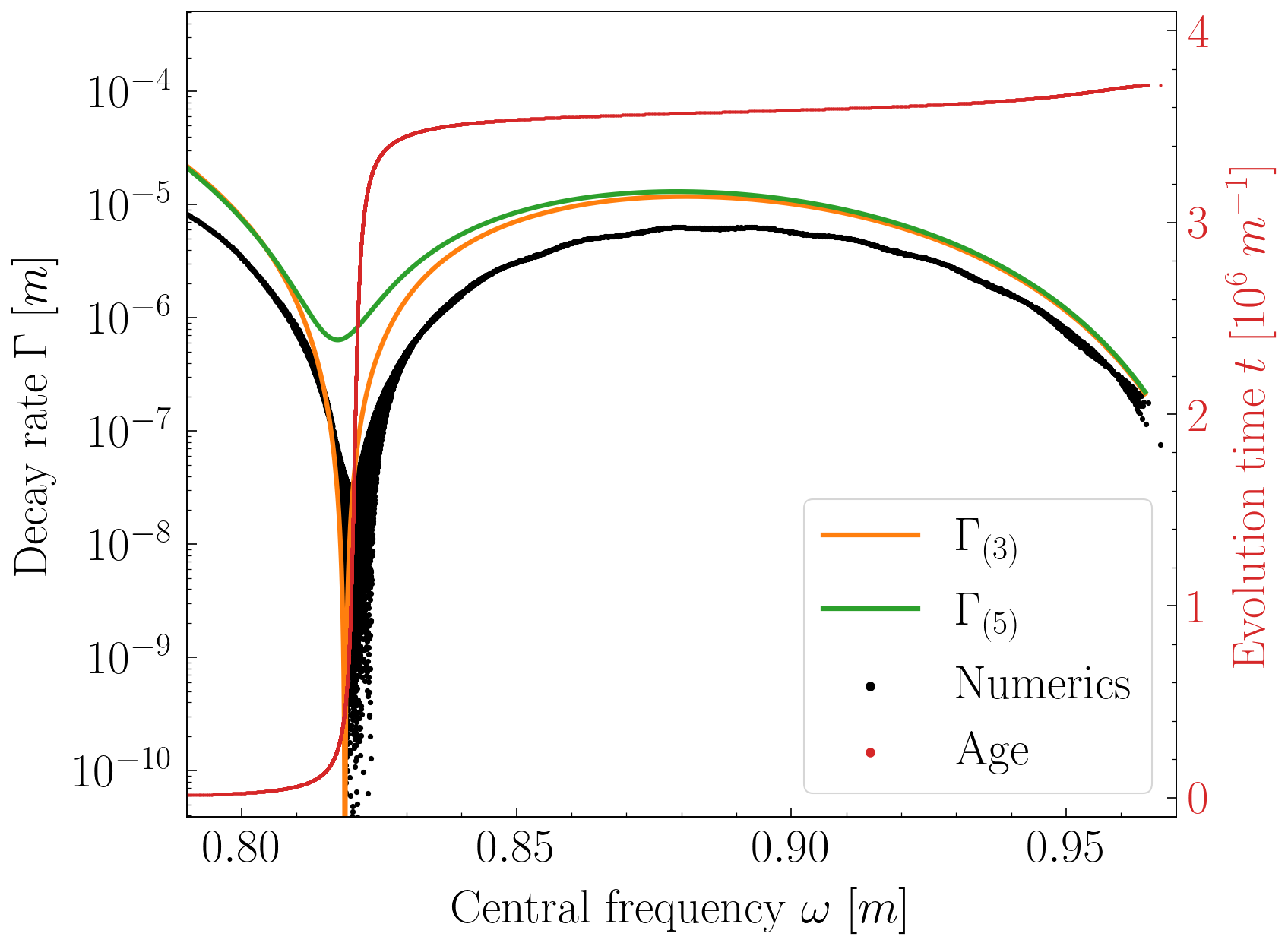}
	\end{minipage}
	\caption{{$V(\phi)=(1/2)m^2F^2\tanh^2(\phi/F)$: {\it Left}: Analytical (colored) and numerical (black) calculations for the oscillon amplitude and energy as a function of the fundamental oscillon frequency $\omega$. The numerical calculation includes time evolution (moving from left to right), whereas the analytical one assumes a stationary configuration for each $\omega$. {\it Right}: Decay rates of oscillons as a function of $\omega$. Black dots show the numerical evolution of the decay rate. With time flowing from left to right, the oscillons disappear quickly after $\omega_\rm{crit}\approx 0.964m$, where $\omega_\rm{crit}$ is defined in \eqref{stability_condition2}. The orange and green curves show the analytic expectation for the decay rate at each quasi-stable oscillon configuration. The orange curve includes the $3\omega$ radiation contribution, whereas the green includes the $3\omega$ and $5\omega$ modes. An accurate prediction of the decay rate, including the dip where the $3\omega$ radiation is vanishing, is correctly provided by our calculations. Finally, the red line shows that the oscillon spends most of its lifetime near the dip at $\omega_\rm{dip}\approx 0.82m$.}}
	\label{fig:tanhfreqenergyamp}
\end{figure}

Let us first consider the $\alpha$-attractor T-model of inflation \cite{Kallosh:2013hoa,Lozanov:2017hjm} with a potential
\begin{align}
	V(\f) = \frac{m^2F^2}{2}\tanh^2 \frac{\f}{F} ~.
\end{align}
The numerical and analytical results for the field amplitude, energy and decay rate as a function of $\omega$ are presented in figure \ref{fig:tanhfreqenergyamp}.

\paragraph{Amplitude and energy:} In the left panel of figure \ref{fig:tanhfreqenergyamp}, we show the central amplitude and total energy of the oscillon configurations as a function of $\omega$. Note that the amplitudes $\f_1(r=0)/F\gtrsim \mathcal{O}[1]$. The upper-limit of the frequency corresponds to $\omega_\rm{crit}$, above which the oscillons are unstable against long-wavelength perturbations. The black dots indicate the numerically obtained energies and amplitudes as the configurations evolve from low to high $\omega$. The agreement between the colored lines (analytic) and the black dots (numerical) indicates that our single frequency ansatz works reasonably well in the range displayed -- conservatively, it is consistent with the numerical solutions at a few~\% level.

\paragraph{Decay rate:} In the right panel of figure \ref{fig:tanhfreqenergyamp}, we show the numerically calculated decay rate (black dots) as the oscillon evolves with time (from low to high $\omega$) until its eventual demise at $\omega=\omega_\rm{crit}$ at the right edge of the panel. Notice the significant ``dip" in decay rate around $\omega_\rm{dip}\approx0.82m$. The solid red line shows that most of the lifetime of the oscillons is spent in the dip. We compare these numerically obtained results with the analytic expectation of our calculations. Note that $\Gamma_{(3)}$ (orange curve), where radiation modes with frequency $3\omega$ were included, beautifully captures the location of the dip in $\Gamma$ as a function of $\omega$. In particular, $\tilde S_3(\kappa_3)=0$ and hence $\Gamma_{(3)}=\Gamma_3=0$ at $\omega_\rm{dip}\approx0.82m$. The $\Gamma_{(5)}=\Gamma_3+\Gamma_5$ calculation (green) barely corrects the $\Gamma_{(3)}$ anywhere, except in the dip, making the decay rate small but finite there. This is to be expected. As we discussed in section \ref{sec:soliton_oscillon_decayrate}, we expect $[\tilde S_3(\kappa_3)]^2\gg [\tilde S_5(\kappa_5)]^2$, except when $[\tilde S_3(\kappa_3)]^2$ vanishes. 

\begin{figure}
	\centering
	\begin{minipage}{0.45\linewidth}
		\includegraphics[width=\linewidth]{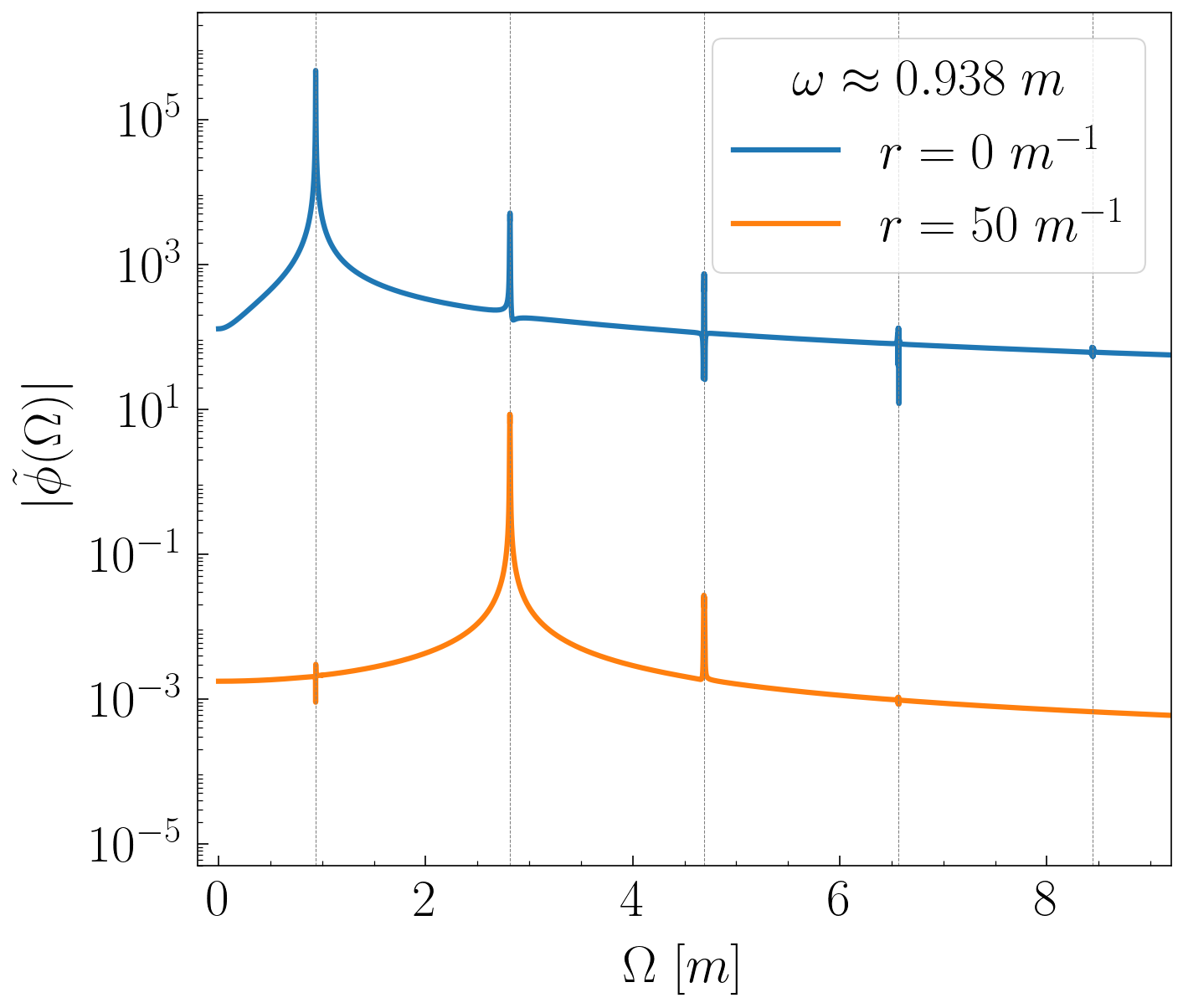}
	\end{minipage}
	\qquad
	\begin{minipage}{0.45\linewidth}
		\includegraphics[width=\linewidth]{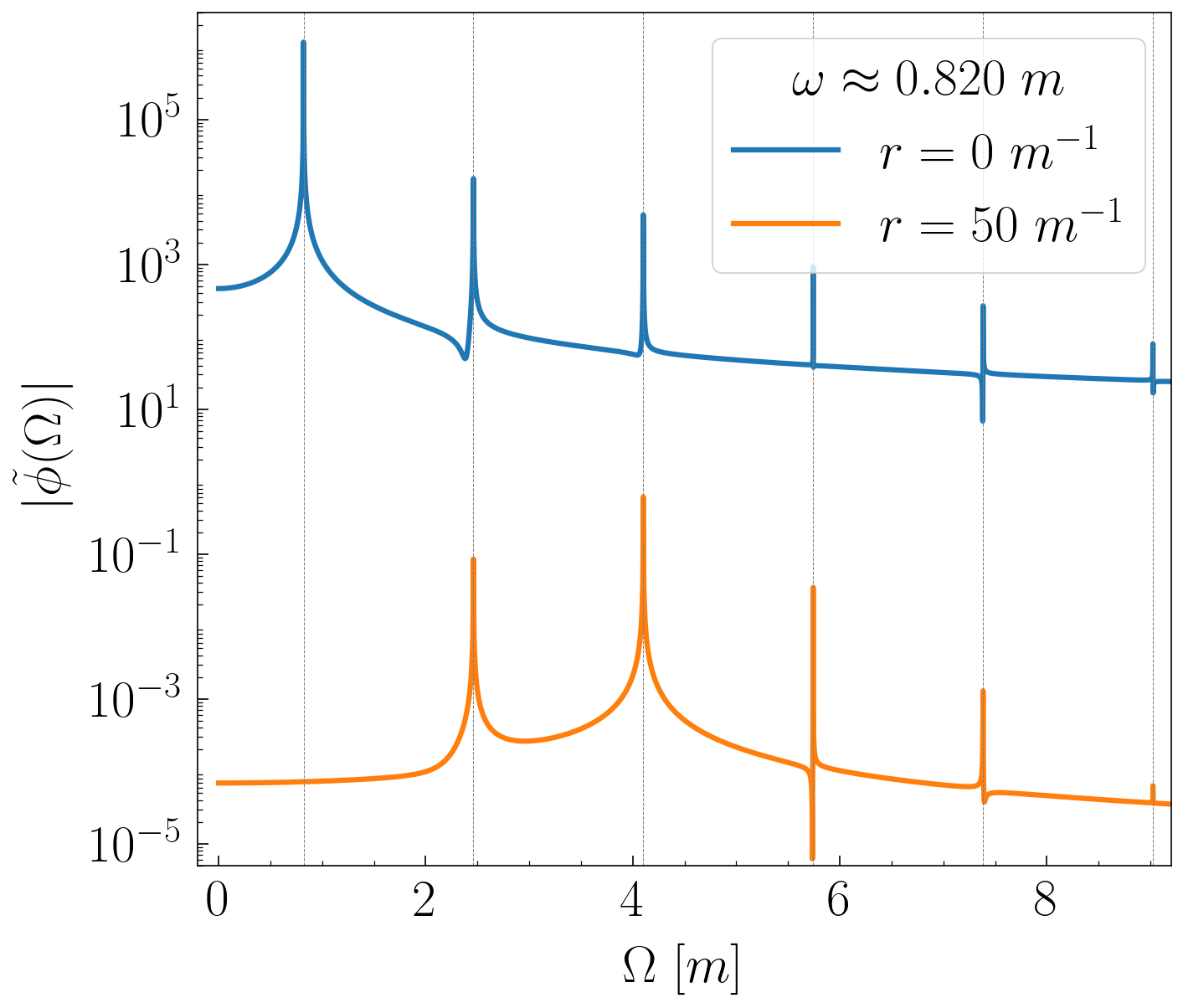}
	\end{minipage}
	\caption{Fourier analysis of the field amplitude at the centre of the oscillon $r=0m^{-1}$ (blue) and far from the center where radiation dominates (orange) [units are arbitrary on the vertical axes]. For both panels, note that the frequency content of the oscillon is dominated by a single fundamental frequency $\omega$, although higher harmonics of $\omega$ are present (blue curves). For the left panel, we have chosen $\omega=0.938m$. In this case the radiation content (orange) is dominated by the $3\omega$ mode as expected, with subdominant content in higher multiples of $\omega$. In contrast, we chose $\omega=\omega_\rm{dip}\approx 0.82m$ for the right panel which is the location of the dip in the decay rate in figure \ref{fig:tanhfreqenergyamp}. As expected, in this case, the $3\omega$ mode is subdominant in the radiation, with the $5\omega$ mode determining the decay rate. These plots provide a verification of our underlying assumptions and confirm the results of our analytic calculation.}
	\label{fig:tanhrootfourieranalysis}
\end{figure}

\paragraph{Frequency content:} It is useful to calculate frequency content of the oscillon as well as the radiation -- this calculation allows us to verify some of the assumptions inherent in our analytic calculation. We take the Fourier transform of $\phi(t,r=0)$ and $\phi(t,r=r_\rm{rad})$. We provide these Fourier transforms for $\omega=0.938m$ as well as $\omega=\omega_\rm{dip}\approx 0.82m$ in figure \ref{fig:tanhrootfourieranalysis}. Consistent with our assumptions, note that a single frequency does dominate the profile near the origin (blue). Similarly, for the radiation (orange), there is a clear hierarchy of power in multiples of $\omega$. This hierarchy is broken at the dip for the $\omega_\rm{dip}\approx 0.82m$ case, with $5\omega$ contribution becoming larger than the $3\omega$ one. We caution that given the finite window for the Fourier transform and numerical uncertainties, the absolute amplitude of the peaks are not quite robust, however, the trends can be trusted.

\subsection{Axion monodromy model}
\begin{figure}[t]
	\centering
	\begin{minipage}{0.45\linewidth}
		\includegraphics[width=\linewidth]{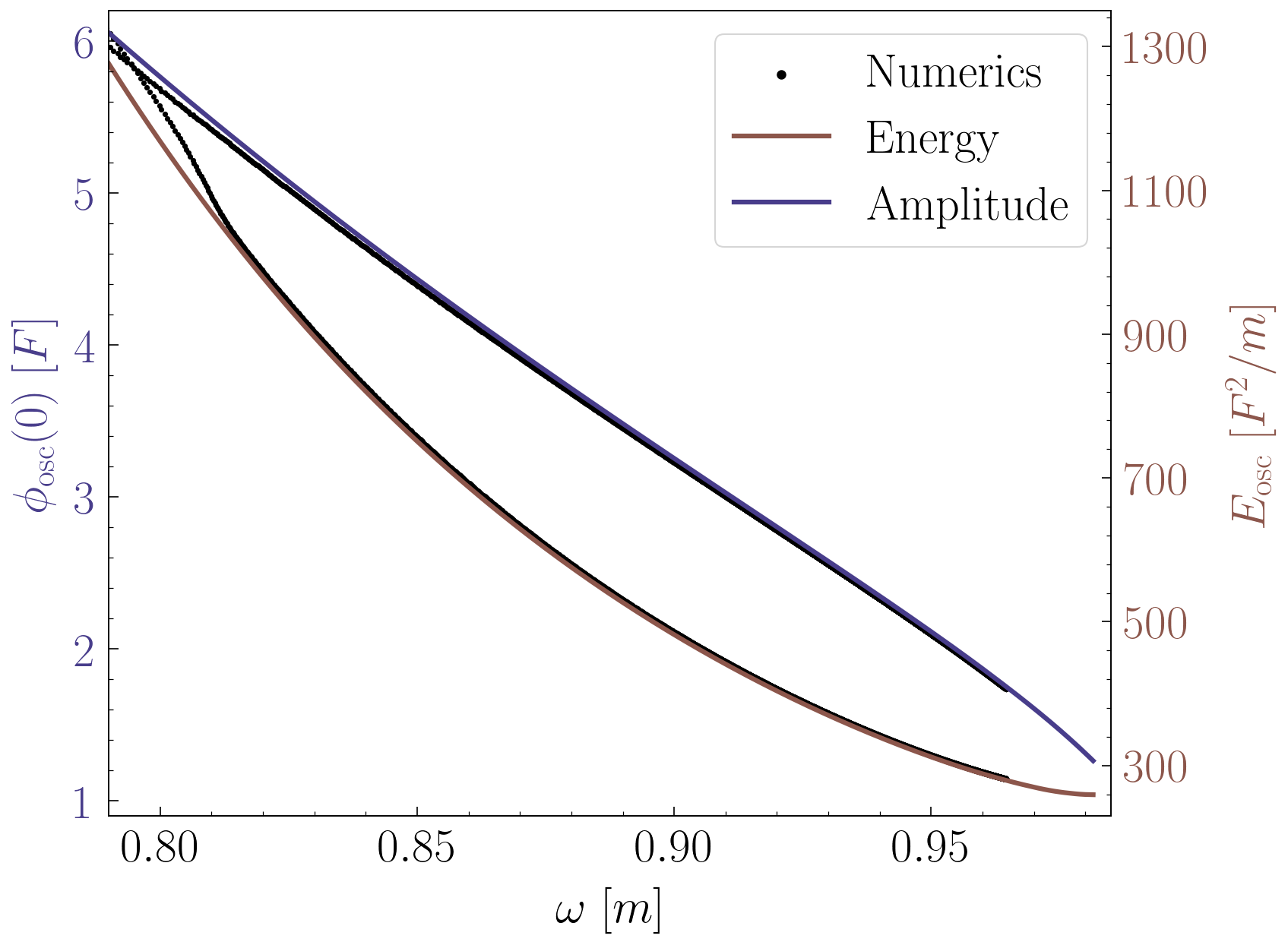}
	\end{minipage}
	\qquad
	\begin{minipage}{0.45\linewidth}
		\includegraphics[width=\linewidth]{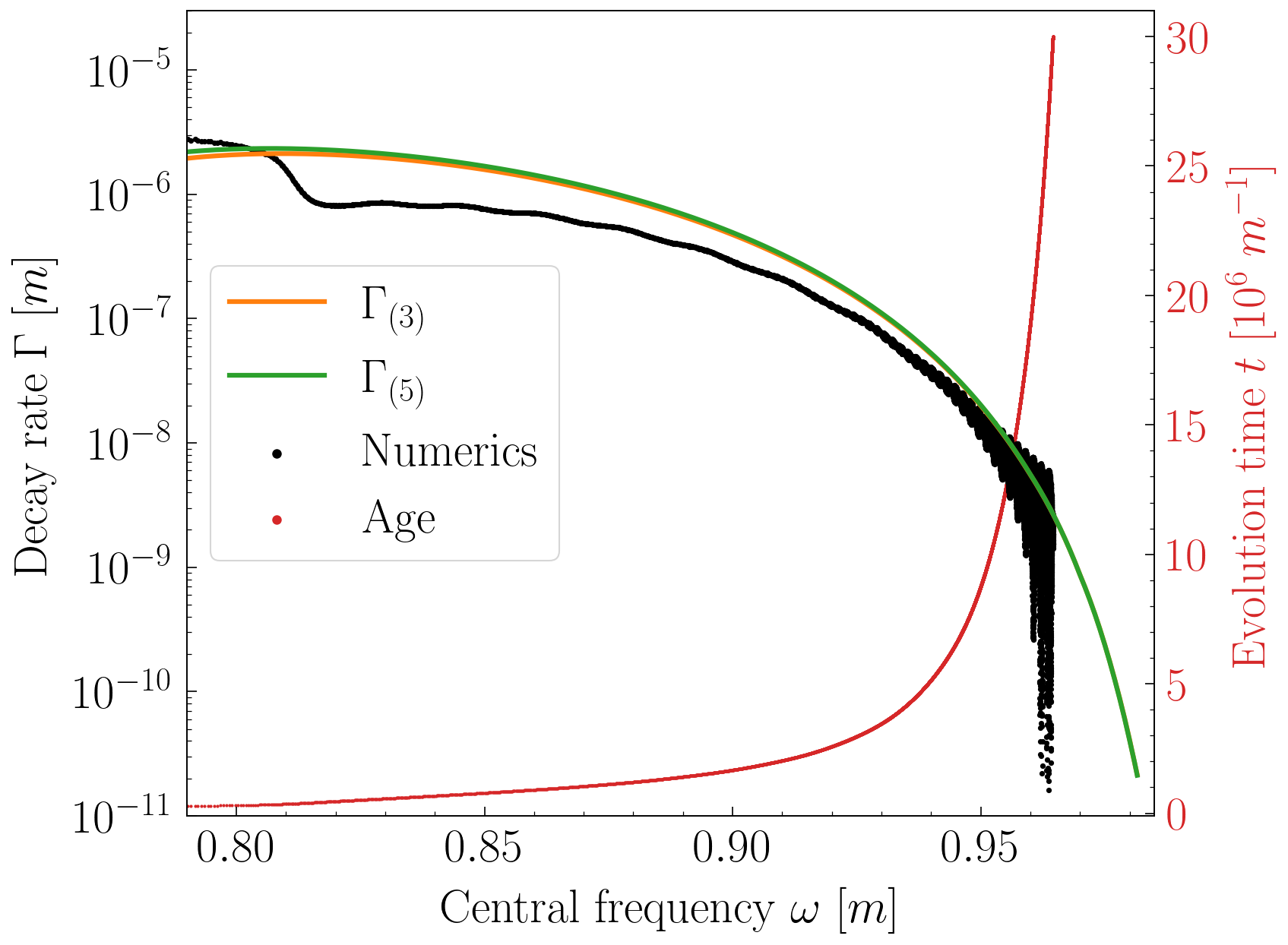}
	\end{minipage}
	\caption{$V(\phi)=m^2F^2\left[\sqrt{1+\phi^2/F^2}-1\right]$: For general description, see the caption of figure \ref{fig:tanhfreqenergyamp}. Once again, the analytics and numerics agree quite well. In this case the behavior of the decay rate is monotonic. Note that most of the oscillon's lifetime arises from the configuration with a frequency close to the $\omega_\rm{crit}\approx 0.982m$, with a lifetime that is longer than $3\times 10^{7}\,m^{-1}$ (potentially much longer). For all cases where the behavior of the decay rate is monotonic up to the critical frequency, we expect that the oscillon tends to spend most of its lifetime near the critical frequency. }
	\label{fig:squarerootfreqenergyamp}
\end{figure}

Now let us consider the axion monodromy model \cite{Silverstein:2008sg,Amin:2011hj,McAllister:2014mpa} 
\begin{align}
	V(\f) = m^2F^2\( \sqrt{1+ \frac{\phi^2}{F^2} } - 1 \) ~.
\end{align}
In figure \ref{fig:squarerootfreqenergyamp}, we show the comparison between the analytical and numerical results for this potential. Apart from the excellent match between theory and numerics, it is worth noting that the numerics do not show any non-monotonic behavior in the decay rate. Our analytics agree with this behavior (green and orange curves). Note that we did not simulate the eventual demise of these oscillons. Their lifetime is longer than $10^{8}m^{-1}$ \cite{Amin:2011hj,Olle:2019kbo}.

\section{Gravitational effects on oscillon lifetimes}
\label{sec:soliton_gravity}
So far our analysis does not include gravity. In this section, the scalar field is assumed to be minimally coupled to gravity. We introduce a lagrangian mechanism that can convert the Hilbert action into one that contains only the first derivative of the metric in section \ref{sec:soliton_gravity_lagrangian}, then we use it to generalize the definition of energy and mass into curved spacetime in section \ref{sec:soliton_gravity_mass}. In section \ref{sec:soliton_gravity_linear}, we study oscillons with linearized gravity. The results for the $\alpha$-attractor T-model of inflation and the axion monodromy model are presented in section \ref{sec:soliton_gravity_case}.

\subsection{Lagrangian mechanism}
\label{sec:soliton_gravity_lagrangian}
The simplest choice of metric to describe oscillons is the spherical coordinates
\begin{align}\label{spherical_metric}
	ds^2 = - e^{2\Phi(t,r)} d t^2 + e^{-2\Psi(t,r)} d r^2 + r^2 (d\theta^2 + \sin^2\theta~ d\j^2) ~,
\end{align}
where $\theta,\j$ are the polar and azimuthal angles, $r$ is $(2\pi)^{-1}$ times the circumference of a two-sphere. The action of our theory is composed of the action of gravity and matter $S= S_\mathrm{G} + S_\mathrm{M}$, specifically
\begin{align}
	S_\mathrm{G} &= \frac{1}{16\pi G} \( \int_\Omega R \sqrt{-g}~ d^4x+ \int_{\pd\Omega}  K \sqrt{|h|}~ d^3x \) ~,\\
	S_\mathrm{M} &= \int \[ -\frac{1}{2}g^{\mu\nu} \f_{,\mu}\f_{,\nu} - V(\f) \] \sqrt{-g} ~d^4x ~,
\end{align}
where $\f_{,\mu}\equiv \pd_\mu\f$ is defined for notation convenience, $R$ is the Ricci scalar and $g$ is the determinant of $g_{\mu\nu}$, i.e.
\begin{align}
	R = & 2 r^{-2} - 2 e^{-2\Phi}\[ \Psi_{,00} - \Psi_{,0}(\Phi + \Psi )_{,0} \] \\
	&- 2e^{2\Psi}\[ r^{-2} + 2 r^{-1}( \Phi + \Psi )_{,1} + \Phi_{,1}(\Phi + \Psi)_{,1} + \Phi_{,11} \] ~,\\
	\sqrt{-g}~ d^4x =& e^{\Phi-\Psi} r^2 \sin\theta ~dt~dr~d\theta~d\j ~.
\end{align}
Apart from the standard Hilbert action, $K$ is a surface term and $h$ is the induced metric on the boundary $\pd\Omega$ \cite{Gibbons:1976ue} that can be appropriately chosen \cite{Lee:1988av}, i.e.
\begin{align}
	K &= 2e^{\Psi} \Phi_{,1} + 4r^{-1} (e^\Psi - 1) ~,\\
	\sqrt{|h|}~d^3x &= e^\Phi r^2 \sin\theta ~dt~d\theta~ d\j ~,
\end{align}
for the three dimensional surface at $r=r_0$ and
\begin{align}
	K &= 2e^{-\Phi} \Psi_{,0} ~,\\
	\sqrt{|h|}~d^3x &= e^{-\Psi} r^2 \sin\theta ~dr~d\theta~ d\j ~,
\end{align}
for that which is bounded by $t=\pm T$. The inclusion of surface terms will not change the Einstein equations, but can convert the Hilbert action into one that contains only the first derivative of the metric so that the usual lagrangian mechanics can be applied. After setting $r_0$ and $T\rightarrow\infty$, the lagrangian, i.e. $S = \int L ~dt$, becomes
\begin{align}
	\label{lagrangian_gravity}
	L_\mathrm{G} &= (2G)^{-1} \int_0^\infty \[ e^{\Phi-\Psi} + e^{\Phi+\Psi}(1+2r\Phi_{,1}) - 2e^\Phi(1+r\Phi_{,1}) \] dr ~,\\
	\label{lagrangian_field}
	L_\mathrm{M} &=  \int_0^\infty ( X-Y-V) ~e^{\Phi-\Psi} 4\pi r^2 dr ~,
\end{align}
where we have defined
\begin{align}
	X \equiv \frac{1}{2}e^{-2\Phi}\f_{,0}^2 \sep 
	Y \equiv \frac{1}{2}e^{2\Psi}\f_{,1}^2 ~.
\end{align}

The gravity and matter are related by the Einstein equation
\begin{align}
	\t{G}{_\mu_\nu} = \t{R}{_\mu_\nu} - \frac{1}{2}g_{\mu\nu} R = 8\pi G ~\t{T}{_\mu_\nu} ~,
\end{align}
where $R_{\mu\nu}$ is the Ricci tensor and the energy-momentum tensor $\t T{^\mu_\nu}$ is given by
\begin{align}
	T_{\mu\nu} = -2 \frac{1}{\sqrt{-g}} \frac{\delta S_\mathrm{M}}{\delta g^{\mu\nu}} = \f_{,\mu}\f_{,\nu} - \frac{1}{2}g_{\mu\nu} g^{\rho\sigma} \f_{,\rho}\f_{,\sigma} - g_{\mu\nu} V(\f) ~.
\end{align}
More specifically
\begin{align}
	\label{G00_full}
	\t{G}{^0_0} &= r^{-2}\( e^{2\Psi} - 1 + 2 r \Psi_{,1} e^{2\Psi}  \) = -8\pi G (X+Y+V) ~,\\
	\label{G10_full}
	\t{G}{^1_0} &= -2 r^{-1} \Psi_{,0} e^{2\Psi} = 8\pi G ~e^{2\Psi}\f_{,0}\f_{,1} ~,\\
	\label{G11_full}
	\t{G}{^1_1} &= r^{-2} \( e^{2\Psi} - 1 + 2 r \Phi_{,1} e^{2\Psi} \) = 8\pi G (X+Y-V) ~,\\
	\label{G22_full}
    \nonumber
	\t{G}{^2_2}&= e^{-2\Phi} \[ \Psi_{,00} - \Psi_{,0} (\Phi + \Psi )_{,0} \] + e^{2\Psi} \[ \Phi_{,11} + (\Phi_{,1}+r^{-1}) (\Phi+\Psi )_{,1} \] \\
    &= 8\pi G(X-Y-V) ~,
\end{align}
where $\t{G}{^3_3}=\t{G}{^2_2}$ and all other components vanish. The $\t{G}{^0_0}$ and $\t{G}{^1_1}$ equations can be alternatively obtained by varying the lagrangian with respect to $\Phi$ and $\Psi$. The others can be derived using the contracted Bianchi identity $\t{G}{^\mu_{\nu;\mu}}=0$, i.e. $\nu=0$ gives \eqref{G10_full}, $\nu=2$ gives $\t{G}{^2_2}=\t{G}{^3_3}$ and $\nu=1$ gives \eqref{G22_full}. Some combanitions will be useful, for example, equations \eqref{G00_full} and \eqref{G11_full} give
\begin{align}\label{G11-G00}
	r^{-1} e^{2\Psi} (\Phi-\Psi)_{,1} = 8\pi G(X+Y) ~,
\end{align}
and equations \eqref{G22_full} and \eqref{G11-G00} give
\begin{align}\label{G11+G22-G00}
	e^{-2\Phi} \[ \Psi_{,00} - \Psi_{,0} (\Phi + \Psi )_{,0} \] + e^{2\Psi} \[ \nabla^2\Phi + \Phi_{,1} (\Phi+\Psi )_{,1} \] = 8\pi G(2X-V) ~,
\end{align}
where $\nabla^2\equiv \pd_r^2 + (2/r)\pd_r$. The equation of motion of $\f$ is
\begin{align}\label{EOM_phi_full_GR}
	e^{-2\Phi}\[-\f_{,00} + (\Phi+\Psi)_{,0} \f_{,0} \] + e^{2\Psi} \[ \nabla^2\f +  (\Phi + \Psi)_{,1} \f_{,1} \] - V'(\f) = 0 ~,
\end{align}
which is obtained by varying the lagrangian with respect to $\f$.

\subsection{Mass and energy}
\label{sec:soliton_gravity_mass}
Following \cite{Lee:1988av}, we distinguish between the mass and energy of oscillons. At large radius the mass density vanishes exponentially, hence $\Phi$ and $\Psi$ scale as $r^{-1}$. The mass then must satisfy
\begin{align}\label{mass_Schwarzschild}
	M_\mathrm{osc} = -G^{-1} \lim\limits_{r\rightarrow\infty} r\Psi = -G^{-1} \lim\limits_{r\rightarrow\infty} r\Phi ~,
\end{align}
to be consistent with the static Schwarzschild solution. There are other ways to express the same mass. For example, the LHS of 00 component of Einstein equation can be rewritten into $r^{-2} [ r\(e^{2\Psi}-1\) ]_{,1}$ hence the mass is also given by
\begin{align}\label{mass_T00}
	M_\mathrm{osc} = \int_0^\infty ( X+Y+V) ~4\pi r^2 dr = -\int_{0}^{\infty} \t{T}{^0_0}~ 4\pi r^2 dr ~,
\end{align}
which is in agreement with the Schwarzschild mass \eqref{mass_Schwarzschild}.

A more enlightening way to describe the mass is to use the Hamiltonian formalism. The lagrangian of matter \eqref{lagrangian_field} indicates that the \emph{energy} of the oscillon is
\begin{align}\label{energy_oscillon_gravity}
	E_\mathrm{osc} =  \int_0^\infty ( X+Y+V) ~e^{\Phi-\Psi} 4\pi r^2 dr = -\int \t{T}{^0_0} \sqrt{-g}~ d^3x  ~.
\end{align}
There is no kinetic term in the lagrangian of gravity \eqref{lagrangian_gravity} thus the energy of gravity is $E_\mathrm{G} = -L_\mathrm{G}$. Then we define the \emph{mass} of oscillons
\begin{align}
	M_\mathrm{osc} \equiv E_\mathrm{osc} + E_G ~.
\end{align}
Combining the energy expression and the 00 component of Einstein equations, we find
\begin{align}
	[r(e^\Phi-e^{\Phi+\Psi})]_{,1} = G (E_\mathrm{M} + E_\mathrm{G})_{,1} ~,
\end{align}
in agreement with the Schwarzschild mass \eqref{mass_Schwarzschild} and the ADM mass \cite{Arnowitt:1962hi}.

\subsection{Oscillons with linearized gravity}
\label{sec:soliton_gravity_linear}
The typical central amplitude, radius and mass of dense oscillons in the NR limit are
\begin{align}
	\f_1(0) \sim F \sep
	R_\mathrm{osc} \sim 5 m^{-1} \sep
	M_\mathrm{osc} \sim 100 F^2/m ~.
\end{align}
Nonlinear effects of gravitational interactions are not important if the size of oscillons is much smaller than their Schwarzschild radius
\begin{align}\label{epsilon_F}
	R_\mathrm{osc} \ll GM_\mathrm{osc}  \Rightarrow
	\epsilon_\phi \sim \f_1/\MP \ll 1  ~,
\end{align}
which is satisfied by a number of cosmological models. In this section, therefore, we study the decay rate and lifetime of oscillons in the NR limit and weak-field limit of gravity, specifically those with $\epsilon_r \lesssim 0.1$ and $\epsilon_\phi \lesssim 0.1$.\footnote{Examples of the EFT that focuses on such low-energy phenomena includes \cite{Mukaida:2016hwd, Eby:2018ufi, Braaten:2018lmj, Namjoo:2017nia, Salehian:2020bon}.} The basic idea is similar to what we have done in section \ref{sec:soliton_oscillon_decayrate}.

In the weak-field approximation, the spherical metric \eqref{spherical_metric} reduces to
\begin{align}
	ds^2 = -(1+2\Phi) d t^2 + (1-2\Psi)d r^2 + r^2(d\theta^2 + \sin^2\theta ~d\j^2) ~.
\end{align}
So far we have encountered five sets of small dimensionless parameters, recall equations \eqref{small_quantity} and \eqref{epsilon_F}, i.e. the spatial derivative parameter $\epsilon_r$, the nonlinear potential parameter $\epsilon_V$, the radiation parameter $\epsilon_\xi$, the amplitude parameter $\epsilon_\phi$ and the gravitational potentials $\Phi$ and $\Psi$ (denoted by $\epsilon_g$). To be consistent, we will keep all small quantities to 1st order, and to 2nd order if spatial derivatives of small parameters are involved. Then the equation of motion of $\f$ \eqref{EOM_phi_full_GR} becomes
\begin{align}\label{EOM_phi}
	-(1-2\Phi)\f_{,00} + (\Phi+\Psi)_{,0}\f_{,0} + \nabla^2\f - V'(\f) = \cal O (\epsilon^2m^2\f) ~.
\end{align}

To be consistent with the field expansion \eqref{profile_single_freq}, we expand the gravitational potentials in terms of a Fourier cosine series
\begin{align}\label{expansion_Phi_Psi}
	\Phi = \frac{1}{2}\Phi_0 + \sum_{j=2}^{\infty}\Phi_j\cos(j\omega t) \sep
	\Psi = \frac{1}{2}\Psi_0 + \sum_{j=2}^{\infty}\Psi_j\cos(j\omega t) ~,
\end{align}
where $j$ is even. We can solve these radial modes by plugging the expansion into the Einstein equations and collecting the coefficient for each Fourier mode. Then equations \eqref{G10_full} and \eqref{G11_full} give
\begin{align}
	\Psi_0 = - r\Phi_{0,1} + \cal O(\epsilon^2) \sep
	\Psi_2 = \cal O(\epsilon^2) \sep
	\Psi_{j\geq 4} = \cal O(\epsilon^3) ~.
\end{align}
Equations \eqref{G11-G00} and \eqref{G11+G22-G00} give
\begin{align}
	\nabla^2\Phi_0 = 4\pi G ~m^2\f_1^2 + \cal O(\epsilon^3 m^2) \sep
	\Phi_{2,1}/r = -2\pi G ~m^2\f_1^2 + \cal O(\epsilon^3 m^2)  ~,
\end{align}
and $\Phi_{j\geq 4} =  \cal O(\epsilon^2)$. Therefore, $-\Phi_0\sim -\Psi_0\sim \Phi_2\sim \epsilon_g\sim \epsilon_\f^2/\epsilon_r$.\footnote{For the quadratic potential, there is no mass scale $F$ and the central amplitude of oscillons satisfies $\phi_1\sim \cal O(\epsilon_r \MP)$ \cite{Lee:1991ax}, hence $\epsilon_\f\sim \epsilon_g\sim \epsilon_r$.} Somewhat surprisingly, we find $\Phi_2$ and $\Psi_2$ are not the same order of magnitude, in constrast with the common results of the isotropic coordinates (Newtonian gauge) where the oscillating part of gravitational potentials is insignificant. This distinction is due to that the notion of time is different in these two coordinates, where $t$ is related to the time in isotropic coordinates $\eta$ by $\eta \simeq t - \frac{\Phi_2}{2\omega} \sin(2\omega t)$ \cite{Salehian:2021khb}. As a result, we will ignore $\Phi_{j\geq 4}$ and $\Psi_{j\geq 2}$ in future calculations.

In order to find the profile of oscillons, we plug the field expansion \eqref{profile_single_freq} and \eqref{expansion_Phi_Psi} into equation \eqref{EOM_phi} and collect the coefficient of $\cos(\omega t)$ to get
\begin{align}\label{profile_gravity}
	[ \nabla^2 + \omega^2(1-\Phi_0) -m^2 ]\f_1 = J_1 + \cal O(\epsilon^2m^2\f_1) ~.
\end{align}
Here $\omega(1-\Phi_0)^{1/2}$ can be regarded as an effective frequency and is larger than $\omega$. This equation can be solved by numerical shooting method.\footnote{The boundary conditions are $\f_1(\infty)\rightarrow 0$, $\Phi_0(\infty)\propto 1/r$ and $\Phi_2(\infty)\rightarrow 0$.} The time-averaged formula for the oscillon energy \eqref{energy_oscillon_gravity} is, up to $\cal O(\epsilon^{-1/2} \f_1^2/m)$,
\begin{align}
	E_\mathrm{osc} = \int_0^\infty \[ \frac{1}{4}(\pd_r\f_1)^2 + \frac{1}{4}(\omega^2+m^2)\(1-\frac{1}{2}\Psi_0 + \frac{1}{2}\Phi_2 \) \f_1^2 + \frac{1}{2}U_0 \] 4\pi r^2dr ~.
\end{align}
As long as the oscillating part of the gravitational potentials is not too important, namely $\epsilon_g\lesssim 0.1$, oscillons share great similarities with mini-boson stars \cite{Friedberg:1986tp}, and we assume the stability condition is still valid \cite{Lee:1988av}
\begin{align}\label{stability}
	\frac{dE_\mathrm{osc}}{d\omega} < 0 ~.
\end{align}
This is confirmed by comparing analytical predictions of $\omega_\mathrm{crit}$ with numerical results in the left panel of figure \ref{fig:lifetime}.

Plugging the field expansion \eqref{profile_single_freq} and \eqref{expansion_Phi_Psi} into equation \eqref{EOM_phi} and collecting the coefficient of $\cos(j\omega t)$, we obtain the radial equation of radiating modes
\begin{align}\label{radiation_eq}
	\[\nabla^2 + \kappa_j^2\] \xi_j(r) = S_j(r) + \cal O(\epsilon^3m^2\f_1) ~,
\end{align}
where $j\geq 3$ and $j$ is odd, and $S_j$ is the effective source. We can keep finite terms of $\xi_j$ as we did in Minkowski spacetime. In particular, if we keep only $\xi_3$ and $\xi_5$, the effective source becomes
\begin{align}
	\label{S3}
	S_3 &= J_3 + \frac{1}{2}\xi_3(M_0+M_6) + \frac{1}{2}\xi_5(M_2+M_{8}) + 2 \omega^2\f_1 \Phi_2 + 9\omega^2\xi_3 \Phi_0 + 20\omega^2 \xi_5 \Phi_2~,\\
	\label{S5}
	S_5 &= J_5 + \frac{1}{2}\xi_3(M_2+M_8) + \frac{1}{2}\xi_5(M_0+M_{10}) + 12\omega^2\xi_3\Phi_2 + 25\omega^2 \xi_5 \Phi_0 ~,
\end{align}
where we have kept higher-order perturbations $\xi_5\Phi_j$ because of their large coefficients. Comparing \eqref{S3} with the corresponding expression in Minkowski spacetime \eqref{S3_Minkowski}, new corrections are introduced due to the coupling of gravity to oscillons and their radiation. The radiation equation can be solved by the iterative method \cite{Zhang:2020bec}.\footnote{The participation of $\Phi_0$ is possible to make the iteration divergent (since both $\Phi_0,\xi_3\propto 1/r$ at large radius), in which case $\xi_j$ can be easily found by adjusting initial values and matching the central amplitudes calculated by equations \eqref{xi_initial} and \eqref{radiation_eq}.}

The conservation law of the energy-momentum tensor is
\begin{align}
	\t{T}{^{\mu}_{\nu;\mu}} = 0 \Rightarrow (\t{T}{^\mu_\nu} \sqrt{-g})_{,\mu} = 0 ~,
\end{align}
which implies
\begin{align}
	\frac{dE_\mathrm{osc}}{dt} = \int  (\t{T}{^i_{0}}\sqrt{-g})_{,i} ~d^3x = \l. -4\pi r^2 \t{T}{^i_0} (1+\Phi-\Psi) \r|_{r\rightarrow\infty} = \l. -4 \pi r^2 \xi_{,0}\xi_{,1} \r|_{r\rightarrow\infty} ~, 
\end{align}
where we have used Gauss's divergence theorem to obtain the second equal sign. The $\xi$ at infinity is calculated in \eqref{xi_inf}, then the decay rate expression is just the same as the one in Minkowski spacetime \eqref{Gamma_Minkowski}.

\subsection{Case study}
\label{sec:soliton_gravity_case}
Now we explore gravitational effects on oscillon lifetimes by studying the $\alpha$-attractor T-model of inflation and the axion monodromy model.

\paragraph{The $\alpha$-attractor T-model of inflation:} The longevity of oscillons in this model (whose lifetime $\sim 10^{6}\,m^{-1}$ in Minkowski spacetime) is characterized by a dip structure in decay rates, see figure \ref{fig:tanhfreqenergyamp}. In the dip, the leading channel of decay rates $\Gamma_3$ vanishes and the scalar radiation is dominated by the subleading channel $\Gamma_5$. Now we argue that the existence of gravity reduces the lifetime of oscillons.

\begin{figure}
	\centering
	\begin{minipage}{0.45\linewidth}
		\includegraphics[width=\linewidth]{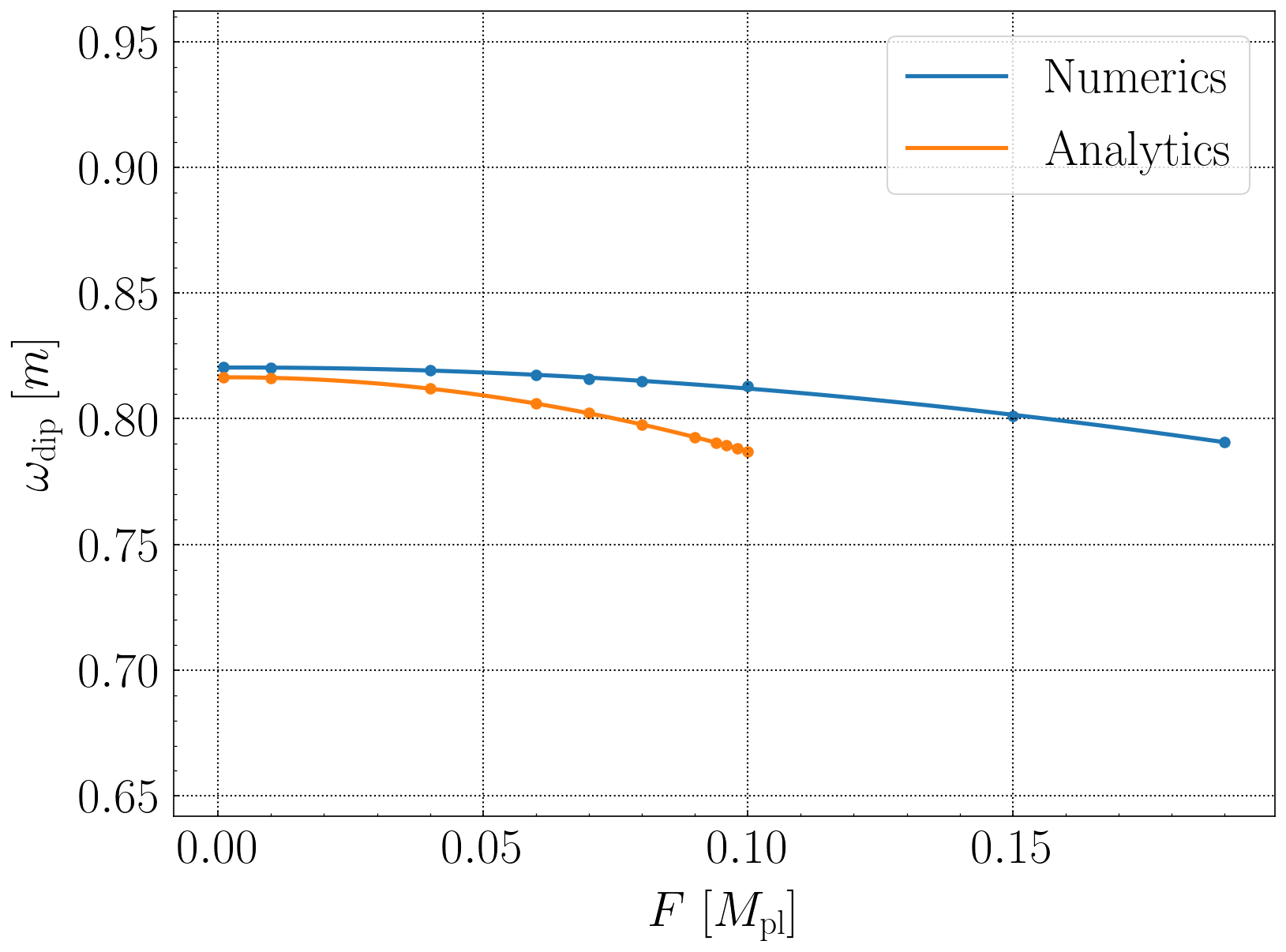}
	\end{minipage}
	\begin{minipage}{0.45\linewidth}
		\includegraphics[width=\linewidth]{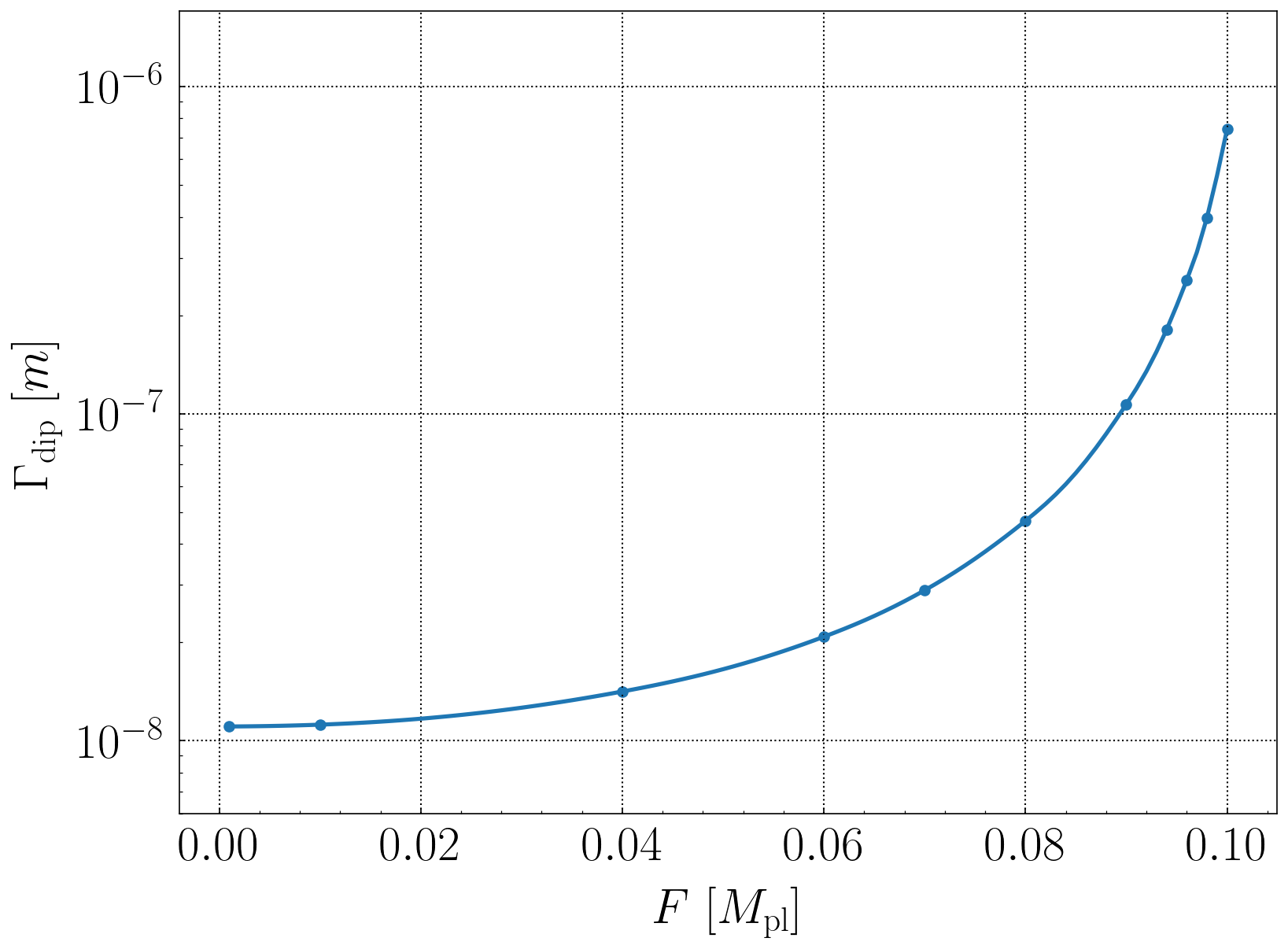}
	\end{minipage}
	\caption{Evidence that the existence of gravity reduces the lifetime of oscillons in the $\alpha$-attractor T-model. In the left panel, we show the dependence of dip locations on values of the characteristic mass scale $F$. The smaller value of $\omega_\mathrm{dip}$ is an indication of a shorter lifetime of oscillons, since the amplitude of radiating modes (in unit of $F$) is inversely related to frequencies and thus larger decay rates are expected. In the right panel, we confirm this expectation by explicitly calculating the decay rate at $\omega_\mathrm{dip}$.}
	\label{fig:tanh}
\end{figure}

We first study how the location of the dip is affected both analytically and numerically in figure \ref{fig:tanh} (left panel). The value of $\omega_\mathrm{dip}$ is significantly reduced when the mass scale $F$ approaches the reduced Planck mass. A smaller value of $\omega_\mathrm{dip}$ is an indication of shorter lifetimes, since the amplitude of radiating modes (in unit of $F$) is inversely related to frequencies and thus larger decay rates are expected. The comparison between analytics and numerics also provides a chance to test our formalism, which correctly captures the gravitational effect on $\omega_\mathrm{dip}$ as long as the assumptions of weak-field gravity and NR limit remain valid. 

\begin{figure}
	\centering
	\includegraphics[width=0.5\linewidth]{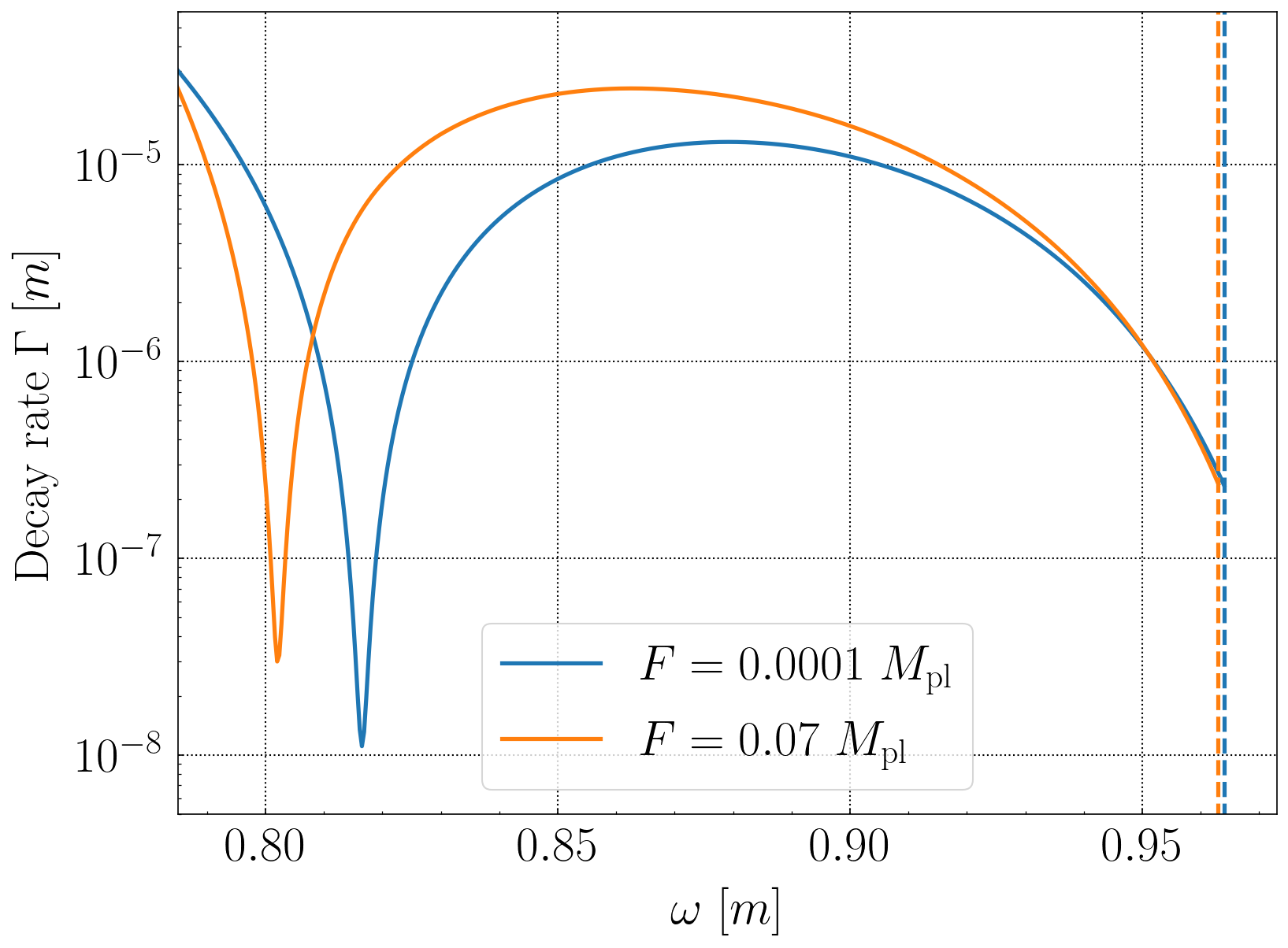}
	\caption{Decay rates for two different values of $F$ in the $\alpha$-attractor T-model.}
	\label{fig:tanhdecayrate}
\end{figure}

To exclude the possibility that the subleading channel of decay rates also vanishes, we explicitly calculate $\Gamma_5$ around $\omega_\mathrm{dip}$ as shown in figure \ref{fig:tanh} (right panel). As a determinant factor, the increasing of $\Gamma_\mathrm{dip}\approx \Gamma_5$ is a clear evidence that the existence of gravity reduces the lifetime of oscillons, which has also been confirmed numerically. For convenience, we present a direct visualization of these two factors in figure \ref{fig:tanhdecayrate}.

\paragraph{The axion monodromy model:} In constrast with the $\alpha$-attractor T-model, the subleading channel of decay rates $\Gamma_5$ is never comparable with the leading channel $\Gamma_3$ in the axion monodromy model. The longevity of oscillons in this case (whose lifetime $\sim 10^{8}\,m^{-1}$ in Minkowski spacetime) is due to the dramatic suppresion of $\Gamma_3$ just before their final collapse at $\omega_\mathrm{crit}$. Consequently, there are two factors that determine the lifetime of oscillons, the value of $\omega_\mathrm{crit}$ and the decay rate around $\omega_\mathrm{crit}$.

\begin{figure}
	\centering
	\begin{minipage}{0.45\linewidth}
		\includegraphics[width=\linewidth]{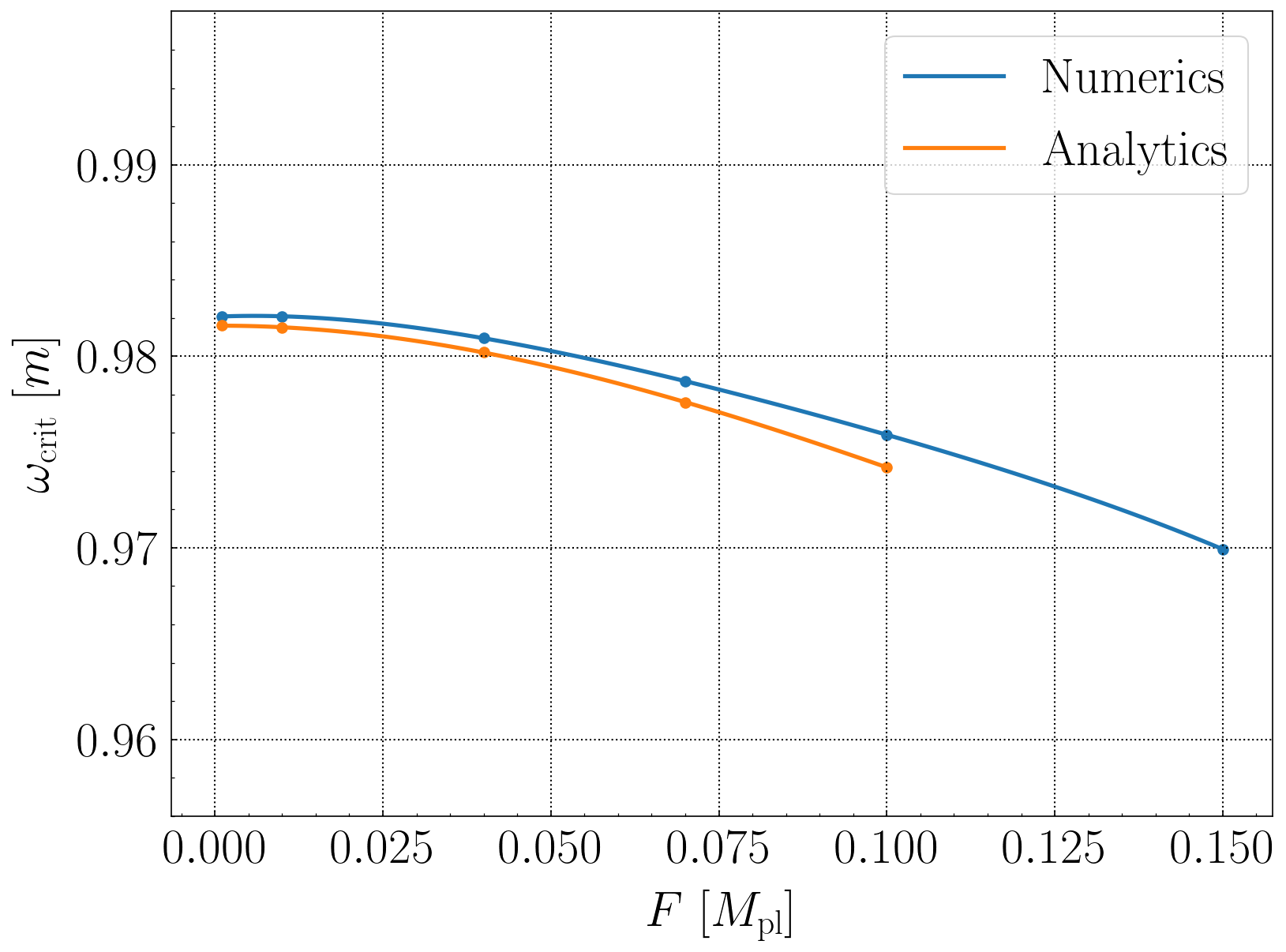}
	\end{minipage}
	\begin{minipage}{0.45\linewidth}
		\includegraphics[width=\linewidth]{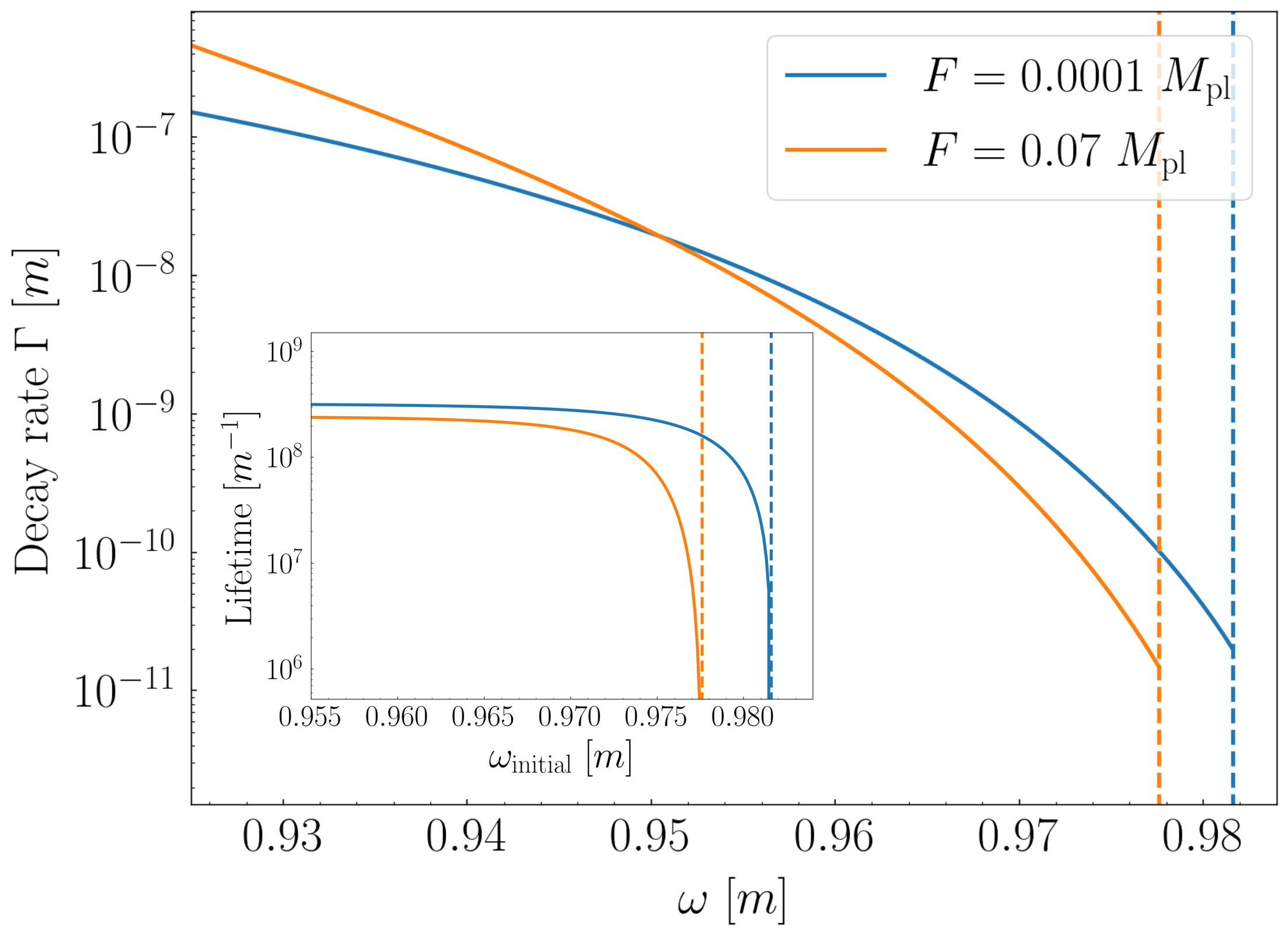}
	\end{minipage}
	\caption{Two determinant factors of oscillon lifetimes in the axion monodromy model, values of the critical frequency (left panel) and decay rates around $\omega_\mathrm{crit}$ (right panel). It is shown that gravitational effects decrease the value of $\omega_\mathrm{crit}$, which tends to reduce the lifetime, while suppressing the decay rate around $\omega_\mathrm{crit}$, which tends to stablize oscillons. In the inset, we show that oscillon lifetimes are reduced slightly for the stronger-field gravity.}
	\label{fig:lifetime}
\end{figure}
In figure \ref{fig:lifetime} (left panel), we show that the values of $\omega_\mathrm{crit}$ are inversely related to $F$. This seems an indication of shorter lifetimes for stronger gravitational effects because oscillons now spend less time around $\omega_\mathrm{crit}$. A good match between analytics and numerics implies that the stability condition \eqref{stability} is still valid as long as gravity is not too important.

Based on our semi-analytical framework, we calculate oscillon decay rates for $F=0.0001 \MP$ and $F=0.07 \MP$ in the right panel. Compared with the very weak-field gravity, it is shown that the decay rate around $\omega_\mathrm{crit}$ is more suppressed for $F=0.07 \MP$. This can be qualitatively understood by inspecting equations \eqref{S3_Minkowski} and \eqref{S3}. Since $J_3$ at $\omega_\mathrm{crit}$ typically has a Gaussian-like shape with a negative amplitude, the introduction of the positive term $2\omega_2\f_1\Phi_2$ diminishes the magnitude of the effective source, and thus tends to reduce the decay rate and increase the total lifetime.

To determine which factor dominates, we integrate out the decay rate and show the lifetime in an inset of the right panel of figure \ref{fig:lifetime}. Our results imply that the first factor plays a more important role and the oscillon lifetime is shorter for stronger-field gravity. Nevertheless, the oscillon is still too long-lived to be simulated with our current numerical algorithms. And we leave this as a testable prediction for a future numerical experiment.

\section{Vector oscillons}
\label{sec:soliton_vector}
\begin{figure}
	\centering
	\begin{minipage}{0.31\linewidth}
		\includegraphics[width=\linewidth]{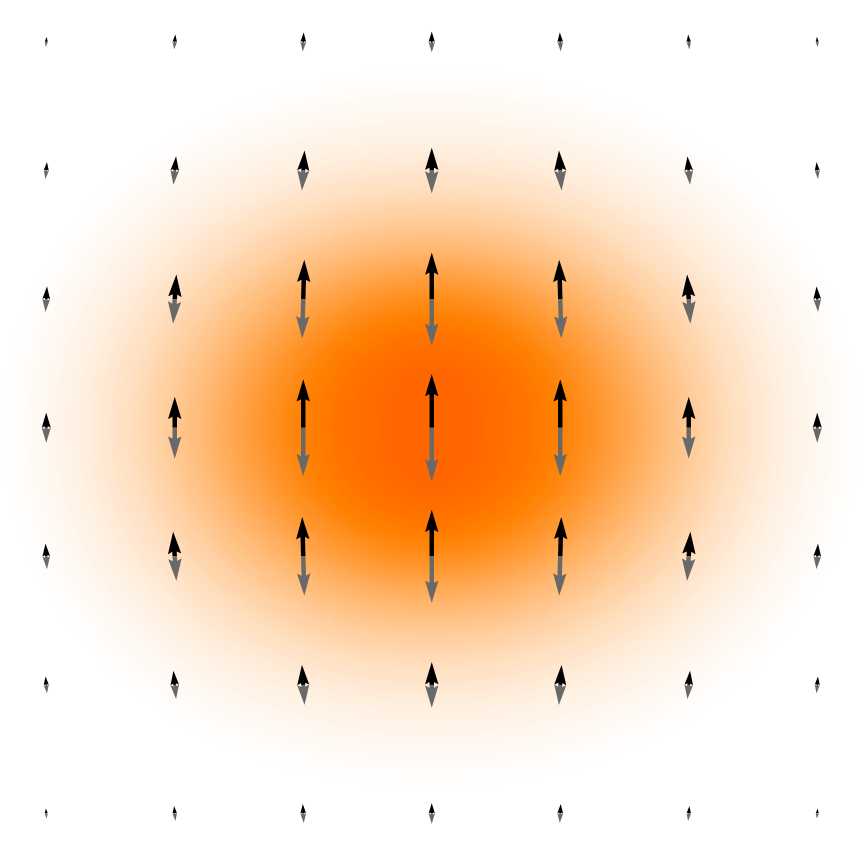}
	\end{minipage}~
	\begin{minipage}{0.3\linewidth}
		\includegraphics[width=\linewidth]{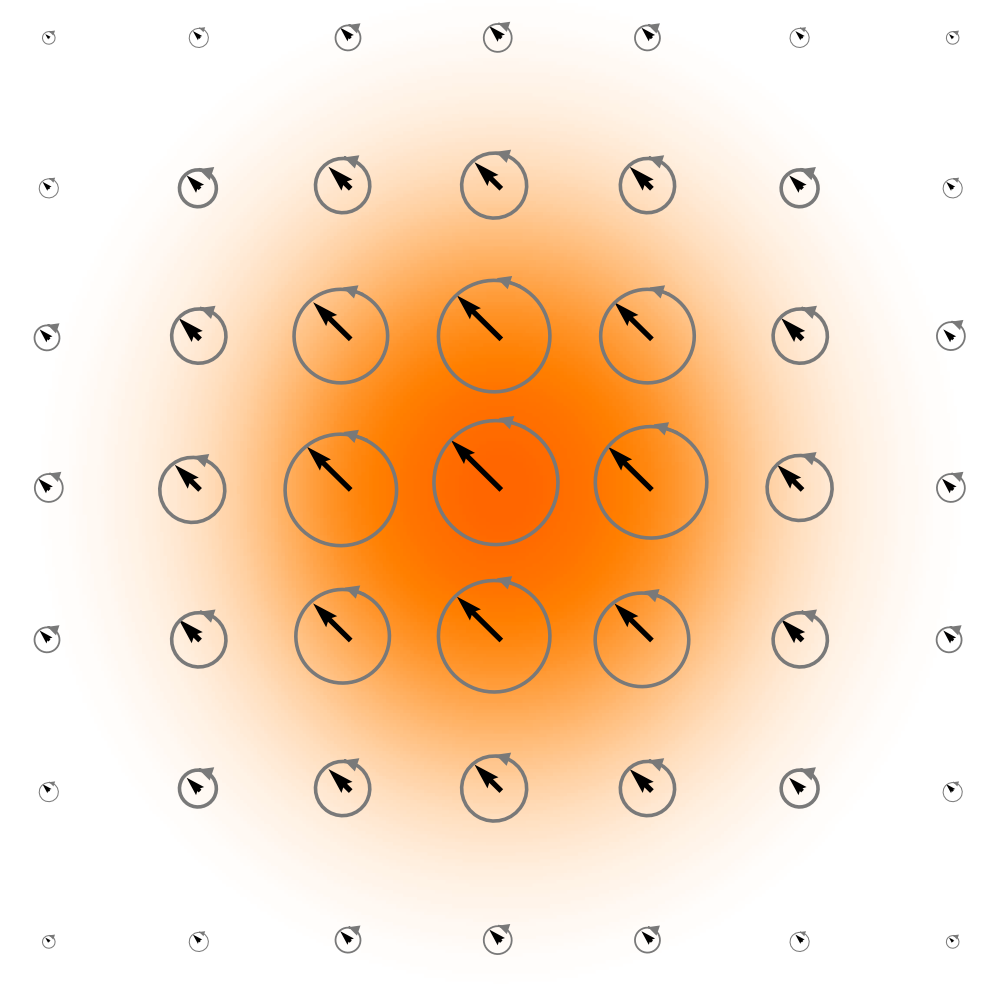}
	\end{minipage}~
	\begin{minipage}{0.29\linewidth}
		\includegraphics[width=\linewidth]{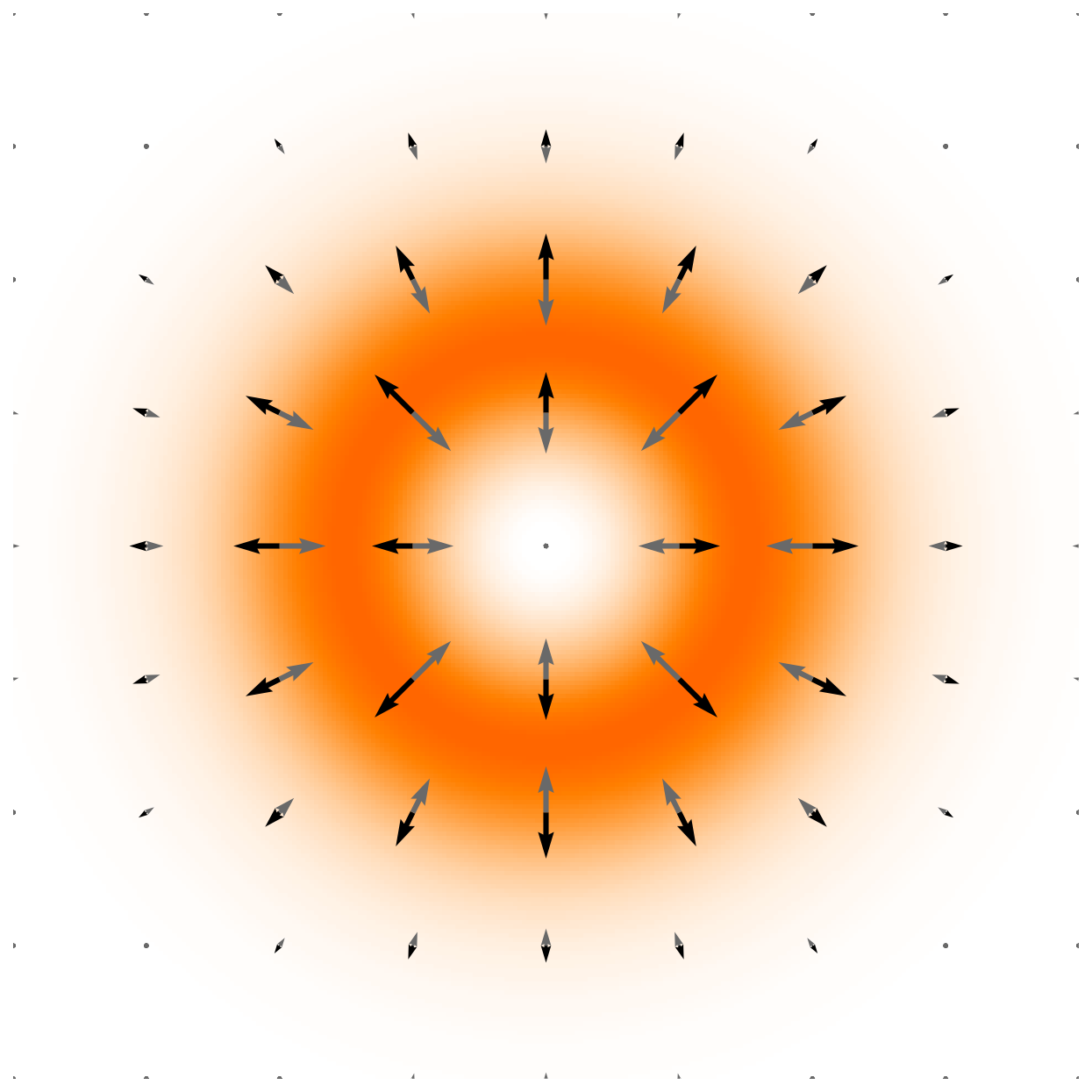}
	\end{minipage}
	\caption{Schematic diagrams of an equatorial cut for directional (left), spinning (middle) and hedgehog (right) oscillons. Arrows denote magnitude and direction of the 3-vector $\b W$, and lighter colors represent higher energy density. Roughly speaking, $\b W$ in directional or hedgehog oscillons oscillates periodically between opposite directions along one specific axis or the radial axis, while in spinning oscillons it rotates around the axis perpendicular to a plane.}
	\label{fig:vectoroscillon}
\end{figure}

In cases where DM is primarily composed of massive vector fields, vector oscillons can be excellent targets to probe DM's self-interactions. To study the property of vector oscillons, we consider a phenomenological real-valued massive spin 1 field $W_\mu$ with the lagrangian
\begin{align}
	\cal L = -\frac{1}{4} F_{\mu\nu}F^{\mu\nu} - V(W_\mu W^\mu) ~,
\end{align}
where $F_{\mu\nu} = \pd_\mu W_\nu - \pd_\nu W_\mu$ and the potential
\begin{align}
	\label{potential}
	V(W_\mu W^\mu) = \frac{m^2}{2} W_\mu W^\mu + \frac{\lambda}{4} (W_\mu W^\mu)^2 + \frac{h}{6} (W_\mu W^\mu)^3 ~.
\end{align}
This type of potential may arise from the interaction of $W_\mu$ to other matter and/or nonminimal coupling of $W_\mu$ to gravity. Without loss of generality, we will set $m=1$ and $\lambda=\pm 1$.\footnote{This can be achieved by redefining the field and spacetime coordinates as $x^\mu \rightarrow x^\mu/m$, $W_\mu \rightarrow m W_\mu/\sqrt{|\lambda|}$ and $h \rightarrow h\lambda^2/m^2$.} The equations of motion are
\begin{align}
	\label{EOM1}
	-\nabla^2 W_0 + \pd_t \nabla\cdot \b W + 2V'(W_\mu W^\mu) W_0 = 0 ~,\\
	\label{EOM2}
	\pd_t^2\b W - \pd_t \nabla W_0 + \nabla\times\nabla\times \b W + 2V'(W_\mu W^\mu) \b W = 0 ~,
\end{align}
where $\nabla\times\nabla\times\b W = \nabla(\nabla\cdot\b W) - \nabla^2\b W$. In order to find localized configurations with the lowest-energy for a fixed particle number, we consider oscillons with some sort of spherical symmetry, i.e. either some components of $W_i$ are (approximately) radially symmetric or the entire vector field $W_\mu$ is spherically symmetric. As illustrated in figure \ref{fig:vectoroscillon}, we call these localized clumps directional, spinning and hedgehog oscillons.

Once a solution is found, the energy can be given by the Noether's current associated with spacetime translations
\begin{align}
	\label{Tmunu}
	T^{\mu\nu} = \pd^\nu W_\sigma F^{\mu\sigma} + g^{\mu\nu}\cal L ~.
\end{align}
And the angular momentum is given by that associated with Lorentz transformations, i.e. $\cal M^{\mu\nu\sigma} = \cal L^{\mu\nu\sigma} + \cal S^{\mu\nu\sigma}$, where 
\begin{align}
	\cal L^{\mu\nu\sigma} &= x^\nu T^{\mu\sigma} - x^\sigma T^{\mu\nu} ~,\\
	\cal S^{\mu\nu\sigma} &= F^{\mu\nu}W^\sigma - F^{\mu\sigma}W^\nu ~.
\end{align}
The density of orbital and spin angular momentum is thus $\cal L_i = (1/2) \epsilon_{ijk} \cal L^{0jk}$ and $\cal S_i = (1/2) \epsilon_{ijk} \cal S^{0jk}$. We will see that spinning oscillons have intrinsic spin and are the composition of particles of spin 1 in the NR limit, thus they should \emph{not} be regarded as rotating directional oscillons, which are made up of spin 0 particles.

In what follows, we will first derive oscillon solutions by doing small-amplitude expansions in section \ref{sec:soliton_vector_leading}--\ref{sec:soliton_vector_subleading}. In section \ref{sec:soliton_vector_eft}, we derive a NR EFT for the the original theory and prove that vector oscillons are lowest-energy states of the field. By carrying out fully relativistic simulations, we study the stability and lifetimes of vector oscillons by carrying out fully relativistic simulations in section \ref{sec:soliton_vector_large}. Although we study properties of vector oscillons with a specific type of self-interactions, the small-amplitude expansion and many of our conclusions (e.g. the lowest energy state, spin, stability and longevity) should remain valid for more general interactions including gravitational ones.

\subsection{Small-amplitude expansions}
\label{sec:soliton_vector_leading}
Consider a vector field in the NR regime that is characterized by a frequency $\omega \approx 1$ and has a small amplitude $|W_\mu|\ll 1$. In this case it approximately obeys the Klein-Gordon equation and a localized solution has the form $W_\mu \propto e^{-\sqrt{1-\omega^2} r}/r$, where $\omega<1$. This motivates us to define a small positive quantity $\epsilon\equiv \sqrt{1-\omega^2}$ and rescale spacetime coordinates by
\begin{align}
	\label{expansion_coordinates}
	\b x' = \epsilon \b x \sep
	t' = \omega t ~,
\end{align}
such that $|\pd_{\mu'} W_\nu| \sim W_\nu$. Then from \eqref{EOM1} and \eqref{EOM2} one sees that $W_0\sim \epsilon W_i$ and $\nabla^2 W_i\sim W_i^3$, which suggests an expansion of the vector field in terms of $\epsilon$, i.e.
\begin{align}
	\label{expansion_field}
	W_\mu = \epsilon W_\mu^{(1)} + \epsilon^2 W_\mu^{(2)} + \epsilon^3 W_\mu^{(3)} + \cdots ~.
\end{align}
Such a scheme was used to study oscillons in scalar field theory \cite{Fodor:2008es, Amin:2010jq, Amin:2013ika}. By plugging these expansions into equation \eqref{EOM2} and collecting terms up to the order $\cal O(\epsilon^4)$, we obtain
\begin{align}
	\label{EOM_Wi_order1}
	\ddot{\b W}^{(1)}+{\b W}^{(1)} =& 0 ~,\\
	\label{EOM_Wi_order2}
	\ddot{\b W}^{(2)}+{\b W}^{(2)} =& 0 ~,\\
	\label{EOM_Wi_order3}
	\ddot{\b W}^{(3)}+{\b W}^{(3)} =& [ \nabla^2 - (1 + \lambda {\b W}^{(1)}\cdot{\b W}^{(1)} )] {\b W}^{(1)} ~,\\
	\label{EOM_Wi_order4}
	\ddot{\b W}^{(4)}+{\b W}^{(4)} =& [ \nabla^2 - (1 + \lambda {\b W}^{(1)}\cdot{\b W}^{(1)} )] {\b W}^{(2)} - 2\lambda ({\b W}^{(1)}\cdot{\b W}^{(2)}) {\b W}^{(1)} ~,
\end{align}
where we have used $W_0^{(n)}=-\nabla\cdot\dot{\b W}^{(n-1)}$ for $n\le 3$ based on \eqref{EOM1} and $\b W^{(0)}=0$. Equations \eqref{EOM_Wi_order1} and \eqref{EOM_Wi_order2} then imply
\begin{align}
	\label{solution_Wi_order1}
	W_i^{(1)} &= v_i^{(1)}(\b x') \sin[t' + \f_i^{(1)}(\b x)] ~,\\
	\label{solution_Wi_order2}
	W_i^{(2)} &= v_i^{(2)}(\b x') \sin[t' + \f_i^{(2)}(\b x)] ~.
\end{align}

To find localized solutions for $v_i^{(1)}$ and $v_i^{(2)}$, we will plug equations \eqref{solution_Wi_order1} and \eqref{solution_Wi_order2} into \eqref{EOM_Wi_order3} and \eqref{EOM_Wi_order4}. The RHS of the resulting equations contain harmonics of $\sin(nt')$ and $\cos(nt')$ with $n$ a positive integer. However, $\b W^{(3)}$ and $\b W^{(4)}$ will grow linearly with time if the coefficients of $\sin(t')$ and $\cos(t')$ do not vanish, since the driving frequency matches the natural frequency of the system. Thus by requiring the coefficients to be 0, we obtain spatial profile equations of $v_i^{(1)}$ and $v_i^{(2)}$, which can be solved by numerical shooting method. By repeating this game, we can find the profiles to arbitrary order.

A significant simplification can be made by noting that the profile equation of $v_i^{(2)}$ resulting from the equation \eqref{EOM_Wi_order4} is linear in $v_i^{(2)}$. This means that once a solution is found, we can construct infinite number of solutions by multiplying it with constants while the leading profile $v_i^{(1)}$ is fixed. Such a continuous degree of freedom (if exists) should not be observed for oscillon solutions, thus we must require $v_i^{(2)}$ to be 0.\footnote{\label{footnote:even_order_terms} This is because: (1) The vector oscillon profile is probably unique for each $\omega$ as suggested by numerical simulations, and it is proved true for scalar oscillons \cite{Zhang:2020bec}. (2) We find that the profile equations of $v_i^{(2)}$ actually do not yield any localized solutions for oscillons that will be studied in this letter and (3) We are always free to multiply the profile by 0 for lowest-energy solutions.} Similarly for higher-order terms, we must require all \emph{even}-order terms of the spatial component $\b W^{(2n)}$ and \emph{odd}-order terms of the temporal component $W_0^{(2n-1)}$ to be 0. Then the system possesses a constant of motion thanks to equation \eqref{EOM_Wi_order1}, i.e.
\begin{align}
	\label{particle_number}
	N = \frac{1}{2\epsilon} \int d^3x' ~(\dot W_i^{(1)}  \dot W_i^{(1)} + W_i^{(1)} W_i^{(1)}) ~,
\end{align}
which may be identified as the particle number. We define the particle number in a more general way in section \ref{sec:soliton_vector_eft} and show that $dE/dN=\omega$ approximately holds, where $E=\int d^3x ~T^{00}$ is energy of an oscillon. Hence $\omega$ may be identified as the chemical potential.

Since we are most interested in lower-energy states for a fixed particle number, we assume that the phase of leading modes is spatially independent, namely $\phi_i^{(1)}(\b x)\equiv \phi_i^{(1)}$, which would otherwise lead to extra energy. And we set $0\leq \phi_i^{(1)}< \pi$ since other choices for $\phi_i^{(1)}$ are mere a redefinition of $v_i^{(1)}$. On the other hand, the gradient of $v_i^{(1)}$ will make positive contributions to the oscillon energy proportional to its surface area, and thus we expect some sort of \emph{spherical symmetry} existing in the solution. For example, we assume that either some $v_i^{(1)}$ are radially symmetric or the entire vector field is spherically symmetric.

By pluging equation \eqref{solution_Wi_order1} into \eqref{EOM_Wi_order3} and by collecting the coefficients of $\cos(t')$ and $\sin(t')$ on the RHS of \eqref{EOM_Wi_order3}, generally we will obtain 6 profile equations that contain $\sin(\f_i^{(1)})$ and $\cos(\f_i^{(1)})$. However, we only need $3$ of them to determine $v_i^{(1)}$, and the others are just over-constraints that may not be satisfied simultaneously. The only way to get rid of these over-constraints is to make $\f_i^{(1)}$ either $0$ or $\pi/2$. Hence we classify the oscillon solutions as follows:
\begin{itemize}
	\item Directional oscillons: $W_1^{(1)}=W_2^{(1)}=0$ and $W_3^{(1)}=v_\rm{d}(r') \sin (t')$, where the phase in $W_3^{(1)}$ has been absorbed by a time shift. Through a spatial rotation, we may obtain other solutions with non-vanishing $W_1^{(1)}, W_2^{(1)}$ and time dependence in the form of $\sin(t')$. 
	\item Spinning oscillons: $W_1^{(1)} = v_\rm{s}(r') \cos(t')$, $W_2^{(1)} = v_\rm{s}(r') \sin(t')$ and $W_3^{(1)}=0$, where the phase in $W_2^{(1)}$ has been absorbed by a time shift and the other phase is fixed to $\pi/2$ in order to get different time dependence from directional oscillons.  Through a spatial rotation, we may obtain other solutions with non-vanishing $W_3^{(1)}$ and time dependence in the form of either $\sin(t')$ or $\cos(t')$.
	\item Hedgehog oscillons: $W_i^{(1)} = (x_i'/r') v_\rm{h}(r')\sin(t')$, which says that the vector field is spherically symmetric $\b W = v_\rm{h}(r') \sin(t') \hat {\b r'}$ and invariant under rotations. Due to the symmetry, $\b W$ must vanish at the center.
\end{itemize}
Thus to the leading order, we conclude that any oscillon solutions that have radially symmetric component $W_i^{(1)}$ can be obtained by either directional or spinning oscillons by a spatial rotation. Note that even though $W_i^{(1)}$ is radially symmetric, generally the oscillon profile with higher-order terms included is not.

\begin{figure}
	\centering
	\includegraphics[width=0.6\linewidth]{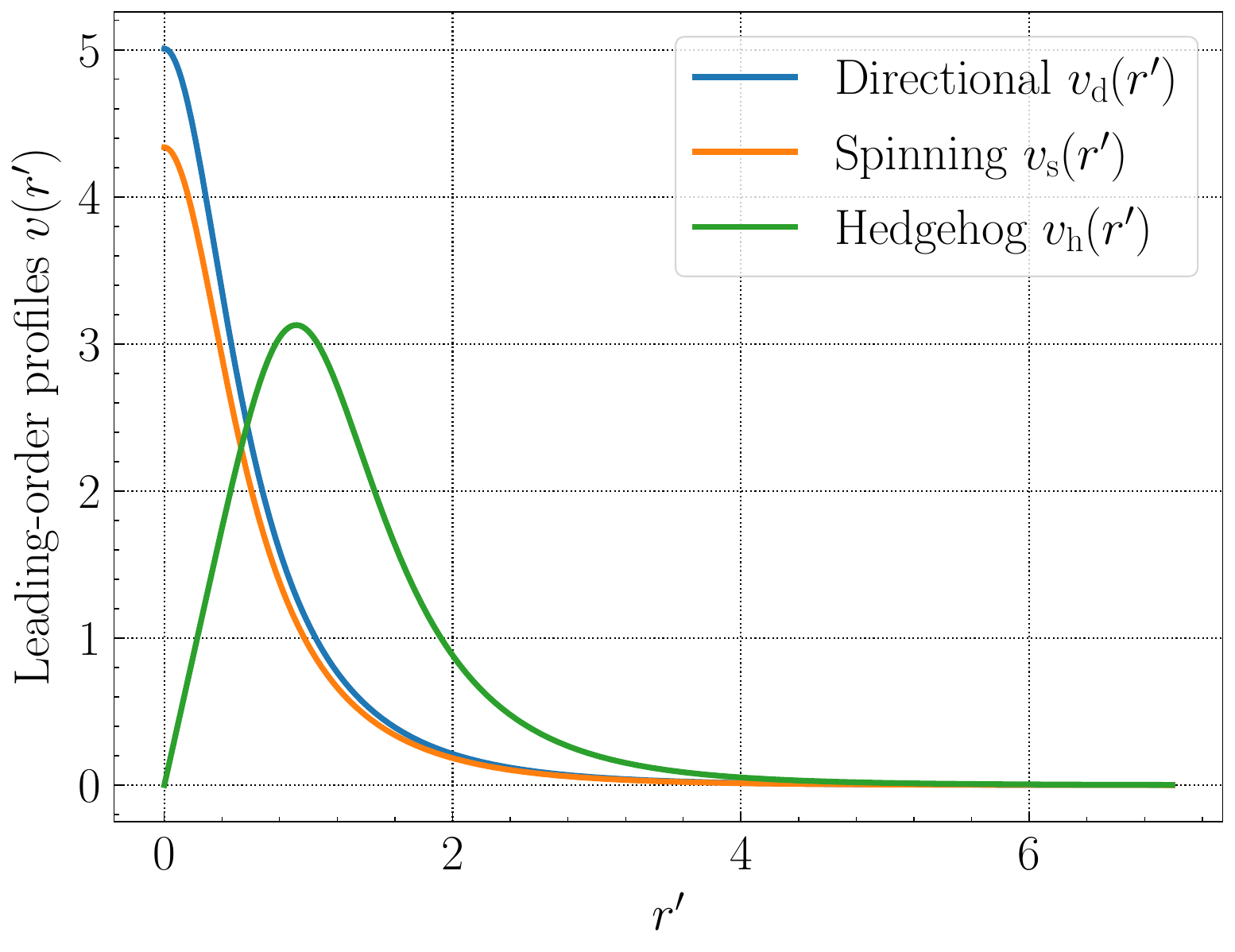}
	\caption{Leading-order profiles of small-amplitude directional, spinning and hedgehog oscillons. These are localized solutions of equations \eqref{EOM_vd}-\eqref{EOM_vh}.}
	\label{fig:smallprofile}
\end{figure}

By plugging the field ansatz into equation \eqref{EOM_Wi_order3} and by collecting the coefficient of $\sin(t')$, we obtain profile equations of the leading mode for directional, spinning and hedgehog oscillons respectively, i.e.
\begin{align}
	\label{EOM_vd}
	\pd_{r'}^2 v_\rm{d} + \frac{2}{r'}\pd_{r'} v_\rm{d} - v_\rm{d} - \frac{3}{4}\lambda v_\rm{d}^3 &= 0 ~,\\
	\label{EOM_vs}
	\pd_{r'}^2 v_\rm{s} + \frac{2}{r'}\pd_{r'} v_\rm{s} - v_\rm{s} - \lambda v_\rm{s}^3 &= 0 ~,\\
	\label{EOM_vh}
	\pd_{r'}^2 v_\rm{h} + \frac{2}{r'}\pd_{r'} v_\rm{h} - \( 1+\frac{2}{{r'}^2} \)v_\rm{h} - \frac{3}{4}\lambda v_\rm{h}^3 &= 0 ~.
\end{align}
Equations \eqref{EOM_vd} and \eqref{EOM_vs} resemble the profile equation for scalar oscillons \cite{Amin:2013ika}, and a localized solution can be obtained only if $\lambda<0$. The solutions of directional and spinning oscillons are related by $v_\rm{s} = \sqrt{3/4}v_\rm{d}$. As for equation \eqref{EOM_vh}, if we consider $v_\rm{h}(r')>0$, this means that $\partial_{r'}v_\rm{h}=0$ and $\partial^2_{r'}v_\rm{h}<0$ at some finite $r'$, namely $\left. \pd^2_{r'} v_\rm{h} \right|_{\pd_{r'}v_\rm{h}=0} = ( 1+2/{r'}^2 ) v_\rm{h} + \frac{3}{4}\lambda v_\rm{h}^3 < 0$. This equation is true only if $\lambda<0$. Hence \emph{attractive} self-interactions are required for all type of oscillon solutions to exist. These profile equations can be solved by numerical shooting method, and the solution is shown in figure \ref{fig:smallprofile}. In section \ref{sec:soliton_vector_subleading}, we go beyond the leading order and obtain the 3rd-order profiles for these oscillons.

\subsection{Energy and angular momentum}
\label{sec:soliton_vector_energy}
Plugging the small-amplitude expansions \eqref{expansion_coordinates} and \eqref{expansion_field} into the energy-momentum tensor \eqref{Tmunu}, we obtain the energy
\begin{align}
	E = \frac{1}{2\epsilon} \int d^3x' ~(\dot W_i^{(1)}  \dot W_i^{(1)} + W_i^{(1)} W_i^{(1)}) + \cal O(\epsilon) ~.
\end{align}
Since the first-order solution $W_i^{(1)}$ is obtained in the last subsection, we calculate the energy to be $5.01/\epsilon$, $18.9/\epsilon$ and $73.4/\epsilon$ for directional, spinning and hedgehog oscillons respectively. Although the energy at this order is the same for all three types of oscillons for a fixed particle number \eqref{particle_number}, in fact it is possible to lift the degeneracy and select the lowest-energy solution by appealing to the relation $dE/dN\approx \omega$.

\begin{figure}
	\centering
	\includegraphics[width=0.5\linewidth]{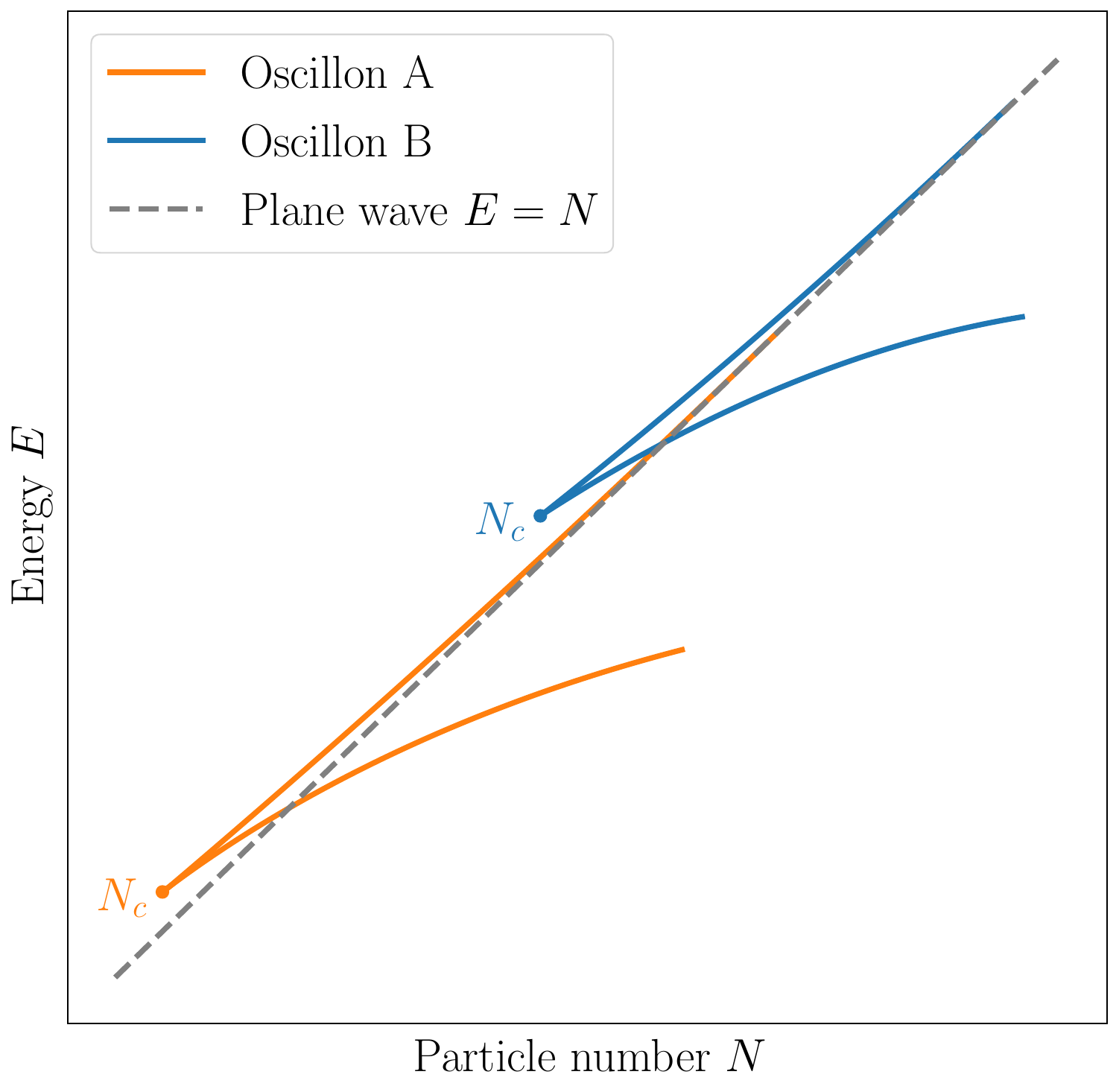}
	\caption{A schematic diagram for the relation between energy and particle numbers. Here the oscillon A has lower energy for a fixed particle number.}
	\label{fig:energyparticlen}
\end{figure}
A schematic diagram of $E$ vs $N$ for vector oscillons is shown in figure \ref{fig:energyparticlen}. There is a cusp $N_c$ at which $dE/d\omega = dN/d\omega = 0$, indicating a classical instability against fission, i.e. oscillons in the upper branch is not stable due to the curvature $\pd^2 E/\pd N^2 = \pd \omega/\pd N >0$. Suppose that we have two oscillons A and B with two different curves. As a rule of thumb, the one that has lower energy for the same frequency is the lower-energy state, to wit, oscillon A has lower energy for a fixed particle number. This is because we can usually extropolate one of the curve starting from the point such that they can be compared at the same particle number, by noting that the slope is $dE/dN=\omega<1$ everywhere and $\omega$ is monotonic when extending the curve towards a specific direction. Therefore, among the three type of configurations, directional and hedgehog oscillons are the lowest- and highest-energy state respectively, which is not surprising since hedgehog profiles have nodes which will contribute extra energy. We have verified this conclusion by explicitly calculating subleading profiles and energy.

The orbital and spin angular momentum to the leading order is
\begin{align}
	L_i &= -\frac{\epsilon_{ijk}}{\epsilon} \int d^3x' ~x'_j \pd_k W_l^{(1)} \dot W_l^{(1)} + \cal O(\epsilon) ~,\\
	S_i &= \frac{\epsilon_{ijk}}{\epsilon} \int d^3x' ~ W_j^{(1)} \dot W_k^{(1)} + \cal O(\epsilon) ~.
\end{align}
At this order the orbital and spin angular momentum actually are conserved separately, and the only non-zero component is the spin of spinning oscillons along the three-axis, i.e.
\begin{align}
	S_{\rm{s}3} = \frac{1}{\epsilon} \int d^3x' ~ v_\rm{s}^2 ~,
\end{align}
which equals its particle number \eqref{particle_number}. Thus we may interpret the spinning oscillon in the NR limit as a composition of particles of spin $1$. Nevertheless one should keep in mind that the spin within this definition can be any values between $S_{\rm{s}3}$ and $-S_{\rm{s}3}$ by performing a spatial rotation while the total spin is fixed. Similarly, directional and hedgehog oscillons may be recognized as collections of spin 0 particles.

\subsection{Beyond the leading order}
\label{sec:soliton_vector_subleading}
In this section, we follow the small-amplitude expansions and go beyond the leading order. Since solutions of $\b W^{(1)}$ have already been obtained, we can simply solve equation \eqref{EOM_Wi_order3} for $\b W^{(3)}$ and require coefficients of $\sin(t')$ and $\cos(t')$ in equation \eqref{EOM2} at order $\cal O(\epsilon^5)$ to be 0 in order to avoid the linear resonance. We can then solve these profiles equations by numerical shooting method.

\begin{figure}
	\centering
	\includegraphics[width=0.6\linewidth]{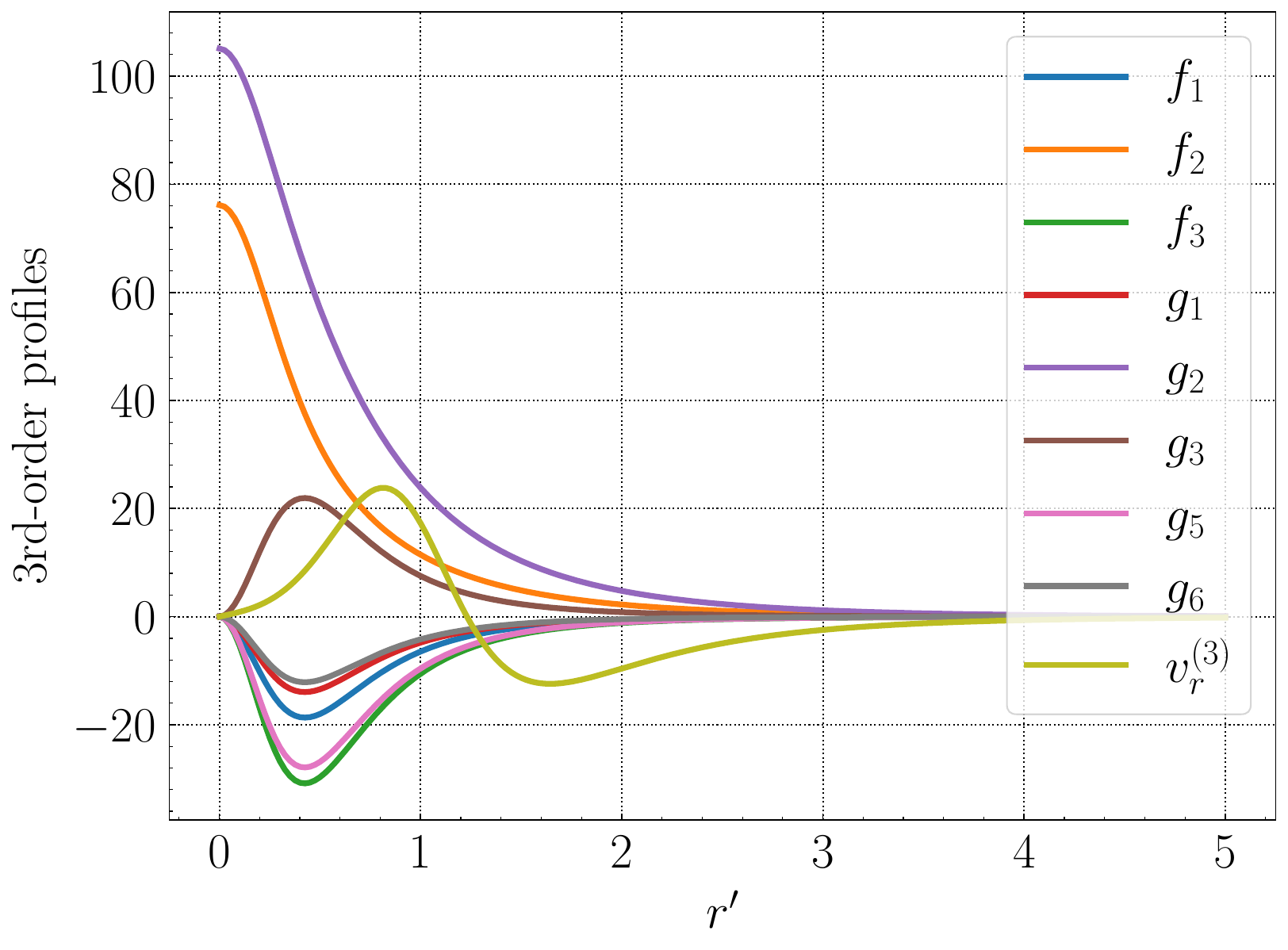}
	\caption{The 3rd-order profiles of small-amplitude oscillons. Here $f_n$, $g_n$ and $v_r^{(3)}$ are defined in equations \eqref{directional_3rd_order3}, \eqref{spinning_3rd_order3}, \eqref{spinning_3rd_order4} and \eqref{hedgehog_3rd_order1}.}
	\label{fig:subleadingprofile}
\end{figure}

For directional oscillons, equation \eqref{EOM_Wi_order3} gives
\begin{align}
	\label{Wm3_directional}
	W_m^{(3)} &= \frac{{x'}^m}{\rho'} [ u_\rho^{(3)} \cos (t') + v_\rho^{(3)} \sin(t') ] ~,\\
	W_3^{(3)} &= u_z^{(3)} \cos(t') + v_z^{(3)} \sin(t') - \frac{1}{32}\lambda v_\mathrm{d}^3 \sin(3t') ~,
\end{align}
where $m=1,2$ and $\rho'=\sqrt{{x'}^2+{y'}^2}$, and we have implemented the cylindrical symmetry in \eqref{Wm3_directional}. However, the profile equations of $u_\rho^{(3)}$ and $u_z^{(3)}$ given by the coefficient of $\cos(t')$ in equation \eqref{EOM2} at order $\cal O(\epsilon^5)$ are linear in $u_\rho^{(3)}$ and $u_z^{(3)}$, hence we must require $u_\rho^{(3)}=u_z^{(3)}=0$ (recall footnote \ref{footnote:even_order_terms}). Then the profile equations of $v_\rho^{(3)}$ and $v_z^{(3)}$ become
\small
\begin{align}
	\label{directional_3rd_order1}
	\(\nabla^2 - 1 - \frac{1}{{\rho'}^2} - \frac{3}{4}\lambda v_\rm{d}^2 \) v_\rho^{(3)} + \[ 4(\pd_{\rho'}v_\rm{d}) (\pd_{z'} v_\rm{d}) + 2v_\rm{d} \pd_{\rho'}\pd_{z'}v_\rm{d} \]\lambda v_\rm{d} = 0 ~,\\
	\label{directional_3rd_order2}
	\( \nabla^2 - 1 - \frac{9}{4}\lambda v_\rm{d}^2 \) v_z^{(3)} - \( \frac{3}{128}{\lambda}^2 + \frac{5}{8}h \) v_\rm{d}^5 + \[ \frac{17}{4} (\pd_{z'} v_\rm{d})^2 + 2 v_\rm{d} \pd_{z'}^2 v_\rm{d} \] \lambda v_\rm{d} = 0 ~,
\end{align}\normalsize
where $\nabla^2 = \pd_{\rho'}^2 + (1/\rho')\pd_{\rho'} + \pd_{z'}^2$. These equations are PDEs and can not be directly solved by numerical shooting. However, by redefining the field 
\begin{align}
	\label{directional_3rd_order3}
	v_\rho^{(3)}\equiv \frac{\rho' z'}{{r'}^2} f_1 \sep
	v_z^{(3)} \equiv \frac{{z'}^2}{{3r'}^2} (f_2+2f_3) + \frac{{\rho'}^2}{{3r'}^2} (f_2-f_3) ~,
\end{align}
we find that $f_n$ can be radially symmetric.  Equations \eqref{directional_3rd_order1} and \eqref{directional_3rd_order2} now become three ODEs, i.e.
\small
\begin{align}
	\( \nabla^2 - 1 - \frac{6}{{r'}^2} - \frac{3}{4}\lambda v_\rm{d}^2 \) f_1 + \[ 4(\pd_{r'}v_\rm{d})^2 + 2v_\rm{d}\pd_{r'}^2 v_\rm{d} -\frac{2}{r'} v_\rm{d}\pd_{r'} v_\rm{d} \] \lambda v_\rm{d} = 0 ~,\\
	\( \nabla^2 - 1- \frac{9}{4}\lambda v_\rm{d}^2 \) f_2 - \( \frac{9 {\lambda}^2}{128} + \frac{15 h}{8} \)v_\rm{d}^5 + \[ \frac{17}{4} (\pd_{r'}v_\rm{d})^2 + 2v_\rm{d} \pd_{r'}^2 v_\rm{d} + \frac{4}{r'} v_\rm{d} \pd_{r'}v_\rm{d} \] \lambda v_\rm{d} = 0 ~,\\
	\( \nabla^2 - 1 - \frac{6}{r^2} - \frac{9}{4} \lambda v_\rm{d}^2 \) f_3 + \[ \frac{17}{4} (\pd_{r'}v_\rm{d})^2 + 2v_\rm{d} \pd_{r'}^2 v_\rm{d} - \frac{2}{r'} v_\rm{d} \pd_{r'}v_\rm{d} \] \lambda v_\rm{d} = 0 ~,
\end{align}\normalsize
where $\nabla^2 = \pd_{r'}^2 + (2/r)\pd_{r'}$. The localized solutions of these equations for $h=1$ are shown in figure \ref{fig:subleadingprofile}.

For spinning oscillons, equation \eqref{EOM_Wi_order3} gives
\begin{align}
	W_i^{(3)} = u_i^{(3)} \cos(t') + v_i^{(3)} \sin(t') ~,
\end{align}
where $i=1,2,3$. By collecting coefficients of $\cos(t')$ and $\sin(t')$ in equation \eqref{EOM2} at order $\cal O(\epsilon^5)$, we obtain
\small
\begin{align}
	\label{spinning_3rd_order1}
	\( \nabla^2 -1 - \frac{5}{2}\lambda v_\rm{s}^2 \) u_1^{(3)} - \frac{1}{2} \lambda v_\rm{s}^2 v_2^{(3)} + \[ \frac{3}{4} (\pd_{y'} v_\rm{s})^2 +  \frac{17}{4} (\pd_{x'} v_\rm{s})^2 + 2 v_\rm{s} \pd_{x'}^2 v_\rm{s} \] \lambda v_\rm{s} - h v_\rm{s}^5 = 0 ~,\\
	\( \nabla^2 - 1 - \frac{5}{2}\lambda v_\rm{s}^2 \) v_2^{(3)} - \frac{1}{2} \lambda v_\rm{s}^2 u_1^{(3)} + \[ \frac{3}{4} (\pd_{x'} v_\rm{s})^2 +  \frac{17}{4} (\pd_{y'} v_\rm{s})^2 + 2 v_\rm{s} \pd_{y'}^2 v_\rm{s} \] \lambda v_\rm{s} - h v_\rm{s}^5 = 0 ~,\\
	\( \nabla^2 - 1 - \frac{3}{2}\lambda v_\rm{s}^2 \) v_1^{(3)} - \frac{1}{2}\lambda v_\rm{s}^2 u_2^{(3)} + \[ \frac{7}{2}(\pd_{x'} v_\rm{s})(\pd_{y'} v_\rm{s}) + 2 v_\rm{s} \pd_{x'}\pd_{y'} v_\rm{s} \] \lambda v_\rm{s} = 0 ~,\\
	\( \nabla^2 - 1 - \frac{3}{2}\lambda v_\rm{s}^2 \) u_2^{(3)} - \frac{1}{2}\lambda v_\rm{s}^2 v_1^{(3)} + \[ \frac{7}{2}(\pd_{x'} v_\rm{s})(\pd_{y'} v_\rm{s}) + 2 v_\rm{s} \pd_{x'}\pd_{y'} v_\rm{s} \] \lambda v_\rm{s} = 0 ~,\\
	\( \nabla^2 - 1 - \lambda v_\rm{s}^2 \) u_3^{(3)} + [ 4(\pd_{x'}v_\rm{s}) (\pd_{z'}v_\rm{s}) + 2 v_\rm{s} \pd_{x'}\pd_{z'} v_\rm{s} ] \lambda v_\rm{s} = 0 ~,\\
	\label{spinning_3rd_order2}
	\( \nabla^2 - 1 - \lambda v_\rm{s}^2 \) v_3^{(3)} + [ 4(\pd_{y'}v_\rm{s}) (\pd_{z'}v_\rm{s}) + 2 v_\rm{s} \pd_{y'}\pd_{z'} v_\rm{s} ] \lambda v_\rm{s} = 0 ~,
\end{align}\normalsize
where $\nabla^2=\pd_{x'}^2 + \pd_{y'}^2 + \pd_{z'}^2$. These equations are PDEs and can not be directly solved by numerical shooting. However, by redefining the field 
\begin{align}
	\label{spinning_3rd_order3}
	u_1^{(3)} - v_2^{(3)} \equiv \frac{{x'}^2-{y'}^2}{{r'}^2} g_1 \sep
	u_1^{(3)} + v_2^{(3)} \equiv \frac{{z'}^2}{3{r'}^2} (g_2 + 2g_3) + \frac{{x'}^2 + {y'}^2}{3{r'}^2} (g_2 - g_3) ~,\\
	\label{spinning_3rd_order4}
	u_2^{(3)} - v_1^{(3)} \equiv \frac{{x'}{y'}}{{r'}^2} g_4 \sep
	u_2^{(3)} + v_1^{(3)} \equiv \frac{x'y'}{{r'}^2} g_5 \sep
	u_3^{(3)} \equiv \frac{x'z'}{{r'}^2} g_6 \sep
	v_3^{(3)} \equiv \frac{y'z'}{{r'}^2} g_6 ~,
\end{align}
we find that $g_n$ can be radially symmetric and then equations \eqref{spinning_3rd_order1}-\eqref{spinning_3rd_order2} become six ODEs, i.e.
\begin{align}
	\( \nabla^2 - 1 - \frac{6}{{r'}^2} - 2\lambda v_\rm{s}^2 \) g_1 + 2 \lambda v_\rm{s}^2 \( \pd_{r'}^2 - \frac{2}{r'}\pd_{r'} \) v_\rm{s} = 0 ~,\\
	\( \nabla^2 - 1 - 3\lambda v_\rm{s}^2 \) g_2 + \[ 10 (\pd_{r'} v_\rm{s})^2 + 4 v_\rm{s} \pd_{r'}^2 v_\rm{s} + \frac{8}{r'} v_\rm{s}\pd_{r'} v_\rm{s} \] \lambda v_\rm{s} - 6 hv_\rm{s}^5 = 0 ~,\\
	\( \nabla^2 - 1 - \frac{6}{{r'}^2} - 3\lambda v_\rm{s}^2 \) g_3 - \[ 5 (\pd_{r'} v_\rm{s})^2 + 2 v_\rm{s} \pd_{r'}^2 v_\rm{s} - \frac{2}{r'} v_\rm{s}\pd_{r'} v_\rm{s}\] \lambda v_\rm{s} = 0 ~,\\
	\label{spinning_3rd_order5}
	\( \nabla^2 - 1 - \frac{6}{{r'}^2} - \lambda v_\rm{s}^2 \) g_4 = 0 ~,\\
	\( \nabla^2 - 1 - \frac{6}{{r'}^2} - 2\lambda v_\rm{s}^2 \) g_5 + \[ 7 (\pd_{r'} v_\rm{s})^2 + 4 v_\rm{s} \pd_{r'}^2 v_\rm{s} -\frac{4}{r'} v_\rm{s}\pd_{r'} v_\rm{s} \] \lambda v_\rm{s} = 0 ~,\\
	\( \nabla^2 - 1 - \frac{6}{{r'}^2} - \lambda v_\rm{s}^2 \) g_6 + \[ 4 (\pd_{r'} v_\rm{s})^2 + 2 v_\rm{s} \pd_{r'}^2 v_\rm{s} -\frac{2}{r'} v_\rm{s}\pd_{r'} v_\rm{s}\] \lambda v_\rm{s} = 0 ~,
\end{align}
where $\nabla^2 = \pd_{r'}^2 + (2/r')\pd_{r'}$. The equation \eqref{spinning_3rd_order5} is linear in $g_4$ hence we must require $g_4=0$, i.e. $u_2^{(3)} = v_1^{(3)}$. The localized solutions of non-vanished $g_n$ for $h=1$ are shown in figure \ref{fig:subleadingprofile}.

For hedgehog oscillons, equation \eqref{EOM_Wi_order3} gives
\begin{align}
	\label{hedgehog_3rd_order1}
	W_i^{(3)} = \frac{{x'}^i}{r'} [ u_r^{(3)} \cos(t') + v_r^{(3)} \sin(t') ] ~,
\end{align}
where $i=1,2,3$ and we have implemented the spherical symmetry. However, the profile equation of $u_r^{(3)}$ given by the coefficient of $\cos(t')$ in equation \eqref{EOM2} at $\cal O(\epsilon^5)$ is linear in $u_r^{(3)}$, hence we must require $u_r^{(3)}=0$. Then the profile equation of $v_r^{(3)}$ becomes
\small
\begin{align}
	\( \nabla^2 - 1 - \frac{2}{{r'}^2} - \frac{9}{4}\lambda v_\rm{h}^2 \) v_r^{(3)} + \( \frac{189}{128}\lambda^2 - \frac{5}{8}h \) v_\rm{h}^5 + \[ \frac{2}{{r'}^2} v_\rm{h}^2 + \frac{17}{4}(\pd_{r'}v_\rm{h})^2 + 2v_\rm{h}^2 \] \lambda v_\rm{h} = 0 ~,
\end{align}\normalsize
where $\nabla^2 = \pd_{r'}^2 + (2/r')\pd_{r'}$. The localized solution of this equation for $h=1$ is shown in figure \ref{fig:subleadingprofile}.

\subsection{Nonrelativistic effective field theory}
\label{sec:soliton_vector_eft}
In the previous sections, we show that directional oscillons are the lowest-energy state among the three types of vector oscillons. In this section, we prove that vector oscillons are ground states of the vector field for a fixed particle number by deriving a NR EFT. The approach we adopt here is similar to that in section \ref{sec:eft}. The resulting EFT comes with an approximate global $U(1)$ symmetry, which reflects the conservation of particle numbers in the NR regime. Therefore, vector oscillons may be regarded as a projection of Proca Q-balls onto the real space, and conversely Proca Q-balls are quasi-static approximations of vector oscillons.

We proceed by expanding the real vector field $W_\mu$ in terms of a complex field $Z_\mu$,
\begin{align}
	\label{NR_expansion}
	W_\mu(t,\b x) \equiv \frac{1}{\sqrt{2m}}\[ Z_\mu(t,\b x) e^{-imt} + Z_\mu^*(t,\b x) e^{imt} \] ~,
\end{align}
where the dependence of $Z_\mu$ on time is weak. Note that unlike the approach for deriving the NR EFT in section \ref{sec:eft}, here we do not assume other small parameters. By plugging this expansion into the action and by integrating out fast-oscillating modes, we can obtain anEFT for the NR field $Z_\mu$. To leading order, the effective lagrangian for slow modes is
\begin{align}
	\label{effective_lagrangian}
	\cal L_\mathrm{NR} = \frac{1}{2m} \( B_{0i}^* B_{0i} + i m Z_i^* B_{0i} - i m Z_i B_{0i}^* + m^2 Z_i^* Z_i - \frac{1}{2} B_{ij}^* B_{ij} \) - V_\mathrm{eff}(Z_\mu^* Z^\mu) ~,
\end{align}
where $B_{\mu\nu}\equiv \pd_\mu Z_\nu - \pd_\nu Z_\mu$. In fact the full lagrangian also contains terms with $e^{i nm t}$ where $n\neq 0$, but for our purpose here, these terms are neglected inasmuch as $Z_\mu$ is slowly varying and thus the time integral in action gives a delta function $\delta(nm)$. From \eqref{effective_lagrangian}, we find the conjugate momentum of $Z_\mu$ and $Z_\mu^*$ to be
\begin{align}
	\Pi_0 = 0 \sep
	\Pi_i = \frac{1}{2\omega}(B_{0i}^* + im Z_i^*) \sep
	\Pi_0^* = 0 \sep
	\Pi_i^* = \frac{1}{2\omega}(B_{0i} - im Z_i) ~.
\end{align}
The Hamiltonian density is then given by $\cal H_\mathrm{NR} = \Pi_i Z_i + \Pi_i^* Z_i^* - \cal L_\mathrm{NR}$, specifically
\small
\begin{align}
	\cal H_\mathrm{NR} =\frac{1}{2\omega} &\[ \pd_0 Z_i^* \pd_0 Z_i -  (\pd_i Z_0^* - im Z_i^*)(\pd_i Z_0 + im Z_i) + \frac{1}{2}B_{ij}^* B_{ij} \] + V_\mathrm{eff}(Z_\mu^* Z^\mu) ~.
\end{align}\normalsize
and the Hamilton's equation reads
\begin{align}
	\pd_0 Z_i = \frac{\delta H_\mathrm{NR}}{\delta \Pi_i} \sep
	\pd_0 \Pi_i = -\frac{\delta H_\mathrm{NR}}{\delta Z_i} ~,
\end{align}
where $H_\mathrm{NR} = \int d^3x ~\cal H_\mathrm{NR}$. Note that the Hamiltonian in this NR effective theory $\cal H_\mathrm{NR}$ is different from the one $\cal H$ that is obtained by plugging equation \eqref{NR_expansion} directly into \eqref{Tmunu}, i.e.
\begin{align}
	\la T^{00}\ra_T - \cal H_\mathrm{NR} = im \Pi_i^* Z_i^* - im \Pi_i Z_i ~,
\end{align}
where $\la\cdots\ra_T$ denotes time averaging over an oscillation period.

Since the effective lagrangian \eqref{effective_lagrangian} is invariant under a global $U(1)$ transformation $Z_\mu \rightarrow e^{-i\alpha} Z_\mu$ and $Z_\mu^* \rightarrow e^{i\alpha} Z_\mu^*$, there exists a conserved charge that may be identified as the particle number
\begin{align}
	\label{particle_number_Zi}
	N = \int d^3x ~j^0 \sep \text{where}\quad
	j^0 = i \Pi_i^* Z_i^* - i \Pi_i Z_i ~.
\end{align}
This definition of particle number is more general than equation \eqref{particle_number}, because in the derivations we only assume that $Z_\mu$ is weakly dependent on time, to wit, the oscillon frequency $\omega$ is close to $m$. This expression reduces to equation \eqref{particle_number} in the NR limit. In order to find the lowest-energy solution, we minimize energy by fixing the particle number
\begin{align}
	\label{variation}
	\nonumber
	\delta (E-\omega N) =  \int d^3 x ~ &[ \delta Z_i (-\pd_0 \Pi_i - i\mu \Pi_i) 
	+ \delta Z_i^* (-\pd_0 \Pi_i^* + i\mu \Pi_i^*) \\
	&+ \delta \Pi_i(\pd_0 Z_i - i\mu Z_i)
	+ \delta \Pi_i^*(\pd_0 Z_i^* + i\mu Z_i^*) ] = 0 ~,
\end{align}
where $\omega$ is a Lagrange multiplier, $E = \int d^3x ~\la\cal H\ra_T$ and $\mu\equiv m-\omega$. Since $\delta Z_i$, $\delta Z_i^*$, $\delta \Pi_i$ and $\delta \Pi_i^*$ are arbitrary small variations, we must require each bracket to vanish and this implies that $Z_i$ has a time denpendence $e^{i\mu t}$. This is the case for the vector oscillon solutions. From \eqref{variation}, we can see
\begin{align}
	\frac{dE}{dN} = \omega ~,
\end{align} 
which implies that $\omega$ is the chemical potential.

\subsection{Large-amplitude oscillons}
\label{sec:soliton_vector_large}
Oscillons with very small amplitudes are actually not stable, since they locate in the upper branch curve in figure \ref{fig:energyparticlen}. One way to avoid the instability is to consider flat-top vector oscillons by formulating small-amplitude expansions in terms of a large $W^6$ coupling $h$, analogous to the scalar case \cite{Amin:2010jq}. In what follows we will instead move on to the \emph{non-perturbative} regime and study large-amplitude oscillons by carrying out fully relativistic simulations in $3+1$ dimensions. For definiteness we set $h=1$ in potential \eqref{potential}, and vector oscillon properties should be insensitive to this parameter. See \cite{Zhang:2021xxa} for the detail of the numerical algorithm.

In order to see that the existence of vector oscillons is not too sensitive to the choice of initial conditions, we use a Gaussian ansatz $F\equiv C e^{-r^2/R^2}$ with $C\lesssim m/\sqrt{\lambda}$ and $R\sim 10\,m^{-1}$ to initialize vector field components for our two different oscillons. Depending on the choice of $C$ and $R$,  the fields latch on to oscillon configurations with different dominant frequency $\omega$ (after an initial transient).  For ease of comparison, we intentionally pick $C$ and $R$ so that in each case we get an oscillon with approximately the same $\omega\approx 0.975\,m$.

For the directional solitons, we start with an initial profile $\b W(t,\bx)|_{t=0}=F(r)\hat{\bz}$ and $\dot{\b W}(t,\bx)|_{t=0}=0$.  Within $t= \mathcal{O}(10^2)\,m^{-1}$, this initial Gaussian profile settles into an oscillon configuration with frequency $\omega \approx 0.975\,m$ and the energy $E_{\rm d}\approx 164\,m/\lambda$.  For this $\omega$, the energy of the oscillon from the NR approximation is $E_{\rm d}=mN+\mathcal{E}\approx 171 m/\lambda$ with a radius $R_{1/e}\approx 6m^{-1}$.

\begin{figure}
	\centering
	\includegraphics[width=\linewidth]{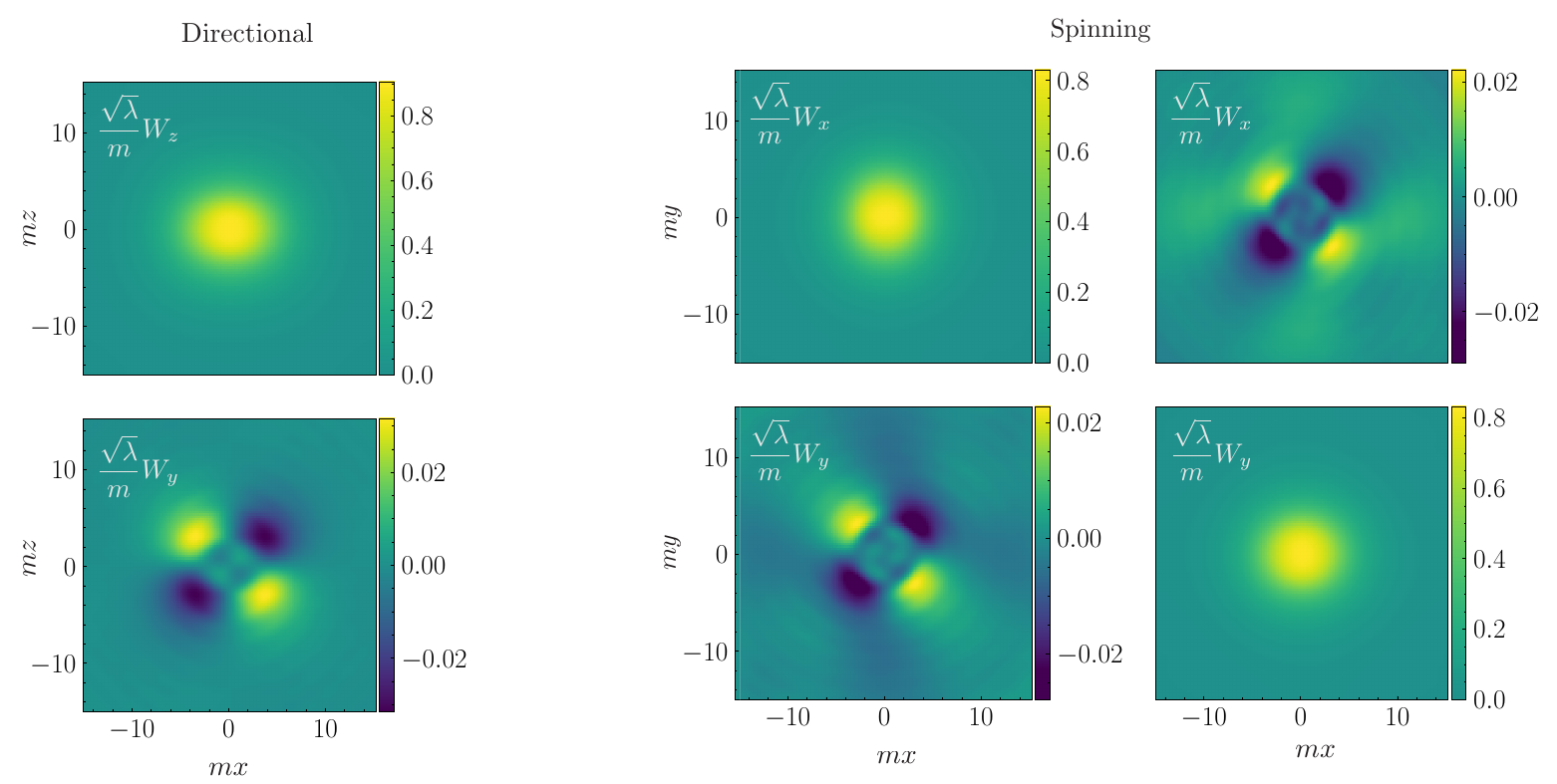}
	\caption{Left: The top panel is a snapshot of the profile of the $z$ component of a directional oscillon (with $\b W$ pointing predominantly in the $\hat{\bz}$ direction). The time is chosen so that the field component has a maximal central value. Bottom panel is the profile for the $y$ component of the field at this same time. Note that these profiles are provided on the $y=0$ plane. On the $z=0$ plane, the $x,y$ components vanish. Right: Snapshot of spatial profiles of the $x$(top) and $y$(bottom) components of $\b W$ for spinning oscillons, with $\b W$ rotating predominantly in the $x$-$y$ plane. In the first column, the time is chosen so that the $W_x$ is at its maximum in the center, whereas for the second column $W_y$ is at its maximum. Note that the deviation from spherical symmetry of the profile of dominant component is small. The subdominant component is small, and does not have spherically symmetric profiles. They are qualitatively consistent with the small-amplitude expansions in section \ref{sec:soliton_vector_leading} and \ref{sec:soliton_vector_subleading}, in particular, the relative amplitude and shape of the subdominant components.}
	\label{fig:largeprofiles}
\end{figure}
Due to relativistic effects, a small deviation of the field configuration from the $\hat{\bz}$ direction is expected, which is indeed observed in our simulations. See figure \ref{fig:largeprofiles} for snapshots of numerical profiles. In the quantities we have checked, such as profiles, energy etc., there is typically a few percent fractional difference between the results of the simulations and the NR solutions. This difference is consistent with our expectation that relativistic corrections should be of order  $|\nabla^2/m^2|\sim 1/(mR_{1/e})^2=\mathcal{O}(10^{-2})$.

Taking advantage of a cylindrical symmetry exhibited by directinal oscillons, we carry out long-time simulations in effectively $2+1$ dimensions with absorbing boundary conditions. After an initial transient, the oscillon does not show significant energy loss for the duration of the simulations ($\sim 10^5 m^{-1}$). We note that the lifetimes may be longer because of non-trivial suppression in the decay rates as seen in the case of scalar oscillons \cite{Zhang:2020bec, Zhang:2020ntm}.

In order to obtain spinning oscillons, we start the simulation with $\b W(t,\bx)|_{t=0} = F(r)\hat{\bx}$, $\dot{\b W}(t,\bx)|_{t=0} = F(r)\hat{\by}$. With these initial conditions, the field quickly settles into a spinning oscillon configuration with frequency $\omega\approx 0.975\,m$ and the energy $E_{\rm{s}}\approx 216\,m/\lambda$. Our analytic estimates yield $E_{\rm{s}}\approx 225\,m/\lambda$. Along with dominant components in the $x-y$ plane, we see small components in the $\hat{\bz}$ direction. Moreover, the energy density deviates slightly from spherical symmetry. Once again, the analytic estimates from our NR theory differ from the results from relativistic simulations by a few percent, consistent with our expectations.

Unlike the directional case, we cannot take advantage of symmetries to do a long-time simulation in effectively lower dimensions. However, we have verified that with absorbing boundary conditions, the spinning oscillon does not decay away for at least $\sim 10^3m^{-1}$. 

\begin{figure}
	\centering
	\includegraphics[width=0.6\linewidth]{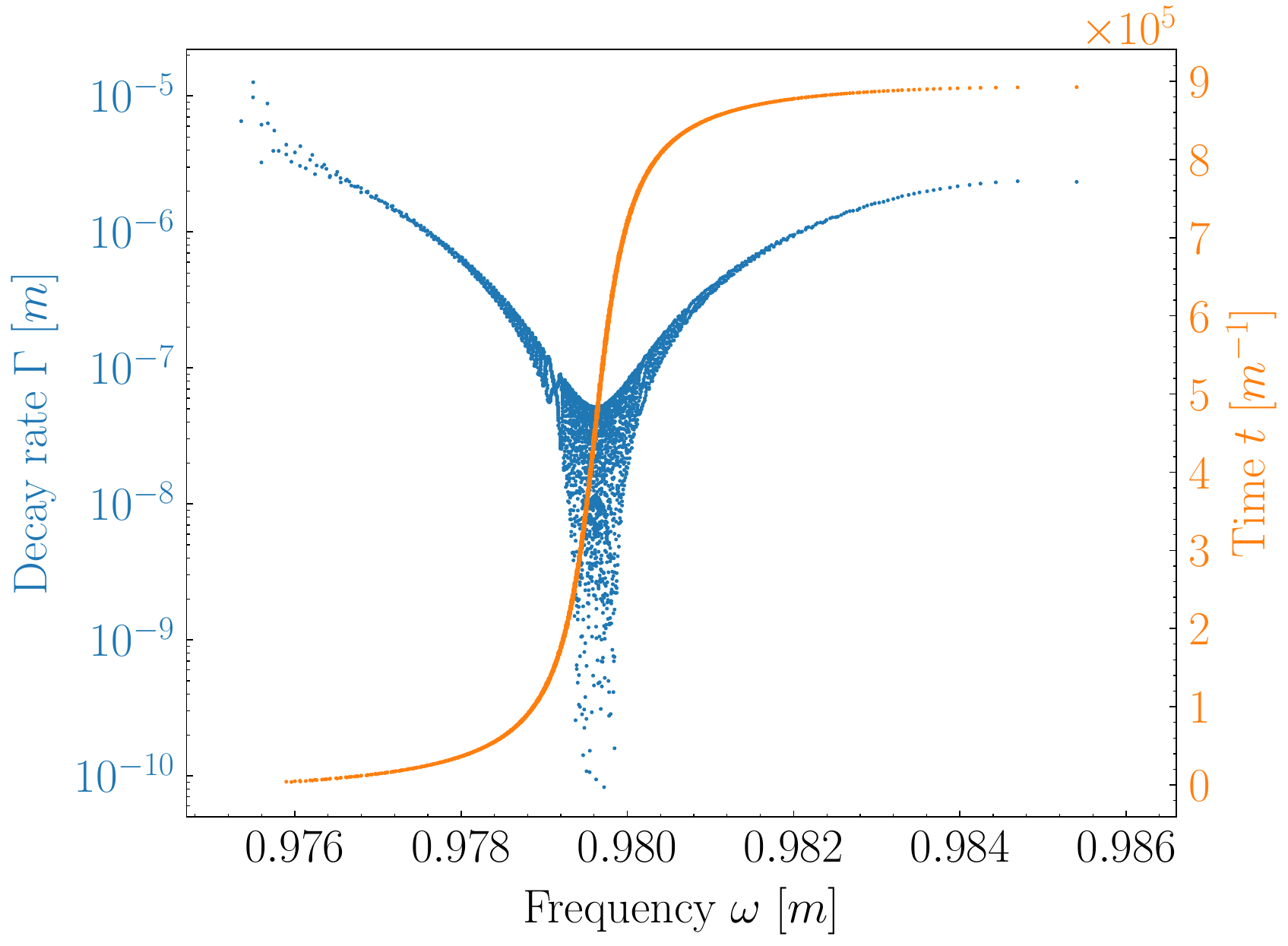}
	\caption{The decay rate (blue) and evolution time (orange) of hedgehog oscillons in terms of their central frequencies $\omega$. They evolve from small to large $\omega$ and have a long lifetime $\sim 10^{6} m^{-1}$ due to the existence of a dip in decay rates.}
	\label{fig:freqdecayratetime}
\end{figure}
As for hedgehog oscillons, the spherical symmetry of $\b W$ requires $\b W=0$ at the center and $\b W$ is an odd function of $x,y,z$. Hence initially we set $W_i(t,x,y,z)|_{t=0} = (x^i/r) C \sin\( \frac{\pi r}{2 R} \) e^{-r^2/R^2}$ while the time derivative is 0. The spherical symmetry makes the simulation of lifetimes for hedgehog oscillons much less CPU-intensive, since it reduces to an $1+1$-dimensional problem. With absorbing boundary conditions and with $C=1 ~m/\sqrt{|\lambda|}$ and $R=9 m^{-1}$ as initial conditions, we find the lifetime of hedgehog oscillons is $\sim 10^6 m^{-1}$, see figure \ref{fig:freqdecayratetime}.
\chapter{Neutron star cooling with lepton-flavor-violating axions}
\label{sec:ns}

\section{Introduction}
As an extension of the SM, there is no strong reason for the ultraviolet theory of axions to respect the lepton flavor symmetry, an accidental one of the SM broken by tiny neutrino masses. The axions whose ultraviolet theory is responsible for the breaking of the flavor symmetry are known as flavons or familons \cite{Davidson:1981zd, Wilczek:1982rv, Feng:1997tn, Bauer:2016rxs}, which can also explain the strong CP problem if they have a coupling to gluons \cite{Ema:2016ops, Calibbi:2016hwq}. Even if the underlying theory preserves lepton flavor, LFV effects can arise from radiative corrections \cite{Choi:2017gpf, Chala:2020wvs, Bauer:2020jbp, Bonilla:2021ufe}. It has been shown that LFV interactions can account for the production of dark matter through thermal freeze-in \cite{Panci:2022wlc}.\footnote{The proposal of using LFV axions to explain the anomalies related to the muon and electron magnetic moments \cite{Bauer:2019gfk} is ruled out by muonium-antimuonium oscillation constraints \cite{Endo:2020mev, Bauer:2021mvw}.} Tests of lepton flavor conservation thus provide important information about new physics.

Laboratory tests of lepton-flavor violation serve as an indirect probe of the axion's LFV interactions. Notably, charged lepton flavor violation would lead to rare lepton decays \cite{Calibbi:2017uvl}. If the axion were heavier than the muon, an effective field theory approach could be used to study decays such as $\mu \rightarrow e \gamma$, $\mu \rightarrow 3e$ and $\mu-e$ conversion, being the best process to detect LFV in the $e\mu$ sector.\footnote{In the SM, LFV decays are suppressed by the neutrino mass-squared difference and $\rm{Br}(\mu\rightarrow e\gamma) \sim \rm{Br}(\mu\rightarrow 3e) \sim 10^{-54}$ \cite{Petcov:1976ff, Calibbi:2017uvl, Hernandez-Tome:2018fbq}, far below the current experimental limits $\rm{Br}(\mu\rightarrow e\gamma) < 4.2 \times 10^{-13}$ \cite{MEG:2016leq} and $\rm{Br}(\mu\rightarrow 3e) < 1.0\times 10^{-12}$ \cite{SINDRUM:1987nra}.} For lighter axions, $\mu\rightarrow ea$ could be the dominating channel and the current limit on $\rm{Br}(\mu\rightarrow ea)$ is of order $10^{-6}$ \cite{Jodidio:1986mz} or $10^{-5}$ \cite{TWIST:2014ymv} depending on the axion mass and chirality of the interaction. The limit will be improved in the future experiments MEG II \cite{MEGII:2018kmf, Jho:2022snj} and Mu3e \cite{Perrevoort:2018ttp} by up to two orders of magnitude \cite{Calibbi:2020jvd}.

\begin{figure}
	\centering
	\includegraphics[width=0.4\linewidth]{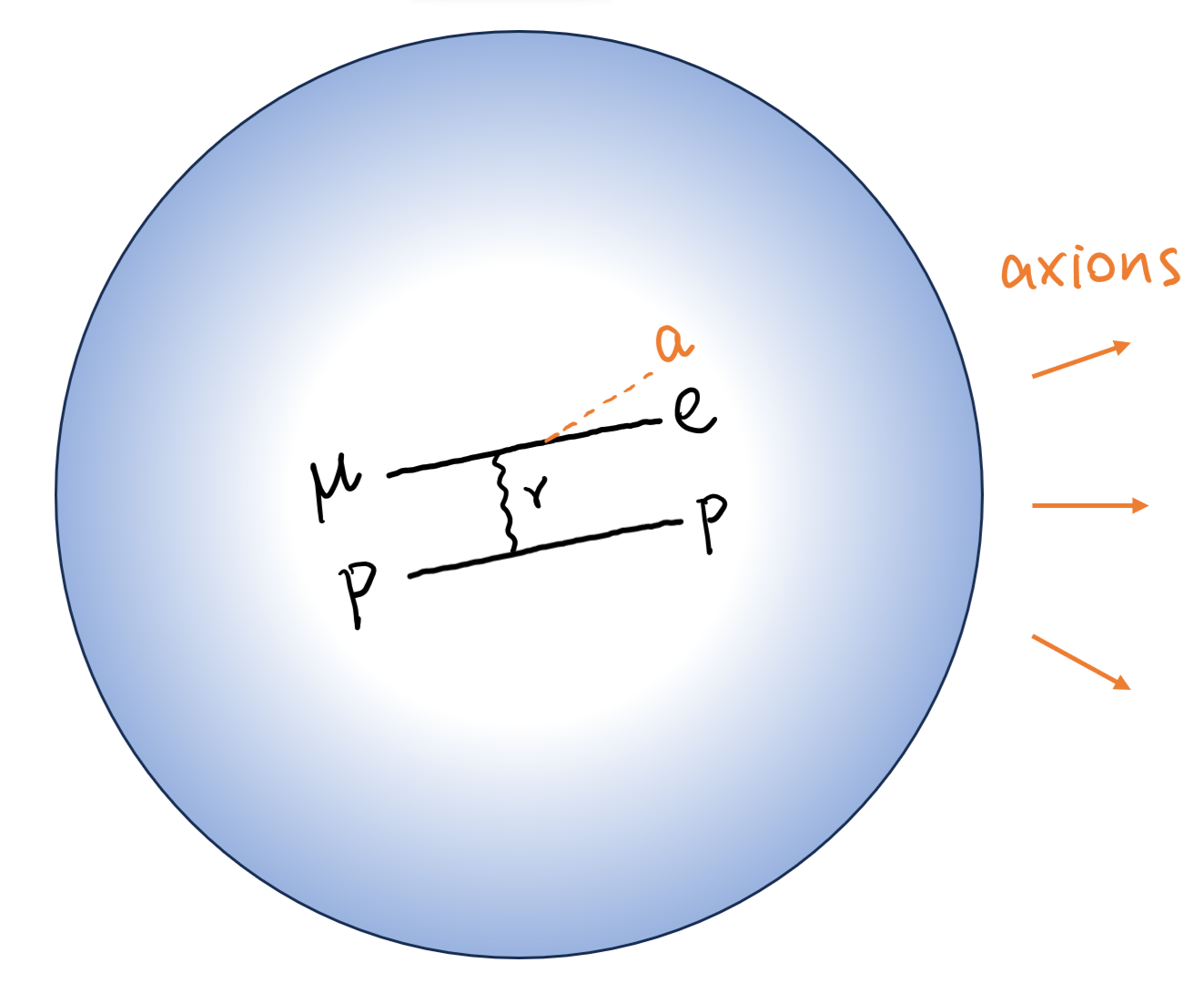}
	\caption{If axions are produced in NS cores, they will carry energy out of the star and make the NS cool down more efficiently than expected.}
	\label{fig:lfvemission}
\end{figure}
In this section, we aim to establish an astrophysical limit on the axion's LFV interactions based on NS cooling arguments, as a complement to current lab limits. The basic idea is illustrated in figure \ref{fig:lfvemission}; if axions are produced in NS cores, they must not carry energy out of the star more efficiently than standard neutrino-mediated cooling channels \cite{Raffelt:1990yz}. In a NS core, unlike nondegenerate stars or even white dwarf stars, the particle densities are so high that the electron Fermi energy exceeds the muon mass, and an appreciable population of muons is present \cite{Haensel:2007yy}. As such, NSs provide a unique opportunity to probe the axion's LFV coupling with muons and electrons.

\section{Neutron stars}
\label{sec:ns_ns}
When the density of a stellar object is high enough, electrons can combine with protons and form neutrons, and the gravitational force is balanced by the neutron degeneracy pressure. Such a compact object is called NSs. A typical NS is about 1 solar mass and $10 \rm{km}$, and the General Relativity effects become important since $GM/R \simeq0.15$. The average mass density is $\rho \sim 4.8\times 10^{14} \rm{g/cm^3}$.

\subsection{\texorpdfstring{$npe\mu$}{npeu} matter}
\label{sec:ns_ns_npemu}
For our purpose, it is sufficient to consider NSs with $npe\mu$ matter. The beta equilibrium and conservation of the baryon number and electric charge indicate \cite{Haensel:2007yy}
\begin{align}
	E_{F,\mu} = E_F \sep
	E_{F,n} = E_{F,p} + E_{F,e} \sep
	n_p = n_e + n_\mu ~,
\end{align}
where we have replaced the chemical potential by the Fermi energy by assuming all particles are degenerate. Given $E_{F,i} = \sqrt{m_i^2 +	 p_{F,i}^2} = \sqrt{m_i^2 + (3\pi^2 n_i)^{2/3}}$, if one of $\rho, n_n, n_p, n_e, n_\mu$ is chosen, the other four can be fully determined. These quantities are calculated in table \ref{tab:nscomposition}, where the nucleon effective mass (due to nuclear interactions) is taken as $0.8 m_N$.
\begin{table}
	\scalebox{0.51}{
	\begin{tabular}{|c|c|c|c|c|c|c|c|c|c|c|c|c|}
		\hline$\rho [\rm{g/cm^{-3}}]$ & $n_n [\rm{cm^{-3}}]$ & $p_{F,n} [\rm{MeV}]$ & $T_{F,n} [\rm{MeV}]$ & $n_p [\rm{cm^{-3}}]$ & $p_{F,p} [\rm{MeV}]$ & $T_{F,p} [\rm{MeV}]$ & $n_e [\rm{cm^{-3}}]$ & $p_{F,e} [\rm{MeV}]$ & $T_{F,e} [\rm{MeV}]$ & $n_\mu [\rm{cm^{-3}}]$ & $p_{F,\mu} [\rm{MeV}]$ & $T_{F,\mu} [\rm{MeV}]$ \\
		\hline $6.66 \times 10^{13}$ & $4.96 \times 10^{37}$ & 224 & 32.7 & $1.58 \times 10^{35}$ & 33.0 & 0.727 & $1.58 \times 10^{35}$ & 33.0 & 32.5 & 0 & 0 & 0 \\
		\hline $2.14 \times 10^{14}$ & $1.59 \times 10^{38}$ & 330 & 69.4 & $1.35 \times 10^{36}$ & 67.5 & 3.02 & $1.35 \times 10^{36}$ & 67.5 & 66.9 & 0 & 0 & 0 \\
		\hline $4.60 \times 10^{14}$ & $3.38 \times 10^{38}$ & 425 & 112 & $5.18 \times 10^{36}$ & 106 & 7.40 & $5.18 \times 10^{36}$ & 106 & 105 & 0 & 0 & 0 \\
		\hline $4.68 \times 10^{14}$ & $3.44 \times 10^{38}$ & 428 & 113 & $5.34 \times 10^{36}$ & 107 & 7.55 & $5.33 \times 10^{36}$ & 107 & 106 & $1.31 \times 10^{34}$ & 14.4 & 0.977 \\
		\hline $4.75 \times 10^{14}$ & $3.49 \times 10^{38}$ & 430 & 114 & $5.51 \times 10^{36}$ & 108 & 7.71 & $5.47 \times 10^{36}$ & 108 & 107 & $3.69 \times 10^{34}$ & 20.3 & 1.94 \\
		\hline $4.83 \times 10^{14}$ & $3.55 \times 10^{38}$ & 432 & 115 & $5.69 \times 10^{36}$ & 109 & 7.87 & $5.62 \times 10^{36}$ & 109 & 108 & $6.73 \times 10^{34}$ & 24.8 & 2.88 \\
		\hline $4.90 \times 10^{14}$ & $3.60 \times 10^{38}$ & 434 & 116 & $5.87 \times 10^{36}$ & 110 & 8.03 & $5.77 \times 10^{36}$ & 109 & 109 & $1.03 \times 10^{35}$ & 28.6 & 3.81 \\
		\hline $4.98 \times 10^{14}$ & $3.66 \times 10^{38}$ & 436 & 118 & $6.05 \times 10^{36}$ & 111 & 8.20 & $5.91 \times 10^{36}$ & 110 & 110 & $1.43 \times 10^{35}$ & 31.9 & 4.72 \\
		\hline $5.05 \times 10^{14}$ & $3.71 \times 10^{38}$ & 439 & 119 & $6.24 \times 10^{36}$ & 112  & 8.37 & $6.06 \times 10^{36}$ & 111 & 111 & $1.87 \times 10^{35}$ & 34.9 & 5.61 \\
		\hline $5.13 \times 10^{14}$ & $3.76 \times 10^{38}$ & 441 & 120 & $6.43 \times 10^{36}$ & 114 & 8.54 & $6.20 \times 10^{36}$ & 112 & 112 & $2.34 \times 10^{35}$ & 37.6 & 6.50 \\
		\hline $5.14 \times 10^{14}$ & $3.77 \times 10^{38}$ & 441 & 120 & $6.46 \times 10^{36}$ & 114 & 8.56 & $6.22 \times 10^{36}$ & 112 & 112 & $2.41 \times 10^{35}$ & 38.0 & 6.62 \\
		\hline $6.09 \times 10^{14}$ & $4.45 \times 10^{38}$ & 466 & 133 & $9.25 \times 10^{36}$ & 128 & 10.9 & $8.16 \times 10^{36}$ & 123 & 122 & $1.09 \times 10^{36}$ & 62.8 & 17.3 \\
		\hline $7.24 \times 10^{14}$ & $5.27 \times 10^{38}$ & 493 & 147 & $1.33 \times 10^{37}$ & 144 & 13.8 & $1.07 \times 10^{37}$ & 135 & 134 & $2.54 \times 10^{36}$ & 83.3 & 28.9 \\
		\hline $8.65 \times 10^{14}$ & $6.26 \times 10^{38}$ & 522 & 164 & $1.88 \times 10^{37}$ & 162 & 17.4 & $1.41 \times 10^{37}$ & 147 & 147 & $4.76 \times 10^{36}$ & 103 & 41.7 \\
		\hline $1.04 \times 10^{15}$ & $7.46 \times 10^{38}$ & 554 & 182 & $2.64 \times 10^{37}$ & 182 & 21.7 & $1.84 \times 10^{37}$ & 161 & 161 & $7.96 \times 10^{36}$ & 122 & 55.6 \\
		\hline $1.25 \times 10^{15}$ & $8.92 \times 10^{38}$ & 588 & 202 & $3.67 \times 10^{37}$ & 203 & 26.9 & $2.42 \times 10^{37}$ & 177 & 176 & $1.25 \times 10^{37}$ & 141 & 70.9 \\
		\hline $1.50 \times 10^{15}$ & $1.07 \times 10^{39}$ & 624 & 225  & $5.04 \times 10^{37}$ & 226 & 33.1 & $3.18 \times 10^{37}$ & 193 & 193 & $1.87 \times 10^{37}$ & 162 & 87.7 \\
		\hline $1.82 \times 10^{15}$ & $1.29 \times 10^{39}$ & 664 & 251 & $6.88 \times 10^{37}$ & 250 & 40.6 & $4.17 \times 10^{37}$ & 212 & 211 & $2.71 \times 10^{37}$ & 183 & 106 \\
		\hline $2.21 \times 10^{15}$ & $1.55 \times 10^{39}$ & 707 & 280 & $9.32 \times 10^{37}$ & 277  & 49.4 & $5.47 \times 10^{37}$ & 232 & 231 & $3.86 \times 10^{37}$ & 206 & 126 \\
		\hline $2.70 \times 10^{15}$ & $1.88 \times 10^{39}$ & 753 & 313 & $1.26 \times 10^{38}$ & 306 & 59.9 & $7.18 \times 10^{37}$ & 254 & 253 & $5.39 \times 10^{37}$ & 231 & 148 \\
		\hline $3.30 \times 10^{15}$ & $2.28 \times 10^{39}$ & 804 & 349 & $1.69 \times 10^{38}$ & 337 & 72.3 & $9.42 \times 10^{37}$ & 278 & 277 & $7.45 \times 10^{37}$ & 257 & 172 \\
		\hline $4.06 \times 10^{15}$ & $2.79 \times 10^{39}$ & 859 & 390 & $2.25 \times 10^{38}$ & 372 & 86.9 & $1.24 \times 10^{38}$ & 304 & 304 & $1.02 \times 10^{38}$ & 285 & 198 \\
		\hline
	\end{tabular}}
	\caption{NS composition in the $npe\mu$ model. Here $T_F$ is the Fermi temperature. No muons are present if $E_{F}<m_\mu$, i.e. $\rho<4.6\times 10^{14}\rm{g/cm^3}$.}
	\label{tab:nscomposition}
\end{table}
In the following we will consider NSs with energy density $\rho>4.6\times 10^{14} \rm{g/cm^3}$ such that muons are produced. All $npe\mu$ particles are degenerate if the temperature $T<10^{10} \rm{K}$. The incorporation of nuclear interactions would slightly change the result. For example, the table I of \cite{cohen1970neutron} indicates a muon threshold at $\rho\approx 2.2 \times 10^{14} \rm{g/cm^3}$.

\subsection{Cooling}
\label{sec:ns_ns_cooling}
Once formed, NSs lose energy through the emission of particles and the temperature decreases with time. Eventually thermal radiation of photons will dominate the energy loss of the stars. Here we summarize the cooling stories presented in \cite{Lattimer:1991ib, Pethick:1991mk}. Also see section 4.8 and 4.9 in \cite{Raffelt:1996wa}.

In the simplest scenario, NSs cool down by emitting neutrinos. If the proton density is bigger than $11.1\%$--$14.8\%$, for which we take the upper value if $\mu_e\gg m_\mu$ or the lower value if no muons are present, the neutrino emission is dominated by the direct Urca process \cite{Lattimer:1991ib},
\begin{align}
	\label{ns_urca}
	n\rightarrow p + e^- + \bar\nu_e \sep
	p + e^- \rightarrow n + \nu_e ~.
\end{align}
The threshold proton concentration is higher for direct Urca process with muons. If the direct Urca processes can happen, the emissivity for \eqref{ns_urca} is given by\footnote{The numerical value given in \cite{Lattimer:1991ib} is with respect to $n_0=0.16 \rm{fm^{-3}}$. Here we have used a more recent value $n_0=0.15 \rm{fm^{-3}}$ \cite{Horowitz:2020evx}.} \cite{Lattimer:1991ib}
\begin{align}
	\nonumber
	\ve_\rm{URCA} &= \frac{457\pi}{10080} G_F^2 \cos^2\theta_C (1+3g_A^2) m_n m_p \mu_e T^6 \\
	&= 3.9\times 10^{27} \(\frac{Y_e n}{n_0}\)^{1/3} \(\frac{T}{10^9 \rm{K}}\)^6 \rm{erg~cm^{-3}~s^{-1}} ~,
\end{align}
where $G_F\simeq 1.436\times 10^{-49} \rm{erg~cm^{-3}}$ is the Fermi's constant, $\theta_C\simeq 0.239$ is the Cabibbo angle, $g_A\simeq -1.261$ is the axial-vector coupling constant, $n=n_n+n_p$ is the baryon density, $Y_e$ is the number density per baryon, and $n_0=0.15 \rm{fm^{-3}}$ is the nuclear saturation density. The numerical value here is obtained by using the physical quantities in the vacuum, and interactions will change the neutron and proton mass, $g_A$ and weak-interaction matrix element \cite{Lattimer:1991ib}.

In case where the direct Urca processes are Pauli blocked, one needs to add a bystander particle and obtain the modified Urca processes,
\begin{align}
	n+n\rightarrow n + p + e^- + \bar\nu_e \sep
	n+p+e^-\rightarrow n+n+\nu_e ~.
\end{align}
Without nucleon superfluidity, the modified Urca process is the dominating cooling channel at temperatures down to $2\times 10^8\rm{K}$ \cite{SOYEUR1979464}, below which the photon radiation takes over. The emissivity is given by (57), (64c) and (75) in \cite{Friman:1979ecl}\footnote{The numerical value given in \cite{Friman:1979ecl} is with respect to $n_0=0.17 \rm{fm^{-3}}$. Here we have used a more recent value $n_0=0.15 \rm{fm^{-3}}$ \cite{Horowitz:2020evx}.}
\begin{align}
	\nonumber
	\ve_\rm{MURCA} &= \frac{21}{16} \times \frac{11513}{60480} \frac{G_F^2 g_A^2 {m_n^*}^3 m_p^*}{2\pi} \( \frac{f}{m_\pi} \)^4 p_F \alpha T^8 \\
	\label{ns_murca}
	&= 4.4 \times 10^{21} \( \frac{m_n^*}{0.8 m_n} \)^3 \( \frac{m_p^*}{0.8 m_p} \) \( \frac{\rho}{6\rho_0} \)^{2/3} \( \frac{T}{10^9 \rm{K}} \)^8 \rm{erg~cm^{-3}~s^{-1}} ~,
\end{align}
where $21/16$ accounts for the inclusion of the exchange diagrams (i.e. equation (75)), $\alpha\sim 1$ is a numerical factor, the superscript $*$ stands for effective masses.

The Urca processes can also happen for exotic states such as pion and kaon condensate and quark matter, if they exist in NSs. The emissivities are given by (19), (20), (26) in \cite{Pethick:1991mk}, and are smaller compared to that of nucleons.

Dense NS matter ($n>2 n_0$) may contain a significant fraction of hyperons. The emissivity of the hyperon Urca processes, if they are allowed, are comparable to that expected for exotic states. The threshold concentration of $\Lambda$'s is quite small, and thus minute traces of A hyperons would be a very effective refrigerant. For processes without strangeness change, such as $\Sigma^-\rightarrow \Lambda + e^- + \bar\nu_e$, the emissivity is not Cabibbo suppressed and would be comparable to that of the nucleon Urca process. Whether any of these processes can occur in NSs is uncertain because of our ignorance of interactions among hyperons, isobars, and nucleons at densities well above nuclear density.

At some densities and low temperatures neutrons may be superfluid and/or protons be superconducting (analogous to Cooper pairs of electrons in metallic superconductors). Because neutrinos are produced only by thermal excitations and not by paired nucleons, superfluidity and superconductivity reduce neutrino production rates. The direct Urca rate is then reduced by a factor of $e^{-\Delta/T}$ and the modified Urca rate by $e^{-2\Delta/T}$, where $\Delta$ is the larger of the neutron and proton paring gaps and typically a few hundred $\rm{keV}$. In this case, the Cooper pair breaking and formation provides the dominant neutrino production channels. Medium effects for neutrino emission processes are discussed in \cite{Potekhin:2015qsa, Voskresensky:1986af, Grigorian:2016leu}.

Eventually photon radiation will dominate the cooling. The emissivity is
\begin{align}
	\ve_\gamma = \frac{4\pi R_\rm{NS}^2 \sigma_\rm{SB} T^4}{4\pi R_\rm{NS}^3/3} = \frac{3\sigma_\rm{SB}}{R_\rm{NS}} T^4 = 1.7 \times 10^{26} \rm{erg~cm^{-3}~s^{-1}} \(\frac{10 \rm{km}}{R_\rm{NS}}\) \( \frac{T}{10^9 K} \)^4 ~,
\end{align} 
where $\sigma_\rm{SB} = \pi^2/60$ is the Stefan-Boltzmann constant and $T$ refers to the surface temperature rather than the core temperature.

\section{Lepton-flavor-violating axions}
\label{sec:ns_lfv}
Due to the particle content of the $npe\mu$ matter, we consider a LFV coupling among the electron, muon, and axion,
\begin{align}
	\label{lfv}
	\cal{L}_\rm{LFV} = \frac{g_{ae\mu}}{m_e + m_\mu} \bar\Psi_e \gamma^\rho \gamma_5 \Psi_\mu \, \pd_\rho a \ + \ \mathrm{h.c.} ~,
\end{align}
where $\Psi_e(x)$ is the electron field, $\Psi_\mu(x)$ is the muon field, $a(x)$ is the axion field, $m_e \approx 0.511 \rm{MeV}$ is the electron mass, $m_\mu \approx 106 \rm{MeV}$ is the muon mass, and $g_{ae\mu}$ is the axion's LFV coupling. The coupling may also be written in terms of the axion decay constant $f_a$ as $g_{ae\mu} = C_{ae\mu}(m_e + m_\mu)/(2f_a)$. This interaction can naturally arise, e.g., in the models of the LFV QCD axion, the LFV axiflavon, the leptonic familon and the majoron (see \cite{Calibbi:2020jvd} and references therein). Past studies of charged lepton flavor violation, from both terrestrial experiments and cosmological/astrophysical observations, furnish constraints on the LFV coupling $g_{ae\mu}$, which we summarize here.

The LFV interaction opens an exotic decay channel for the muon $\mu\rightarrow ea$, as long as the axion mass is not too large $m_a < m_\mu - m_e$.  The branching ratio is predicted to be \cite{Bjorkeroth:2018dzu}
\begin{align}
	\mathrm{Br}(\mu \rightarrow ea) \approx \frac{\Gamma(\mu\rightarrow ea)}{\Gamma(\mu\rightarrow e \nu \bar\nu)} = 7.0 \times 10^{15} g_{ae\mu}^2 ~.
\end{align}
Initial searches for the two-body muon decay were performed by Derenzo using a magnetic spectrometer, resulting in an upper limit on the branching ratio of $2 \times 10^{-4}$ for the mass range $98.1$--$103.5 \rm{MeV}$ \cite{Derenzo:1969za}. Jodidio et al. constrained the branching ratio for a massless familon to be $< 2.6\times 10^{-6}$, which was later extended to massive particles up to $\sim 10 \rm{MeV}$ \cite{Calibbi:2020jvd}. Bryman \& Clifford analyzed data of muon and tauon decays obtained from NaI(Tl) and magnetic spectrometers, concluding an upper limit of $3 \times 10^{-4}$ for masses less than $104 \rm{MeV}$ \cite{Bryman:1986wn}. Bilger et al. studied muon decay in the mass range $103$--$105 \rm{MeV}$ using a high purity germanium detector and established a limit of $5.7 \times 10^{-4}$ \cite{Bilger:1998rp}, while the PIENU collaboration improved the limit in the mass range $87.0$--$95.1 \rm{MeV}$ \cite{PIENU:2020loi}. The TWIST experiment performed a broader search for masses up to $\sim 80 \rm{MeV}$ by accommodating nonzero anisotropies, resulting in an upper limit of $2.1\times 10^{-5}$ for massless axions \cite{TWIST:2014ymv}. These constraints on $\mathrm{Br}(\mu \rightarrow ea)$ translate into upper limits on the LFV coupling $g_{ae\mu}$, and we summarize the current status in Tab.~\ref{tab:lfv_constraint}.
\begin{table*}
	\centering
	\small
	\setlength\tabcolsep{0.8pt}
	\begin{tabular}{|c|c|c|c|c|c|}
		\hline
		$|g_{ae\mu}|$ & $\frac{2f_a}{C_{ae\mu}} ~[\rm{GeV}]$ & $\rm{Br}(\mu\rightarrow ea)$ & $m_a ~ [\rm{MeV}]$ & Experiment & Reference\\
		\hline
		$<3.0 \times 10^{-6}$ & $>3.5\times 10^4$ & $<1.0$ & $\lesssim 1$ & NS cooling & This work \\
		\hline
		$\lesssim 8\times 10^{-10}$ & $\gtrsim 1\times 10^8$ & $\lesssim 4 \times 10^{-3}$ & $ \lesssim 50$ & SN 1987A, $\mu\rightarrow ea$ & \cite{Calibbi:2020jvd} \\
		\hline
		$<4.2\times 10^{-10}$ & $>2.5\times 10^8$ & $<1.3 \times 10^{-3}$ & $\lesssim 10^{-7}$ & Cosmology, $\Delta N_\rm{eff}$ & \cite{DEramo:2021usm} \\
		\hline
		$<2.9\times 10^{-10}$ & $>3.7\times 10^8$ & $<5.7 \times 10^{-4}$ & $103-105$ & Rare muon decay & \cite{Bilger:1998rp} \\
		\hline
		$\lesssim 2\times 10^{-10}$ & $\gtrsim 5\times 10^8$ & $\lesssim 3 \times 10^{-4}$ & $<104$ & Rare muon decay & \cite{Bryman:1986wn} \\
		\hline
		$<2\times 10^{-10}$ & $>6\times 10^8$ & $<2 \times 10^{-4}$ & $98.1 - 103.5$ & Rare muon decay & \cite{Derenzo:1969za} \\
		\hline
		$<1\times 10^{-10}$ & $>9\times 10^8$ & $< 1\times 10^{-4}$ & $47.8 - 95.1$ & Rare muon decay (PIENU)\footnote{The PIENU collaboration obtained upper limits on the branching ratio from $10^{-4}$ to $10^{-5}$ for the considered mass range.} & \cite{PIENU:2020loi} \\
		\hline
		$<5.5\times 10^{-11}$ & $>1.9\times 10^9$ & $<2.1 \times 10^{-5}$ & $< 13$ & Rare muon decay (TWIST) & \cite{TWIST:2014ymv} \\
		\hline
		$\lesssim 4\times 10^{-11}$ & $\gtrsim 3\times 10^9$ & $\lesssim 9 \times 10^{-6}$ & $ \lesssim 50$ & SN 1987A, $lf\rightarrow l'fa$ & This work \\
		\hline
		$<1.9\times 10^{-11}$ & $>5.5\times 10^9$ & $<2.6 \times 10^{-6}$ & $\lesssim 10$ & Rare muon decay & \cite{Jodidio:1986mz, Calibbi:2020jvd} \\
		\hline
	\end{tabular}
	\normalsize
	\caption{A summary of constraints on the axion's LFV coupling in the $e$-$\mu$ sector, where stronger constraints are presented at the bottom. See the main text for more detailed descriptions. For the NS cooling limit, we calculate the axion emissivity via $l+f \to l' + f + a$ and compare with the neutrino emissivity via the modified Urca channels. For the SN 1987A limit, we compare with the upper bound on energy loss rate. }
	\label{tab:lfv_constraint}
\end{table*}

Apart from terrestrial experiments, cosmological and astrophysical observations also constrain the axion's LFV interaction. If this interaction were too strong, relativistic axions would be produced thermally in the early universe; however, the presence of a dark radiation in the universe is incompatible with observations of the cosmic microwave background anisotropies. Constraints on dark radiation are typically expressed in terms of a parameter $N_\mathrm{eff}$ called the effective number of neutrino species.  A recent study of flavor-violating axions in the early universe finds that current observational limits on $N_\mathrm{eff}$ require the LFV coupling to obey $|2f_a/C_{ae\mu}| > 2.5 \times 10^{8} \rm{GeV}$ \cite{DEramo:2021usm}. Astrophysical probes of the axion's LFV interaction have not been extensively explored.  Calibbi et al. considered the bound on $\mathrm{Br}(\mu \to ea)$ from SN 1987A associated with the cooling of the proto-NS \cite{Calibbi:2020jvd}.  Assuming that the dominant energy loss channel is free muon decay $\mu \to ea$, they derive an upper limit on the branching ratio at the level of $4\times 10^{-3}$. We find that a stronger constraint is obtained from the 2-to-3 scattering channels, such as $\mu p \to e p a$, and we discuss this result further below.

To provide a comprehensive overview, we also introduce the constraints on LFV couplings involving $\tau$ leptons. Currently, laboratory limits on the branching ratios of rare tauon decays are $\rm{Br}(\tau\rightarrow ea)<2.7\times 10^{-3}$ and $\rm{Br}(\tau\rightarrow \mu a) < 4.5\times 10^{-3}$ \cite{ARGUS:1995bjh, Calibbi:2020jvd}. Constraints from $N_\rm{eff}$ are more stringent, $\rm{Br}(\tau\rightarrow ea)\lesssim 3\times 10^{-4}$ and $\rm{Br}(\tau\rightarrow \mu a) \lesssim 5\times 10^{-4}$ \cite{DEramo:2021usm}. Each of these limits is expected to improve significantly, by up to three orders of magnitude, in the future Belle II \cite{Belle-II:2018jsg, Calibbi:2020jvd} and CMB-S4 experiment \cite{CMB-S4:2016ple, Abazajian:2019eic, DEramo:2021usm}. However, it remains challenging to impose constraints on $\tau$ leptons from astrophysical systems due to their considerable mass of $1.8 \rm{GeV}$, which far exceeds stellar core temperatures.

\section{Axion emission}
\label{sec:ns_emissivity}

The emission of axions from NS matter via the LFV interaction can proceed through various channels. One might expect the dominant channel to be the decay of free muons $\mu \to ea$; however, since the electrons in NS matter are degenerate, this channel is Pauli blocked, and its rate is suppressed in comparison with scattering channels. Since NS matter consists of degenerate electrons, muons, protons, and neutrons, various scattering channels are available. We denote these collectively as\footnote{We neglect the Compton process for axions, since the number density of photons is low compared to other particles.}
\begin{align}
	\label{lfv1}
	l + f \rightarrow l' + f + a ~,
\end{align}
where a lepton $l=e,\mu$ is converted to another $l'=\mu,e$ with the spectator particle $f=p, e, \mu$. We consider channels in which the neutron star's muon is present in the initial state, and channels in which muons are created thanks to the large electron Fermi momentum. The scattering is mediated by the electromagnetic interaction (photon exchange), and channels involving neutrons are neglected. Assuming that all particles are degenerate, scattering predominantly happens for particles at the Fermi surface. These processes are kinematically allowed if $|p_{F,l} - p_{F,f}| < p_{F,l'} + p_{F,f}$ and $|p_{F,l'} - p_{F,f}| < p_{F,l} + p_{F,f}$, implying the existence of a threshold momentum of the spectator particle
\begin{align}
	\label{threshold}
	p_{F,f} > \frac{p_{F,e} - p_{F,\mu}}{2} ~.
\end{align}
Here we have introduced the Fermi momentum $p_{F,i}$ of the particle species $i$.

The quantities of interest are the axion emissivities $\ve_a^{(lf)}$, which corresponds to the energy released in axions per unit volume per unit time through the channel $lf \to l^\prime f a$. We assign $(E_1,\b p_1)$ and $(E_1',\b p_1')$ for the initial and final four-momenta of the converting leptons $l$ and $l'$, $(E_2,\b p_2)$ and $(E_2',\b p_2')$ for the spectator $f$, and $(E_3',\b p_3')$ for the axion. Then the axion emissivity is calculated as
\begin{align}
	\label{emissivity_integral}
	\nonumber
	\ve_a^{(lf)} & = \frac{(2\pi)^4}{S} \! \int 
	\prod_{i=1}^2 \widetilde{dp_i} \ 
	\prod_{j=1}^3 \widetilde{dp_j'} \ 
	\sum\limits_\rm{spin} \, \bigl| \mathcal{M}^{(lf)} \bigr|^2 
	\\ & \quad 
	\times \delta^{(4)}(p_1+p_2-p_1'-p_2'-p_3') E_3' \, f_1 \, f_2 \, (1-f_1') \, (1-f_2') ~,
\end{align}
where $S$ is the symmetry factor accounting for identical initial and final state particles, $\mathcal{M}^{(lf)}$ is the Lorentz invariant matrix element, $f_i$ and $f_i'$ are the Fermi-Dirac distribution functions, the factor $(1-f_i')$ takes into account the Pauli blocking due to particle degeneracy, and $\widetilde{dp}\equiv d^3 p/[(2\pi)^3 2E]$ is the Lorentz-invariant differential phase space element. We do not include a factor of $(1+f_3')$, since $f_3' \ll 1$ and there is no Bose enhancement of axion production since NSs are essentially transparent to axions for the currently allowed parameter space.

\subsection{Fermi-surface approximation}
\label{sec:ns_emissivity_fermi}
Calculating the emissivity \eqref{emissivity_integral} requires evaluating the 15 momentum integrals along with the 4 constraints from energy and momentum conservation.  We evaluate all but 2 of these integrals analytically using the Fermi surface approximation, and we calculate the last 2 integrals using numerical techniques.  The Fermi surface approximation assumes that the integrals are dominated by momenta near the Fermi surface $|{\bm p}| \approx p_F$; smaller and larger momenta do not contribute because of Pauli blocking or Boltzmann suppression \cite{Harris:2020qim}. To implement the Fermi surface approximation we introduce Dirac delta functions that fix the magnitude of the fermion 3-momenta to equal their respective Fermi momenta, and we promote the fermion energies to integration variables via the prescription: 
\begin{align}
	d^3p \rightarrow d^3p \int \frac{E}{p_F} \delta(p-p_F) dE ~.
\end{align}
This approximation allows the emissivity to be written as 
\begin{align}
	\ve_a^{(lf)} = \frac{1}{2^5(2\pi)^{11} p_{F,1} p_{F,2} p_{F,1'} p_{F,2'} S} J A ~,
\end{align}
which splits the calculation into two parts: an angular integral $A$ and an energy integral $J$, defined by 
\begin{align}
	\label{angular_integral}
	\nonumber
	A & \equiv \int \! d^3p_1 d^3p_2 d^3p_1' d^3p_2' d^2\Omega_3' \delta(p_1-p_{F,1}) \delta(p_2-p_{F,2}) \delta(p_1'-p_{F,1'}) \delta(p_2' - p_{F,2'}) \\
	&\phantom{\equiv \int \! d^3p_1 d^3p_2 d^3p_1' d^3p_2' d^2\Omega_3'} \times \delta^3(\b p_1 + \b p_2 - \b p_1' - \b p_2') \frac{\sum_\rm{spin} \abs{\cal M^{(lf)}}_\rm{Fermi}^2}{{E_3'}^{n}} ~,\\
	\label{energy_integral}
	J & \equiv \int \! dE_1 dE_2 dE_1' dE_2' dE_3' \delta(E_1+E_2-E_1'-E_2'-E_3') f_1 f_2 (1-f_1')(1-f_2') {E_3'}^{n+2} ~.
\end{align}
The matrix element $\abs{\cal M^{(lf)}}_\rm{Fermi}$ is evaluated with fermion 3-momenta and energies fixed to the respective Fermi momenta and Fermi energies.  
The exponent $n$ is chosen such that ${E_3'}^{-n}\sum_\rm{spin} \abs{\cal M^{(lf)}}_\rm{Fermi}^2$ is independent of $E_3'$. 
We have neglected the axion momentum in the momentum conservation delta function since $p_3'\sim T \ll p_{F,\mu}$. The mass dimension of $J$ and $A$ is $6+n$ and $3-n$, and that of $\abs{\cal M^{(lf)}}^2$ is $-2$. For the LFV channels considered in this work, we note that $p_{F,2}=p_{F,2'}$, $n=2$, and $S=1$ for $f$ being a proton and $S=2$ otherwise. 

\subsection{Energy integral}
The energy integral can be written as
\begin{align}
	\nonumber
	J &\approx \int_{-\infty}^\infty \! dx_1 \int_{-\infty}^\infty dx_2 \int_{-\infty}^\infty \! dx_1' \int_{-\infty}^\infty \! dx_2' \int_0^\infty \! dz \frac{T^{6+n} z^{2+n} \delta\( x_1 + x_2 + x_1' + x_2' - z\)}{(e^{x_1}+1) (e^{x_2}+1) (e^{x_1'}+1) (e^{x_2'}+1)} \\
	&= \frac{T^{6+n}}{6} \int_0^\infty dz \frac{z^{3+n}(z^2+4\pi^2)}{e^z - 1} ~,
\end{align}
where 
$x_i \equiv (E_i-E_{F,i})/T$ , 
$x_i' \equiv (E_{F,i}'-E_i')/T$, 
and $z\equiv E_3'/T$.  
The approximation symbols arise from extending the limits of integration to infinity. 
The second equality is derived using the technique in \cite{baym2008landau}. 
For $n=2$, we obtain
\begin{align}
	J = \frac{164\pi^8}{945} T^8 ~.
\end{align}

\subsection{Angular integral}
For the angular integral, we first integrate $d^3p_2'$ with the momentum delta function and $dp_1, dp_2, dp_1'$ with the Fermi surface delta function. It is convenient to align all angles with respect to $\b p_1$, so $\int \! d^2\Omega_1$ simply gives $4\pi$. The angular integral $A$ becomes
\begin{align}
	\label{A1}
	\nonumber
	A &= 4\pi p_{F,1}^2 p_{F,2}^2 p_{F,1'}^2 \int_{-1}^1 dc_{12} \int_{-1}^1 dc_{11'} \int_{-1}^1 dc_{13'} \int_0^{2\pi} d\j_{12} \int_0^{2\pi} d\j_{11'} \int_0^{2\pi} d\j_{13'} \\
	\nonumber
	&\phantom{=} \times \delta(p_2'-p_{F,2'}) {E_3'}^{-n} \sum_\rm{spin} \abs{\cal M^{(lf)}}^2_\rm{Fermi} ~,\\
	&= 32\pi^3 p_{F,1}^2 p_{F,2}^2 p_{F,1'}^2 \int_{-1}^1 dc_{12} \int_{-1}^1 dc_{11'} \int_{-1}^1 dc_{13'} \int_0^\pi dv_\j \\
	\nonumber
	&\phantom{=} \times \delta(p_2'-p_{F,2'}) \la {E_3'}^{-n} \sum_\rm{spin} \abs{\cal M^{(lf)}}^2_\rm{Fermi} \ra _{\j_{13'}} ~,
\end{align}
where $c_{ij}$ denotes the cosine of the angle between $\b p_i$ and $\b p_j$, $u_\j\equiv \j_{11'} + \j_{12}$, $v_\j \equiv \j_{11'}-\j_{12}$, and $\la \cdots \ra_{\j_{13'}}$ stands for an average over $\j_{13'}$. To obtain the second equality, we have assumed that $\la {E_3'}^{-n} \sum_\rm{spin} \abs{\cal M}^2_\rm{Fermi} \ra _{\j_{13'}}$ and $\delta(p_2'-p_{F,2'})$ do not depend on $u_\j$, and may rely on $v_\j$ only through $\cos v_\j$.

To simplify the expression further, we note that $2$ and $2'$ represent identical particle species whereas $1$ and $1'$ represent different particle species, and either $p_{F,2}\geq p_{F,1},p_{F,1'}$ or $p_{F,2}< p_{F,1},p_{F,1'}$. The delta function then becomes
\begin{align}
	\label{A2}
	\delta(p_2'-p_{F,2'}) = \frac{ \delta( v_\j - v_{\j,0} ) }{ p_{F,1'} \sqrt{(1-c_{11'}^2) (1-c_{12}^2) (1-\cos^2 v_{\j,0}) } }  ~,
\end{align}
where
\begin{align}
	\label{A3}
	v_{\j,0} = \arccos\[ \frac{p_{F,1}^2 + p_{F,1'}^2 - 2 p_{F,1} p_{F,1'} c_{11'} + 2 p_{F,2} (p_{F,1} - p_{F,1'} c_{11'}) c_{12} }{2 p_{F,1'} p_{F,2} \sqrt{(1-c_{11'}^2) (1-c_{12}^2) } } \]~.
\end{align}
To have a real-valued $v_{\j,0}$ within the range from $0$ to $\pi$, we must require $\cos^2 v_{\j,0} < 1$. This restricts the range of $dc_{11'}$ and $dc_{12}$ integrals to be within
\begin{align}
	\label{A4}
	c_{11'}^- < c_{11'} < c_{11'}^+ \sep
	c_{12}^- < c_{12} < c_{12}^+ ~,
\end{align}
where
\begin{align}
	\label{A5}
	\nonumber
	c_{11'}^\pm &= \frac{(p_{F,1}+p_{F,2} c_{12}) (p_{F,1}^2 + p_{F,1'}^2 + 2 p_{F,1} p_{F,2} c_{12}) }{2 p_{F,1'} (p_{F,1}^2 + p_{F,2}^2 + 2p_{F,1} p_{F,2} c_{12})}  \\
	&\phantom{=} \pm \frac{ p_{F,2} \sqrt{(c_{12}^2-1) [(p_{F,1}^2 - p_{F,1'}^2 + 2 p_{F,1} p_{F,2} c_{12})^2 - (2p_{F,2} p_{F,1'} )^2] }}{2 p_{F,1'} (p_{F,1}^2 + p_{F,2}^2 + 2p_{F,1} p_{F,2} c_{12})} ~,
\end{align}
and
\begin{align}
	\label{A6}
	c_{12}^+ = \min\[1, \frac{p_{F,1'}^2 - p_{F,1}^2 + 2 p_{F,2} p_{F,1'}}{2 p_{F,1} p_{F,2}} \] \sep
	c_{12}^- = \max\[-1, \frac{p_{F,1'}^2 - p_{F,1}^2 - 2 p_{F,2} p_{F,1'}}{2 p_{F,1} p_{F,2}} \] ~.
\end{align}
Combining equations \eqref{A1}-\eqref{A6}, we find
\small
\begin{align}
	\label{lfv_angular_integral8}
	A = 32\pi^3 p_{F,1}^2 p_{F,2}^2 p_{F,1'} \int_{c_{12}^-}^{c_{12}^+} dc_{12} \int_{c_{11'}^-}^{c_{11'}^+} dc_{11'} \int_{-1}^1 dc_{13'} \frac{ \la {E_3'}^{-n} \sum_\rm{spin} \abs{\cal M^{(lf)}}^2_\rm{Fermi} \ra _{\j_{13'}, v_\j = v_{\j,0}} }{ \sqrt{(1-c_{11'}^2) (1-c_{12}^2) (1-\cos^2 v_{\j,0}) } } ~.
\end{align}
\normalsize
We need to calculate the matrix element at the Fermi surface to evaluate this integral.

\subsection{Matrix element}
\begin{figure}
	\centering
	\includegraphics[width=0.6\linewidth]{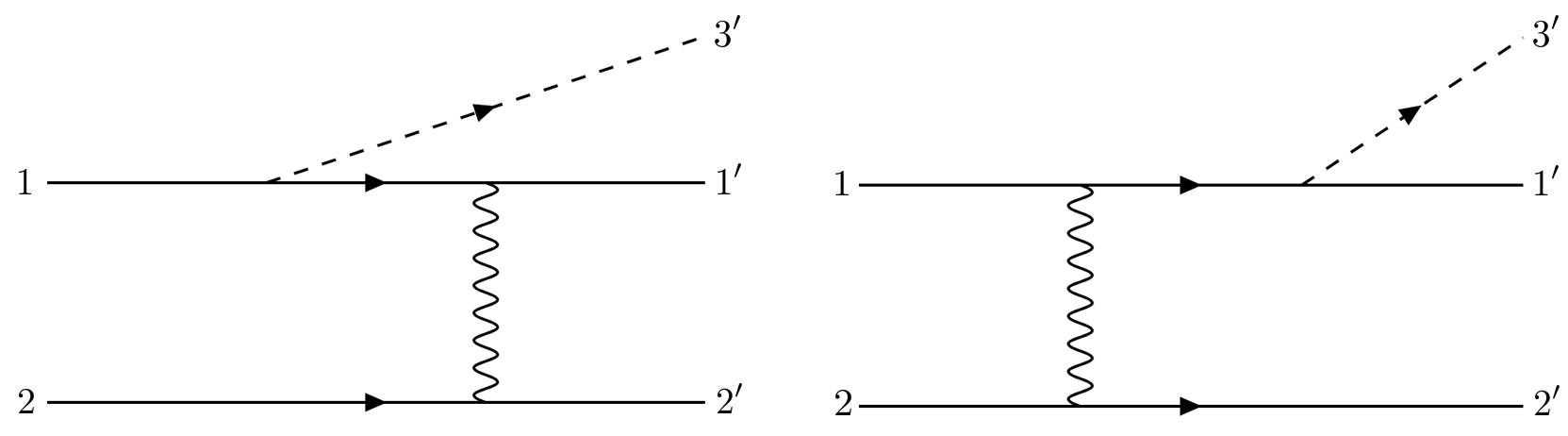}
	\caption{Feynman diagrams for the LFV process $l + f \rightarrow l' + f + a$. If $f$ is a lepton, there occur two more graphs which can be obtained by exchanging $(1\leftrightarrow 2)$ for $f$ being identical to $l$ or $(1'\leftrightarrow 2')$ for $f$ being identical to $l'$.}
	\label{fig:lfvfeynmandiagram}
\end{figure}

Now we evaluate the matrix element. It is convenient to use the LFV coupling
\begin{align}
	\cal L_\rm{LFV} = -i g_{ae\mu} a ( \bar\Psi_e \gamma_5 \Psi_\mu + \bar\Psi_\mu \gamma_5 \Psi_e ) ~,
\end{align}
which is equivalent to the use of the pseudovector (derivative) form \eqref{lfv} if each fermion line is attached to at most one axion line \cite{Raffelt:1987yt}. Given the two Feynman diagrams in figure \ref{fig:lfvfeynmandiagram}, the matrix elements are
\begin{align}
	i\cal M^{(1)} = \pm e^2 g_{ae\mu} \[ \bar u_1' \gamma^\mu \frac{-\sl r + m_1'}{r^2 + {m_1'}^2} \gamma_5 u_1 \] \frac{-g_{\mu\nu}}{k^2} [\bar u_2' \gamma^\nu u_2] ~,\\
	i\cal M^{(2)} = \pm e^2 g_{ae\mu} \[ \bar u_1' \gamma_5 \frac{-\sl s + m_1}{s^2 + m_1^2} \gamma^\mu  u_1 \] \frac{-g_{\mu\nu}}{k^2} [\bar u_2' \gamma^\nu u_2] ~,
\end{align}
where $k\equiv p_2-p_2'$, $r\equiv p_1-p_3'$, $s\equiv p_1'+p_3'$ and $\pm$ refers to the sign of the spectator particle's electric charge. In NSs we have $|m_1^2 - {m_1'}^2| \approx m_\mu^2 \gg E_F E_3'$, thus $r^2 + {m_1'}^2 \approx -m_1^2 + {m_1'}^2$ and $s^2 + m_1^2 \approx -{m_1'}^2 + m_1^2$. The matrix element for exchange diagrams can be obtained by $(1\leftrightarrow 2)$ or $(1'\leftrightarrow 2')$, with an additional factor of $-1$ included. 

The spin-summed squared matrix element is
\begin{align}
	\label{Mlp}
	\sum_\rm{spin} \abs{\cal M^{(lp)}}^2 &= -\frac{128 g_{ae\mu}^2 e^4}{(p_2-p_2')^4} \frac{(p_1\cdot p_1' + m_1 m_1') (p_2\cdot p_3') (p_2'\cdot p_3')}{(m_1^2 - {m_1'}^2)^2} ~, \\
	\label{Mll}
	\sum_\rm{spin} \abs{\cal M^{(ll)}}^2 &= \sum_\rm{spin} \abs{\cal M^{(lp)}}^2 + (1\leftrightarrow 2) + \cal T^{(ll)} ~, \\
	\label{Mll'}
	\sum_\rm{spin} \abs{\cal M^{(ll')}}^2 &= \sum_\rm{spin} \abs{\cal M^{(lp)}}^2 + (1'\leftrightarrow 2') + \cal T^{(ll')} ~,
\end{align}
where $l=e,\mu$ and $l'=\mu,e$. The second term in \eqref{Mll} and \eqref{Mll'} is the contribution solely from the exchange diagrams given by the first term but with $(1\leftrightarrow 2)$. The third term in \eqref{Mll} is the interference between prototype and exchange diagrams given by
\small
\begin{align}
	\nonumber
	\cal T^{(ll)} &= \frac{64 g_{ae\mu}^2 e^4}{(p_1-p_2')^2 (p_2-p_2')^2} \frac{p_2' \cdot p_3'}{(m_1^2 - {m_1'}^2)^2} \\
	&\phantom{=}\times [(p_2\cdot p_1' + m_1 m_1') (p_1\cdot p_3') + (p_1\cdot p_1' + m_1 m_1') (p_2\cdot p_3') - (p_1\cdot p_2 + m_1^2) (p_1'\cdot p_3')] ~,
	\label{Tll}
\end{align}
\normalsize
and $T^{(ll')}$ in \eqref{Mll'} by $\cal T^{(ll)}$ but with $(1\leftrightarrow 1')$ and $(2\leftrightarrow 2')$. Here we evaluate the traces of products of gamma matrices and spinors with the help of the Mathematica package FeynCalc \cite{Shtabovenko:2020gxv}.

At the Fermi surface, the spin-summed squared matrix element becomes
\begin{align}
	\sum_\rm{spin} \abs{\cal M^{(lf)}}^2_{\rm{Fermi}} = \frac{32 e^4 g_{ae\mu}^2 {E_3'}^2}{E_{F,1}^2 E_{F,2}^2 \beta_2^4 (\beta_1^2 - {\beta_1'}^2)^2} G^{(lf)} ~,
\end{align}
where $f=p,e,\mu$. The $G^{(lf)}$ factor is found to be
\begin{align}
	\label{Glp}
	G^{(lp)} &= \frac{(1-\beta_{F,2} c_{23'}) (1-\beta_{F,2} c_{2'3'}) (1- \beta_{F,1} \beta_{F,1'} c_{11'})}{(1-c_{22'})^2} ~,\\
	\label{Gll}
	G^{(ll)} &= G^{(lp)} + (1\leftrightarrow 2) + H^{(ll)} ~,\\
	\label{Gll'}
	G^{(ll')} &= G^{(lp)} + (1'\leftrightarrow 2') + H^{(ll')} ~,
\end{align}
where we have assumed that electrons are ultra relativistic so $\beta_{F,e}=1$. The second term in \eqref{Gll} and \eqref{Gll'} is the contribution solely from the exchange diagrams given by the first term but with $(1\leftrightarrow 2)$. The third term in \eqref{Gll} is the interference between prototype and exchange diagrams given by
\begin{align}
	\nonumber
	H^{(ll)} = \frac{(1-\beta_{F,1} c_{2'3'}) }{2 (1-c_{12'}) (1-c_{22'})} &\[  \beta_{F,1}\big( c_{13'}+c_{23'} + \beta_{F,1}(1-c_{12}) + \beta_{F,1'}(c_{11'} + c_{21'}) \right. \\
	&\left. + \beta_{F,1}\beta_{F,1'} ( c_{12} c_{1'3'} - c_{11'} c_{23'} - c_{13'} c_{21'} - c_{1'3'} ) \big) - 2 \] ~,
\end{align}
and $H^{(ll')}$ in \eqref{Gll'} by $H^{(ll)}$ but with $(1\leftrightarrow 1')$ and $(2\leftrightarrow 2')$.

\subsection{Axion emissivity}
In summary, the axion emissivity is given by
\begin{align}
	\label{emissivity}
	\ve_a^{(lf)} &= \frac{328\pi^2 \alpha^2 g_{ae\mu}^2 }{945 m_\mu^4} \frac{\beta_{F,1} E_{F,1}^3 }{\beta_{F,2}^2 p_{F,2}^2} F^{(lf)} T^8 ~,\\
	F^{(lf)} &\equiv \frac{1}{8S} \int_{c_{12}^-}^{c_{12}^+} dc_{12} \int_{c_{11'}^-}^{c_{11'}^+} dc_{11'} \int_{-1}^1 dc_{13'} \frac{ \la G^{(lf)}\ra_{\j_{13'}, v_\j = v_{\j,0}} }{ \sqrt{(1-c_{11'}^2) (1-c_{12}^2) (1-\cos^2 v_{\j,0}) } } ~.
\end{align}
To derive \eqref{emissivity}, we have assumed that the axion mass is small compared to the NS temperature $m_a\ll T$, muons and electrons are in the beta equilibrium (i.e., $E_{F,e} \approx E_{F,\mu}$), electrons are ultra relativistic but muons are not (i.e., $p_{F,\mu}\lesssim m_\mu$), and $T\ll m_\mu^2/E_{F,e}$.

The temperature dependence of the axion emissivity \eqref{emissivity} is especially interesting and important for understanding the limits from neutron star cooling.  For comparison, note that axion bremstrahlung via lepton-flavor-preserving (LFP) interactions (such as $ep \to epa$ or $\mu p \to \mu p a$) goes as $\ve_a \propto T^6$.  In other words, the LFV interaction leads to an emissivity that's suppressed by an additional factor of $T^2 E_{F,e}^2 / (m_\mu^2 - m_e^2)^2 \sim T^2/m_\mu^2$, which is of order $(100 \rm{keV}/100 \rm{MeV})^2 \sim 10^{-6}$ for $T\sim 10^{9} \rm{K}$. A detailed discussion appears in the Supplemental Material, but the essential idea can be understood as follows. The phase-space integrals over momenta can be converted to energy integrals, and each integral for degenerate leptons and protons is restricted to the Fermi surface of thickness $\sim T$, giving a factor of $T^4$. The phase-space integral of axions (i.e., $d^3p_3'/E_3'$) gives a factor of $T^2$. The axions are emitted thermally and have an energy $\sim T$. The energy conservation delta function gives $T^{-1}$. The squared matrix element has a temperature dependence $T^2$. Putting all these together, we see that the emissivity is proportional to $T^8$. In comparison, the squared matrix element for the LFP interactions has no temperature dependence since one power of $T$ from the coupling vertex is canceled by $T^{-1}$ from the lepton propagator.

\begin{figure}
	\centering
	\begin{minipage}{0.45\linewidth}
		\includegraphics[width=\linewidth]{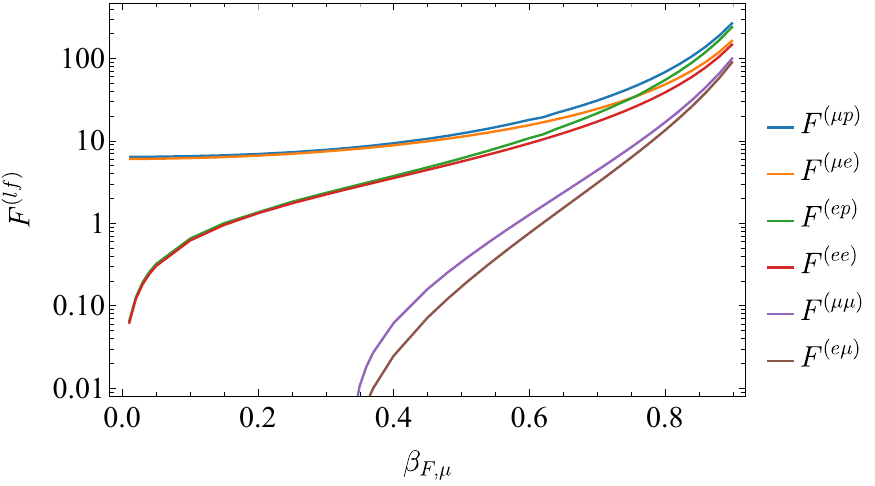}
	\end{minipage}\quad
	\begin{minipage}{0.45\linewidth}
		\includegraphics[width=\linewidth]{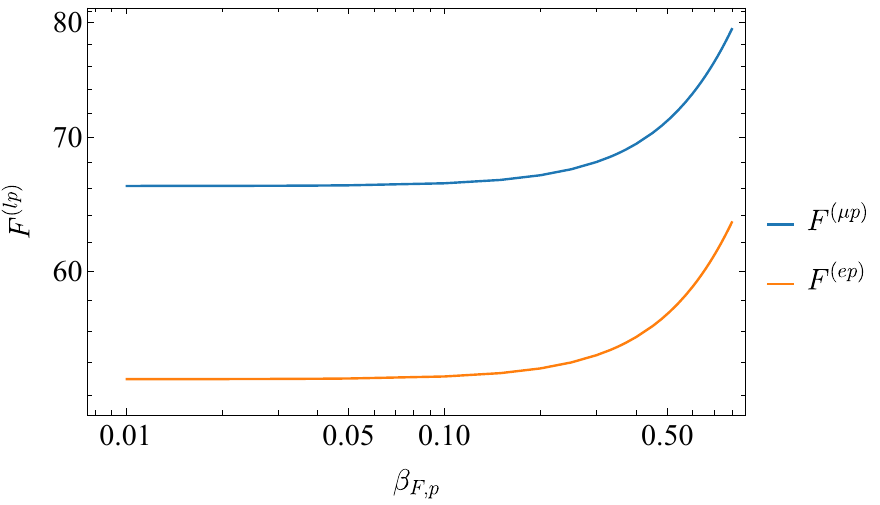}
	\end{minipage}
	\caption{The factor $F^{(lf)}$ as a function of the Fermi velocity of muons (left) and protons (right). Here we have set $\beta_{F,p}=0.3$ and $\beta_{F,\mu}=0.8$ for the left and right panels respectively for the $f=p$ processes.}
	\label{fig:lfvf}
\end{figure}

The $dc_{13'}$ integral in \eqref{emissivity} can be evaluated analytically. We calculate the other integrals using numerical techniques and present the result for $F^{(lf)}$ in figure \ref{fig:lfvf}. In the left panel we vary the muon Fermi velocity $\beta_{F,\mu} = p_{F,\mu} / E_{F,\mu}$. From the right panel we see that $F^{(lp)}$ is not sensitive to $\beta_{F,p}$ if protons are NR, i.e., $\beta_{F,p} \lesssim 0.5$, which is expected in NSs. Therefore, we may use the values of $F^{(lf)}$ shown in the left panel to calculate the emissivity.

\begin{figure}
	\centering
	\includegraphics[width=0.5\linewidth]{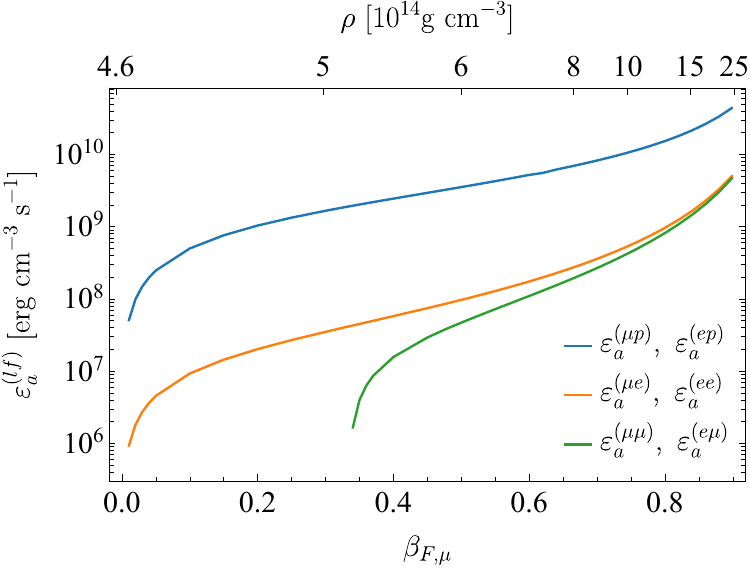}
	\caption{Axion emissivities $\ve_a^{(lf)}$ for the LFV process $l+f\rightarrow l' + f + a$, given by equation \eqref{emissivity}, as a function of the muon Fermi velocity $\beta_{F,\mu}$. The top axis, in a nonlinear scale, represents the corresponding mass density of a NS assuming the $npe\mu$ matter. Here we take $g_{ae\mu}=10^{-11}$ and $T=10^9\rm{K}$, and more generally $\ve_a^{(lf)} \propto g_{ae\mu}^2 T^8$.}
	\label{fig:lfvemissivity}
\end{figure}

The numerical values of axion emissivity \eqref{emissivity} are shown in figure \ref{fig:lfvemissivity} for the six channels $lf \to l'fa$, where the effective mass of protons is taken to be $0.8 m_p$ (see \cite{Li:2018lpy} and references therein).\footnote{Thanks to the electric charge neutrality and the beta equilibrium condition $E_{F,e}\approx E_{F,\mu}$, the emissivity can be fully determined once the effective proton mass and $\beta_{F,\mu}$ are given.} The emissivities are equal for the channels $ef \to \mu fa$ and $\mu f \to efa$ due to the strong degeneracy of particles and the beta equilibrium condition $E_{F,e}\approx E_{F,\mu}$, so the plot only shows three curves corresponding to different spectator particles $f=p,e,\mu$. The channels with a spectator proton ($f=p$) have the largest emissivity across the range of muon Fermi momenta shown here; this is a consequence of the enhanced matrix element and the larger available phase space for these scatterings. For the channels with a spectator muon ($f=\mu$), the emissivity drops to zero below $\beta_{F,\mu} \approx 0.34$; this corresponds to a violation of the kinematic threshold in \eqref{threshold}.  For all channels, the emissivity decreases with decreasing muon Fermi velocity due to the reduced kinematically allowed phase space. On the other hand, for larger muon Fermi velocity, the channels with spectator electrons and muons coincide, since both particles can be regarded as massless. For the top axis in figure \ref{fig:lfvemission}, we show the corresponding mass density of a NS assuming the $npe\mu$ model; see section \ref{sec:ns_ns_npemu} for more details.

The total axion emissivity is obtained by summing over the six channels.  For this estimate we set $\beta_{F,\mu}=0.84$, corresponding to 
\begin{align}
	\label{fermi_momenta_npemu}
	p_{F,n} \simeq 624 \rm{MeV} \sep
	p_{F,p} \simeq 226 \rm{MeV} \sep
	p_{F,e} \simeq 193 \rm{MeV} \sep
	p_{F,\mu} \simeq 162 \rm{MeV} ~.
\end{align}
We find the axion emissivity via LFV interactions to be
\begin{align}
	\label{emissivity_lfv}
	\ve_a^\rm{LFV} \simeq 4.8\times 10^{32} g_{ae\mu}^2 T_9^8 ~\rm{erg~cm^{-3}~s^{-1}} ~,
\end{align}
where $T_9\equiv T/(10^9\rm{K})$ and $10^9\rm{K} \approx 86.2 \rm{keV}$.

\section{Implications for neutron star cooling}
\label{ns_observation}

In low-mass NSs, slow cooling could occur via neutrino emission by the modified Urca (Murca) processes $nn\rightarrow npe\bar\nu$, $npe\rightarrow nn\nu$ or slightly less efficient processes such as the nucleon bremsstrahlung \cite{Yakovlev:2004iq, Potekhin:2015qsa}. At the density $\rho=6\rho_0$, where $\rho_0=2.5\times 10^{14} \, \rm{g~cm^{-3}}$ is the nuclear saturation density \cite{Horowitz:2020evx}, and with the effective nucleon mass taken to be $0.8 m_N$ \cite{Li:2018lpy}, the emissivity of the Murca process is given by $\ve_\nu = 4.4 \times 10^{21} T_9^8 ~ \rm{erg~cm^{-3}~s^{-1}}$ \cite{Friman:1979ecl}. Comparing this rate with \eqref{emissivity_lfv}, one finds that the axion emission from LFV couplings dominates the neutrino emission unless
\begin{align}
	|g_{ae\mu}| \lesssim 3.0 \times 10^{-6} ~,
\end{align}
which is indeed the case based on the existing constraints. In heavier NSs, the LFV emission of axions tends to have a less significant impact. This is because fast neutrino emission could occur via the direct Urca processes \cite{Lattimer:1991ib}. In the presence of superfluidity, the formation of Cooper pairs can dominate over the Murca process \cite{Leinson:2009mq, Leinson:2009nu}, further diminishing the role of LFV axion emission.

Axions are predominantly produced in NSs through the nucleon bremsstrahlung process $nn\rightarrow nna$. At the same core conditions, its emissivity is given by $\ve_a^{(nn)} \simeq 2.8\times 10^{38} g_{ann}^2 T_9^6 ~\rm{erg~cm^{-3}~s^{-1}}$ \cite{Iwamoto:1984ir, Brinkmann:1988vi}. The nucleon bremsstrahlung process dominate the LFV processes if 
\begin{align}
	|g_{ae\mu}| \lesssim 7.6 \times 10^2 |g_{ann}| T_9^{-1} ~.
\end{align}
The current best constraint on the axion-neutron coupling is $|g_{ann}|\lesssim 2.8\times 10^{-10}$ \cite{Beznogov:2018fda}. Therefore, it is unlikely for the LFV couplings to play a significant role in NSs with an age $\gtrsim 1 \rm{yr}$, where the temperature has cooled to $10^9 \rm{K}$ \cite{Pethick:1991mk}. 

These limits on the axion's LFV coupling are relatively weak, and this is a consequence of the $\ve_a^\rm{LFV} \propto T^8$ scaling, which is suppressed compared to LFP channels by a factor of $(T/m_\mu)^2$, which is tiny in old NSs. However, in the proto-NS that forms just after a supernova, this ratio can be order one, which suggests that stronger limits can be obtained by considering the effect of axion emission on supernova rather than neutron stars.  Since our analysis has focused on neutron star environments, adapting our results to the more complex proto-NS system requires some extrapolation.  We estimate the axion emissivity from a supernova by extrapolating \eqref{emissivity_lfv} to high temperatures. By imposing the bound on the energy loss of SN 1987A, $\ve_a/\rho \lesssim 10^{19} \, \rm{erg~g^{-1}~s^{-1}}$ \cite{Raffelt:1990yz}, one finds that at a typical core condition $\rho\sim 8\times 10^{14} \, \rm{g~cm^{-3}}$,
\begin{align}
	\label{constraint_sn}
	\abs{g_{ae\mu}} \lesssim 4\times 10^{-11} \( \frac{50 \rm{MeV}}{T} \)^4 ~,
\end{align}
which is to be evaluated at $T\sim (30-60) ~\rm{MeV}$.
This constraint is more stringent than that obtained from considering $\mu\rightarrow ea$ in a supernova and is comparable to the current best terrestrial limit. One should note that at typical core conditions of a proto-NS, nucleons and muons are at the borderline between degeneracy and nondegeneracy, and we expect a similar constraint if a nondegenerate emission rate is used.\footnote{For the axion bremsstrahlung by nucleons, the emission rates with degenerate and nondegenerate nucleons coincide at typical supernova core conditions \cite{Raffelt:1990yz}.}

Incorporating the effect of nuclear interactions, we have used an effective proton mass of $0.8 \, m_p$ where $m_p$ denotes the proton mass in vacuum \cite{Li:2018lpy}. If protons are nonrelativistic, a reduced effective mass diminishes axion emission and consequently weakens constraints on $g_{ae\mu}$. For an effective mass around $0.5 \, m_N$, a correction factor of $\sim 0.8/0.5$ arises based on equation \eqref{emissivity}. However, the emission rate remains unaffected for relativistic protons. Additionally, in next section, we discuss the screening effect of electric fields. By introducing an effective mass for photon propagators of the order $k_{\rm{TF}}$, where $k_{\rm{TF}}$ represents the Thomas-Fermi wavenumber, we estimate a correction factor of $\sim 2$ due to the screening effect.\footnote{While this methodology isn't apt for strongly coupled plasmas like NSs and white dwarfs, it does furnish reasonably accurate estimates of the screening effect in axion bremsstrahlung processes within white dwarfs \cite{Raffelt:1990yz}.} Our numerical analysis indicates that extrapolating the degenerate rate \eqref{emissivity_lfv} to $T\sim 50\rm{MeV}$ tends to overestimate the axion emission, introducing a correction factor of $\sim 3$ in the supernova constraint.

\section{Brute force calculations}
In foregoing sections we calculate the emissivity of LFV axions using the Fermi surface approximation and use it to explore the impact of the LFV coupling on NS cooling. This is appropriate at low temperatures, where all involved particles are degenerate. However, as we can see in table \ref{tab:nscomposition}, protons and muons start to become nondegenerate at $T\sim 10\rm{MeV}$, at which point the Fermi surface approximation starts to break down. Thus in section, we aim to provide a complementary method for evaluating the axion emissivity -- brute-force Monte Carlo integration technique. The numerical details are provided in \cite{Zhang:2023vva}.

\begin{figure}
	\centering
	\includegraphics[width=0.6\linewidth]{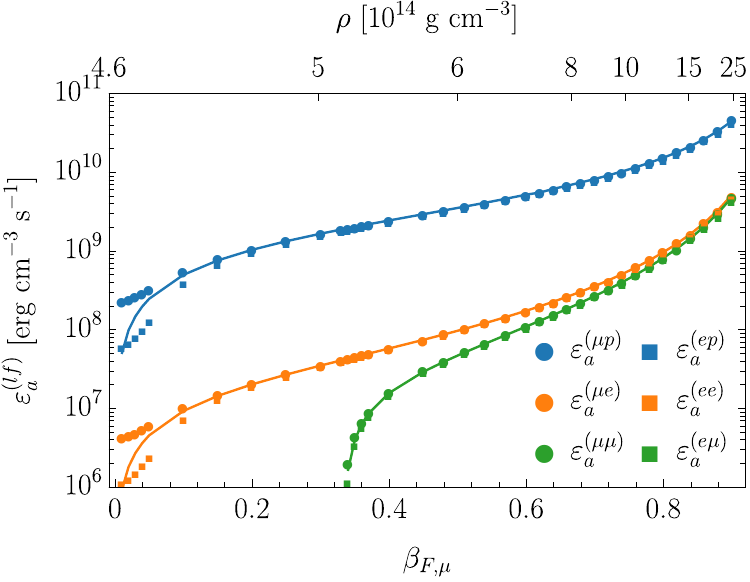}
	\caption{Axion emissivity computed by brute force (dots and squares) vs. by using the Fermi surface approximation (lines). The results agree well for $\beta_{F,\mu} \gtrsim 0.1$ and the agreement is good within about $10\%$ at $\beta_{F,\mu} \approx 0.8$. Data shown in this figure was generated using a NS core temperature of $T = 10^9~\mathrm{K}$ and a axion mass $m_a = 0$.}
	\label{fig:numerical-emissivity}
\end{figure}

The results of brute-force evaluations for the various axion emission channels are shown in figure \ref{fig:numerical-emissivity}. The numerical results (dots and squares) agree very well with the analytical results (lines) for a wide range of $\beta_{F,\mu}$. For small $\beta_{F,\mu}$ the numerical results tend to diverge from the analytical results, which is expected because in this regime the number density of muons is small, which means that the degenerate matter approximation breaks down. In addition, we observe that the emissivities are paired by channel such that $\varepsilon_a^{(\mu p)} \approx \varepsilon_a^{(ep)}$, $\varepsilon_a^{(\mu e)} \approx \varepsilon_a^{(e \mu)}$, and $\varepsilon_a^{(\mu \mu)} \approx \varepsilon_a^{(e e)}$. This is a consequence of the strong particle degeneracy and the beta equilibrium condition $E_{F,e}\approx E_{F,\mu}$.

In degenerate NS matter, electric fields are screened because of the polarizability of charged particles. To estimate this effect, we replace the photon propagator $k^{-2}$ in the matrix element by $(k^2+ k_\rm{TF}^2)^{-1}$ \cite{Raffelt:1996wa}, where $k_\rm{TF}^2 = \sum_{i} 4\alpha p_{F,i} E_{F,i}/\pi$ is the Thomas-Fermi screening scale contributed by electrons, muons and protons. Noting that $k^2 \sim (p_{F,e}-p_{F,\mu})^2 \sim E_{F,e}^2 (1-\beta_{F,\mu})^2$ at low temperatures, the screening effect is small if $\beta_{F,\mu} \lesssim 1-k_\rm{TF}/E_{F,e}$, which becomes $\beta_{F,\mu} \lesssim 0.75$ at the core condition given by \eqref{fermi_momenta_npemu}. Therefore, for midly relativistic muons, the emissivity of LFV axions without including the screening effect is subject to $\cal O(1)$ corrections. On the other hand, incorporating the screening effect in axion emissivities is crucial at high temperatures since $k_\rm{TF}^2$ dominates over $k^2$, especially near the pole $k^2=0$.

\begin{figure}
	\centering
	\includegraphics[width=0.6\linewidth]{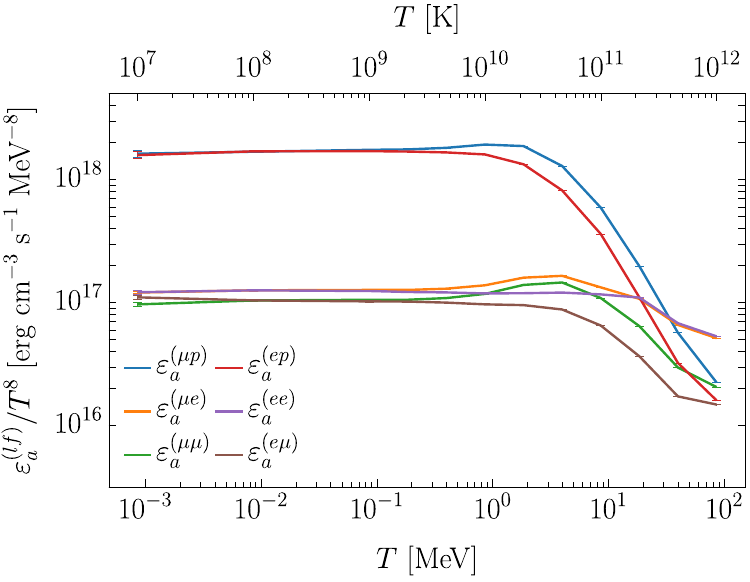}
	\caption{Axion emissivities as a function of temperature. To generate the data we set $\beta_{F, \mu} = 0.836788$, $m_a = 0$, and $g_{ae\mu} = 10^{-11}$.}
	\label{fig:numerical-emissivity-vs-T}
\end{figure}

Employing this change in photon propagators, we present the temperature dependence of the axion emissivity is in figure \ref{fig:numerical-emissivity-vs-T}.
Since we expect the emissivity to scale as $\varepsilon^{(lf)} \propto T^8$ for low temperatures we normalize the emissivity by $T^8$ so that a $T^8$ scaling would be a constant line in this figure. As expected, the emissivity diverges from the scaling $T^8$ for temperatures higher than $\sim 10\rm{MeV}$, where protons and muons are at the borderline between degeneracy and nondegeneracy. We can see that extrapolating the degenerate rate \eqref{emissivity_lfv} to $T\sim 50\rm{MeV}$ tends to overestimate the axion emission, introducing a correction factor of $\sim 0.1$ in axion emissivities and, consequently, a factor of $\sim 3$ in the supernova constraint.
\chapter{Nonminimal gravitational interactions of dark matter}
\label{sec:nonminimal}

\section{Introduction}
Vector dark matter (VDM) has been drawing increasing attention of theorists and experimentalists in recent years \cite{Fabbrichesi:2020wbt, Caputo:2021eaa}. Proposals for VDM cover a large range of mass, with a division occurring at $30\mathrm{eV}$ (as we demonstrate in the introduction section \ref{sec:intro_ultralight}), below or over which the vector bosons behave like waves \cite{Freitas:2021cfi, Adshead:2021kvl, Gorghetto:2022sue, Amin:2022pzv, Antypas:2022asj} or particles \cite{Lebedev:2011iq, Djouadi:2011aa, Baek:2012se, Arcadi:2020jqf, Baouche:2021wwa, Ghorbani:2021yiw}. VDM can be produced through, for example, gravitational particle production \cite{Graham:2015rva, Ahmed:2019mjo, Ema:2019yrd, Ahmed:2020fhc, Kolb:2020fwh}, freeze-in mechanism \cite{Duch:2017khv, Barman:2020ifq, Barman:2021qds}, energy transfer from scalar fields \cite{Agrawal:2017eqm, Agrawal:2018vin, Dror:2018pdh, Co:2018lka, Bastero-Gil:2018uel, Salehian:2020asa, Co:2021rhi} and misalignment mechanism \cite{Nelson:2011sf, Arias:2012az, Nakayama:2019rhg, Kitajima:2023fun, Alonso-Alvarez:2019ixv}.\footnote{The misalignment mechanism proposed in \cite{Nelson:2011sf} can not produce sufficient DM, which can be fixed by adding a nonminimal coupling to gravity \cite{Arias:2012az}. However, the added nonminimal coupling induces a ghost instability and a quadratic divergence \cite{Himmetoglu:2009qi, Himmetoglu:2008zp, Himmetoglu:2008hx, Esposito-Farese:2009wbc, Graham:2015rva}. A viable misalignment mechanism for VDM can be constructed by including a nonminimal kinetic coupling to the inflaton \cite{Nakayama:2019rhg, Kitajima:2023fun}.}

While most of the studies on VDM consider vector fields that are minimally coupled to gravity, additional nonminimal couplings are well motivated as they naturally arise as quantum corrections to a minimally coupled classical theory \cite{Birrell:1982ix}. It has been demonstrated that the inclusion of the nonminimal coupling is integral to the renormalization of field theories in curved spacetime \cite{Weinberg:1995mt, callan1970new, freedman1974energy, FREEDMAN1974354}. Moreover, the nonminimal terms are phenomenologically relevant and have been employed in the context of inflation \cite{Turner:1987bw, Ford:1989me, Faraoni:2000wk, Golovnev:2008cf, Golovnev:2008hv, Golovnev:2009ks, Golovnev:2009rm}, modified gravity \cite{Moffat:2005si, Brownstein:2005zz, Tasinato:2014eka, Heisenberg:2014rta, DeFelice:2016yws, deFelice:2017paw}, DM \cite{Ji:2021rrn, Sankharva:2021spi, Barman:2021qds, Ivanov:2019iec} and dark energy \cite{Kouwn:2015cdw, Koivisto:2008xf}.

In this section, we study ultralight VDM that is nonminimally coupled to gravity in the wave regime. Although the DM mass $m\sim 10^{-22}\mathrm{eV}$ is preferred in some galaxies \cite{Schive:2014dra, Chen:2016unw, Khelashvili:2022ffq}, a lower bound $m \gtrsim 10^{-19} \mathrm{eV}$ has been reported by constraining velocity dispersion in Segue 1 and Segue 2 galaxies \cite{Dalal:2022rmp}, and it further increases to $m\gtrsim 10^{-18}\mathrm{eV}$ if the DM is produced after inflation via a process with a finite-correlation length \cite{Amin:2022nlh}. The impact of nonminimal couplings on these constraints warrants a detailed analysis, and for generality we focus on $m\gtrsim 10^{-22}\mathrm{eV}$. Remaining agnostic about the ultraviolet physics, in section \ref{sec:nonminimal_eft}, we derive a NR EFT and the modified version of the SP equations that include the nonminimal couplings. With this tool, we analyze the phenomenology of the nonminimal couplings in several contexts in section \ref{sec:nonminimal_pheno}, such as the mass-radius relation of vector solitons, the growth of linear perturbations, and the propagation of GWs.

\section{Nonrelativistic effective field theory}
\label{sec:nonminimal_eft}
As discussed in section \ref{sec:eft}, the NR dynamics of wave DM can be described by the SP equations, which are the leading-order approximation of the KGE system. To investigate the impacts of nonminimal couplings to gravity, in this section we derive a modified version of the SP equations by identifying small, dimensionless quantities and expanding the relativistic equations to the leading order \cite{Salehian:2020bon, Salehian:2021khb}. We also convert the modified SP equations to a fluid description, which is more commonly used in the study of linear perturbations.

\subsection{Wave description}
\label{sec:nonminimal_eft_wave}
For the purpose of describing the low-energy limit of VDM, we can start without loss of generality from the covariant action $S$ that contains nonminimal couplings to gravity with the lowest mass dimension. The progenitor action we consider is 
\begin{align}
	\label{eq:full_action}
	S = S_G + S_M ~,
\end{align}
where the gravity and matter parts are given by
\begin{align}
	S_G &= \int \dd[4]{x} \sqrt{-g} \left[ \frac{1}{2} \MP^2 R \right] ~, \\
	S_M &= \int \dd[4]{x} \sqrt{-g} \left[ -\frac{1}{4} X_{\mu\nu} X^{\mu\nu} - \frac{1}{2}m^2 X_\mu X^\mu + \frac{1}{2} \xi_1 R X_\mu X^\mu + \frac{1}{2} \xi_2 R^{\mu\nu} X_\mu X_\nu + \cdots \right] ~.
\end{align}
Here, $\xi_1$ and $\xi_2$ characterize the nonminimal coupling to gravity, $R$ and $R_{\mu\nu}$ are the Ricci scalar and Ricci tensor, $X_\mu$ is the VDM field, and $X_{\mu\nu} \equiv \partial_\mu X_\nu - \partial_\nu X_\mu$. In the low-energy limit, an expanding universe containing VDM can be described by the perturbed FRW metric \cite{Cembranos:2016ugq}
\begin{align}
	\label{metric}
	\dd{s}^2 = -(1+2\Phi) \dd{t}^2 + a^2(t) (1-2\Phi) \delta\indices{_i_j} \dd{x}^i \dd{x}^j ~.
\end{align}
The action \eqref{eq:full_action} and metric \eqref{metric} are valid if the VDM has become the dominant and NR component of the universe.  After the universe enters the matter-dominated era, the dynamics of $X_\mu$ is dominated by oscillations of frequency $\omega\sim m$, thus it is motivated to redefine the vector field in terms of a new, complex NR field $\psi_\mu$ by
\begin{align}
	\label{eq:nr_expansion}
	X_\mu(t,\vb{x}) = \frac{1}{\sqrt{2ma}} \left[ e^{-imt} \psi_\mu(t,\vb{x}) + e^{imt} \psi_\mu^*(t,\vb{x}) \right] ~.
\end{align}
The power of the scale factor is chosen such that the amplitude of $\psi_\mu$ does not change significantly with the expansion of the universe, since the energy density scales like $\rho\sim m^2 a^{-2} X_i X_i \propto a^{-3}$ during matter domination. To ensure that the field redefinition preserves the number of propagating degrees of freedom, one could employ a constraint that implies a field equation of $\psi_\mu$ remaining first order in time derivatives \cite{Salehian:2020bon},
\begin{align}
	\label{eq:nr_constraint}
	e^{-imt} \dot\psi_i + e^{imt} \dot\psi_i^* = 0 ~.
\end{align}
An alternative nonlocal field redefinition was exploited in \cite{Namjoo:2017nia}.

To derive a NR EFT of VDM, we follow the prescription in section \ref{sec:eft} and identify the following small, dimensionless parameters
\begin{align}
	\epsilon_H \sim \frac{H}{m} ~,\quad
	\epsilon_t \sim \abs{ \frac{\dot Q}{m Q} } ~,\quad
	\epsilon_k \sim \abs{ \frac{\nabla^2 Q}{a^2 m^2 Q} } ~,\quad
	\epsilon_\psi \sim \frac{|\psi_i|}{\MP \sqrt{m}} ~,\quad
	\epsilon_g \sim |\Phi| ~,
\end{align}
where $Q$ can be any of the slowly varying variables including $a, H, \psi_\mu, \Phi$. These parameters respectively characterize the smallness of the expansion rate, time variation, spatial gradient, field amplitude and gravity strength. In addition, the nonminimal coupling terms in \eqref{eq:full_action} should not play too significant a role in order to retain the success of General Relativity and to avoid the ghost instability of longitudinal modes discussed in refs.~\cite{Himmetoglu:2009qi, Himmetoglu:2008zp, Himmetoglu:2008hx, Esposito-Farese:2009wbc, Graham:2015rva}. Hence one must have another small parameter much less than unity in terms of $\xi_a ~(a=1,2)$,
\begin{align}
	\epsilon_\xi \sim \abs{ \frac{\xi_a R}{m^2} } ~,
\end{align}
where $R\sim a^{-2} \nabla^2\Phi \sim \rho/\MP^2$ and $\rho$ is the local DM density. This sets an upper limit on $\xi_a$,
\begin{align}
	\label{xi_range}
	\abs{\xi_a} \ll \frac{m^2 \MP^2}{\rho} = 1.5\times 10^{15} \left( \frac{m}{10^{-20} \mathrm{eV}} \right)^2 \left( \frac{5 \times 10^{10} ~\mathrm{GeV/m^3}}{\rho} \right) ~.
\end{align}
Here the reference value of DM density is taken to be its average value at the matter-radiation equality \cite{ParticleDataGroup:2022pth}. If one is interested in the dynamics of local DM today, one should take the DM density to be around $0.3~\mathrm{GeV/cm^3}$ \cite{deBoer:2010eh, Bovy:2012tw, mckee2015stars, Sivertsson:2017rkp}, which implies $\abs{\xi_a} \ll 2.6\times 10^{20} (m / 10^{-20} \mathrm{eV})^2$.\footnote{In section \ref{sec:nonminimal_pheno_gw}, we will take this value as the upper limit of $\xi_a$ when studying the variation of gravitational wave speed.} In general, these small parameters are not independent from each other and we do not know a priori the relative magnitudes between them, thus a reliable expansion strategy would require the system of equations to be expanded up to a specific order that includes every parameter. That is, each order is a homogeneous function of all small parameters. In the following discussions, we will denote all small parameters collectively by $\epsilon = \{\epsilon_H, \epsilon_t, \epsilon_k, \epsilon_\psi, \epsilon_g, \epsilon_\xi \}$ for notation convenience and work up to the leading order in $\epsilon$.

By plugging the field redefinition \eqref{eq:nr_expansion} into the action \eqref{eq:full_action} and integrating out the fast oscillating terms, we obtain a NR effective action
\begin{align}
	\label{eq:nr_action}
	S = \int \dd[4]{x} & \bigg[ \MP^2 a\left( -3 \dot a^2 + \Phi\nabla^2\Phi - 6a \ddot a\Phi \right) \nonumber \\
	& + \frac{1}{2} a^2 m \abs{\psi_0 - \frac{i}{a^2m}\nabla\cdot\vb*{\psi}}^2 + i \dot{\vb*{\psi}}\cdot\vb*{\psi}^\ast + \frac{1}{2a^2m} (\nabla^2\vb*{\psi})\cdot\vb*{\psi}^* - m\Phi \abs{\vb*{\psi}}^2  \nonumber \\
	& + \frac{\abs{\vb*{\psi}}^2}{2a^2m} \left[ 2 \xi_1 (\nabla^2 \Phi + 3 \dot{a}^2 + 3a\ddot a ) + \xi_2 (\nabla^2\Phi + 2 \dot{a}^2 + a\ddot a) \right]\bigg] \times \left[ 1 + \order{\epsilon} \right] ~,
\end{align}
where $\vb*{\psi}$ is the spatial part of $\psi_\mu$, the overdot stands for time derivative, and $\order{\epsilon}$ includes all relativistic corrections.\footnote{Higher-order time derivatives will emerge if one takes the relativistic corrections into account, but they do not introduce unphysical degrees of freedom because they can be systematically removed by applying field equations at lower orders \cite{Namjoo:2017nia}.} To the leading order, the constraint equation on $\psi_0$ is not affected by the nonminimal coupling, implying that the existence of nonminimal couplings does not bring about the singularity problem discussed in \cite{Mou:2022hqb, Clough:2022ygm, Coates:2022qia}. As promised, \eqref{eq:nr_action} is also free from the ghost instability of longitudinal modes \cite{Himmetoglu:2009qi, Himmetoglu:2008zp, Himmetoglu:2008hx, Esposito-Farese:2009wbc, Graham:2015rva}.

By varying \eqref{eq:nr_action} with respect to $a,\psi_i^*,\Phi$, in the NR limit, we obtain the field equations
\begin{align}
	\label{EOM_psi}
	i\partial_t \psi_i = - \frac{\nabla^2}{2m a^2} \psi_i + m\Phi \psi_i - \frac{(2\xi_1+\xi_2) 2\pi G}{m} \rho_\xi \psi_i ~,\\
	\label{EOM_Phi}
	\frac{\nabla^2}{a^2}\Phi = 4\pi G (\rho_\xi-\bar\rho_\xi) \sep
	H^2 = \frac{8\pi G}{3} \bar\rho_\xi ~,\\
	\label{EOM_rho}
	\rho_\xi \equiv \rho - \frac{2\xi_1 + \xi_2}{2m^2 a^2} \nabla^2\rho \sep
	\rho = \frac{1}{a^3} m \abs{\boldsymbol\psi}^2 \equiv \sum_{i=1}^{3} \rho_i ~,
\end{align}
where $\Phi$ can be split into the conventional Newtonian part $\Phi_N$ and a nonminimal part $\Phi_\xi = -(2\xi_1+\xi_2) 2\pi G \rho/m^2$, the overline stands for spatial averaging and we have broken the energy density into its component parts $\rho_i \equiv a^{-3} m \abs{\psi_i}^2$. Besides the Newtonian potential $\Phi_N$, the nonminimal coupling results in a nonminimal potential $\Phi_\xi$, which represents attractive (repulsive) self-interactions for $2\xi_1+\xi_2>0$ ($<0$). These self-interactions do not depend on specific polarization states, and are distinct from those due to terms of the form $(X_\mu X^\mu)^2$, which would result in both $\psi_j \psi_j^\ast \psi_i$ and $\psi_j \psi_j \psi_i^\ast$ terms in \eqref{EOM_psi} and can break the degeneracy in energy of polarized vector solitons \cite{Jain:2021pnk, Zhang:2021xxa}. The modified SP equations and Friedmann equation \eqref{EOM_psi}-\eqref{EOM_rho} are our master equations in the wave description for VDM.

\subsection{Fluid description}
The fluid description is related to the SP equations through the Madelung transformation \cite{madelung1927quantentheorie}. To do the transform, let us define the fluid velocity ${\boldsymbol v}_i$ in terms of the phase of each field component $\psi_i$,
\begin{align}
	\psi_i \equiv \sqrt{\frac{\rho_i a^3}{m}} e^{i\theta} ~,\quad
	{\boldsymbol v}_i \equiv \frac{1}{ma} \nabla\theta_i ~.
\end{align}
For convenience, we do not use the Einstein summation convention in and only in this subsection. The equations \eqref{EOM_psi} become
\begin{align}
	\label{EOM_continuity}
	\dot\rho_i + 3 H \rho_i +  \frac{1}{a} \nabla\cdot(\rho_i {\boldsymbol v}_i) &= 0 ~,\\
	\label{EOM_Euler}
	\dot{{\boldsymbol v}_i} + H{\boldsymbol v}_i + \frac{1}{a} ({\boldsymbol v}_i\cdot\nabla) {\boldsymbol v}_i &= - \frac{1}{a}\nabla \left(\Phi_N + \Phi_{Q,i} + 2\Phi_\xi - \frac{(2\xi_1+\xi_2)}{2m^2a^2}\nabla^2\Phi_\xi \right) ~,
\end{align}
where $\Phi_\xi$ is the nonminimal part in \eqref{EOM_Phi} and
\begin{align}
	\label{quantum_potential}
	\Phi_{Q,i} \equiv - \frac{1}{2a^2m^2} \frac{\nabla^2\sqrt{\rho_i}}{\sqrt{\rho_i} } ~.
\end{align}
Physically, the Newtonian potential $\Phi_N$ attracts matter, the quantum potential $\Phi_Q$ repulses matter, and the property of interactions induced by nonminimal couplings depends on the sign of $2\xi_1+\xi_2$. In the large mass limit, the equations can be used to describe particle-like CDM \cite{Widrow:1993qq}. In section \ref{sec:nonminimal_pheno_perturbation}, we will use \eqref{EOM_continuity} and \eqref{EOM_Euler} to study the growth of linear perturbations and the Jeans scale.

\section{Phenomenological implications}
\label{sec:nonminimal_pheno}

With the wave and fluid description of VDM, we study the mass-radius relation of vector solitons and the growth of linear perturbations in section \ref{sec:nonminimal_pheno_soliton} and \ref{sec:nonminimal_pheno_perturbation} respectively. Moving beyond scalar perturbations, we will investigate the effects of nonminimal couplings on GW speed in section \ref{sec:nonminimal_pheno_gw}.

\subsection{Mass-radius relation of vector solitons}
\label{sec:nonminimal_pheno_soliton}
In section \ref{sec:soliton_vector}, we find that vector field theories admit three types of classically stable solutions with zero orbital angular momentum, also see \cite{Loginov:2015rya, Brito:2015pxa, Sanchis-Gual:2017bhw, Adshead:2021kvl, Jain:2021pnk, Zhang:2021xxa, Gorghetto:2022sue, Amin:2022pzv, Amin:2023imi, Jain:2023ojg}. The ground state solitons are polarized, not spherically symmetric in field configuration but are spherically symmetric in energy/mass density \cite{Jain:2021pnk, Zhang:2021xxa}. It is useful to note how the nonminimal couplings affect their mass-radius relation. Once again, we neglect the universe expansion since the size of solitons is much smaller than the Hubble patch.

The mass-radius relation can be found by minimizing the energy of the soliton at a fixed particle number. We define the total particle number $N$ and mass $M$ of a vector soliton by
\begin{align}
	\label{soliton_number_mass}
	N \equiv \int \dd[3]{x} \psi_i \psi_i^* ~,\quad
	M \equiv \int \dd[3]{x} \rho = m N ~,
\end{align}
where $N$ is the conserved charge associated with the global $U(1)$ symmetry $\psi_i\rightarrow e^{i\alpha} \psi_i$ in the NR action \eqref{eq:nr_action}. The radius $R$ is defined as that of the ball enclosing $99\%$ of the total mass. The NR action \eqref{eq:nr_action} is also invariant under time translation, and the associated conserved charge is identified as the energy $E_\psi$,
\begin{align}
	\label{eq:soliton_energy}
	E_\psi = \int \dd[3]{x} \left[ \frac{1}{2m} \partial_j \psi_i \partial_j \psi_i^* + m\Phi_N \abs{\boldsymbol\psi}^2 + 2m\Phi_\xi \abs{\boldsymbol\psi}^2 - \frac{(2\xi_1 + \xi_2) \nabla^2\Phi_\xi}{2m} \abs{\boldsymbol\psi}^2 \right] ~.
\end{align}
This energy does not include the rest mass energy of the original vector field $X_\mu$. In the NR limit, the total energy of $X_\mu$ is $E_X=E_\psi + mN$. Using the thin-wall approximation and replacing $\nabla$ with $1/R$ and $\int \dd[3]{x}$ with $R^3$ \cite{Lee:1991ax, Coleman:1985ki, Visinelli:2017ooc}, the energy can be written as\footnote{An alternative method is to assume a Gaussian ansatz for soliton profiles \cite{Chavanis:2011zm}. However, numerically Gaussian profiles are bad approximations if the nonminimal coupling becomes important.}
\begin{align}
	\label{eq:E_M_R_relation}
	E_\psi \propto \frac{\beta_1}{2} \frac{M}{m^2 R^2} - \frac{M^2}{\MP^2 R} - \frac{\beta_2}{3} \frac{(2\xi_1+\xi_2) M^2}{m^2 \MP^2 R^3} - \frac{\beta_3}{5} \frac{(2\xi_1+\xi_2)^2 M^2}{m^4 \MP^2 R^5} ~,
\end{align}
where $\beta_1,\beta_2,\beta_3$ are numerical coefficients to be determined, and we have replaced center field amplitudes $\psi_c$ with $M\sim m\psi_c^2 R^3$. This expression of energy allows us to minimize the energy at a fixed particle number, to wit, $\delta(E_\psi+\mu N)=0$ where $\mu$ is a Lagrangian multiplier \cite{Lee:1991ax}. By varying $E_\psi+\mu N$ with respect to $R$ for a fixed $M$, we obtain
\begin{align}
	\label{eq:MR}
	M = \frac{\beta_1 m^3R^3}{m^4R^4 + \beta_2(2\xi_1+\xi_2) m^2R^2 + \beta_3 (2\xi_1+\xi_2)^2} \frac{\MP^2}{m} ~.
\end{align}
The numerical coefficients $\beta_1,\beta_2,\beta_3$ can be determined by comparing \eqref{eq:MR} with the mass-radius relation obtained using the numerical shooting method.
\begin{figure}
	\centering
	\includegraphics[width=0.6\linewidth]{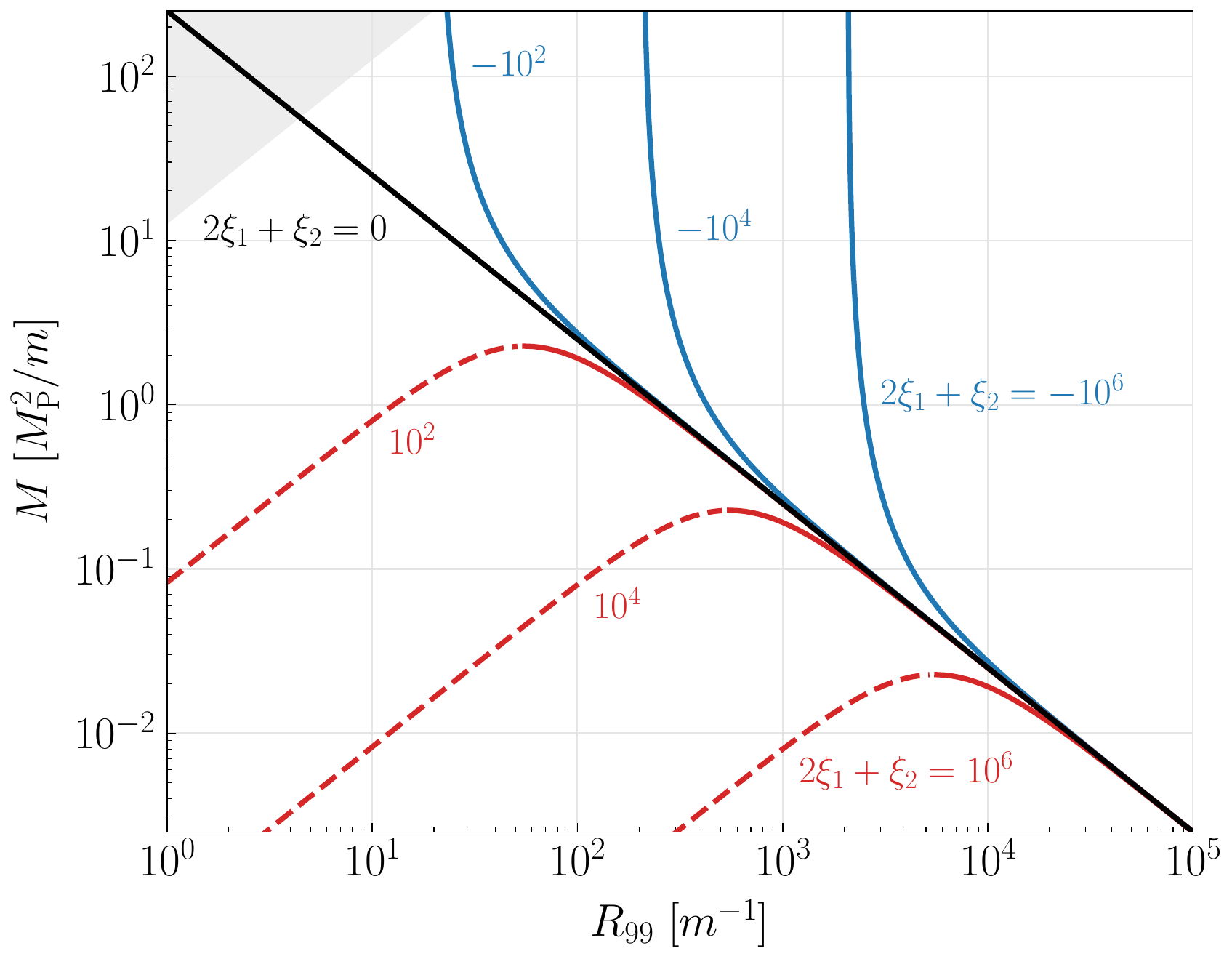}
	\caption{Mass-radius relation of the ground-state vector solitons, given by equation \eqref{eq:MR}. The dashed line indicates a classical instability against small perturbations. The gray region at the upper left corner corresponds to the regime in which $R$ is comparable with the Schwarzschild radius, $mM/\MP^2 \leq 4\pi mR$, where gravity becomes strong and the NR approximation breaks down.}
	\label{fig:mass_radius_relation}
\end{figure}
We find the above formula to be a good approximation with $\beta_1=250$, $\beta_2=30.25$, $\beta_3=0$ for $2\xi_1+\xi_2>0$ and with $\beta_1=250$, $\beta_2=8.5$, $\beta_3=18.06$ for $2\xi_1+\xi_2<0$.\footnote{In finding these numbers, we have assumed that the size of solitons is much larger than the strength of the nonminimal couplings, $m^2R^2 \gg 2\xi_1+\xi_2$, and thus approximated $\rho_\xi\approx \rho$ in \eqref{EOM_rho}.} The mass-radius relation \eqref{eq:MR} is shown in figure \ref{fig:mass_radius_relation}, which manifests a similar qualitative behavior compared to that from covariant quartic self-interactions \cite{Chavanis:2011zi, Chavanis:2011zm, Salehian:2021khb}. One can see that the low-radius part of the curve is dramatically changed compared to the minimal case, signifying the fact that the nonminimal coupling becomes more important than Newtonian gravity at small scales. Also, if one wants to investigate nonlinear dynamics of vector solitons, one must carry out rigorous lattice simulations (See \cite{Gorghetto:2022sue,Amin:2022pzv,Zhang:2021xxa,Brito:2015pxa,Sanchis-Gual:2017bhw,Amin:2023imi,Jain:2023ojg} for examples.).

We end this subsection with a brief discussion on the possibility of inferring soliton configurations from the density profile of galaxies. Solitons offer a natural solution to the core-cusp problem, which refers to the mismatch between the cuspidal density profile predicted by CDM simulations and the flatter ones observed in galactic centers \cite{Ferreira:2020fam}. By hosting a solitonic core, the galaxy can exhibit a smooth central density profile, thereby preventing gravitational clustering of matter and resolving the core-cusp problem. While a proliferation of favorable evidence for such solitonic cores has emerged in recent years, the existence of solitonic cores also pose strong constraints on the mass of DM bosons, which were obtained by comparing the predicted density profiles of solitonic galaxy cores and the observed ones \cite{Schive:2014hza, Marsh:2015wka, Chen:2016unw, Gonzalez-Morales:2016yaf, Broadhurst:2019fsl, Bar:2018acw, Veltmaat:2018dfz, DeMartino:2018zkx, Pozo:2023zmx, ParticleDataGroup:2022pth}. Nonminimal couplings to gravity could weaken such constraints and alleviate tensions between the halo mass and soliton mass in some galaxies \cite{Robles:2018fur, Desjacques:2019zhf, Safarzadeh:2019sre}.  Further investigation is needed to explore this possibility.

\subsection{Growth of linear perturbations}
\label{sec:nonminimal_pheno_perturbation}

In this section, we will use the fluid description to study the growth of density perturbations for wave VDM.\footnote{For scalar DM, the density perturbation has been calculated in the fluid description up to the third order in $\delta$ and $v$ to obtain the one-loop power spectrum \cite{Li:2018kyk}. One may also use the wave perturbations $\delta\psi\equiv \psi-\bar\psi$ to study structure growth; that said, this approach breaks down at higher redshifts compared to the fluid description \cite{Li:2018kyk}.} We find that significant nonminimal couplings may lead to distinct evolution of perturbations, where both the small and large scale perturbations can grow significantly. As the universe expands, the importance of the nonminimal coupling decreases and eventually the standard picture of wave DM \cite{Hu:2000ke} is recovered, albeit with a modified Jeans scale.

Linearizing the fluid equations \eqref{EOM_continuity} and \eqref{EOM_Euler}, we obtain the equations for perturbations around a homogeneous background
\begin{align}
	\dot\delta_i + \frac{1}{a}\nabla\cdot{\boldsymbol v}_i &= 0 ~,\\
	\dot{{\boldsymbol v}_i} + H{\boldsymbol v}_i  &= -\frac{1}{a}\nabla \left(\Phi_N + \Phi_{Q,i} + 2 \Phi_\xi - \frac{(2\xi_1+\xi_2)}{2m^2a^2}\nabla^2\Phi_\xi \right) ~,
\end{align}
where $\Phi_N,\Phi_{Q,i},\Phi_\xi$ are given by \eqref{EOM_Phi} and \eqref{quantum_potential}, $\rho_i \equiv \bar\rho_i (1+\delta_i)$ and the background densities satisfy $\dot{\bar\rho}_i + 3H\bar\rho_i = 0$. These equations can be combined to yield a set of second-order differential equations for the overdensity $\delta_i$. In Fourier space, it can be written as
\begin{align}
	\label{overdensity_eq}
	\left[\partial_t^2 + 2H\partial_t +
	\begin{pmatrix}
		\frac{k^4}{4a^4m^2} + c_1 & c_2 & c_3 \\
		0 & \frac{k^4}{4a^4m^2} & 0 \\
		0 & 0 & \frac{k^4}{4a^4m^2}
	\end{pmatrix}
	\right]
	\begin{pmatrix}
		\Delta_1 \\
		\Delta_2 \\
		\Delta_3
	\end{pmatrix} = 0 ~,
\end{align}
where
\begin{align*}
	\begin{pmatrix}
		\Delta_1 \\
		\Delta_2 \\
		\Delta_3
	\end{pmatrix}
	\equiv
	Q
	\begin{pmatrix}
		\delta_1 \\
		\delta_2 \\
		\delta_3
	\end{pmatrix} ~,\quad
	Q = \frac{1}{\sqrt{6}} 
	\begin{pmatrix}
		\sqrt{2} & \sqrt{2} & \sqrt{2} \\
		2 & -1 & -1 \\
		0 & \sqrt{3} & -\sqrt{3}
	\end{pmatrix} ~,
\end{align*}
and we have defined
\begin{align*}
	c_1 &\equiv - \left[ \frac{k^2}{2m^2a^2} (2\xi_1+\xi_2) \sqrt{4\pi G\bar\rho} + \sqrt{4\pi G\bar\rho} \right]^2 ~,\\
	c_2 &\equiv -\frac{\pi G (2a^2m^2 + k^2(2\xi_1+\xi_2))^2}{\sqrt{2} a^4m^4} (2\bar\rho_1 - \bar\rho_2 - \bar\rho_3) ~,\\
	c_3 &\equiv -\frac{\sqrt{3}\pi G (2a^2m^2 + k^2(2\xi_1+\xi_2))^2}{\sqrt{2} a^4m^4} (\bar\rho_2 - \bar\rho_3) ~.
\end{align*}
Assuming $\bar\rho_1=\bar\rho_2=\bar\rho_3$, which is expected unless there is a mechanism that favors VDM with a particular polarization state at the time of production and throughout its subsequent evolution,\footnote{In fact, this assumption holds as long as there is no statistically preferred direction. In particular, helicities (in Fourier space) produced by mechanisms described in refs.~\cite{Agrawal:2018vin, Co:2018lka, Bastero-Gil:2018uel, Co:2021rhi, Graham:2015rva, Dror:2018pdh, Salehian:2020asa} do not invalidate this assumption.} the differential operator of \eqref{overdensity_eq} becomes diagonal and one can study $\Delta_i$ separately.\footnote{Since $\Delta_1$ corresponds to the direction $\delta_1 = \delta_2 = \delta_3$ (which does not distinguish between spatial indices), $\Delta_1$ should be interpreted as the isotropic component of the density contrast; accordingly, $\Delta_2$ and $\Delta_3$ are the anisotropic components. If $\bar{\rho}_1 = \bar{\rho}_2 = \bar{\rho}_3$, then equation \eqref{overdensity_eq} is symmetric under permutations of the spatial indices $i=1,2,3$, and the isotropic and anisotropic components correspond to orthogonal invariant subspaces of the solution space. Note that the orthogonal transform $Q$ is not to be understood as a spatial rotation since the $\delta_i$'s do not form a vector under spatial rotation.} While both $\Delta_2$ and $\Delta_3$ oscillate and do not experience significant growth due to their positive effective mass, the growth of $\Delta_1$ depends on the sign of 
\begin{align}
	\label{jeans_effective_mass}
	\nonumber
	\Omega^2 &\equiv \frac{k^4}{4a^4m^2} + c_1 \\
	\nonumber
	&= -4\pi G \bar\rho \left[ 1 + (2\xi_1 + \xi_2) \frac{k^2}{a^2m^2}  + \left( (2\xi_1 + \xi_2)^2 - \frac{m^2}{4\pi G \bar\rho} \right) \frac{k^4}{4a^4m^4} \right] \\
	&= \frac{1}{4a^4m^2} \left[ 1 -(2\xi_1+\xi_2)^2 \frac{4\pi G \bar{\rho}}{m^2} \right] (k^2 + k_-^2)(k^2 - k_+^2) ~,
\end{align}
where
\begin{align}
	k_\pm^2 \equiv k_{J,0}^2 \frac{1}{1 \mp (2\xi_1 + \xi_2)\sqrt{4\pi G\bar\rho/m^2}} ~,\quad
	k_{J,0} = (16\pi G m^2 \bar\rho a^4)^{1/4} ~,
\end{align}
with $k_{J,0}$ being the comoving Jeans scale without nonminimal couplings \cite{Hu:2000ke, Gorghetto:2022sue}. The sign of $\Omega^2$ is undetermined, because the upper limit of $\xi_a$ $(a=1,2)$ is given by $\xi_a\ll m^2 \MP^2 / \bar\rho$, i.e. equation \eqref{xi_range}, which does not fix the relative magnitude between $\xi_a^2$ and $m^2 \MP^2 / \bar\rho$.

The nonminimal couplings can play an important role if $(2\xi_1 + \xi_2)^2\gg m^2\MP^2/\bar\rho$. In this case, we have $k_+^2<0, k_-^2>0$ for positive $2\xi_1+\xi_2$, hence $\Omega^2<0$ and perturbations grow for all $k$ modes.  For negative $2\xi_1+\xi_2$, we have $k_+^2>0, k_-^2<0$, thus perturbations with $\abs{k_+}<k<\abs{k_-}$ oscillate whereas others grow. In contrast, the small-scale perturbations with $k>k_{J,0}$ are suppressed for minimally coupled wave DM \cite{Hu:2000ke, Gorghetto:2022sue} -- the presence of significant nonminimal couplings could enhance small-scale structure.

The evolution of CDM perturbations $\delta \propto a$ for $k<k_\rm{obs} \sim 10 ~h\mathrm{Mpc^{-1}}\sim 10^3 k_\rm{eq}$ is consistent with current observations \cite{Irsic:2017yje, Bechtol:2022koa}. To retain the success of CDM on large-scale perturbations, we demand $\Omega^2 \approx -4\pi G\bar\rho$ for $k<k_\rm{obs}$ since the matter-radiation equality, which yields
\begin{align}
	\label{xi_large_scale_perturbation}
	\abs{2\xi_1 + \xi_2} \ll \frac{a_\rm{eq}^2m^2}{k_\rm{obs}^2} = 10^{10} \left( \frac{m}{10^{-20} \mathrm{eV}} \right)^2 \left( \frac{10^{-28} \rm{eV}}{H_\mathrm{eq}} \right)^2 ~,
\end{align}
This bound is more stringent than equation \eqref{xi_range}, but weaker than the constraint we will obtain for $\xi_2$ in section \ref{sec:nonminimal_pheno_gw} based current limits on GW speed.

As the energy density decreases with the expansion of the universe, eventually $m^2\MP^2/\bar\rho$ will dominate over $(2\xi_1+\xi_2)^2$ and we recover the standard evolution of perturbations for wave DM \cite{Hu:2000ke, Gorghetto:2022sue}. The comoving Jeans scale for the subsequent evolution is
\begin{align}
	\label{jeans_scale}
	k_J = k_+ \simeq a (16 \pi G m^2 \bar\rho)^{1/4} \left[ 1 + (2\xi_1 + \xi_2) \sqrt{\frac{\pi G \bar\rho}{m^2}} \right] ~.
\end{align}
Perturbations grow at large length scales with $k\ll k_J$ and oscillate at small scales with $k\gg k_J$.

It has been shown that minimally coupled wave DM has a sharp break in the power spectrum of density perturbations with $k$ larger than $k_{J,0,\mathrm{eq}}$, the comoving Jeans scale at matter-radiation equality~\cite{Hu:2000ke}. The existence of nonminimal couplings is expected to leave imprints on the spectrum at small scales and also shift the break toward higher or lower $k$ depending on the value of $2\xi_1+\xi_2$. The small-scale structure could be enhanced if the nonminimal coupling is significant in early times. It might be interesting to explore the related phenomenology in detail in the future.

\subsection{Speed of gravitational waves}
\label{sec:nonminimal_pheno_gw}
The presence of nonminimal couplings for VDM changes the propagation speed of GWs. In this subsection, we investigate this effect and put constraints on $\xi_2$ and the VDM mass based on current limits on GW speed.\footnote{In contrast, the nonminimal coupling of the form $\phi^2R$ for scalar DM \cite{Ji:2021rrn} effectively modifies the Planck constant and has no impact on GW speed, unless interactions suppressed by higher-order gradients are considered \cite{Ivanov:2019iec}.}

Since the propagation distance of GWs is typically much less than the Hubble horizon, we will neglect the expansion of the universe and consider a metric with tensor perturbations $g_{\mu\nu} = \eta_{\mu\nu} + h_{\mu\nu}$, where $\eta_{\mu\nu}$ is the Minkowski metric by choosing locally inertial coordinates of VDM, and $h_{\mu\nu}$ is traceless and transverse. The action for linearized GWs is
\begin{align}
	\label{action_tensor}
	S^{(2)} = \frac{1}{2} \sum_{\lambda=+,\times} \int d^4x~ M_*^2 \[ \dot h_\lambda^2 - c_T^2 (\nabla h_\lambda)^2 \] ~,
\end{align}
where $h_{\lambda}$ is the amplitude of polarization states, and the effective reduced Planck mass $M_*$ and the speed of tensor modes $c_T$ are\footnote{To obtain the expression, we have integrated out the fast oscillating modes of VDM, as we did when deriving the NR EFT.}
\begin{align}
	M_*^2 &= \MP^2  \[ 1 + \xi_1 \frac{\rho}{m^2\MP^2} + \xi_2 \frac{\rho-\rho_{\hat n}}{m^2\MP^2} + \cal O(\epsilon^2) \] ~,\\
	M_*^2 c_T^2 &= \MP^2 \[ 1 + (\xi_1 + \xi_2)\frac{\rho}{m^2\MP^2} + \cal O(\epsilon^2) \]~.
\end{align}
Here, $\hat n$ is a unit vector with the same direction as the propagation of GWs, and $\rho_{\hat n} \equiv m\abs{\b \psi \cdot \hat n}^2$ represents the energy density contributed by the component of the field $\b \psi$ parallel to $\hat n$. Since $M_*^2>0$ and $c_T^2>0$, the tensor-mode action \eqref{action_tensor} is free from ghost and gradient instabilities \cite{DeFelice:2011bh}. 

The deviation of the GW speed from the light speed can be characterized by $\alpha_T \equiv c_T^2-1 \approx \xi_2 \rho_{\hat n} / (m^2\MP^2)$. If the VDM can be regarded as homogeneous and unpolarized (i.e. $\b\psi$ does not have a preferred direction) at the length scales comparable to the propagating distance of GWs, then $\rho_{\hat n}\simeq \rho/3$ and we obtain
\begin{align}
	\label{alpha_T}
	\alpha_T \simeq 1.7 \times 10^{-11} \(\frac{\xi_2}{10^{10}}\) \( \frac{\rho}{0.4 \rm{GeV/cm^3}} \) \(\frac{10^{-20} \rm{eV}}{m}\)^{2} ~.
\end{align}
For GW events detected by LIGO, we may use a ``separate-region approximation'' to describe the GW speed -- assuming that GWs propagate through two separate homogeneous regions of galactic halos and intergalactic medium. Denoting the total propagating distance and the distance propagated in galactic halos as $d$ and $d_\rm{halo}$, one can estimate the time differences between photons and GWs accumulated in these two regions
\begin{align}
	\label{time_delay_halo}
	\Delta t_\rm{halo} &\sim \frac{1}{2} d_\rm{halo} \alpha_T \simeq 8.89 \rm{s} ~ \( \frac{d_\rm{halo}}{10 \rm{kpc}} \) \(\frac{\alpha_T}{1.7 \times 10^{-11}}\) ~,\\
	\label{time_delay_intga}
	\Delta t_\rm{intga} &\sim \frac{1}{2}(d-d_\rm{halo}) \alpha_T \simeq 0.09 \rm{s} ~ \( \frac{d-d_\rm{halo}}{40 \rm{Mpc}} \)\(\frac{\alpha_T}{4.3 \times 10^{-17}}\)  ~,
\end{align}
where the reference values of $\alpha_T$ in \eqref{time_delay_halo} and \eqref{time_delay_intga} are those given by \eqref{alpha_T} with the DM density being $0.4\rm{GeV/cm^3}$ (local DM density)  \cite{deBoer:2010eh, Bovy:2012tw, mckee2015stars, Sivertsson:2017rkp} and $1\rm{GeV/m^3}$ (DM density at cosmological scales) \cite{ParticleDataGroup:2022pth}. Thus for a leading-order calculation of arrival time differences between photons and GWs, one may neglect the contribution due to the intergalactic propagation.

\begin{figure}
	\centering
	\begin{minipage}{0.49\linewidth}
		\includegraphics[width=\linewidth]{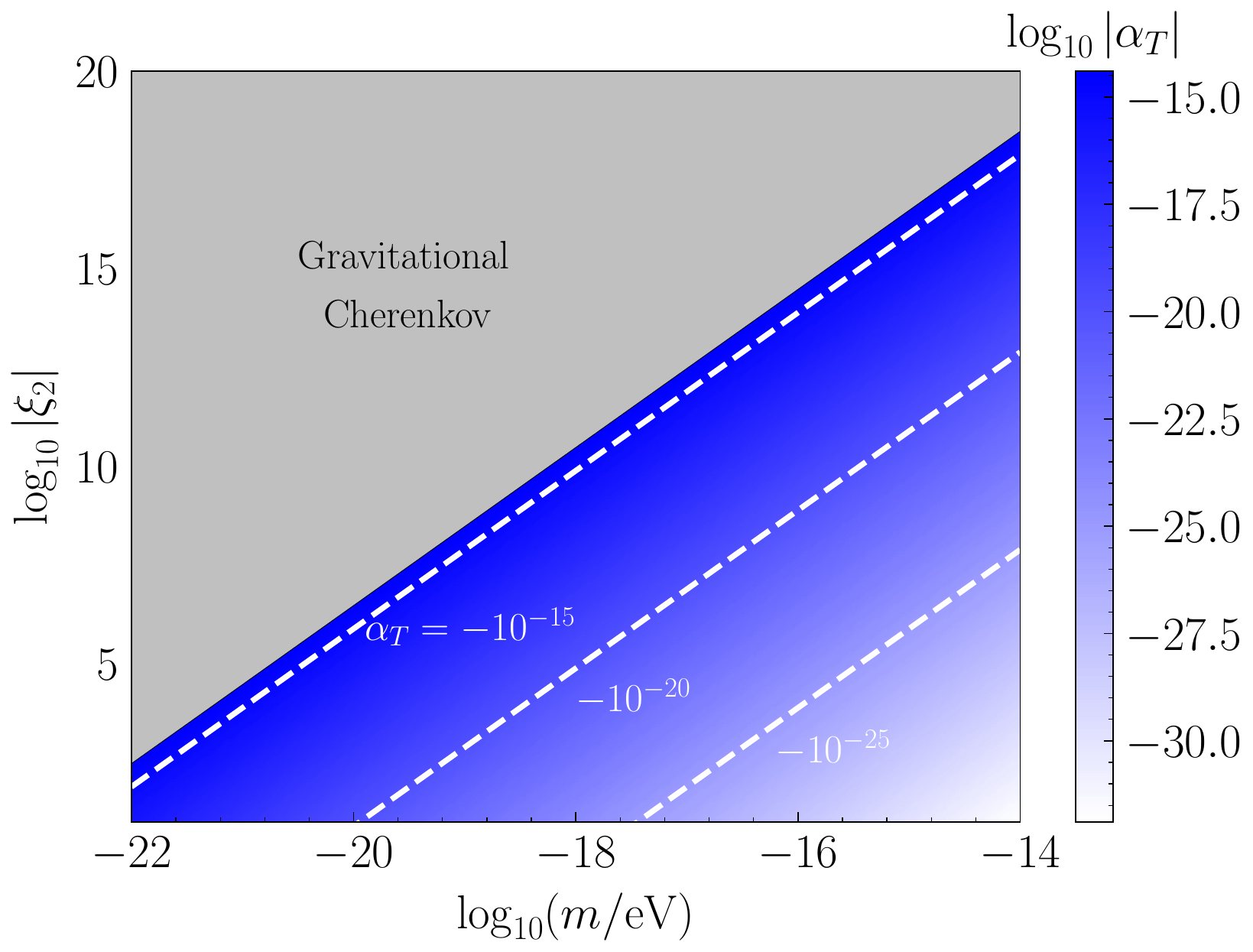}
	\end{minipage}
	\begin{minipage}{0.49\linewidth}
		\includegraphics[width=\linewidth]{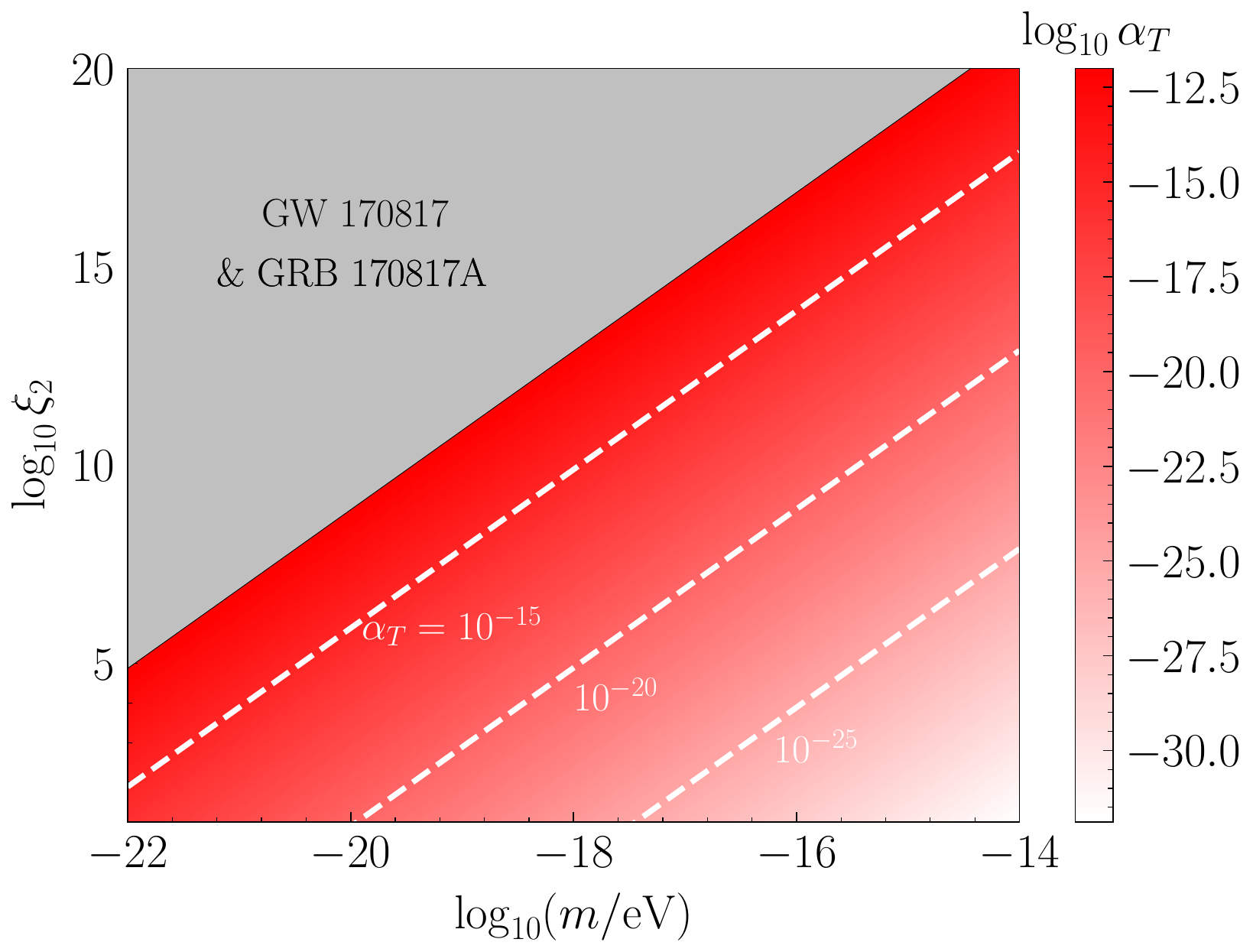}
	\end{minipage}
	\caption{Constraints on the speed of gravitational waves, characterized by $\alpha_T = c_T^2 - 1 \approx 2(c_T-1)$, for the negative (left) and positive (right) nonminimal coupling $\xi_2$. The gray regions are excluded by the lack of observation for gravitational Cherenkov radiation of cosmic rays with a galactic origin or by the difference in arrival time between GW 170817 \& GRB 170817A.}
	\label{fig:gwspeed}
\end{figure}

A strong constraint on $\alpha_T$ can be extracted with the fact that GW 170817 and GRB 170817A, which were emitted by the same binary neutron stars, arrived on the Earth at nearly the same time (the coalescence signal is $1.7\rm{s}$ prior to the peak of GRB) after propagating a distance of $d \simeq 40~\rm{Mpc}$ \cite{LIGOScientific:2017zic, Baker:2017hug}. As a conservative estimation, we assume that the GRB peak was emitted after the coalescence within $10\rm{s}$, then the time difference between the GRB and GW accumulated during the propagation satisfies $-8.3 \rm{s} < \Delta t < 1.7\rm{s}$. Replacing $\Delta t_\rm{halo}$ in equation \eqref{time_delay_halo} with $\Delta t$, the $\alpha_T$ is constrained to be within $-1.6\times 10^{-11} \lesssim \alpha_T \lesssim 3.3\times 10^{-12}$. A stronger lower bound $\alpha_T \gtrsim -4\times 10^{-15}$ can be put due to a lack of observation for gravitational Cherenkov radiation from cosmic rays with a galactic origin \cite{Moore:2001bv}. Combining these considerations, we obtain
\begin{align}
	\label{alphaT_constraint}
	-4\times 10^{-15} \lesssim \alpha_T \lesssim 3\times 10^{-12} \( \frac{10 ~\rm{kpc}}{d_\rm{halo}} \) ~.
\end{align}
In terms of the $\xi_2$ parameter and VDM mass, equations \eqref{alpha_T} and \eqref{alphaT_constraint} yield
\begin{align}
	\label{xi2_constraint}
	-2\times 10^{46} \( \frac{0.4 \rm{GeV/cm^3}}{\rho_\rm{halo}} \) \rm{eV^{-2}} \lesssim \frac{\xi_2}{m^2} \lesssim 2\times 10^{49} \( \frac{10 \rm{kpc}}{d_\rm{halo}} \) \( \frac{0.4 \rm{GeV/cm^3}}{\rho_\rm{halo}} \) \rm{eV^{-2}} ~.
\end{align}
The constraints \eqref{alphaT_constraint} and \eqref{xi2_constraint} are shown in figure \ref{fig:gwspeed}.



\bibliographystyle{jhep}
\bibliography{ref}

\providecommand{\href}[2]{#2}\begingroup\raggedright\begin{thebibliography}{100}

\bibitem{LZ:2022lsv}
{\scshape LZ} collaboration, \emph{{First Dark Matter Search Results from the
  LUX-ZEPLIN (LZ) Experiment}},
  \href{https://doi.org/10.1103/PhysRevLett.131.041002}{\emph{Phys. Rev. Lett.}
  {\bfseries 131} (2023) 041002}
  [\href{https://arxiv.org/abs/2207.03764}{{\ttfamily 2207.03764}}].

\bibitem{PandaX-4T:2021bab}
{\scshape PandaX-4T} collaboration, \emph{{Dark Matter Search Results from the
  PandaX-4T Commissioning Run}},
  \href{https://doi.org/10.1103/PhysRevLett.127.261802}{\emph{Phys. Rev. Lett.}
  {\bfseries 127} (2021) 261802}
  [\href{https://arxiv.org/abs/2107.13438}{{\ttfamily 2107.13438}}].

\bibitem{XENON:2018voc}
{\scshape XENON} collaboration, \emph{{Dark Matter Search Results from a One
  Ton-Year Exposure of XENON1T}},
  \href{https://doi.org/10.1103/PhysRevLett.121.111302}{\emph{Phys. Rev. Lett.}
  {\bfseries 121} (2018) 111302}
  [\href{https://arxiv.org/abs/1805.12562}{{\ttfamily 1805.12562}}].

\bibitem{LUX:2016ggv}
{\scshape LUX} collaboration, \emph{{Results from a search for dark matter in
  the complete LUX exposure}},
  \href{https://doi.org/10.1103/PhysRevLett.118.021303}{\emph{Phys. Rev. Lett.}
  {\bfseries 118} (2017) 021303}
  [\href{https://arxiv.org/abs/1608.07648}{{\ttfamily 1608.07648}}].

\bibitem{DEAP:2019yzn}
{\scshape DEAP} collaboration, \emph{{Search for dark matter with a 231-day
  exposure of liquid argon using DEAP-3600 at SNOLAB}},
  \href{https://doi.org/10.1103/PhysRevD.100.022004}{\emph{Phys. Rev. D}
  {\bfseries 100} (2019) 022004}
  [\href{https://arxiv.org/abs/1902.04048}{{\ttfamily 1902.04048}}].

\bibitem{AxionLimits}
C.~O'Hare, ``cajohare/axionlimits: Axionlimits.''
  \url{https://cajohare.github.io/AxionLimits/}, July, 2020.
\newblock 10.5281/zenodo.3932430.

\bibitem{zwicky1933rotshift}
F.~Zwicky, \emph{The redshift of extragalactic nebulae}, {\emph{Helvetica
  Physica Acta, Vol. 6, p. 110-127} {\bfseries 6} (1933) 110}.

\bibitem{Rubin:1970zza}
V.~C. Rubin and W.~K. Ford, Jr., \emph{{Rotation of the Andromeda Nebula from a
  Spectroscopic Survey of Emission Regions}},
  \href{https://doi.org/10.1086/150317}{\emph{Astrophys. J.} {\bfseries 159}
  (1970) 379}.

\bibitem{Bertone:2016nfn}
G.~Bertone and D.~Hooper, \emph{{History of dark matter}},
  \href{https://doi.org/10.1103/RevModPhys.90.045002}{\emph{Rev. Mod. Phys.}
  {\bfseries 90} (2018) 045002}
  [\href{https://arxiv.org/abs/1605.04909}{{\ttfamily 1605.04909}}].

\bibitem{dodelson2020modern}
S.~Dodelson and F.~Schmidt, \emph{Modern cosmology}. Academic press, 2020.

\bibitem{Baumann:2022mni}
D.~Baumann, \emph{{Cosmology}}. Cambridge University Press, 7, 2022,
  \href{https://doi.org/10.1017/9781108937092}{10.1017/9781108937092}.

\bibitem{Planck:2018vyg}
{\scshape Planck} collaboration, \emph{{Planck 2018 results. VI. Cosmological
  parameters}},
  \href{https://doi.org/10.1051/0004-6361/201833910}{\emph{Astron. Astrophys.}
  {\bfseries 641} (2020) A6}
  [\href{https://arxiv.org/abs/1807.06209}{{\ttfamily 1807.06209}}].

\bibitem{Roszkowski:2017nbc}
L.~Roszkowski, E.~M. Sessolo and S.~Trojanowski, \emph{{WIMP dark matter
  candidates and searches\textemdash{}current status and future prospects}},
  \href{https://doi.org/10.1088/1361-6633/aab913}{\emph{Rept. Prog. Phys.}
  {\bfseries 81} (2018) 066201}
  [\href{https://arxiv.org/abs/1707.06277}{{\ttfamily 1707.06277}}].

\bibitem{Lin:2019uvt}
T.~Lin, \emph{{Dark matter models and direct detection}},
  \href{https://doi.org/10.22323/1.333.0009}{\emph{PoS} {\bfseries 333} (2019)
  009} [\href{https://arxiv.org/abs/1904.07915}{{\ttfamily 1904.07915}}].

\bibitem{Arcadi:2017kky}
G.~Arcadi, M.~Dutra, P.~Ghosh, M.~Lindner, Y.~Mambrini, M.~Pierre et~al.,
  \emph{{The waning of the WIMP? A review of models, searches, and
  constraints}},
  \href{https://doi.org/10.1140/epjc/s10052-018-5662-y}{\emph{Eur. Phys. J. C}
  {\bfseries 78} (2018) 203}
  [\href{https://arxiv.org/abs/1703.07364}{{\ttfamily 1703.07364}}].

\bibitem{Safdi:2022xkm}
B.~R. Safdi, \emph{{TASI Lectures on the Particle Physics and Astrophysics of
  Dark Matter}},  \href{https://arxiv.org/abs/2303.02169}{{\ttfamily
  2303.02169}}.

\bibitem{Flores:1994gz}
R.~A. Flores and J.~R. Primack, \emph{{Observational and theoretical
  constraints on singular dark matter halos}},
  \href{https://doi.org/10.1086/187350}{\emph{Astrophys. J. Lett.} {\bfseries
  427} (1994) L1} [\href{https://arxiv.org/abs/astro-ph/9402004}{{\ttfamily
  astro-ph/9402004}}].

\bibitem{Moore:1994yx}
B.~Moore, \emph{{Evidence against dissipationless dark matter from observations
  of galaxy haloes}}, \href{https://doi.org/10.1038/370629a0}{\emph{Nature}
  {\bfseries 370} (1994) 629}.

\bibitem{Hui:2021tkt}
L.~Hui, \emph{{Wave Dark Matter}},
  \href{https://doi.org/10.1146/annurev-astro-120920-010024}{\emph{Ann. Rev.
  Astron. Astrophys.} {\bfseries 59} (2021) 247}
  [\href{https://arxiv.org/abs/2101.11735}{{\ttfamily 2101.11735}}].

\bibitem{Ferreira:2020fam}
E.~G.~M. Ferreira, \emph{{Ultra-light dark matter}},
  \href{https://doi.org/10.1007/s00159-021-00135-6}{\emph{Astron. Astrophys.
  Rev.} {\bfseries 29} (2021) 7}
  [\href{https://arxiv.org/abs/2005.03254}{{\ttfamily 2005.03254}}].

\bibitem{Colin:2000dn}
P.~Colin, V.~Avila-Reese and O.~Valenzuela, \emph{{Substructure and halo
  density profiles in a warm dark matter cosmology}},
  \href{https://doi.org/10.1086/317057}{\emph{Astrophys. J.} {\bfseries 542}
  (2000) 622} [\href{https://arxiv.org/abs/astro-ph/0004115}{{\ttfamily
  astro-ph/0004115}}].

\bibitem{Spergel:1999mh}
D.~N. Spergel and P.~J. Steinhardt, \emph{{Observational evidence for
  selfinteracting cold dark matter}},
  \href{https://doi.org/10.1103/PhysRevLett.84.3760}{\emph{Phys. Rev. Lett.}
  {\bfseries 84} (2000) 3760}
  [\href{https://arxiv.org/abs/astro-ph/9909386}{{\ttfamily
  astro-ph/9909386}}].

\bibitem{Jacobs:2014yca}
D.~M. Jacobs, G.~D. Starkman and B.~W. Lynn, \emph{{Macro Dark Matter}},
  \href{https://doi.org/10.1093/mnras/stv774}{\emph{Mon. Not. R. Astron. Soc.}
  {\bfseries 450} (2015) 3418}
  [\href{https://arxiv.org/abs/1410.2236}{{\ttfamily 1410.2236}}].

\bibitem{Khlopov:2019qcr}
M.~Y. Khlopov, \emph{{Direct and Indirect Probes for Composite Dark Matter}},
  \href{https://doi.org/10.3389/fphy.2019.00004}{\emph{Front. in Phys.}
  {\bfseries 7} (2019) 4}.

\bibitem{Carr:2020gox}
B.~Carr, K.~Kohri, Y.~Sendouda and J.~Yokoyama, \emph{{Constraints on
  primordial black holes}},
  \href{https://doi.org/10.1088/1361-6633/ac1e31}{\emph{Rept. Prog. Phys.}
  {\bfseries 84} (2021) 116902}
  [\href{https://arxiv.org/abs/2002.12778}{{\ttfamily 2002.12778}}].

\bibitem{Carr:2020xqk}
B.~Carr and F.~Kuhnel, \emph{{Primordial Black Holes as Dark Matter: Recent
  Developments}},
  \href{https://doi.org/10.1146/annurev-nucl-050520-125911}{\emph{Annu. Rev.
  Nucl. Part. Sci.} (2020) }
  [\href{https://arxiv.org/abs/2006.02838}{{\ttfamily 2006.02838}}].

\bibitem{deBoer:2010eh}
W.~de~Boer and M.~Weber, \emph{{The Dark Matter Density in the Solar
  Neighborhood reconsidered}},
  \href{https://doi.org/10.1088/1475-7516/2011/04/002}{\emph{JCAP} {\bfseries
  04} (2011) 002} [\href{https://arxiv.org/abs/1011.6323}{{\ttfamily
  1011.6323}}].

\bibitem{Bovy:2012tw}
J.~Bovy and S.~Tremaine, \emph{{On the local dark matter density}},
  \href{https://doi.org/10.1088/0004-637X/756/1/89}{\emph{Astrophys. J.}
  {\bfseries 756} (2012) 89} [\href{https://arxiv.org/abs/1205.4033}{{\ttfamily
  1205.4033}}].

\bibitem{mckee2015stars}
C.~F. McKee, A.~Parravano and D.~J. Hollenbach, \emph{Stars, gas, and dark
  matter in the solar neighborhood}, {\emph{The Astrophysical Journal}
  {\bfseries 814} (2015) 13}.

\bibitem{Sivertsson:2017rkp}
S.~Sivertsson, H.~Silverwood, J.~I. Read, G.~Bertone and P.~Steger, \emph{{The
  localdark matter density from SDSS-SEGUE G-dwarfs}},
  \href{https://doi.org/10.1093/mnras/sty977}{\emph{Mon. Not. Roy. Astron.
  Soc.} {\bfseries 478} (2018) 1677}
  [\href{https://arxiv.org/abs/1708.07836}{{\ttfamily 1708.07836}}].

\bibitem{Guth:2014hsa}
A.~H. Guth, M.~P. Hertzberg and C.~Prescod-Weinstein, \emph{{Do Dark Matter
  Axions Form a Condensate with Long-Range Correlation?}},
  \href{https://doi.org/10.1103/PhysRevD.92.103513}{\emph{Phys. Rev. D}
  {\bfseries 92} (2015) 103513}
  [\href{https://arxiv.org/abs/1412.5930}{{\ttfamily 1412.5930}}].

\bibitem{Hertzberg:2016tal}
M.~P. Hertzberg, \emph{{Quantum and Classical Behavior in Interacting Bosonic
  Systems}}, \href{https://doi.org/10.1088/1475-7516/2016/11/037}{\emph{JCAP}
  {\bfseries 11} (2016) 037}
  [\href{https://arxiv.org/abs/1609.01342}{{\ttfamily 1609.01342}}].

\bibitem{Tremaine:1979we}
S.~Tremaine and J.~E. Gunn, \emph{{Dynamical Role of Light Neutral Leptons in
  Cosmology}}, \href{https://doi.org/10.1103/PhysRevLett.42.407}{\emph{Phys.
  Rev. Lett.} {\bfseries 42} (1979) 407}.

\bibitem{Davoudiasl:2020uig}
H.~Davoudiasl, P.~B. Denton and D.~A. McGady, \emph{{Ultralight fermionic dark
  matter}}, \href{https://doi.org/10.1103/PhysRevD.103.055014}{\emph{Phys. Rev.
  D} {\bfseries 103} (2021) 055014}
  [\href{https://arxiv.org/abs/2008.06505}{{\ttfamily 2008.06505}}].

\bibitem{Hu:2000ke}
W.~Hu, R.~Barkana and A.~Gruzinov, \emph{{Cold and fuzzy dark matter}},
  \href{https://doi.org/10.1103/PhysRevLett.85.1158}{\emph{Phys. Rev. Lett.}
  {\bfseries 85} (2000) 1158}
  [\href{https://arxiv.org/abs/astro-ph/0003365}{{\ttfamily
  astro-ph/0003365}}].

\bibitem{Schive:2014hza}
H.-Y. Schive, M.-H. Liao, T.-P. Woo, S.-K. Wong, T.~Chiueh, T.~Broadhurst
  et~al., \emph{{Understanding the Core-Halo Relation of Quantum Wave Dark
  Matter from 3D Simulations}},
  \href{https://doi.org/10.1103/PhysRevLett.113.261302}{\emph{Phys. Rev. Lett.}
  {\bfseries 113} (2014) 261302}
  [\href{https://arxiv.org/abs/1407.7762}{{\ttfamily 1407.7762}}].

\bibitem{Marsh:2015wka}
D.~J.~E. Marsh and A.-R. Pop, \emph{{Axion dark matter, solitons and the
  cusp\textendash{}core problem}},
  \href{https://doi.org/10.1093/mnras/stv1050}{\emph{Mon. Not. Roy. Astron.
  Soc.} {\bfseries 451} (2015) 2479}
  [\href{https://arxiv.org/abs/1502.03456}{{\ttfamily 1502.03456}}].

\bibitem{Chen:2016unw}
S.-R. Chen, H.-Y. Schive and T.~Chiueh, \emph{{Jeans Analysis for Dwarf
  Spheroidal Galaxies in Wave Dark Matter}},
  \href{https://doi.org/10.1093/mnras/stx449}{\emph{Mon. Not. Roy. Astron.
  Soc.} {\bfseries 468} (2017) 1338}
  [\href{https://arxiv.org/abs/1606.09030}{{\ttfamily 1606.09030}}].

\bibitem{Gonzalez-Morales:2016yaf}
A.~X. Gonz\'alez-Morales, D.~J.~E. Marsh, J.~Pe\~narrubia and L.~A. Ure\~na
  L\'opez, \emph{{Unbiased constraints on ultralight axion mass from dwarf
  spheroidal galaxies}},
  \href{https://doi.org/10.1093/mnras/stx1941}{\emph{Mon. Not. Roy. Astron.
  Soc.} {\bfseries 472} (2017) 1346}
  [\href{https://arxiv.org/abs/1609.05856}{{\ttfamily 1609.05856}}].

\bibitem{Broadhurst:2019fsl}
T.~Broadhurst, I.~de~Martino, H.~N. Luu, G.~F. Smoot and S.~H.~H. Tye,
  \emph{{Ghostly Galaxies as Solitons of Bose-Einstein Dark Matter}},
  \href{https://doi.org/10.1103/PhysRevD.101.083012}{\emph{Phys. Rev. D}
  {\bfseries 101} (2020) 083012}
  [\href{https://arxiv.org/abs/1902.10488}{{\ttfamily 1902.10488}}].

\bibitem{Bar:2018acw}
N.~Bar, D.~Blas, K.~Blum and S.~Sibiryakov, \emph{{Galactic rotation curves
  versus ultralight dark matter: Implications of the soliton-host halo
  relation}}, \href{https://doi.org/10.1103/PhysRevD.98.083027}{\emph{Phys.
  Rev. D} {\bfseries 98} (2018) 083027}
  [\href{https://arxiv.org/abs/1805.00122}{{\ttfamily 1805.00122}}].

\bibitem{Veltmaat:2018dfz}
J.~Veltmaat, J.~C. Niemeyer and B.~Schwabe, \emph{{Formation and structure of
  ultralight bosonic dark matter halos}},
  \href{https://doi.org/10.1103/PhysRevD.98.043509}{\emph{Phys. Rev. D}
  {\bfseries 98} (2018) 043509}
  [\href{https://arxiv.org/abs/1804.09647}{{\ttfamily 1804.09647}}].

\bibitem{DeMartino:2018zkx}
I.~De~Martino, T.~Broadhurst, S.~H.~H. Tye, T.~Chiueh and H.-Y. Schive,
  \emph{{Dynamical Evidence of a Solitonic Core of $10^{9}M_\odot$ in the Milky
  Way}}, \href{https://doi.org/10.1016/j.dark.2020.100503}{\emph{Phys. Dark
  Univ.} {\bfseries 28} (2020) 100503}
  [\href{https://arxiv.org/abs/1807.08153}{{\ttfamily 1807.08153}}].

\bibitem{Pozo:2023zmx}
A.~Pozo, T.~Broadhurst, G.~F. Smoot and T.~Chiueh, \emph{{Dwarf Galaxies United
  by Dark Bosons}},  \href{https://arxiv.org/abs/2302.00181}{{\ttfamily
  2302.00181}}.

\bibitem{ParticleDataGroup:2022pth}
{\scshape Particle Data Group} collaboration, \emph{{Review of Particle
  Physics}}, \href{https://doi.org/10.1093/ptep/ptac097}{\emph{PTEP} {\bfseries
  2022} (2022) 083C01}.

\bibitem{Schive:2014dra}
H.-Y. Schive, T.~Chiueh and T.~Broadhurst, \emph{{Cosmic Structure as the
  Quantum Interference of a Coherent Dark Wave}},
  \href{https://doi.org/10.1038/nphys2996}{\emph{Nature Phys.} {\bfseries 10}
  (2014) 496} [\href{https://arxiv.org/abs/1406.6586}{{\ttfamily 1406.6586}}].

\bibitem{Khelashvili:2022ffq}
M.~Khelashvili, A.~Rudakovskyi and S.~Hossenfelder, \emph{{Dark matter profiles
  of SPARC galaxies: a challenge to fuzzy dark matter}},
  \href{https://doi.org/10.1093/mnras/stad1595}{\emph{Mon. Not. Roy. Astron.
  Soc.} {\bfseries 523} (2023) 3393}
  [\href{https://arxiv.org/abs/2207.14165}{{\ttfamily 2207.14165}}].

\bibitem{Irsic:2017yje}
V.~Ir\v{s}i\v{c}, M.~Viel, M.~G. Haehnelt, J.~S. Bolton and G.~D. Becker,
  \emph{{First constraints on fuzzy dark matter from Lyman-$\alpha$ forest data
  and hydrodynamical simulations}},
  \href{https://doi.org/10.1103/PhysRevLett.119.031302}{\emph{Phys. Rev. Lett.}
  {\bfseries 119} (2017) 031302}
  [\href{https://arxiv.org/abs/1703.04683}{{\ttfamily 1703.04683}}].

\bibitem{Nori:2018pka}
M.~Nori, R.~Murgia, V.~Ir\v{s}i\v{c}, M.~Baldi and M.~Viel, \emph{{Lyman
  $\alpha$ forest and non-linear structure characterization in Fuzzy Dark
  Matter cosmologies}}, \href{https://doi.org/10.1093/mnras/sty2888}{\emph{Mon.
  Not. Roy. Astron. Soc.} {\bfseries 482} (2019) 3227}
  [\href{https://arxiv.org/abs/1809.09619}{{\ttfamily 1809.09619}}].

\bibitem{Rogers:2020ltq}
K.~K. Rogers and H.~V. Peiris, \emph{{Strong Bound on Canonical Ultralight
  Axion Dark Matter from the Lyman-Alpha Forest}},
  \href{https://doi.org/10.1103/PhysRevLett.126.071302}{\emph{Phys. Rev. Lett.}
  {\bfseries 126} (2021) 071302}
  [\href{https://arxiv.org/abs/2007.12705}{{\ttfamily 2007.12705}}].

\bibitem{Dalal:2022rmp}
N.~Dalal and A.~Kravtsov, \emph{{Excluding fuzzy dark matter with sizes and
  stellar kinematics of ultrafaint dwarf galaxies}},
  \href{https://doi.org/10.1103/PhysRevD.106.063517}{\emph{Phys. Rev. D}
  {\bfseries 106} (2022) 063517}
  [\href{https://arxiv.org/abs/2203.05750}{{\ttfamily 2203.05750}}].

\bibitem{Amin:2022nlh}
M.~A. Amin and M.~Mirbabayi, \emph{{A lower bound on dark matter mass}},
  \href{https://arxiv.org/abs/2211.09775}{{\ttfamily 2211.09775}}.

\bibitem{DelPopolo:2016emo}
A.~Del~Popolo and M.~Le~Delliou, \emph{{Small scale problems of the
  $\Lambda$CDM model: a short review}},
  \href{https://doi.org/10.3390/galaxies5010017}{\emph{Galaxies} {\bfseries 5}
  (2017) 17} [\href{https://arxiv.org/abs/1606.07790}{{\ttfamily 1606.07790}}].

\bibitem{Marsh:2015xka}
D.~J.~E. Marsh, \emph{{Axion Cosmology}},
  \href{https://doi.org/10.1016/j.physrep.2016.06.005}{\emph{Phys. Rept.}
  {\bfseries 643} (2016) 1} [\href{https://arxiv.org/abs/1510.07633}{{\ttfamily
  1510.07633}}].

\bibitem{DiLuzio:2020wdo}
L.~Di~Luzio, M.~Giannotti, E.~Nardi and L.~Visinelli, \emph{{The landscape of
  QCD axion models}},
  \href{https://doi.org/10.1016/j.physrep.2020.06.002}{\emph{Phys. Rept.}
  {\bfseries 870} (2020) 1} [\href{https://arxiv.org/abs/2003.01100}{{\ttfamily
  2003.01100}}].

\bibitem{Sikivie:2020zpn}
P.~Sikivie, \emph{{Invisible Axion Search Methods}},
  \href{https://doi.org/10.1103/RevModPhys.93.015004}{\emph{Rev. Mod. Phys.}
  {\bfseries 93} (2021) 015004}
  [\href{https://arxiv.org/abs/2003.02206}{{\ttfamily 2003.02206}}].

\bibitem{Kim:1986ax}
J.~E. Kim, \emph{{Light Pseudoscalars, Particle Physics and Cosmology}},
  \href{https://doi.org/10.1016/0370-1573(87)90017-2}{\emph{Phys. Rept.}
  {\bfseries 150} (1987) 1}.

\bibitem{Peccei:1977hh}
R.~D. Peccei and H.~R. Quinn, \emph{{CP Conservation in the Presence of
  Instantons}}, \href{https://doi.org/10.1103/PhysRevLett.38.1440}{\emph{Phys.
  Rev. Lett.} {\bfseries 38} (1977) 1440}.

\bibitem{Weinberg:1977ma}
S.~Weinberg, \emph{{A New Light Boson?}},
  \href{https://doi.org/10.1103/PhysRevLett.40.223}{\emph{Phys. Rev. Lett.}
  {\bfseries 40} (1978) 223}.

\bibitem{Wilczek:1977pj}
F.~Wilczek, \emph{{Problem of Strong $P$ and $T$ Invariance in the Presence of
  Instantons}}, \href{https://doi.org/10.1103/PhysRevLett.40.279}{\emph{Phys.
  Rev. Lett.} {\bfseries 40} (1978) 279}.

\bibitem{Kim:1979if}
J.~E. Kim, \emph{{Weak Interaction Singlet and Strong CP Invariance}},
  \href{https://doi.org/10.1103/PhysRevLett.43.103}{\emph{Phys. Rev. Lett.}
  {\bfseries 43} (1979) 103}.

\bibitem{Shifman:1979if}
M.~A. Shifman, A.~I. Vainshtein and V.~I. Zakharov, \emph{{Can Confinement
  Ensure Natural CP Invariance of Strong Interactions?}},
  \href{https://doi.org/10.1016/0550-3213(80)90209-6}{\emph{Nucl. Phys. B}
  {\bfseries 166} (1980) 493}.

\bibitem{Zhitnitsky:1980tq}
A.~R. Zhitnitsky, \emph{{On Possible Suppression of the Axion Hadron
  Interactions. (In Russian)}}, {\emph{Sov. J. Nucl. Phys.} {\bfseries 31}
  (1980) 260}.

\bibitem{Dine:1981rt}
M.~Dine, W.~Fischler and M.~Srednicki, \emph{{A Simple Solution to the Strong
  CP Problem with a Harmless Axion}},
  \href{https://doi.org/10.1016/0370-2693(81)90590-6}{\emph{Phys. Lett. B}
  {\bfseries 104} (1981) 199}.

\bibitem{Borsanyi:2016ksw}
S.~Borsanyi et~al., \emph{{Calculation of the axion mass based on
  high-temperature lattice quantum chromodynamics}},
  \href{https://doi.org/10.1038/nature20115}{\emph{Nature} {\bfseries 539}
  (2016) 69} [\href{https://arxiv.org/abs/1606.07494}{{\ttfamily 1606.07494}}].

\bibitem{Gorghetto:2018ocs}
M.~Gorghetto and G.~Villadoro, \emph{{Topological Susceptibility and QCD Axion
  Mass: QED and NNLO corrections}},
  \href{https://doi.org/10.1007/JHEP03(2019)033}{\emph{JHEP} {\bfseries 03}
  (2019) 033} [\href{https://arxiv.org/abs/1812.01008}{{\ttfamily
  1812.01008}}].

\bibitem{Svrcek:2006yi}
P.~Svrcek and E.~Witten, \emph{{Axions In String Theory}},
  \href{https://doi.org/10.1088/1126-6708/2006/06/051}{\emph{JHEP} {\bfseries
  06} (2006) 051} [\href{https://arxiv.org/abs/hep-th/0605206}{{\ttfamily
  hep-th/0605206}}].

\bibitem{Arvanitaki:2009fg}
A.~Arvanitaki, S.~Dimopoulos, S.~Dubovsky, N.~Kaloper and J.~March-Russell,
  \emph{{String Axiverse}},
  \href{https://doi.org/10.1103/PhysRevD.81.123530}{\emph{Phys. Rev. D}
  {\bfseries 81} (2010) 123530}
  [\href{https://arxiv.org/abs/0905.4720}{{\ttfamily 0905.4720}}].

\bibitem{Ringwald:2012cu}
A.~Ringwald, \emph{{Searching for axions and ALPs from string theory}},
  \href{https://doi.org/10.1088/1742-6596/485/1/012013}{\emph{J. Phys. Conf.
  Ser.} {\bfseries 485} (2014) 012013}
  [\href{https://arxiv.org/abs/1209.2299}{{\ttfamily 1209.2299}}].

\bibitem{Raffelt:1990yz}
G.~G. Raffelt, \emph{{Astrophysical methods to constrain axions and other novel
  particle phenomena}},
  \href{https://doi.org/10.1016/0370-1573(90)90054-6}{\emph{Phys. Rept.}
  {\bfseries 198} (1990) 1}.

\bibitem{Hamaguchi:2018oqw}
K.~Hamaguchi, N.~Nagata, K.~Yanagi and J.~Zheng, \emph{{Limit on the Axion
  Decay Constant from the Cooling Neutron Star in Cassiopeia A}},
  \href{https://doi.org/10.1103/PhysRevD.98.103015}{\emph{Phys. Rev. D}
  {\bfseries 98} (2018) 103015}
  [\href{https://arxiv.org/abs/1806.07151}{{\ttfamily 1806.07151}}].

\bibitem{Beznogov:2018fda}
M.~V. Beznogov, E.~Rrapaj, D.~Page and S.~Reddy, \emph{{Constraints on
  Axion-like Particles and Nucleon Pairing in Dense Matter from the Hot Neutron
  Star in HESS J1731-347}},
  \href{https://doi.org/10.1103/PhysRevC.98.035802}{\emph{Phys. Rev. C}
  {\bfseries 98} (2018) 035802}
  [\href{https://arxiv.org/abs/1806.07991}{{\ttfamily 1806.07991}}].

\bibitem{Buschmann:2021juv}
M.~Buschmann, C.~Dessert, J.~W. Foster, A.~J. Long and B.~R. Safdi,
  \emph{{Upper Limit on the QCD Axion Mass from Isolated Neutron Star
  Cooling}}, \href{https://doi.org/10.1103/PhysRevLett.128.091102}{\emph{Phys.
  Rev. Lett.} {\bfseries 128} (2022) 091102}
  [\href{https://arxiv.org/abs/2111.09892}{{\ttfamily 2111.09892}}].

\bibitem{Burrows:1988ah}
A.~Burrows, M.~S. Turner and R.~P. Brinkmann, \emph{{Axions and SN 1987a}},
  \href{https://doi.org/10.1103/PhysRevD.39.1020}{\emph{Phys. Rev. D}
  {\bfseries 39} (1989) 1020}.

\bibitem{Burrows:1990pk}
A.~Burrows, M.~T. Ressell and M.~S. Turner, \emph{{Axions and SN1987A: Axion
  trapping}}, \href{https://doi.org/10.1103/PhysRevD.42.3297}{\emph{Phys. Rev.
  D} {\bfseries 42} (1990) 3297}.

\bibitem{Keil:1996ju}
W.~Keil, H.-T. Janka, D.~N. Schramm, G.~Sigl, M.~S. Turner and J.~R. Ellis,
  \emph{{A Fresh look at axions and SN-1987A}},
  \href{https://doi.org/10.1103/PhysRevD.56.2419}{\emph{Phys. Rev. D}
  {\bfseries 56} (1997) 2419}
  [\href{https://arxiv.org/abs/astro-ph/9612222}{{\ttfamily
  astro-ph/9612222}}].

\bibitem{Hanhart:2000ae}
C.~Hanhart, D.~R. Phillips and S.~Reddy, \emph{{Neutrino and axion emissivities
  of neutron stars from nucleon-nucleon scattering data}},
  \href{https://doi.org/10.1016/S0370-2693(00)01382-4}{\emph{Phys. Lett. B}
  {\bfseries 499} (2001) 9}
  [\href{https://arxiv.org/abs/astro-ph/0003445}{{\ttfamily
  astro-ph/0003445}}].

\bibitem{Fischer:2016cyd}
T.~Fischer, S.~Chakraborty, M.~Giannotti, A.~Mirizzi, A.~Payez and A.~Ringwald,
  \emph{{Probing axions with the neutrino signal from the next galactic
  supernova}}, \href{https://doi.org/10.1103/PhysRevD.94.085012}{\emph{Phys.
  Rev. D} {\bfseries 94} (2016) 085012}
  [\href{https://arxiv.org/abs/1605.08780}{{\ttfamily 1605.08780}}].

\bibitem{Carenza:2019pxu}
P.~Carenza, T.~Fischer, M.~Giannotti, G.~Guo, G.~Mart\'\i{}nez-Pinedo and
  A.~Mirizzi, \emph{{Improved axion emissivity from a supernova via
  nucleon-nucleon bremsstrahlung}},
  \href{https://doi.org/10.1088/1475-7516/2019/10/016}{\emph{JCAP} {\bfseries
  10} (2019) 016} [\href{https://arxiv.org/abs/1906.11844}{{\ttfamily
  1906.11844}}].

\bibitem{Carenza:2020cis}
P.~Carenza, B.~Fore, M.~Giannotti, A.~Mirizzi and S.~Reddy, \emph{{Enhanced
  Supernova Axion Emission and its Implications}},
  \href{https://doi.org/10.1103/PhysRevLett.126.071102}{\emph{Phys. Rev. Lett.}
  {\bfseries 126} (2021) 071102}
  [\href{https://arxiv.org/abs/2010.02943}{{\ttfamily 2010.02943}}].

\bibitem{Lella:2023bfb}
A.~Lella, P.~Carenza, G.~Co', G.~Lucente, M.~Giannotti, A.~Mirizzi et~al.,
  \emph{{Getting the most on supernova axions}},
  \href{https://arxiv.org/abs/2306.01048}{{\ttfamily 2306.01048}}.

\bibitem{Salehian:2021khb}
B.~Salehian, H.-Y. Zhang, M.~A. Amin, D.~I. Kaiser and M.~H. Namjoo,
  \emph{{Beyond Schr\"odinger-Poisson: nonrelativistic effective field theory
  for scalar dark matter}},
  \href{https://doi.org/10.1007/JHEP09(2021)050}{\emph{JHEP} {\bfseries 09}
  (2021) 050} [\href{https://arxiv.org/abs/2104.10128}{{\ttfamily
  2104.10128}}].

\bibitem{Mou:2022hqb}
Z.-G. Mou and H.-Y. Zhang, \emph{{A singularity problem for interacting massive
  vectors}}, \href{https://doi.org/10.1103/PhysRevLett.129.151101}{\emph{Phys.
  Rev. Lett.} {\bfseries 129} (2022) 151101}
  [\href{https://arxiv.org/abs/2204.11324}{{\ttfamily 2204.11324}}].

\bibitem{Zhang:2020bec}
H.-Y. Zhang, M.~A. Amin, E.~J. Copeland, P.~M. Saffin and K.~D. Lozanov,
  \emph{{Classical Decay Rates of Oscillons}},
  \href{https://doi.org/10.1088/1475-7516/2020/07/055}{\emph{JCAP} {\bfseries
  07} (2020) 055} [\href{https://arxiv.org/abs/2004.01202}{{\ttfamily
  2004.01202}}].

\bibitem{Zhang:2020ntm}
H.-Y. Zhang, \emph{{Gravitational effects on oscillon lifetimes}},
  \href{https://doi.org/10.1088/1475-7516/2021/03/102}{\emph{JCAP} {\bfseries
  03} (2021) 102} [\href{https://arxiv.org/abs/2011.11720}{{\ttfamily
  2011.11720}}].

\bibitem{Zhang:2021xxa}
H.-Y. Zhang, M.~Jain and M.~A. Amin, \emph{{Polarized vector oscillons}},
  \href{https://doi.org/10.1103/PhysRevD.105.096037}{\emph{Phys. Rev. D}
  {\bfseries 105} (2022) 096037}
  [\href{https://arxiv.org/abs/2111.08700}{{\ttfamily 2111.08700}}].

\bibitem{Zhang:2023vva}
H.-Y. Zhang, R.~Hagimoto and A.~J. Long, \emph{{Neutron star cooling with
  lepton-flavor-violating axions}},
  \href{https://arxiv.org/abs/2309.03889}{{\ttfamily 2309.03889}}.

\bibitem{Zhang:2023fhs}
H.-Y. Zhang and S.~Ling, \emph{{Phenomenology of wavelike vector dark matter
  nonminimally coupled to gravity}},
  \href{https://doi.org/10.1088/1475-7516/2023/07/055}{\emph{JCAP} {\bfseries
  07} (2023) 055} [\href{https://arxiv.org/abs/2305.03841}{{\ttfamily
  2305.03841}}].

\bibitem{Cheung:2007st}
C.~Cheung, P.~Creminelli, A.~Fitzpatrick, J.~Kaplan and L.~Senatore, \emph{{The
  Effective Field Theory of Inflation}},
  \href{https://doi.org/10.1088/1126-6708/2008/03/014}{\emph{JHEP} {\bfseries
  03} (2008) 014} [\href{https://arxiv.org/abs/0709.0293}{{\ttfamily
  0709.0293}}].

\bibitem{Carrasco:2012cv}
J.~J.~M. Carrasco, M.~P. Hertzberg and L.~Senatore, \emph{{The Effective Field
  Theory of Cosmological Large Scale Structures}},
  \href{https://doi.org/10.1007/JHEP09(2012)082}{\emph{JHEP} {\bfseries 09}
  (2012) 082} [\href{https://arxiv.org/abs/1206.2926}{{\ttfamily 1206.2926}}].

\bibitem{Namjoo:2017nia}
M.~H. Namjoo, A.~H. Guth and D.~I. Kaiser, \emph{{Relativistic Corrections to
  Nonrelativistic Effective Field Theories}},
  \href{https://doi.org/10.1103/PhysRevD.98.016011}{\emph{Phys. Rev. D}
  {\bfseries 98} (2018) 016011}
  [\href{https://arxiv.org/abs/1712.00445}{{\ttfamily 1712.00445}}].

\bibitem{Eby:2018ufi}
J.~Eby, K.~Mukaida, M.~Takimoto, L.~C.~R. Wijewardhana and M.~Yamada,
  \emph{{Classical nonrelativistic effective field theory and the role of
  gravitational interactions}},
  \href{https://doi.org/10.1103/PhysRevD.99.123503}{\emph{Phys. Rev.}
  {\bfseries D99} (2019) 123503}
  [\href{https://arxiv.org/abs/1807.09795}{{\ttfamily 1807.09795}}].

\bibitem{Braaten:2018lmj}
E.~Braaten, A.~Mohapatra and H.~Zhang, \emph{{Classical Nonrelativistic
  Effective Field Theories for a Real Scalar Field}},
  \href{https://doi.org/10.1103/PhysRevD.98.096012}{\emph{Phys. Rev. D}
  {\bfseries 98} (2018) 096012}
  [\href{https://arxiv.org/abs/1806.01898}{{\ttfamily 1806.01898}}].

\bibitem{Salehian:2020bon}
B.~Salehian, M.~H. Namjoo and D.~I. Kaiser, \emph{{Effective theories for a
  nonrelativistic field in an expanding universe: Induced self-interaction,
  pressure, sound speed, and viscosity}},
  \href{https://doi.org/10.1007/JHEP07(2020)059}{\emph{JHEP} {\bfseries 07}
  (2020) 059} [\href{https://arxiv.org/abs/2005.05388}{{\ttfamily
  2005.05388}}].

\bibitem{Adamek:2013wja}
J.~Adamek, D.~Daverio, R.~Durrer and M.~Kunz, \emph{{General Relativistic
  $N$-body simulations in the weak field limit}},
  \href{https://doi.org/10.1103/PhysRevD.88.103527}{\emph{Phys. Rev. D}
  {\bfseries 88} (2013) 103527}
  [\href{https://arxiv.org/abs/1308.6524}{{\ttfamily 1308.6524}}].

\bibitem{Adamek:2015eda}
J.~Adamek, D.~Daverio, R.~Durrer and M.~Kunz, \emph{{General relativity and
  cosmic structure formation}},
  \href{https://doi.org/10.1038/nphys3673}{\emph{Nature Phys.} {\bfseries 12}
  (2016) 346} [\href{https://arxiv.org/abs/1509.01699}{{\ttfamily
  1509.01699}}].

\bibitem{Mocz:2019pyf}
P.~Mocz et~al., \emph{{First star-forming structures in fuzzy cosmic
  filaments}},
  \href{https://doi.org/10.1103/PhysRevLett.123.141301}{\emph{Phys. Rev. Lett.}
  {\bfseries 123} (2019) 141301}
  [\href{https://arxiv.org/abs/1910.01653}{{\ttfamily 1910.01653}}].

\bibitem{Mocz:2019uyd}
P.~Mocz et~al., \emph{{Galaxy formation with BECDM \textendash{} II. Cosmic
  filaments and first galaxies}},
  \href{https://doi.org/10.1093/mnras/staa738}{\emph{Mon. Not. Roy. Astron.
  Soc.} {\bfseries 494} (2020) 2027}
  [\href{https://arxiv.org/abs/1911.05746}{{\ttfamily 1911.05746}}].

\bibitem{Amin:2019ums}
M.~A. Amin and P.~Mocz, \emph{{Formation, gravitational clustering, and
  interactions of nonrelativistic solitons in an expanding universe}},
  \href{https://doi.org/10.1103/PhysRevD.100.063507}{\emph{Phys. Rev. D}
  {\bfseries 100} (2019) 063507}
  [\href{https://arxiv.org/abs/1902.07261}{{\ttfamily 1902.07261}}].

\bibitem{Musoke:2019ima}
N.~Musoke, S.~Hotchkiss and R.~Easther, \emph{{Lighting the Dark: Evolution of
  the Postinflationary Universe}},
  \href{https://doi.org/10.1103/PhysRevLett.124.061301}{\emph{Phys. Rev. Lett.}
  {\bfseries 124} (2020) 061301}
  [\href{https://arxiv.org/abs/1909.11678}{{\ttfamily 1909.11678}}].

\bibitem{Schwabe:2016rze}
B.~Schwabe, J.~C. Niemeyer and J.~F. Engels, \emph{{Simulations of solitonic
  core mergers in ultralight axion dark matter cosmologies}},
  \href{https://doi.org/10.1103/PhysRevD.94.043513}{\emph{Phys. Rev. D}
  {\bfseries 94} (2016) 043513}
  [\href{https://arxiv.org/abs/1606.05151}{{\ttfamily 1606.05151}}].

\bibitem{Glennon:2020dxs}
N.~Glennon and C.~Prescod-Weinstein, \emph{{Modifying PyUltraLight to model
  scalar dark matter with self-interactions}},
  \href{https://doi.org/10.1103/PhysRevD.104.083532}{\emph{Phys. Rev. D}
  {\bfseries 104} (2021) 083532}
  [\href{https://arxiv.org/abs/2011.09510}{{\ttfamily 2011.09510}}].

\bibitem{Hertzberg:2020dbk}
M.~P. Hertzberg, Y.~Li and E.~D. Schiappacasse, \emph{{Merger of Dark Matter
  Axion Clumps and Resonant Photon Emission}},
  \href{https://doi.org/10.1088/1475-7516/2020/07/067}{\emph{JCAP} {\bfseries
  07} (2020) 067} [\href{https://arxiv.org/abs/2005.02405}{{\ttfamily
  2005.02405}}].

\bibitem{Amin:2020vja}
M.~A. Amin and Z.-G. Mou, \emph{{Electromagnetic Bursts from Mergers of
  Oscillons in Axion-like Fields}},
  \href{https://doi.org/10.1088/1475-7516/2021/02/024}{\emph{JCAP} {\bfseries
  02} (2020) 024} [\href{https://arxiv.org/abs/2009.11337}{{\ttfamily
  2009.11337}}].

\bibitem{Lancaster:2019mde}
L.~Lancaster, C.~Giovanetti, P.~Mocz, Y.~Kahn, M.~Lisanti and D.~N. Spergel,
  \emph{{Dynamical Friction in a Fuzzy Dark Matter Universe}},
  \href{https://doi.org/10.1088/1475-7516/2020/01/001}{\emph{JCAP} {\bfseries
  01} (2020) 001} [\href{https://arxiv.org/abs/1909.06381}{{\ttfamily
  1909.06381}}].

\bibitem{Bar-Or:2018pxz}
B.~Bar-Or, J.-B. Fouvry and S.~Tremaine, \emph{{Relaxation in a Fuzzy Dark
  Matter Halo}},
  \href{https://doi.org/10.3847/1538-4357/aaf28c}{\emph{Astrophys. J.}
  {\bfseries 871} (2019) 28}
  [\href{https://arxiv.org/abs/1809.07673}{{\ttfamily 1809.07673}}].

\bibitem{Mocz:2017wlg}
P.~Mocz, M.~Vogelsberger, V.~H. Robles, J.~Zavala, M.~Boylan-Kolchin,
  A.~Fialkov et~al., \emph{{Galaxy formation with BECDM \textendash{} I.
  Turbulence and relaxation of idealized haloes}},
  \href{https://doi.org/10.1093/mnras/stx1887}{\emph{Mon. Not. Roy. Astron.
  Soc.} {\bfseries 471} (2017) 4559}
  [\href{https://arxiv.org/abs/1705.05845}{{\ttfamily 1705.05845}}].

\bibitem{Du:2016zcv}
X.~Du, C.~Behrens and J.~C. Niemeyer, \emph{{Substructure of fuzzy dark matter
  haloes}}, \href{https://doi.org/10.1093/mnras/stw2724}{\emph{Mon. Not. Roy.
  Astron. Soc.} {\bfseries 465} (2017) 941}
  [\href{https://arxiv.org/abs/1608.02575}{{\ttfamily 1608.02575}}].

\bibitem{May:2021wwp}
S.~May and V.~Springel, \emph{{Structure formation in large-volume cosmological
  simulations of fuzzy dark matter: impact of the non-linear dynamics}},
  \href{https://doi.org/10.1093/mnras/stab1764}{\emph{Mon. Not. Roy. Astron.
  Soc.} {\bfseries 506} (2021) 2603}
  [\href{https://arxiv.org/abs/2101.01828}{{\ttfamily 2101.01828}}].

\bibitem{Levkov:2018kau}
D.~Levkov, A.~Panin and I.~Tkachev, \emph{{Gravitational Bose-Einstein
  condensation in the kinetic regime}},
  \href{https://doi.org/10.1103/PhysRevLett.121.151301}{\emph{Phys. Rev. Lett.}
  {\bfseries 121} (2018) 151301}
  [\href{https://arxiv.org/abs/1804.05857}{{\ttfamily 1804.05857}}].

\bibitem{Kirkpatrick:2020fwd}
K.~Kirkpatrick, A.~E. Mirasola and C.~Prescod-Weinstein, \emph{{Relaxation
  times for Bose-Einstein condensation in axion miniclusters}},
  \href{https://doi.org/10.1103/PhysRevD.102.103012}{\emph{Phys. Rev. D}
  {\bfseries 102} (2020) 103012}
  [\href{https://arxiv.org/abs/2007.07438}{{\ttfamily 2007.07438}}].

\bibitem{Hui:2020hbq}
L.~Hui, A.~Joyce, M.~J. Landry and X.~Li, \emph{{Vortices and waves in light
  dark matter}},
  \href{https://doi.org/10.1088/1475-7516/2021/01/011}{\emph{JCAP} {\bfseries
  01} (2021) 011} [\href{https://arxiv.org/abs/2004.01188}{{\ttfamily
  2004.01188}}].

\bibitem{Veltmaat:2016rxo}
J.~Veltmaat and J.~C. Niemeyer, \emph{{Cosmological particle-in-cell
  simulations with ultralight axion dark matter}},
  \href{https://doi.org/10.1103/PhysRevD.94.123523}{\emph{Phys. Rev. D}
  {\bfseries 94} (2016) 123523}
  [\href{https://arxiv.org/abs/1608.00802}{{\ttfamily 1608.00802}}].

\bibitem{Edwards:2018ccc}
F.~Edwards, E.~Kendall, S.~Hotchkiss and R.~Easther, \emph{{PyUltraLight: A
  Pseudo-Spectral Solver for Ultralight Dark Matter Dynamics}},
  \href{https://doi.org/10.1088/1475-7516/2018/10/027}{\emph{JCAP} {\bfseries
  10} (2018) 027} [\href{https://arxiv.org/abs/1807.04037}{{\ttfamily
  1807.04037}}].

\bibitem{Co:2018lka}
R.~T. Co, A.~Pierce, Z.~Zhang and Y.~Zhao, \emph{{Dark Photon Dark Matter
  Produced by Axion Oscillations}},
  \href{https://doi.org/10.1103/PhysRevD.99.075002}{\emph{Phys. Rev. D}
  {\bfseries 99} (2019) 075002}
  [\href{https://arxiv.org/abs/1810.07196}{{\ttfamily 1810.07196}}].

\bibitem{Agrawal:2018vin}
P.~Agrawal, N.~Kitajima, M.~Reece, T.~Sekiguchi and F.~Takahashi, \emph{{Relic
  Abundance of Dark Photon Dark Matter}},
  \href{https://doi.org/10.1016/j.physletb.2019.135136}{\emph{Phys. Lett. B}
  {\bfseries 801} (2020) 135136}
  [\href{https://arxiv.org/abs/1810.07188}{{\ttfamily 1810.07188}}].

\bibitem{Ford:1989me}
L.~H. Ford, \emph{{INFLATION DRIVEN BY A VECTOR FIELD}},
  \href{https://doi.org/10.1103/PhysRevD.40.967}{\emph{Phys. Rev. D} {\bfseries
  40} (1989) 967}.

\bibitem{Golovnev:2008cf}
A.~Golovnev, V.~Mukhanov and V.~Vanchurin, \emph{{Vector Inflation}},
  \href{https://doi.org/10.1088/1475-7516/2008/06/009}{\emph{JCAP} {\bfseries
  06} (2008) 009} [\href{https://arxiv.org/abs/0802.2068}{{\ttfamily
  0802.2068}}].

\bibitem{Baryakhtar:2017ngi}
M.~Baryakhtar, R.~Lasenby and M.~Teo, \emph{{Black Hole Superradiance
  Signatures of Ultralight Vectors}},
  \href{https://doi.org/10.1103/PhysRevD.96.035019}{\emph{Phys. Rev. D}
  {\bfseries 96} (2017) 035019}
  [\href{https://arxiv.org/abs/1704.05081}{{\ttfamily 1704.05081}}].

\bibitem{Fukuda:2019ewf}
H.~Fukuda and K.~Nakayama, \emph{{Aspects of Nonlinear Effect on Black Hole
  Superradiance}}, \href{https://doi.org/10.1007/JHEP01(2020)128}{\emph{JHEP}
  {\bfseries 01} (2020) 128}
  [\href{https://arxiv.org/abs/1910.06308}{{\ttfamily 1910.06308}}].

\bibitem{Wang:2022hra}
Z.~Wang, T.~Helfer, K.~Clough and E.~Berti, \emph{{Superradiance in massive
  vector fields with spatially varying mass}},
  \href{https://doi.org/10.1103/PhysRevD.105.104055}{\emph{Phys. Rev. D}
  {\bfseries 105} (2022) 104055}
  [\href{https://arxiv.org/abs/2201.08305}{{\ttfamily 2201.08305}}].

\bibitem{Clough:2022ygm}
K.~Clough, T.~Helfer, H.~Witek and E.~Berti, \emph{{Ghost Instabilities in
  Self-Interacting Vector Fields: The Problem with Proca Fields}},
  \href{https://doi.org/10.1103/PhysRevLett.129.151102}{\emph{Phys. Rev. Lett.}
  {\bfseries 129} (2022) 151102}
  [\href{https://arxiv.org/abs/2204.10868}{{\ttfamily 2204.10868}}].

\bibitem{East:2022ppo}
W.~E. East, \emph{{Vortex String Formation in Black Hole Superradiance of a
  Dark Photon with the Higgs Mechanism}},
  \href{https://doi.org/10.1103/PhysRevLett.129.141103}{\emph{Phys. Rev. Lett.}
  {\bfseries 129} (2022) 141103}
  [\href{https://arxiv.org/abs/2205.03417}{{\ttfamily 2205.03417}}].

\bibitem{Tasinato:2014eka}
G.~Tasinato, \emph{{Cosmic Acceleration from Abelian Symmetry Breaking}},
  \href{https://doi.org/10.1007/JHEP04(2014)067}{\emph{JHEP} {\bfseries 04}
  (2014) 067} [\href{https://arxiv.org/abs/1402.6450}{{\ttfamily 1402.6450}}].

\bibitem{Heisenberg:2014rta}
L.~Heisenberg, \emph{{Generalization of the Proca Action}},
  \href{https://doi.org/10.1088/1475-7516/2014/05/015}{\emph{JCAP} {\bfseries
  05} (2014) 015} [\href{https://arxiv.org/abs/1402.7026}{{\ttfamily
  1402.7026}}].

\bibitem{DeFelice:2016yws}
A.~De~Felice, L.~Heisenberg, R.~Kase, S.~Mukohyama, S.~Tsujikawa and Y.-l.
  Zhang, \emph{{Cosmology in generalized Proca theories}},
  \href{https://doi.org/10.1088/1475-7516/2016/06/048}{\emph{JCAP} {\bfseries
  06} (2016) 048} [\href{https://arxiv.org/abs/1603.05806}{{\ttfamily
  1603.05806}}].

\bibitem{deFelice:2017paw}
A.~de~Felice, L.~Heisenberg and S.~Tsujikawa, \emph{{Observational constraints
  on generalized Proca theories}},
  \href{https://doi.org/10.1103/PhysRevD.95.123540}{\emph{Phys. Rev. D}
  {\bfseries 95} (2017) 123540}
  [\href{https://arxiv.org/abs/1703.09573}{{\ttfamily 1703.09573}}].

\bibitem{Heisenberg:2020xak}
L.~Heisenberg and H.~Villarrubia-Rojo, \emph{{Proca in the sky}},
  \href{https://doi.org/10.1088/1475-7516/2021/03/032}{\emph{JCAP} {\bfseries
  03} (2021) 032} [\href{https://arxiv.org/abs/2010.00513}{{\ttfamily
  2010.00513}}].

\bibitem{Fodor:2008du}
G.~Fodor, P.~Forgacs, Z.~Horvath and M.~Mezei, \emph{{Computation of the
  radiation amplitude of oscillons}},
  \href{https://doi.org/10.1103/PhysRevD.79.065002}{\emph{Phys. Rev. D}
  {\bfseries 79} (2009) 065002}
  [\href{https://arxiv.org/abs/0812.1919}{{\ttfamily 0812.1919}}].

\bibitem{Fodor:2019ftc}
G.~Fodor, \emph{{A review on radiation of oscillons and oscillatons}}, Ph.D.
  thesis, Wigner RCP, Budapest, 2019.
\newblock \href{https://arxiv.org/abs/1911.03340}{{\ttfamily 1911.03340}}.

\bibitem{Bogolyubsky:1976yu}
I.~Bogolyubsky and V.~Makhankov, \emph{{Lifetime of Pulsating Solitons in Some
  Classical Models}}, {\emph{Pisma Zh. Eksp. Teor. Fiz.} {\bfseries 24} (1976)
  15}.

\bibitem{Gleiser:1993pt}
M.~Gleiser, \emph{{Pseudostable bubbles}},
  \href{https://doi.org/10.1103/PhysRevD.49.2978}{\emph{Phys. Rev.} {\bfseries
  D49} (1994) 2978} [\href{https://arxiv.org/abs/hep-ph/9308279}{{\ttfamily
  hep-ph/9308279}}].

\bibitem{Copeland:1995fq}
E.~J. Copeland, M.~Gleiser and H.-R. Muller, \emph{{Oscillons: Resonant
  configurations during bubble collapse}},
  \href{https://doi.org/10.1103/PhysRevD.52.1920}{\emph{Phys. Rev. D}
  {\bfseries 52} (1995) 1920}
  [\href{https://arxiv.org/abs/hep-ph/9503217}{{\ttfamily hep-ph/9503217}}].

\bibitem{Kasuya:2002zs}
S.~Kasuya, M.~Kawasaki and F.~Takahashi, \emph{{I-balls}},
  \href{https://doi.org/10.1016/S0370-2693(03)00344-7}{\emph{Phys. Lett.}
  {\bfseries B559} (2003) 99}
  [\href{https://arxiv.org/abs/hep-ph/0209358}{{\ttfamily hep-ph/0209358}}].

\bibitem{Amin:2010jq}
M.~A. Amin and D.~Shirokoff, \emph{{Flat-top oscillons in an expanding
  universe}}, \href{https://doi.org/10.1103/PhysRevD.81.085045}{\emph{Phys.
  Rev.} {\bfseries D81} (2010) 085045}
  [\href{https://arxiv.org/abs/1002.3380}{{\ttfamily 1002.3380}}].

\bibitem{Amin:2013ika}
M.~A. Amin, \emph{{K-oscillons: Oscillons with noncanonical kinetic terms}},
  \href{https://doi.org/10.1103/PhysRevD.87.123505}{\emph{Phys. Rev.}
  {\bfseries D87} (2013) 123505}
  [\href{https://arxiv.org/abs/1303.1102}{{\ttfamily 1303.1102}}].

\bibitem{Seidel:1991zh}
E.~Seidel and W.~Suen, \emph{{Oscillating soliton stars}},
  \href{https://doi.org/10.1103/PhysRevLett.66.1659}{\emph{Phys. Rev. Lett.}
  {\bfseries 66} (1991) 1659}.

\bibitem{Visinelli:2017ooc}
L.~Visinelli, S.~Baum, J.~Redondo, K.~Freese and F.~Wilczek, \emph{{Dilute and
  dense axion stars}},
  \href{https://doi.org/10.1016/j.physletb.2017.12.010}{\emph{Phys. Lett. B}
  {\bfseries 777} (2018) 64}
  [\href{https://arxiv.org/abs/1710.08910}{{\ttfamily 1710.08910}}].

\bibitem{Chavanis:2017loo}
P.-H. Chavanis, \emph{{Phase transitions between dilute and dense axion
  stars}}, \href{https://doi.org/10.1103/PhysRevD.98.023009}{\emph{Phys. Rev.
  D} {\bfseries 98} (2018) 023009}
  [\href{https://arxiv.org/abs/1710.06268}{{\ttfamily 1710.06268}}].

\bibitem{Eby:2019ntd}
J.~Eby, M.~Leembruggen, L.~Street, P.~Suranyi and L.~R. Wijewardhana,
  \emph{{Global view of QCD axion stars}},
  \href{https://doi.org/10.1103/PhysRevD.100.063002}{\emph{Phys. Rev. D}
  {\bfseries 100} (2019) 063002}
  [\href{https://arxiv.org/abs/1905.00981}{{\ttfamily 1905.00981}}].

\bibitem{Amin:2010xe}
M.~A. Amin, \emph{{Inflaton fragmentation: Emergence of pseudo-stable inflaton
  lumps (oscillons) after inflation}},
  \href{https://arxiv.org/abs/1006.3075}{{\ttfamily 1006.3075}}.

\bibitem{Amin:2011hj}
M.~A. Amin, R.~Easther, H.~Finkel, R.~Flauger and M.~P. Hertzberg,
  \emph{{Oscillons After Inflation}},
  \href{https://doi.org/10.1103/PhysRevLett.108.241302}{\emph{Phys. Rev. Lett.}
  {\bfseries 108} (2012) 241302}
  [\href{https://arxiv.org/abs/1106.3335}{{\ttfamily 1106.3335}}].

\bibitem{Gleiser:2011xj}
M.~Gleiser, N.~Graham and N.~Stamatopoulos, \emph{{Generation of Coherent
  Structures After Cosmic Inflation}},
  \href{https://doi.org/10.1103/PhysRevD.83.096010}{\emph{Phys. Rev.}
  {\bfseries D83} (2011) 096010}
  [\href{https://arxiv.org/abs/1103.1911}{{\ttfamily 1103.1911}}].

\bibitem{Grandclement:2011wz}
P.~Grandclement, G.~Fodor and P.~Forgacs, \emph{{Numerical simulation of
  oscillatons: extracting the radiating tail}},
  \href{https://doi.org/10.1103/PhysRevD.84.065037}{\emph{Phys. Rev. D}
  {\bfseries 84} (2011) 065037}
  [\href{https://arxiv.org/abs/1107.2791}{{\ttfamily 1107.2791}}].

\bibitem{Lozanov:2017hjm}
K.~D. Lozanov and M.~A. Amin, \emph{{Self-resonance after inflation: oscillons,
  transients and radiation domination}},
  \href{https://doi.org/10.1103/PhysRevD.97.023533}{\emph{Phys. Rev.}
  {\bfseries D97} (2018) 023533}
  [\href{https://arxiv.org/abs/1710.06851}{{\ttfamily 1710.06851}}].

\bibitem{Hong:2017ooe}
J.-P. Hong, M.~Kawasaki and M.~Yamazaki, \emph{{Oscillons from Pure Natural
  Inflation}}, \href{https://doi.org/10.1103/PhysRevD.98.043531}{\emph{Phys.
  Rev.} {\bfseries D98} (2018) 043531}
  [\href{https://arxiv.org/abs/1711.10496}{{\ttfamily 1711.10496}}].

\bibitem{Ikeda:2017qev}
T.~Ikeda, C.-M. Yoo and V.~Cardoso, \emph{{Self-gravitating oscillons and new
  critical behavior}},
  \href{https://doi.org/10.1103/PhysRevD.96.064047}{\emph{Phys. Rev.}
  {\bfseries D96} (2017) 064047}
  [\href{https://arxiv.org/abs/1708.01344}{{\ttfamily 1708.01344}}].

\bibitem{Bond:2015zfa}
J.~R. Bond, J.~Braden and L.~Mersini-Houghton, \emph{{Cosmic bubble and domain
  wall instabilities III: The role of oscillons in three-dimensional bubble
  collisions}},
  \href{https://doi.org/10.1088/1475-7516/2015/09/004}{\emph{JCAP} {\bfseries
  09} (2015) 004} [\href{https://arxiv.org/abs/1505.02162}{{\ttfamily
  1505.02162}}].

\bibitem{Antusch:2017flz}
S.~Antusch, F.~Cefala, S.~Krippendorf, F.~Muia, S.~Orani and F.~Quevedo,
  \emph{{Oscillons from String Moduli}},
  \href{https://doi.org/10.1007/JHEP01(2018)083}{\emph{JHEP} {\bfseries 01}
  (2018) 083} [\href{https://arxiv.org/abs/1708.08922}{{\ttfamily
  1708.08922}}].

\bibitem{Arvanitaki:2019rax}
A.~Arvanitaki, S.~Dimopoulos, M.~Galanis, L.~Lehner, J.~O. Thompson and
  K.~Van~Tilburg, \emph{{The Large-Misalignment Mechanism for the Formation of
  Compact Axion Structures: Signatures from the QCD Axion to Fuzzy Dark
  Matter}}, \href{https://doi.org/10.1103/PhysRevD.101.083014}{\emph{Phys. Rev.
  D} {\bfseries 101} (2019) 083014}
  [\href{https://arxiv.org/abs/1909.11665}{{\ttfamily 1909.11665}}].

\bibitem{Brax:2020oye}
P.~Brax, J.~A.~R. Cembranos and P.~Valageas, \emph{{Nonrelativistic formation
  of scalar clumps as a candidate for dark matter}},
  \href{https://doi.org/10.1103/PhysRevD.102.083012}{\emph{Phys. Rev. D}
  {\bfseries 102} (2020) 083012}
  [\href{https://arxiv.org/abs/2007.04638}{{\ttfamily 2007.04638}}].

\bibitem{Kawasaki:2020jnw}
M.~Kawasaki, W.~Nakano, H.~Nakatsuka and E.~Sonomoto, \emph{{Oscillons of
  Axion-Like Particle: Mass distribution and power spectrum}},
  \href{https://doi.org/10.1088/1475-7516/2021/01/061}{\emph{JCAP} {\bfseries
  01} (2020) 061} [\href{https://arxiv.org/abs/2010.09311}{{\ttfamily
  2010.09311}}].

\bibitem{Seidel:1993zk}
E.~Seidel and W.-M. Suen, \emph{{Formation of solitonic stars through
  gravitational cooling}},
  \href{https://doi.org/10.1103/PhysRevLett.72.2516}{\emph{Phys. Rev. Lett.}
  {\bfseries 72} (1994) 2516}
  [\href{https://arxiv.org/abs/gr-qc/9309015}{{\ttfamily gr-qc/9309015}}].

\bibitem{UrenaLopez:2001tw}
L.~Urena-Lopez, \emph{{Oscillatons revisited}},
  \href{https://doi.org/10.1088/0264-9381/19/10/307}{\emph{Class. Quant. Grav.}
  {\bfseries 19} (2002) 2617}
  [\href{https://arxiv.org/abs/gr-qc/0104093}{{\ttfamily gr-qc/0104093}}].

\bibitem{Alcubierre:2003sx}
M.~Alcubierre, R.~Becerril, S.~F. Guzman, T.~Matos, D.~Nunez and L.~A.
  Urena-Lopez, \emph{{Numerical studies of Phi**2 oscillatons}},
  \href{https://doi.org/10.1088/0264-9381/20/13/332}{\emph{Class. Quant. Grav.}
  {\bfseries 20} (2003) 2883}
  [\href{https://arxiv.org/abs/gr-qc/0301105}{{\ttfamily gr-qc/0301105}}].

\bibitem{Arnowitt:1962hi}
R.~L. Arnowitt, S.~Deser and C.~W. Misner, \emph{{The Dynamics of general
  relativity}}, \href{https://doi.org/10.1007/s10714-008-0661-1}{\emph{Gen.
  Rel. Grav.} {\bfseries 40} (2008) 1997}
  [\href{https://arxiv.org/abs/gr-qc/0405109}{{\ttfamily gr-qc/0405109}}].

\bibitem{Gourgoulhon:2007ue}
E.~Gourgoulhon, \emph{{3+1 formalism and bases of numerical relativity}},
  \href{https://arxiv.org/abs/gr-qc/0703035}{{\ttfamily gr-qc/0703035}}.

\bibitem{Croon:2018ybs}
D.~Croon, J.~Fan and C.~Sun, \emph{{Boson Star from Repulsive Light Scalars and
  Gravitational Waves}},
  \href{https://doi.org/10.1088/1475-7516/2019/04/008}{\emph{JCAP} {\bfseries
  04} (2019) 008} [\href{https://arxiv.org/abs/1810.01420}{{\ttfamily
  1810.01420}}].

\bibitem{Woodard:2015zca}
R.~P. Woodard, \emph{{Ostrogradsky's theorem on Hamiltonian instability}},
  \href{https://doi.org/10.4249/scholarpedia.32243}{\emph{Scholarpedia}
  {\bfseries 10} (2015) 32243}
  [\href{https://arxiv.org/abs/1506.02210}{{\ttfamily 1506.02210}}].

\bibitem{Cline:2003gs}
J.~M. Cline, S.~Jeon and G.~D. Moore, \emph{{The Phantom menaced: Constraints
  on low-energy effective ghosts}},
  \href{https://doi.org/10.1103/PhysRevD.70.043543}{\emph{Phys. Rev. D}
  {\bfseries 70} (2004) 043543}
  [\href{https://arxiv.org/abs/hep-ph/0311312}{{\ttfamily hep-ph/0311312}}].

\bibitem{Sawicki:2012pz}
I.~Sawicki and A.~Vikman, \emph{{Hidden Negative Energies in Strongly
  Accelerated Universes}},
  \href{https://doi.org/10.1103/PhysRevD.87.067301}{\emph{Phys. Rev. D}
  {\bfseries 87} (2013) 067301}
  [\href{https://arxiv.org/abs/1209.2961}{{\ttfamily 1209.2961}}].

\bibitem{Pagani:1987ue}
E.~Pagani, G.~Tecchiolli and S.~Zerbini, \emph{{On the Problem of Stability for
  Higher Order Derivatives: Lagrangian Systems}},
  \href{https://doi.org/10.1007/BF00402140}{\emph{Lett. Math. Phys.} {\bfseries
  14} (1987) 311}.

\bibitem{Smilga:2004cy}
A.~V. Smilga, \emph{{Benign versus malicious ghosts in higher-derivative
  theories}},
  \href{https://doi.org/10.1016/j.nuclphysb.2004.10.037}{\emph{Nucl. Phys. B}
  {\bfseries 706} (2005) 598}
  [\href{https://arxiv.org/abs/hep-th/0407231}{{\ttfamily hep-th/0407231}}].

\bibitem{Deffayet:2021nnt}
C.~Deffayet, S.~Mukohyama and A.~Vikman, \emph{{Ghosts without Runaway
  Instabilities}},
  \href{https://doi.org/10.1103/PhysRevLett.128.041301}{\emph{Phys. Rev. Lett.}
  {\bfseries 128} (2022) 041301}
  [\href{https://arxiv.org/abs/2108.06294}{{\ttfamily 2108.06294}}].

\bibitem{Nakayama:2019rhg}
K.~Nakayama, \emph{{Vector Coherent Oscillation Dark Matter}},
  \href{https://doi.org/10.1088/1475-7516/2019/10/019}{\emph{JCAP} {\bfseries
  10} (2019) 019} [\href{https://arxiv.org/abs/1907.06243}{{\ttfamily
  1907.06243}}].

\bibitem{Kolb:2020fwh}
E.~W. Kolb and A.~J. Long, \emph{{Completely dark photons from gravitational
  particle production during the inflationary era}},
  \href{https://doi.org/10.1007/JHEP03(2021)283}{\emph{JHEP} {\bfseries 03}
  (2021) 283} [\href{https://arxiv.org/abs/2009.03828}{{\ttfamily
  2009.03828}}].

\bibitem{Weinberg:1995mt}
S.~Weinberg, \emph{{The Quantum theory of fields. Vol. 1: Foundations}}.
  Cambridge University Press, 6, 2005.

\bibitem{Coates:2022qia}
A.~Coates and F.~M. Ramazano\u{g}lu, \emph{{Intrinsic Pathology of
  Self-Interacting Vector Fields}},
  \href{https://doi.org/10.1103/PhysRevLett.129.151103}{\emph{Phys. Rev. Lett.}
  {\bfseries 129} (2022) 151103}
  [\href{https://arxiv.org/abs/2205.07784}{{\ttfamily 2205.07784}}].

\bibitem{castro2011numerical}
M.~J. Castro-D{\'\i}az, E.~D. Fern{\'a}ndez-Nieto, J.~M. Gonz{\'a}lez-Vida and
  C.~Par{\'e}s-Madro{\~n}al, \emph{Numerical treatment of the loss of
  hyperbolicity of the two-layer shallow-water system}, {\emph{Journal of
  Scientific Computing} {\bfseries 48} (2011) 16}.

\bibitem{Frolov:2002rr}
A.~V. Frolov, L.~Kofman and A.~A. Starobinsky, \emph{{Prospects and problems of
  tachyon matter cosmology}},
  \href{https://doi.org/10.1016/S0370-2693(02)02582-0}{\emph{Phys. Lett. B}
  {\bfseries 545} (2002) 8}
  [\href{https://arxiv.org/abs/hep-th/0204187}{{\ttfamily hep-th/0204187}}].

\bibitem{Kroger:2003qh}
H.~Kroger, G.~Melkonian and S.~G. Rubin, \emph{{Cosmological dynamics of scalar
  field with non-minimal kinetic term}},
  \href{https://doi.org/10.1023/B:GERG.0000032157.83125.14}{\emph{Gen. Rel.
  Grav.} {\bfseries 36} (2004) 1649}
  [\href{https://arxiv.org/abs/astro-ph/0310182}{{\ttfamily
  astro-ph/0310182}}].

\bibitem{Hinterbichler:2011tt}
K.~Hinterbichler, \emph{{Theoretical Aspects of Massive Gravity}},
  \href{https://doi.org/10.1103/RevModPhys.84.671}{\emph{Rev. Mod. Phys.}
  {\bfseries 84} (2012) 671} [\href{https://arxiv.org/abs/1105.3735}{{\ttfamily
  1105.3735}}].

\bibitem{Lee:1977eg}
B.~W. Lee, C.~Quigg and H.~B. Thacker, \emph{{Weak Interactions at Very
  High-Energies: The Role of the Higgs Boson Mass}},
  \href{https://doi.org/10.1103/PhysRevD.16.1519}{\emph{Phys. Rev. D}
  {\bfseries 16} (1977) 1519}.

\bibitem{Schwartz:2014sze}
M.~D. Schwartz, \emph{{Quantum Field Theory and the Standard Model}}. Cambridge
  University Press, 3, 2014.

\bibitem{Amin:2010dc}
M.~A. Amin, R.~Easther and H.~Finkel, \emph{{Inflaton Fragmentation and
  Oscillon Formation in Three Dimensions}},
  \href{https://doi.org/10.1088/1475-7516/2010/12/001}{\emph{JCAP} {\bfseries
  1012} (2010) 001} [\href{https://arxiv.org/abs/1009.2505}{{\ttfamily
  1009.2505}}].

\bibitem{Farhi:2007wj}
E.~Farhi, N.~Graham, A.~H. Guth, N.~Iqbal, R.~Rosales and N.~Stamatopoulos,
  \emph{{Emergence of Oscillons in an Expanding Background}},
  \href{https://doi.org/10.1103/PhysRevD.77.085019}{\emph{Phys. Rev. D}
  {\bfseries 77} (2008) 085019}
  [\href{https://arxiv.org/abs/0712.3034}{{\ttfamily 0712.3034}}].

\bibitem{Gleiser:2010qt}
M.~Gleiser, N.~Graham and N.~Stamatopoulos, \emph{{Long-Lived Time-Dependent
  Remnants During Cosmological Symmetry Breaking: From Inflation to the
  Electroweak Scale}},
  \href{https://doi.org/10.1103/PhysRevD.82.043517}{\emph{Phys. Rev. D}
  {\bfseries 82} (2010) 043517}
  [\href{https://arxiv.org/abs/1004.4658}{{\ttfamily 1004.4658}}].

\bibitem{Kolb:1993hw}
E.~W. Kolb and I.~I. Tkachev, \emph{{Nonlinear axion dynamics and formation of
  cosmological pseudosolitons}},
  \href{https://doi.org/10.1103/PhysRevD.49.5040}{\emph{Phys. Rev.} {\bfseries
  D49} (1994) 5040} [\href{https://arxiv.org/abs/astro-ph/9311037}{{\ttfamily
  astro-ph/9311037}}].

\bibitem{Olle:2019kbo}
J.~Ollé, O.~Pujolàs and F.~Rompineve, \emph{{Oscillons and Dark Matter}},
  \href{https://doi.org/10.1088/1475-7516/2020/02/006}{\emph{JCAP} {\bfseries
  02} (2019) 006} [\href{https://arxiv.org/abs/1906.06352}{{\ttfamily
  1906.06352}}].

\bibitem{Kawasaki:2019czd}
M.~Kawasaki, W.~Nakano and E.~Sonomoto, \emph{{Oscillon of Ultra-Light
  Axion-like Particle}},
  \href{https://doi.org/10.1088/1475-7516/2020/01/047}{\emph{JCAP} {\bfseries
  2001} (2020) 047} [\href{https://arxiv.org/abs/1909.10805}{{\ttfamily
  1909.10805}}].

\bibitem{Zhou:2013tsa}
S.-Y. Zhou, E.~J. Copeland, R.~Easther, H.~Finkel, Z.-G. Mou and P.~M. Saffin,
  \emph{{Gravitational Waves from Oscillon Preheating}},
  \href{https://doi.org/10.1007/JHEP10(2013)026}{\emph{JHEP} {\bfseries 10}
  (2013) 026} [\href{https://arxiv.org/abs/1304.6094}{{\ttfamily 1304.6094}}].

\bibitem{Antusch:2016con}
S.~Antusch, F.~Cefala and S.~Orani, \emph{{Gravitational waves from oscillons
  after inflation}}, \href{https://doi.org/10.1103/PhysRevLett.120.219901,
  10.1103/PhysRevLett.118.011303}{\emph{Phys. Rev. Lett.} {\bfseries 118}
  (2017) 011303} [\href{https://arxiv.org/abs/1607.01314}{{\ttfamily
  1607.01314}}].

\bibitem{Liu:2017hua}
J.~Liu, Z.-K. Guo, R.-G. Cai and G.~Shiu, \emph{{Gravitational Waves from
  Oscillons with Cuspy Potentials}},
  \href{https://doi.org/10.1103/PhysRevLett.120.031301}{\emph{Phys. Rev. Lett.}
  {\bfseries 120} (2018) 031301}
  [\href{https://arxiv.org/abs/1707.09841}{{\ttfamily 1707.09841}}].

\bibitem{Lozanov:2019ylm}
K.~D. Lozanov and M.~A. Amin, \emph{{Gravitational perturbations from oscillons
  and transients after inflation}},
  \href{https://doi.org/10.1103/PhysRevD.99.123504}{\emph{Phys. Rev.}
  {\bfseries D99} (2019) 123504}
  [\href{https://arxiv.org/abs/1902.06736}{{\ttfamily 1902.06736}}].

\bibitem{Amin:2018xfe}
M.~A. Amin, J.~Braden, E.~J. Copeland, J.~T. Giblin, C.~Solorio, Z.~J. Weiner
  et~al., \emph{{Gravitational waves from asymmetric oscillon dynamics?}},
  \href{https://doi.org/10.1103/PhysRevD.98.024040}{\emph{Phys. Rev. D}
  {\bfseries 98} (2018) 024040}
  [\href{https://arxiv.org/abs/1803.08047}{{\ttfamily 1803.08047}}].

\bibitem{Cotner:2019ykd}
E.~Cotner, A.~Kusenko, M.~Sasaki and V.~Takhistov, \emph{{Analytic Description
  of Primordial Black Hole Formation from Scalar Field Fragmentation}},
  \href{https://doi.org/10.1088/1475-7516/2019/10/077}{\emph{JCAP} {\bfseries
  10} (2019) 077} [\href{https://arxiv.org/abs/1907.10613}{{\ttfamily
  1907.10613}}].

\bibitem{Kou:2019bbc}
X.-X. Kou, C.~Tian and S.-Y. Zhou, \emph{{Oscillon Preheating in Full General
  Relativity}}, \href{https://doi.org/10.1088/1361-6382/abd09f}{\emph{Class.
  Quant. Grav.} {\bfseries 38} (2021) 045005}
  [\href{https://arxiv.org/abs/1912.09658}{{\ttfamily 1912.09658}}].

\bibitem{Lozanov:2014zfa}
K.~D. Lozanov and M.~A. Amin, \emph{{End of inflation, oscillons, and
  matter-antimatter asymmetry}},
  \href{https://doi.org/10.1103/PhysRevD.90.083528}{\emph{Phys. Rev. D}
  {\bfseries 90} (2014) 083528}
  [\href{https://arxiv.org/abs/1408.1811}{{\ttfamily 1408.1811}}].

\bibitem{Sakstein:2018pfd}
J.~Sakstein and M.~Trodden, \emph{{Oscillons in Higher-Derivative Effective
  Field Theories}},
  \href{https://doi.org/10.1103/PhysRevD.98.123512}{\emph{Phys. Rev.}
  {\bfseries D98} (2018) 123512}
  [\href{https://arxiv.org/abs/1809.07724}{{\ttfamily 1809.07724}}].

\bibitem{Graham:2006vy}
N.~Graham, \emph{{An Electroweak oscillon}},
  \href{https://doi.org/10.1103/PhysRevLett.98.101801,
  10.1103/PhysRevLett.98.189904}{\emph{Phys. Rev. Lett.} {\bfseries 98} (2007)
  101801} [\href{https://arxiv.org/abs/hep-th/0610267}{{\ttfamily
  hep-th/0610267}}].

\bibitem{Gleiser:2008dt}
M.~Gleiser and J.~Thorarinson, \emph{{A Class of Nonperturbative Configurations
  in Abelian-Higgs Models: Complexity from Dynamical Symmetry Breaking}},
  \href{https://doi.org/10.1103/PhysRevD.79.025016}{\emph{Phys. Rev.}
  {\bfseries D79} (2009) 025016}
  [\href{https://arxiv.org/abs/0808.0514}{{\ttfamily 0808.0514}}].

\bibitem{Sfakianakis:2012bq}
E.~I. Sfakianakis, \emph{{Analysis of Oscillons in the SU(2) Gauged Higgs
  Model}},  \href{https://arxiv.org/abs/1210.7568}{{\ttfamily 1210.7568}}.

\bibitem{Coleman:1985ki}
S.~R. Coleman, \emph{{Q Balls}},
  \href{https://doi.org/10.1016/0550-3213(85)90286-X,
  10.1016/0550-3213(86)90520-1}{\emph{Nucl. Phys.} {\bfseries B262} (1985)
  263}.

\bibitem{Nugaev:2019vru}
E.~Y. Nugaev and A.~V. Shkerin, \emph{{Review of Nontopological Solitons in
  Theories with $U(1)$-Symmetry}},
  \href{https://doi.org/10.1134/S1063776120020077}{\emph{J. Exp. Theor. Phys.}
  {\bfseries 130} (2020) 301}
  [\href{https://arxiv.org/abs/1905.05146}{{\ttfamily 1905.05146}}].

\bibitem{Kaup:1968}
D.~J. Kaup, \emph{Klein-gordon geon},
  \href{https://doi.org/10.1103/PhysRev.172.1331}{\emph{Phys. Rev.} {\bfseries
  172} (1968) 1331}.

\bibitem{Liebling:2012fv}
S.~L. Liebling and C.~Palenzuela, \emph{{Dynamical Boson Stars}},
  \href{https://doi.org/10.12942/lrr-2012-6}{\emph{Living Rev. Rel.} {\bfseries
  15} (2012) 6} [\href{https://arxiv.org/abs/1202.5809}{{\ttfamily
  1202.5809}}].

\bibitem{2017Sci...356..422N}
J.~H.~V. {Nguyen}, D.~{Luo} and R.~G. {Hulet}, \emph{{Formation of matter-wave
  soliton trains by modulational instability}},
  \href{https://doi.org/10.1126/science.aal3220}{\emph{Science} {\bfseries 356}
  (2017) 422} [\href{https://arxiv.org/abs/1703.04662}{{\ttfamily
  1703.04662}}].

\bibitem{Niemeyer:2019aqm}
J.~C. Niemeyer, \emph{{Small-scale structure of fuzzy and axion-like dark
  matter}},  \href{https://arxiv.org/abs/1912.07064}{{\ttfamily 1912.07064}}.

\bibitem{Segur:1987mg}
H.~Segur and M.~D. Kruskal, \emph{{Nonexistence of Small Amplitude Breather
  Solutions in $\phi^4$ Theory}},
  \href{https://doi.org/10.1103/PhysRevLett.58.747}{\emph{Phys. Rev. Lett.}
  {\bfseries 58} (1987) 747}.

\bibitem{Fodor:2009kf}
G.~Fodor, P.~Forgacs, Z.~Horvath and M.~Mezei, \emph{{Radiation of scalar
  oscillons in 2 and 3 dimensions}},
  \href{https://doi.org/10.1016/j.physletb.2009.03.054}{\emph{Phys. Lett.}
  {\bfseries B674} (2009) 319}
  [\href{https://arxiv.org/abs/0903.0953}{{\ttfamily 0903.0953}}].

\bibitem{Hertzberg:2010yz}
M.~P. Hertzberg, \emph{{Quantum Radiation of Oscillons}},
  \href{https://doi.org/10.1103/PhysRevD.82.045022}{\emph{Phys. Rev.}
  {\bfseries D82} (2010) 045022}
  [\href{https://arxiv.org/abs/1003.3459}{{\ttfamily 1003.3459}}].

\bibitem{Salmi:2012ta}
P.~Salmi and M.~Hindmarsh, \emph{{Radiation and Relaxation of Oscillons}},
  \href{https://doi.org/10.1103/PhysRevD.85.085033}{\emph{Phys. Rev.}
  {\bfseries D85} (2012) 085033}
  [\href{https://arxiv.org/abs/1201.1934}{{\ttfamily 1201.1934}}].

\bibitem{Mukaida:2016hwd}
K.~Mukaida, M.~Takimoto and M.~Yamada, \emph{{On Longevity of
  I-ball/Oscillon}}, \href{https://doi.org/10.1007/JHEP03(2017)122}{\emph{JHEP}
  {\bfseries 03} (2017) 122}
  [\href{https://arxiv.org/abs/1612.07750}{{\ttfamily 1612.07750}}].

\bibitem{Ibe:2019vyo}
M.~Ibe, M.~Kawasaki, W.~Nakano and E.~Sonomoto, \emph{{Decay of I-ball/Oscillon
  in Classical Field Theory}},
  \href{https://doi.org/10.1007/JHEP04(2019)030}{\emph{JHEP} {\bfseries 04}
  (2019) 030} [\href{https://arxiv.org/abs/1901.06130}{{\ttfamily
  1901.06130}}].

\bibitem{Eby:2015hyx}
J.~Eby, P.~Suranyi and L.~Wijewardhana, \emph{{The Lifetime of Axion Stars}},
  \href{https://doi.org/10.1142/S0217732316500905}{\emph{Mod. Phys. Lett. A}
  {\bfseries 31} (2016) 1650090}
  [\href{https://arxiv.org/abs/1512.01709}{{\ttfamily 1512.01709}}].

\bibitem{Eby:2020ply}
J.~Eby, L.~Street, P.~Suranyi and L.~C.~R. Wijewardhana, \emph{{Global view of
  axion stars with nearly Planck-scale decay constants}},
  \href{https://doi.org/10.1103/PhysRevD.103.063043}{\emph{Phys. Rev. D}
  {\bfseries 103} (2021) 063043}
  [\href{https://arxiv.org/abs/2011.09087}{{\ttfamily 2011.09087}}].

\bibitem{UrenaLopez:2002gx}
L.~Urena-Lopez, T.~Matos and R.~Becerril, \emph{{Inside oscillatons}},
  \href{https://doi.org/10.1088/0264-9381/19/23/320}{\emph{Class. Quant. Grav.}
  {\bfseries 19} (2002) 6259}.

\bibitem{Hui:2016ltb}
L.~Hui, J.~P. Ostriker, S.~Tremaine and E.~Witten, \emph{{Ultralight scalars as
  cosmological dark matter}},
  \href{https://doi.org/10.1103/PhysRevD.95.043541}{\emph{Phys. Rev. D}
  {\bfseries 95} (2017) 043541}
  [\href{https://arxiv.org/abs/1610.08297}{{\ttfamily 1610.08297}}].

\bibitem{Graham:2015rva}
P.~W. Graham, J.~Mardon and S.~Rajendran, \emph{{Vector Dark Matter from
  Inflationary Fluctuations}},
  \href{https://doi.org/10.1103/PhysRevD.93.103520}{\emph{Phys. Rev. D}
  {\bfseries 93} (2016) 103520}
  [\href{https://arxiv.org/abs/1504.02102}{{\ttfamily 1504.02102}}].

\bibitem{Bastero-Gil:2018uel}
M.~Bastero-Gil, J.~Santiago, L.~Ubaldi and R.~Vega-Morales, \emph{{Vector dark
  matter production at the end of inflation}},
  \href{https://doi.org/10.1088/1475-7516/2019/04/015}{\emph{JCAP} {\bfseries
  04} (2019) 015} [\href{https://arxiv.org/abs/1810.07208}{{\ttfamily
  1810.07208}}].

\bibitem{Kolb:2021xfn}
E.~W. Kolb, A.~J. Long and E.~McDonough, \emph{{Catastrophic production of slow
  gravitinos}}, \href{https://doi.org/10.1103/PhysRevD.104.075015}{\emph{Phys.
  Rev. D} {\bfseries 104} (2021) 075015}
  [\href{https://arxiv.org/abs/2102.10113}{{\ttfamily 2102.10113}}].

\bibitem{Kolb:2021nob}
E.~W. Kolb, A.~J. Long and E.~McDonough, \emph{{Gravitino Swampland
  Conjecture}},
  \href{https://doi.org/10.1103/PhysRevLett.127.131603}{\emph{Phys. Rev. Lett.}
  {\bfseries 127} (2021) 131603}
  [\href{https://arxiv.org/abs/2103.10437}{{\ttfamily 2103.10437}}].

\bibitem{Babichev:2016bxi}
E.~Babichev, L.~Marzola, M.~Raidal, A.~Schmidt-May, F.~Urban, H.~Veerm\"ae
  et~al., \emph{{Heavy spin-2 Dark Matter}},
  \href{https://doi.org/10.1088/1475-7516/2016/09/016}{\emph{JCAP} {\bfseries
  09} (2016) 016} [\href{https://arxiv.org/abs/1607.03497}{{\ttfamily
  1607.03497}}].

\bibitem{Alexander:2020gmv}
S.~Alexander, L.~Jenks and E.~McDonough, \emph{{Higher spin dark matter}},
  \href{https://doi.org/10.1016/j.physletb.2021.136436}{\emph{Phys. Lett. B}
  {\bfseries 819} (2021) 136436}
  [\href{https://arxiv.org/abs/2010.15125}{{\ttfamily 2010.15125}}].

\bibitem{ParticleDataGroup:2020ssz}
{\scshape Particle Data Group} collaboration, \emph{{Review of Particle
  Physics}}, \href{https://doi.org/10.1093/ptep/ptaa104}{\emph{PTEP} {\bfseries
  2020} (2020) 083C01}.

\bibitem{Palenzuela:2017kcg}
C.~Palenzuela, P.~Pani, M.~Bezares, V.~Cardoso, L.~Lehner and S.~Liebling,
  \emph{{Gravitational Wave Signatures of Highly Compact Boson Star Binaries}},
  \href{https://doi.org/10.1103/PhysRevD.96.104058}{\emph{Phys. Rev. D}
  {\bfseries 96} (2017) 104058}
  [\href{https://arxiv.org/abs/1710.09432}{{\ttfamily 1710.09432}}].

\bibitem{Helfer:2018vtq}
T.~Helfer, E.~A. Lim, M.~A. Garcia and M.~A. Amin, \emph{{Gravitational Wave
  Emission from Collisions of Compact Scalar Solitons}},
  \href{https://doi.org/10.1103/PhysRevD.99.044046}{\emph{Phys. Rev. D}
  {\bfseries 99} (2019) 044046}
  [\href{https://arxiv.org/abs/1802.06733}{{\ttfamily 1802.06733}}].

\bibitem{Dietrich:2018jov}
T.~Dietrich, F.~Day, K.~Clough, M.~Coughlin and J.~Niemeyer, \emph{{Neutron
  star\textendash{}axion star collisions in the light of multimessenger
  astronomy}}, \href{https://doi.org/10.1093/mnras/sty3158}{\emph{Mon. Not.
  Roy. Astron. Soc.} {\bfseries 483} (2019) 908}
  [\href{https://arxiv.org/abs/1808.04746}{{\ttfamily 1808.04746}}].

\bibitem{Caputo:2021eaa}
A.~Caputo, A.~J. Millar, C.~A.~J. O'Hare and E.~Vitagliano, \emph{{Dark photon
  limits: A handbook}},
  \href{https://doi.org/10.1103/PhysRevD.104.095029}{\emph{Phys. Rev. D}
  {\bfseries 104} (2021) 095029}
  [\href{https://arxiv.org/abs/2105.04565}{{\ttfamily 2105.04565}}].

\bibitem{Levkov:2020txo}
D.~Levkov, A.~Panin and I.~Tkachev, \emph{{Radio-emission of axion stars}},
  \href{https://doi.org/10.1103/PhysRevD.102.023501}{\emph{Phys. Rev. D}
  {\bfseries 102} (2020) 023501}
  [\href{https://arxiv.org/abs/2004.05179}{{\ttfamily 2004.05179}}].

\bibitem{Caldwell:2016dcw}
{\scshape MADMAX Working Group} collaboration, \emph{{Dielectric Haloscopes: A
  New Way to Detect Axion Dark Matter}},
  \href{https://doi.org/10.1103/PhysRevLett.118.091801}{\emph{Phys. Rev. Lett.}
  {\bfseries 118} (2017) 091801}
  [\href{https://arxiv.org/abs/1611.05865}{{\ttfamily 1611.05865}}].

\bibitem{Baryakhtar:2018doz}
M.~Baryakhtar, J.~Huang and R.~Lasenby, \emph{{Axion and hidden photon dark
  matter detection with multilayer optical haloscopes}},
  \href{https://doi.org/10.1103/PhysRevD.98.035006}{\emph{Phys. Rev. D}
  {\bfseries 98} (2018) 035006}
  [\href{https://arxiv.org/abs/1803.11455}{{\ttfamily 1803.11455}}].

\bibitem{Chiles:2021gxk}
J.~Chiles et~al., \emph{{First Constraints on Dark Photon Dark Matter with
  Superconducting Nanowire Detectors in an Optical Haloscope}},  10, 2021.

\bibitem{Chen:2021bdr}
Y.~Chen, M.~Jiang, J.~Shu, X.~Xue and Y.~Zeng, \emph{{Dissecting axion and dark
  photon with a network of vector sensors}},
  \href{https://doi.org/10.1103/PhysRevResearch.4.033080}{\emph{Phys. Rev.
  Res.} {\bfseries 4} (2022) 033080}
  [\href{https://arxiv.org/abs/2111.06732}{{\ttfamily 2111.06732}}].

\bibitem{Kallosh:2013hoa}
R.~Kallosh and A.~Linde, \emph{{Universality Class in Conformal Inflation}},
  \href{https://doi.org/10.1088/1475-7516/2013/07/002}{\emph{JCAP} {\bfseries
  1307} (2013) 002} [\href{https://arxiv.org/abs/1306.5220}{{\ttfamily
  1306.5220}}].

\bibitem{Silverstein:2008sg}
E.~Silverstein and A.~Westphal, \emph{{Monodromy in the CMB: Gravity Waves and
  String Inflation}},
  \href{https://doi.org/10.1103/PhysRevD.78.106003}{\emph{Phys. Rev.}
  {\bfseries D78} (2008) 106003}
  [\href{https://arxiv.org/abs/0803.3085}{{\ttfamily 0803.3085}}].

\bibitem{McAllister:2014mpa}
L.~McAllister, E.~Silverstein, A.~Westphal and T.~Wrase, \emph{{The Powers of
  Monodromy}}, \href{https://doi.org/10.1007/JHEP09(2014)123}{\emph{JHEP}
  {\bfseries 09} (2014) 123} [\href{https://arxiv.org/abs/1405.3652}{{\ttfamily
  1405.3652}}].

\bibitem{Friedberg:1976me}
R.~Friedberg, T.~Lee and A.~Sirlin, \emph{{A Class of Scalar-Field Soliton
  Solutions in Three Space Dimensions}},
  \href{https://doi.org/10.1103/PhysRevD.13.2739}{\emph{Phys. Rev. D}
  {\bfseries 13} (1976) 2739}.

\bibitem{Andersen:2012wg}
E.~A. Andersen and A.~Tranberg, \emph{{Four results on $phi^4$ oscillons in D+1
  dimensions}}, \href{https://doi.org/10.1007/JHEP12(2012)016}{\emph{JHEP}
  {\bfseries 12} (2012) 016} [\href{https://arxiv.org/abs/1210.2227}{{\ttfamily
  1210.2227}}].

\bibitem{Lee:1991ax}
T.~Lee and Y.~Pang, \emph{{Nontopological solitons}},
  \href{https://doi.org/10.1016/0370-1573(92)90064-7}{\emph{Phys. Rept.}
  {\bfseries 221} (1992) 251}.

\bibitem{Gibbons:1976ue}
G.~Gibbons and S.~Hawking, \emph{{Action Integrals and Partition Functions in
  Quantum Gravity}},
  \href{https://doi.org/10.1103/PhysRevD.15.2752}{\emph{Phys. Rev. D}
  {\bfseries 15} (1977) 2752}.

\bibitem{Lee:1988av}
T.~D. Lee and Y.~Pang, \emph{{Stability of Mini-Boson Stars}},
  \href{https://doi.org/10.1016/0550-3213(89)90365-9}{\emph{Nucl. Phys. B}
  {\bfseries 315} (1989) 477}.

\bibitem{Friedberg:1986tp}
R.~Friedberg, T.~Lee and Y.~Pang, \emph{{MINI - SOLITON STARS}},
  \href{https://doi.org/10.1103/PhysRevD.35.3640}{\emph{Phys. Rev. D}
  {\bfseries 35} (1987) 3640}.

\bibitem{Fodor:2008es}
G.~Fodor, P.~Forgacs, Z.~Horvath and A.~Lukacs, \emph{{Small amplitude
  quasi-breathers and oscillons}},
  \href{https://doi.org/10.1103/PhysRevD.78.025003}{\emph{Phys. Rev. D}
  {\bfseries 78} (2008) 025003}
  [\href{https://arxiv.org/abs/0802.3525}{{\ttfamily 0802.3525}}].

\bibitem{Davidson:1981zd}
A.~Davidson and K.~C. Wali, \emph{{MINIMAL FLAVOR UNIFICATION VIA
  MULTIGENERATIONAL PECCEI-QUINN SYMMETRY}},
  \href{https://doi.org/10.1103/PhysRevLett.48.11}{\emph{Phys. Rev. Lett.}
  {\bfseries 48} (1982) 11}.

\bibitem{Wilczek:1982rv}
F.~Wilczek, \emph{{Axions and Family Symmetry Breaking}},
  \href{https://doi.org/10.1103/PhysRevLett.49.1549}{\emph{Phys. Rev. Lett.}
  {\bfseries 49} (1982) 1549}.

\bibitem{Feng:1997tn}
J.~L. Feng, T.~Moroi, H.~Murayama and E.~Schnapka, \emph{{Third generation
  familons, b factories, and neutrino cosmology}},
  \href{https://doi.org/10.1103/PhysRevD.57.5875}{\emph{Phys. Rev. D}
  {\bfseries 57} (1998) 5875}
  [\href{https://arxiv.org/abs/hep-ph/9709411}{{\ttfamily hep-ph/9709411}}].

\bibitem{Bauer:2016rxs}
M.~Bauer, T.~Schell and T.~Plehn, \emph{{Hunting the Flavon}},
  \href{https://doi.org/10.1103/PhysRevD.94.056003}{\emph{Phys. Rev. D}
  {\bfseries 94} (2016) 056003}
  [\href{https://arxiv.org/abs/1603.06950}{{\ttfamily 1603.06950}}].

\bibitem{Ema:2016ops}
Y.~Ema, K.~Hamaguchi, T.~Moroi and K.~Nakayama, \emph{{Flaxion: a minimal
  extension to solve puzzles in the standard model}},
  \href{https://doi.org/10.1007/JHEP01(2017)096}{\emph{JHEP} {\bfseries 01}
  (2017) 096} [\href{https://arxiv.org/abs/1612.05492}{{\ttfamily
  1612.05492}}].

\bibitem{Calibbi:2016hwq}
L.~Calibbi, F.~Goertz, D.~Redigolo, R.~Ziegler and J.~Zupan, \emph{{Minimal
  axion model from flavor}},
  \href{https://doi.org/10.1103/PhysRevD.95.095009}{\emph{Phys. Rev. D}
  {\bfseries 95} (2017) 095009}
  [\href{https://arxiv.org/abs/1612.08040}{{\ttfamily 1612.08040}}].

\bibitem{Choi:2017gpf}
K.~Choi, S.~H. Im, C.~B. Park and S.~Yun, \emph{{Minimal Flavor Violation with
  Axion-like Particles}},
  \href{https://doi.org/10.1007/JHEP11(2017)070}{\emph{JHEP} {\bfseries 11}
  (2017) 070} [\href{https://arxiv.org/abs/1708.00021}{{\ttfamily
  1708.00021}}].

\bibitem{Chala:2020wvs}
M.~Chala, G.~Guedes, M.~Ramos and J.~Santiago, \emph{{Running in the ALPs}},
  \href{https://doi.org/10.1140/epjc/s10052-021-08968-2}{\emph{Eur. Phys. J. C}
  {\bfseries 81} (2021) 181}
  [\href{https://arxiv.org/abs/2012.09017}{{\ttfamily 2012.09017}}].

\bibitem{Bauer:2020jbp}
M.~Bauer, M.~Neubert, S.~Renner, M.~Schnubel and A.~Thamm, \emph{{The
  Low-Energy Effective Theory of Axions and ALPs}},
  \href{https://doi.org/10.1007/JHEP04(2021)063}{\emph{JHEP} {\bfseries 04}
  (2021) 063} [\href{https://arxiv.org/abs/2012.12272}{{\ttfamily
  2012.12272}}].

\bibitem{Bonilla:2021ufe}
J.~Bonilla, I.~Brivio, M.~B. Gavela and V.~Sanz, \emph{{One-loop corrections to
  ALP couplings}}, \href{https://doi.org/10.1007/JHEP11(2021)168}{\emph{JHEP}
  {\bfseries 11} (2021) 168}
  [\href{https://arxiv.org/abs/2107.11392}{{\ttfamily 2107.11392}}].

\bibitem{Panci:2022wlc}
P.~Panci, D.~Redigolo, T.~Schwetz and R.~Ziegler, \emph{{Axion dark matter from
  lepton flavor-violating decays}},
  \href{https://doi.org/10.1016/j.physletb.2023.137919}{\emph{Phys. Lett. B}
  {\bfseries 841} (2023) 137919}
  [\href{https://arxiv.org/abs/2209.03371}{{\ttfamily 2209.03371}}].

\bibitem{Bauer:2019gfk}
M.~Bauer, M.~Neubert, S.~Renner, M.~Schnubel and A.~Thamm, \emph{{Axionlike
  Particles, Lepton-Flavor Violation, and a New Explanation of $a_\mu$ and
  $a_e$}}, \href{https://doi.org/10.1103/PhysRevLett.124.211803}{\emph{Phys.
  Rev. Lett.} {\bfseries 124} (2020) 211803}
  [\href{https://arxiv.org/abs/1908.00008}{{\ttfamily 1908.00008}}].

\bibitem{Endo:2020mev}
M.~Endo, S.~Iguro and T.~Kitahara, \emph{{Probing $e\mu$ flavor-violating ALP
  at Belle II}}, \href{https://doi.org/10.1007/JHEP06(2020)040}{\emph{JHEP}
  {\bfseries 06} (2020) 040}
  [\href{https://arxiv.org/abs/2002.05948}{{\ttfamily 2002.05948}}].

\bibitem{Bauer:2021mvw}
M.~Bauer, M.~Neubert, S.~Renner, M.~Schnubel and A.~Thamm, \emph{{Flavor probes
  of axion-like particles}},
  \href{https://doi.org/10.1007/JHEP09(2022)056}{\emph{JHEP} {\bfseries 09}
  (2022) 056} [\href{https://arxiv.org/abs/2110.10698}{{\ttfamily
  2110.10698}}].

\bibitem{Calibbi:2017uvl}
L.~Calibbi and G.~Signorelli, \emph{{Charged Lepton Flavour Violation: An
  Experimental and Theoretical Introduction}},
  \href{https://doi.org/10.1393/ncr/i2018-10144-0}{\emph{Riv. Nuovo Cim.}
  {\bfseries 41} (2018) 71} [\href{https://arxiv.org/abs/1709.00294}{{\ttfamily
  1709.00294}}].

\bibitem{Petcov:1976ff}
S.~T. Petcov, \emph{{The Processes $\mu \rightarrow e + \gamma, \mu \rightarrow
  e + \overline{e}, \nu' \rightarrow \nu + \gamma$ in the Weinberg-Salam Model
  with Neutrino Mixing}}, {\emph{Sov. J. Nucl. Phys.} {\bfseries 25} (1977)
  340}.

\bibitem{Hernandez-Tome:2018fbq}
G.~Hern\'andez-Tom\'e, G.~L\'opez~Castro and P.~Roig, \emph{{Flavor violating
  leptonic decays of $\tau$ and $\mu$ leptons in the Standard Model with
  massive neutrinos}},
  \href{https://doi.org/10.1140/epjc/s10052-019-6563-4}{\emph{Eur. Phys. J. C}
  {\bfseries 79} (2019) 84} [\href{https://arxiv.org/abs/1807.06050}{{\ttfamily
  1807.06050}}].

\bibitem{MEG:2016leq}
{\scshape MEG} collaboration, \emph{{Search for the lepton flavour violating
  decay $\mu ^+ \rightarrow \mathrm {e}^+ \gamma $ with the full dataset of the
  MEG experiment}},
  \href{https://doi.org/10.1140/epjc/s10052-016-4271-x}{\emph{Eur. Phys. J. C}
  {\bfseries 76} (2016) 434}
  [\href{https://arxiv.org/abs/1605.05081}{{\ttfamily 1605.05081}}].

\bibitem{SINDRUM:1987nra}
{\scshape SINDRUM} collaboration, \emph{{Search for the Decay mu+
  ---\ensuremath{>} e+ e+ e-}},
  \href{https://doi.org/10.1016/0550-3213(88)90462-2}{\emph{Nucl. Phys. B}
  {\bfseries 299} (1988) 1}.

\bibitem{Jodidio:1986mz}
A.~Jodidio et~al., \emph{{Search for Right-Handed Currents in Muon Decay}},
  \href{https://doi.org/10.1103/PhysRevD.34.1967}{\emph{Phys. Rev. D}
  {\bfseries 34} (1986) 1967}.

\bibitem{TWIST:2014ymv}
{\scshape TWIST} collaboration, \emph{{Search for two body muon decay
  signals}}, \href{https://doi.org/10.1103/PhysRevD.91.052020}{\emph{Phys. Rev.
  D} {\bfseries 91} (2015) 052020}
  [\href{https://arxiv.org/abs/1409.0638}{{\ttfamily 1409.0638}}].

\bibitem{MEGII:2018kmf}
{\scshape MEG II} collaboration, \emph{{The design of the MEG II experiment}},
  \href{https://doi.org/10.1140/epjc/s10052-018-5845-6}{\emph{Eur. Phys. J. C}
  {\bfseries 78} (2018) 380}
  [\href{https://arxiv.org/abs/1801.04688}{{\ttfamily 1801.04688}}].

\bibitem{Jho:2022snj}
Y.~Jho, S.~Knapen and D.~Redigolo, \emph{{Lepton-flavor violating axions at MEG
  II}}, \href{https://doi.org/10.1007/JHEP10(2022)029}{\emph{JHEP} {\bfseries
  10} (2022) 029} [\href{https://arxiv.org/abs/2203.11222}{{\ttfamily
  2203.11222}}].

\bibitem{Perrevoort:2018ttp}
{\scshape Mu3e} collaboration, \emph{{The Rare and Forbidden: Testing Physics
  Beyond the Standard Model with Mu3e}},
  \href{https://doi.org/10.21468/SciPostPhysProc.1.052}{\emph{SciPost Phys.
  Proc.} {\bfseries 1} (2019) 052}
  [\href{https://arxiv.org/abs/1812.00741}{{\ttfamily 1812.00741}}].

\bibitem{Calibbi:2020jvd}
L.~Calibbi, D.~Redigolo, R.~Ziegler and J.~Zupan, \emph{{Looking forward to
  lepton-flavor-violating ALPs}},
  \href{https://doi.org/10.1007/JHEP09(2021)173}{\emph{JHEP} {\bfseries 09}
  (2021) 173} [\href{https://arxiv.org/abs/2006.04795}{{\ttfamily
  2006.04795}}].

\bibitem{Haensel:2007yy}
P.~Haensel, A.~Y. Potekhin and D.~G. Yakovlev, \emph{{Neutron stars 1: Equation
  of state and structure}}, vol.~326. Springer, New York, USA, 2007,
  \href{https://doi.org/10.1007/978-0-387-47301-7}{10.1007/978-0-387-47301-7}.

\bibitem{cohen1970neutron}
J.~M. Cohen, W.~D. Langer, L.~C. Rosen and A.~Cameron, \emph{Neutron star
  models based on an improved equation of state}, {\emph{Astrophysics and Space
  Science} {\bfseries 6} (1970) 228}.

\bibitem{Lattimer:1991ib}
J.~M. Lattimer, M.~Prakash, C.~J. Pethick and P.~Haensel, \emph{{Direct URCA
  process in neutron stars}},
  \href{https://doi.org/10.1103/PhysRevLett.66.2701}{\emph{Phys. Rev. Lett.}
  {\bfseries 66} (1991) 2701}.

\bibitem{Pethick:1991mk}
C.~J. Pethick, \emph{{Cooling of neutron stars}},
  \href{https://doi.org/10.1103/RevModPhys.64.1133}{\emph{Rev. Mod. Phys.}
  {\bfseries 64} (1992) 1133}.

\bibitem{Raffelt:1996wa}
G.~G. Raffelt, \emph{{Stars as laboratories for fundamental physics}: {The
  astrophysics of neutrinos, axions, and other weakly interacting particles}}.
  University of Chicago press, 5, 1996.

\bibitem{Horowitz:2020evx}
C.~J. Horowitz, J.~Piekarewicz and B.~Reed, \emph{{Insights into nuclear
  saturation density from parity violating electron scattering}},
  \href{https://doi.org/10.1103/PhysRevC.102.044321}{\emph{Phys. Rev. C}
  {\bfseries 102} (2020) 044321}
  [\href{https://arxiv.org/abs/2007.07117}{{\ttfamily 2007.07117}}].

\bibitem{SOYEUR1979464}
M.~Soyeur and G.~Brown, \emph{Contribution of the crust to the neutrino
  luminosity of neutron stars},
  \href{https://doi.org/https://doi.org/10.1016/0375-9474(79)90595-5}{\emph{Nuclear
  Physics A} {\bfseries 324} (1979) 464}.

\bibitem{Friman:1979ecl}
B.~L. Friman and O.~V. Maxwell, \emph{{Neutron Star Neutrino Emissivities}},
  \href{https://doi.org/10.1086/157313}{\emph{Astrophys. J.} {\bfseries 232}
  (1979) 541}.

\bibitem{Potekhin:2015qsa}
A.~Y. Potekhin, J.~A. Pons and D.~Page, \emph{{Neutron stars - cooling and
  transport}}, \href{https://doi.org/10.1007/s11214-015-0180-9}{\emph{Space
  Sci. Rev.} {\bfseries 191} (2015) 239}
  [\href{https://arxiv.org/abs/1507.06186}{{\ttfamily 1507.06186}}].

\bibitem{Voskresensky:1986af}
D.~N. Voskresensky and A.~V. Senatorov, \emph{{Emission of Neutrinos by Neutron
  Stars}}, {\emph{Sov. Phys. JETP} {\bfseries 63} (1986) 885}.

\bibitem{Grigorian:2016leu}
H.~Grigorian, D.~N. Voskresensky and D.~Blaschke, \emph{{Influence of the
  stiffness of the equation of state and in-medium effects on the cooling of
  compact stars}}, \href{https://doi.org/10.1140/epja/i2016-16067-4}{\emph{Eur.
  Phys. J. A} {\bfseries 52} (2016) 67}
  [\href{https://arxiv.org/abs/1603.02634}{{\ttfamily 1603.02634}}].

\bibitem{Bjorkeroth:2018dzu}
F.~Bj\"orkeroth, E.~J. Chun and S.~F. King, \emph{{Flavourful Axion
  Phenomenology}}, \href{https://doi.org/10.1007/JHEP08(2018)117}{\emph{JHEP}
  {\bfseries 08} (2018) 117}
  [\href{https://arxiv.org/abs/1806.00660}{{\ttfamily 1806.00660}}].

\bibitem{Derenzo:1969za}
S.~E. Derenzo, \emph{{Measurement of the low-energy end of the mu-plus decay
  spectrum}}, \href{https://doi.org/10.1103/PhysRev.181.1854}{\emph{Phys. Rev.}
  {\bfseries 181} (1969) 1854}.

\bibitem{Bryman:1986wn}
D.~A. Bryman and E.~T.~H. Clifford, \emph{{EXOTIC MUON DECAY mu
  ---\ensuremath{>} e x}},
  \href{https://doi.org/10.1103/PhysRevLett.57.2787}{\emph{Phys. Rev. Lett.}
  {\bfseries 57} (1986) 2787}.

\bibitem{Bilger:1998rp}
R.~Bilger, K.~Foehl, H.~Clement, M.~Croni, A.~Erhardt, R.~Meier et~al.,
  \emph{{Search for exotic muon decays}},
  \href{https://doi.org/10.1016/S0370-2693(98)01507-X}{\emph{Phys. Lett. B}
  {\bfseries 446} (1999) 363}
  [\href{https://arxiv.org/abs/hep-ph/9811333}{{\ttfamily hep-ph/9811333}}].

\bibitem{PIENU:2020loi}
{\scshape PIENU} collaboration, \emph{{Improved search for two body muon decay
  ${\mu}^+{\rightarrow}e^+X_H$}},
  \href{https://doi.org/10.1103/PhysRevD.101.052014}{\emph{Phys. Rev. D}
  {\bfseries 101} (2020) 052014}
  [\href{https://arxiv.org/abs/2002.09170}{{\ttfamily 2002.09170}}].

\bibitem{DEramo:2021usm}
F.~D'Eramo and S.~Yun, \emph{{Flavor violating axions in the early Universe}},
  \href{https://doi.org/10.1103/PhysRevD.105.075002}{\emph{Phys. Rev. D}
  {\bfseries 105} (2022) 075002}
  [\href{https://arxiv.org/abs/2111.12108}{{\ttfamily 2111.12108}}].

\bibitem{ARGUS:1995bjh}
{\scshape ARGUS} collaboration, \emph{{A Search for lepton flavor violating
  decays tau ----\ensuremath{>} e alpha, tau ---\ensuremath{>} mu alpha}},
  \href{https://doi.org/10.1007/BF01579801}{\emph{Z. Phys. C} {\bfseries 68}
  (1995) 25}.

\bibitem{Belle-II:2018jsg}
{\scshape Belle-II} collaboration, \emph{{The Belle II Physics Book}},
  \href{https://doi.org/10.1093/ptep/ptz106}{\emph{PTEP} {\bfseries 2019}
  (2019) 123C01} [\href{https://arxiv.org/abs/1808.10567}{{\ttfamily
  1808.10567}}].

\bibitem{CMB-S4:2016ple}
{\scshape CMB-S4} collaboration, \emph{{CMB-S4 Science Book, First Edition}},
  \href{https://arxiv.org/abs/1610.02743}{{\ttfamily 1610.02743}}.

\bibitem{Abazajian:2019eic}
K.~Abazajian et~al., \emph{{CMB-S4 Science Case, Reference Design, and Project
  Plan}},  \href{https://arxiv.org/abs/1907.04473}{{\ttfamily 1907.04473}}.

\bibitem{Harris:2020qim}
S.~P. Harris, J.-F. Fortin, K.~Sinha and M.~G. Alford, \emph{{Axions in neutron
  star mergers}},
  \href{https://doi.org/10.1088/1475-7516/2020/07/023}{\emph{JCAP} {\bfseries
  07} (2020) 023} [\href{https://arxiv.org/abs/2003.09768}{{\ttfamily
  2003.09768}}].

\bibitem{baym2008landau}
G.~Baym and C.~Pethick, \emph{Landau Fermi-liquid theory: concepts and
  applications}. John Wiley \& Sons, 2008.

\bibitem{Raffelt:1987yt}
G.~Raffelt and D.~Seckel, \emph{{Bounds on Exotic Particle Interactions from SN
  1987a}}, \href{https://doi.org/10.1103/PhysRevLett.60.1793}{\emph{Phys. Rev.
  Lett.} {\bfseries 60} (1988) 1793}.

\bibitem{Shtabovenko:2020gxv}
V.~Shtabovenko, R.~Mertig and F.~Orellana, \emph{{FeynCalc 9.3: New features
  and improvements}},
  \href{https://doi.org/10.1016/j.cpc.2020.107478}{\emph{Comput. Phys. Commun.}
  {\bfseries 256} (2020) 107478}
  [\href{https://arxiv.org/abs/2001.04407}{{\ttfamily 2001.04407}}].

\bibitem{Li:2018lpy}
B.-A. Li, B.-J. Cai, L.-W. Chen and J.~Xu, \emph{{Nucleon Effective Masses in
  Neutron-Rich Matter}},
  \href{https://doi.org/10.1016/j.ppnp.2018.01.001}{\emph{Prog. Part. Nucl.
  Phys.} {\bfseries 99} (2018) 29}
  [\href{https://arxiv.org/abs/1801.01213}{{\ttfamily 1801.01213}}].

\bibitem{Yakovlev:2004iq}
D.~G. Yakovlev and C.~J. Pethick, \emph{{Neutron star cooling}},
  \href{https://doi.org/10.1146/annurev.astro.42.053102.134013}{\emph{Ann. Rev.
  Astron. Astrophys.} {\bfseries 42} (2004) 169}
  [\href{https://arxiv.org/abs/astro-ph/0402143}{{\ttfamily
  astro-ph/0402143}}].

\bibitem{Leinson:2009mq}
L.~B. Leinson, \emph{{Superfluid response and the neutrino emissivity of baryon
  matter. Fermi-liquid effects}},
  \href{https://doi.org/10.1103/PhysRevC.79.045502}{\emph{Phys. Rev. C}
  {\bfseries 79} (2009) 045502}
  [\href{https://arxiv.org/abs/0904.0320}{{\ttfamily 0904.0320}}].

\bibitem{Leinson:2009nu}
L.~B. Leinson, \emph{{Neutrino emission from triplet pairing of neutrons in
  neutron stars}},
  \href{https://doi.org/10.1103/PhysRevC.81.025501}{\emph{Phys. Rev. C}
  {\bfseries 81} (2010) 025501}
  [\href{https://arxiv.org/abs/0912.2164}{{\ttfamily 0912.2164}}].

\bibitem{Iwamoto:1984ir}
N.~Iwamoto, \emph{{Axion Emission from Neutron Stars}},
  \href{https://doi.org/10.1103/PhysRevLett.53.1198}{\emph{Phys. Rev. Lett.}
  {\bfseries 53} (1984) 1198}.

\bibitem{Brinkmann:1988vi}
R.~P. Brinkmann and M.~S. Turner, \emph{{Numerical Rates for Nucleon-Nucleon
  Axion Bremsstrahlung}},
  \href{https://doi.org/10.1103/PhysRevD.38.2338}{\emph{Phys. Rev. D}
  {\bfseries 38} (1988) 2338}.

\bibitem{Fabbrichesi:2020wbt}
M.~Fabbrichesi, E.~Gabrielli and G.~Lanfranchi, \emph{{The Dark Photon}},
  \href{https://arxiv.org/abs/2005.01515}{{\ttfamily 2005.01515}}.

\bibitem{Freitas:2021cfi}
F.~F. Freitas, C.~A.~R. Herdeiro, A.~P. Morais, A.~Onofre, R.~Pasechnik,
  E.~Radu et~al., \emph{{Ultralight bosons for strong gravity applications from
  simple Standard Model extensions}},
  \href{https://doi.org/10.1088/1475-7516/2021/12/047}{\emph{JCAP} {\bfseries
  12} (2021) 047} [\href{https://arxiv.org/abs/2107.09493}{{\ttfamily
  2107.09493}}].

\bibitem{Adshead:2021kvl}
P.~Adshead and K.~D. Lozanov, \emph{{Self-gravitating Vector Dark Matter}},
  \href{https://doi.org/10.1103/PhysRevD.103.103501}{\emph{Phys. Rev. D}
  {\bfseries 103} (2021) 103501}
  [\href{https://arxiv.org/abs/2101.07265}{{\ttfamily 2101.07265}}].

\bibitem{Gorghetto:2022sue}
M.~Gorghetto, E.~Hardy, J.~March-Russell, N.~Song and S.~M. West, \emph{{Dark
  photon stars: formation and role as dark matter substructure}},
  \href{https://doi.org/10.1088/1475-7516/2022/08/018}{\emph{JCAP} {\bfseries
  08} (2022) 018} [\href{https://arxiv.org/abs/2203.10100}{{\ttfamily
  2203.10100}}].

\bibitem{Amin:2022pzv}
M.~A. Amin, M.~Jain, R.~Karur and P.~Mocz, \emph{{Small-scale structure in
  vector dark matter}},
  \href{https://doi.org/10.1088/1475-7516/2022/08/014}{\emph{JCAP} {\bfseries
  08} (2022) 014} [\href{https://arxiv.org/abs/2203.11935}{{\ttfamily
  2203.11935}}].

\bibitem{Antypas:2022asj}
D.~Antypas et~al., \emph{{New Horizons: Scalar and Vector Ultralight Dark
  Matter}},  \href{https://arxiv.org/abs/2203.14915}{{\ttfamily 2203.14915}}.

\bibitem{Lebedev:2011iq}
O.~Lebedev, H.~M. Lee and Y.~Mambrini, \emph{{Vector Higgs-portal dark matter
  and the invisible Higgs}},
  \href{https://doi.org/10.1016/j.physletb.2012.01.029}{\emph{Phys. Lett. B}
  {\bfseries 707} (2012) 570}
  [\href{https://arxiv.org/abs/1111.4482}{{\ttfamily 1111.4482}}].

\bibitem{Djouadi:2011aa}
A.~Djouadi, O.~Lebedev, Y.~Mambrini and J.~Quevillon, \emph{{Implications of
  LHC searches for Higgs--portal dark matter}},
  \href{https://doi.org/10.1016/j.physletb.2012.01.062}{\emph{Phys. Lett. B}
  {\bfseries 709} (2012) 65} [\href{https://arxiv.org/abs/1112.3299}{{\ttfamily
  1112.3299}}].

\bibitem{Baek:2012se}
S.~Baek, P.~Ko, W.-I. Park and E.~Senaha, \emph{{Higgs Portal Vector Dark
  Matter : Revisited}},
  \href{https://doi.org/10.1007/JHEP05(2013)036}{\emph{JHEP} {\bfseries 05}
  (2013) 036} [\href{https://arxiv.org/abs/1212.2131}{{\ttfamily 1212.2131}}].

\bibitem{Arcadi:2020jqf}
G.~Arcadi, A.~Djouadi and M.~Kado, \emph{{The Higgs-portal for vector dark
  matter and the effective field theory approach: A reappraisal}},
  \href{https://doi.org/10.1016/j.physletb.2020.135427}{\emph{Phys. Lett. B}
  {\bfseries 805} (2020) 135427}
  [\href{https://arxiv.org/abs/2001.10750}{{\ttfamily 2001.10750}}].

\bibitem{Baouche:2021wwa}
N.~Baouche, A.~Ahriche, G.~Faisel and S.~Nasri, \emph{{Phenomenology of the
  hidden SU(2) vector dark matter model}},
  \href{https://doi.org/10.1103/PhysRevD.104.075022}{\emph{Phys. Rev. D}
  {\bfseries 104} (2021) 075022}
  [\href{https://arxiv.org/abs/2105.14387}{{\ttfamily 2105.14387}}].

\bibitem{Ghorbani:2021yiw}
K.~Ghorbani, \emph{{Light vector dark matter with scalar mediator and muon g-2
  anomaly}}, \href{https://doi.org/10.1103/PhysRevD.104.115008}{\emph{Phys.
  Rev. D} {\bfseries 104} (2021) 115008}
  [\href{https://arxiv.org/abs/2104.13810}{{\ttfamily 2104.13810}}].

\bibitem{Ahmed:2019mjo}
A.~Ahmed, B.~Grzadkowski and A.~Socha, \emph{{Production of Purely
  Gravitational Vector Dark Matter}},
  \href{https://doi.org/10.5506/APhysPolB.50.1809}{\emph{Acta Phys. Polon. B}
  {\bfseries 50} (2019) 1809}.

\bibitem{Ema:2019yrd}
Y.~Ema, K.~Nakayama and Y.~Tang, \emph{{Production of purely gravitational dark
  matter: the case of fermion and vector boson}},
  \href{https://doi.org/10.1007/JHEP07(2019)060}{\emph{JHEP} {\bfseries 07}
  (2019) 060} [\href{https://arxiv.org/abs/1903.10973}{{\ttfamily
  1903.10973}}].

\bibitem{Ahmed:2020fhc}
A.~Ahmed, B.~Grzadkowski and A.~Socha, \emph{{Gravitational production of
  vector dark matter}},
  \href{https://doi.org/10.1007/JHEP08(2020)059}{\emph{JHEP} {\bfseries 08}
  (2020) 059} [\href{https://arxiv.org/abs/2005.01766}{{\ttfamily
  2005.01766}}].

\bibitem{Duch:2017khv}
M.~Duch, B.~Grzadkowski and D.~Huang, \emph{{Strongly self-interacting vector
  dark matter via freeze-in}},
  \href{https://doi.org/10.1007/JHEP01(2018)020}{\emph{JHEP} {\bfseries 01}
  (2018) 020} [\href{https://arxiv.org/abs/1710.00320}{{\ttfamily
  1710.00320}}].

\bibitem{Barman:2020ifq}
B.~Barman, S.~Bhattacharya and B.~Grzadkowski, \emph{{Feebly coupled vector
  boson dark matter in effective theory}},
  \href{https://doi.org/10.1007/JHEP12(2020)162}{\emph{JHEP} {\bfseries 12}
  (2020) 162} [\href{https://arxiv.org/abs/2009.07438}{{\ttfamily
  2009.07438}}].

\bibitem{Barman:2021qds}
B.~Barman, N.~Bernal, A.~Das and R.~Roshan, \emph{{Non-minimally coupled vector
  boson dark matter}},
  \href{https://doi.org/10.1088/1475-7516/2022/01/047}{\emph{JCAP} {\bfseries
  01} (2022) 047} [\href{https://arxiv.org/abs/2108.13447}{{\ttfamily
  2108.13447}}].

\bibitem{Agrawal:2017eqm}
P.~Agrawal, G.~Marques-Tavares and W.~Xue, \emph{{Opening up the QCD axion
  window}}, \href{https://doi.org/10.1007/JHEP03(2018)049}{\emph{JHEP}
  {\bfseries 03} (2018) 049}
  [\href{https://arxiv.org/abs/1708.05008}{{\ttfamily 1708.05008}}].

\bibitem{Dror:2018pdh}
J.~A. Dror, K.~Harigaya and V.~Narayan, \emph{{Parametric Resonance Production
  of Ultralight Vector Dark Matter}},
  \href{https://doi.org/10.1103/PhysRevD.99.035036}{\emph{Phys. Rev. D}
  {\bfseries 99} (2019) 035036}
  [\href{https://arxiv.org/abs/1810.07195}{{\ttfamily 1810.07195}}].

\bibitem{Salehian:2020asa}
B.~Salehian, M.~A. Gorji, H.~Firouzjahi and S.~Mukohyama, \emph{{Vector dark
  matter production from inflation with symmetry breaking}},
  \href{https://doi.org/10.1103/PhysRevD.103.063526}{\emph{Phys. Rev. D}
  {\bfseries 103} (2021) 063526}
  [\href{https://arxiv.org/abs/2010.04491}{{\ttfamily 2010.04491}}].

\bibitem{Co:2021rhi}
R.~T. Co, K.~Harigaya and A.~Pierce, \emph{{Gravitational waves and dark photon
  dark matter from axion rotations}},
  \href{https://doi.org/10.1007/JHEP12(2021)099}{\emph{JHEP} {\bfseries 12}
  (2021) 099} [\href{https://arxiv.org/abs/2104.02077}{{\ttfamily
  2104.02077}}].

\bibitem{Nelson:2011sf}
A.~E. Nelson and J.~Scholtz, \emph{{Dark Light, Dark Matter and the
  Misalignment Mechanism}},
  \href{https://doi.org/10.1103/PhysRevD.84.103501}{\emph{Phys. Rev. D}
  {\bfseries 84} (2011) 103501}
  [\href{https://arxiv.org/abs/1105.2812}{{\ttfamily 1105.2812}}].

\bibitem{Arias:2012az}
P.~Arias, D.~Cadamuro, M.~Goodsell, J.~Jaeckel, J.~Redondo and A.~Ringwald,
  \emph{{WISPy Cold Dark Matter}},
  \href{https://doi.org/10.1088/1475-7516/2012/06/013}{\emph{JCAP} {\bfseries
  06} (2012) 013} [\href{https://arxiv.org/abs/1201.5902}{{\ttfamily
  1201.5902}}].

\bibitem{Kitajima:2023fun}
N.~Kitajima and K.~Nakayama, \emph{{Viable vector coherent oscillation
  dark~matter}},
  \href{https://doi.org/10.1088/1475-7516/2023/07/014}{\emph{JCAP} {\bfseries
  07} (2023) 014} [\href{https://arxiv.org/abs/2303.04287}{{\ttfamily
  2303.04287}}].

\bibitem{Alonso-Alvarez:2019ixv}
G.~Alonso-\'Alvarez, T.~Hugle and J.~Jaeckel, \emph{{Misalignment
  \textbackslash{}\& Co.: (Pseudo-)scalar and vector dark matter with curvature
  couplings}}, \href{https://doi.org/10.1088/1475-7516/2020/02/014}{\emph{JCAP}
  {\bfseries 02} (2020) 014}
  [\href{https://arxiv.org/abs/1905.09836}{{\ttfamily 1905.09836}}].

\bibitem{Himmetoglu:2009qi}
B.~Himmetoglu, C.~R. Contaldi and M.~Peloso, \emph{{Ghost instabilities of
  cosmological models with vector fields nonminimally coupled to the
  curvature}}, \href{https://doi.org/10.1103/PhysRevD.80.123530}{\emph{Phys.
  Rev. D} {\bfseries 80} (2009) 123530}
  [\href{https://arxiv.org/abs/0909.3524}{{\ttfamily 0909.3524}}].

\bibitem{Himmetoglu:2008zp}
B.~Himmetoglu, C.~R. Contaldi and M.~Peloso, \emph{{Instability of anisotropic
  cosmological solutions supported by vector fields}},
  \href{https://doi.org/10.1103/PhysRevLett.102.111301}{\emph{Phys. Rev. Lett.}
  {\bfseries 102} (2009) 111301}
  [\href{https://arxiv.org/abs/0809.2779}{{\ttfamily 0809.2779}}].

\bibitem{Himmetoglu:2008hx}
B.~Himmetoglu, C.~R. Contaldi and M.~Peloso, \emph{{Instability of the ACW
  model, and problems with massive vectors during inflation}},
  \href{https://doi.org/10.1103/PhysRevD.79.063517}{\emph{Phys. Rev. D}
  {\bfseries 79} (2009) 063517}
  [\href{https://arxiv.org/abs/0812.1231}{{\ttfamily 0812.1231}}].

\bibitem{Esposito-Farese:2009wbc}
G.~Esposito-Farese, C.~Pitrou and J.-P. Uzan, \emph{{Vector theories in
  cosmology}}, \href{https://doi.org/10.1103/PhysRevD.81.063519}{\emph{Phys.
  Rev. D} {\bfseries 81} (2010) 063519}
  [\href{https://arxiv.org/abs/0912.0481}{{\ttfamily 0912.0481}}].

\bibitem{Birrell:1982ix}
N.~D. Birrell and P.~C.~W. Davies, \emph{{Quantum Fields in Curved Space}},
  Cambridge Monographs on Mathematical Physics. Cambridge Univ. Press,
  Cambridge, UK, 2, 1984,
  \href{https://doi.org/10.1017/CBO9780511622632}{10.1017/CBO9780511622632}.

\bibitem{callan1970new}
C.~G. Callan~Jr, S.~Coleman and R.~Jackiw, \emph{A new improved energy-momentum
  tensor}, {\emph{Annals of physics} {\bfseries 59} (1970) 42}.

\bibitem{freedman1974energy}
D.~Z. Freedman, I.~J. Muzinich and E.~J. Weinberg, \emph{On the energy-momentum
  tensor in gauge field theories}, {\emph{Annals of Physics} {\bfseries 87}
  (1974) 95}.

\bibitem{FREEDMAN1974354}
D.~Z. Freedman and E.~J. Weinberg, \emph{The energy-momentum tensor in scalar
  and gauge field theories},
  \href{https://doi.org/https://doi.org/10.1016/0003-4916(74)90040-2}{\emph{Annals
  of Physics} {\bfseries 87} (1974) 354}.

\bibitem{Turner:1987bw}
M.~S. Turner and L.~M. Widrow, \emph{{Inflation Produced, Large Scale Magnetic
  Fields}}, \href{https://doi.org/10.1103/PhysRevD.37.2743}{\emph{Phys. Rev. D}
  {\bfseries 37} (1988) 2743}.

\bibitem{Faraoni:2000wk}
V.~Faraoni, \emph{{Inflation and quintessence with nonminimal coupling}},
  \href{https://doi.org/10.1103/PhysRevD.62.023504}{\emph{Phys. Rev. D}
  {\bfseries 62} (2000) 023504}
  [\href{https://arxiv.org/abs/gr-qc/0002091}{{\ttfamily gr-qc/0002091}}].

\bibitem{Golovnev:2008hv}
A.~Golovnev, V.~Mukhanov and V.~Vanchurin, \emph{{Gravitational waves in vector
  inflation}}, \href{https://doi.org/10.1088/1475-7516/2008/11/018}{\emph{JCAP}
  {\bfseries 11} (2008) 018} [\href{https://arxiv.org/abs/0810.4304}{{\ttfamily
  0810.4304}}].

\bibitem{Golovnev:2009ks}
A.~Golovnev and V.~Vanchurin, \emph{{Cosmological perturbations from vector
  inflation}}, \href{https://doi.org/10.1103/PhysRevD.79.103524}{\emph{Phys.
  Rev. D} {\bfseries 79} (2009) 103524}
  [\href{https://arxiv.org/abs/0903.2977}{{\ttfamily 0903.2977}}].

\bibitem{Golovnev:2009rm}
A.~Golovnev, \emph{{Linear perturbations in vector inflation and stability
  issues}}, \href{https://doi.org/10.1103/PhysRevD.81.023514}{\emph{Phys. Rev.
  D} {\bfseries 81} (2010) 023514}
  [\href{https://arxiv.org/abs/0910.0173}{{\ttfamily 0910.0173}}].

\bibitem{Moffat:2005si}
J.~W. Moffat, \emph{{Scalar-tensor-vector gravity theory}},
  \href{https://doi.org/10.1088/1475-7516/2006/03/004}{\emph{JCAP} {\bfseries
  03} (2006) 004} [\href{https://arxiv.org/abs/gr-qc/0506021}{{\ttfamily
  gr-qc/0506021}}].

\bibitem{Brownstein:2005zz}
J.~R. Brownstein and J.~W. Moffat, \emph{{Galaxy rotation curves without
  non-baryonic dark matter}},
  \href{https://doi.org/10.1086/498208}{\emph{Astrophys. J.} {\bfseries 636}
  (2006) 721} [\href{https://arxiv.org/abs/astro-ph/0506370}{{\ttfamily
  astro-ph/0506370}}].

\bibitem{Ji:2021rrn}
L.~Ji, \emph{{Wave Dark Matter Non-minimally Coupled to Gravity}},
  \href{https://arxiv.org/abs/2106.11971}{{\ttfamily 2106.11971}}.

\bibitem{Sankharva:2021spi}
K.~Sankharva and S.~Sethi, \emph{{Nonminimally coupled ultralight axions as
  cold dark matter}},
  \href{https://doi.org/10.1103/PhysRevD.105.103517}{\emph{Phys. Rev. D}
  {\bfseries 105} (2022) 103517}
  [\href{https://arxiv.org/abs/2110.04322}{{\ttfamily 2110.04322}}].

\bibitem{Ivanov:2019iec}
D.~Ivanov and S.~Liberati, \emph{{Testing Non-minimally Coupled BEC Dark Matter
  with Gravitational Waves}},
  \href{https://doi.org/10.1088/1475-7516/2020/07/065}{\emph{JCAP} {\bfseries
  07} (2020) 065} [\href{https://arxiv.org/abs/1909.02368}{{\ttfamily
  1909.02368}}].

\bibitem{Kouwn:2015cdw}
S.~Kouwn, P.~Oh and C.-G. Park, \emph{{Massive Photon and Dark Energy}},
  \href{https://doi.org/10.1103/PhysRevD.93.083012}{\emph{Phys. Rev. D}
  {\bfseries 93} (2016) 083012}
  [\href{https://arxiv.org/abs/1512.00541}{{\ttfamily 1512.00541}}].

\bibitem{Koivisto:2008xf}
T.~Koivisto and D.~F. Mota, \emph{{Vector Field Models of Inflation and Dark
  Energy}}, \href{https://doi.org/10.1088/1475-7516/2008/08/021}{\emph{JCAP}
  {\bfseries 08} (2008) 021} [\href{https://arxiv.org/abs/0805.4229}{{\ttfamily
  0805.4229}}].

\bibitem{Cembranos:2016ugq}
J.~A.~R. Cembranos, A.~L. Maroto and S.~J. N\'u\~nez Jare\~no,
  \emph{{Perturbations of ultralight vector field dark matter}},
  \href{https://doi.org/10.1007/JHEP02(2017)064}{\emph{JHEP} {\bfseries 02}
  (2017) 064} [\href{https://arxiv.org/abs/1611.03793}{{\ttfamily
  1611.03793}}].

\bibitem{Jain:2021pnk}
M.~Jain and M.~A. Amin, \emph{{Polarized solitons in higher-spin wave dark
  matter}},  9, 2021.
\newblock 10.1103/PhysRevD.105.056019.

\bibitem{madelung1927quantentheorie}
E.~Madelung, \emph{Quantentheorie in hydrodynamischer form}, {\emph{Zeitschrift
  f{\"u}r Physik} {\bfseries 40} (1927) 322}.

\bibitem{Widrow:1993qq}
L.~M. Widrow and N.~Kaiser, \emph{{Using the Schrodinger equation to simulate
  collisionless matter}}, {\emph{Astrophys. J. Lett.} {\bfseries 416} (1993)
  L71}.

\bibitem{Loginov:2015rya}
A.~Y. Loginov, \emph{{Nontopological solitons in the model of the
  self-interacting complex vector field}},
  \href{https://doi.org/10.1103/PhysRevD.91.105028}{\emph{Phys. Rev. D}
  {\bfseries 91} (2015) 105028}.

\bibitem{Brito:2015pxa}
R.~Brito, V.~Cardoso, C.~A.~R. Herdeiro and E.~Radu, \emph{{Proca stars:
  Gravitating Bose\textendash{}Einstein condensates of massive spin 1
  particles}},
  \href{https://doi.org/10.1016/j.physletb.2015.11.051}{\emph{Phys. Lett. B}
  {\bfseries 752} (2016) 291}
  [\href{https://arxiv.org/abs/1508.05395}{{\ttfamily 1508.05395}}].

\bibitem{Sanchis-Gual:2017bhw}
N.~Sanchis-Gual, C.~Herdeiro, E.~Radu, J.~C. Degollado and J.~A. Font,
  \emph{{Numerical evolutions of spherical Proca stars}},
  \href{https://doi.org/10.1103/PhysRevD.95.104028}{\emph{Phys. Rev. D}
  {\bfseries 95} (2017) 104028}
  [\href{https://arxiv.org/abs/1702.04532}{{\ttfamily 1702.04532}}].

\bibitem{Amin:2023imi}
M.~A. Amin, A.~J. Long and E.~D. Schiappacasse, \emph{{Photons from dark photon
  solitons via parametric resonance}},
  \href{https://doi.org/10.1088/1475-7516/2023/05/015}{\emph{JCAP} {\bfseries
  05} (2023) 015} [\href{https://arxiv.org/abs/2301.11470}{{\ttfamily
  2301.11470}}].

\bibitem{Jain:2023ojg}
M.~Jain, M.~A. Amin, J.~Thomas and W.~Wanichwecharungruang, \emph{{Kinetic
  relaxation and Bose-star formation in multicomponent dark matter- I}},
  \href{https://arxiv.org/abs/2304.01985}{{\ttfamily 2304.01985}}.

\bibitem{Chavanis:2011zm}
P.~H. Chavanis and L.~Delfini, \emph{{Mass-radius relation of Newtonian
  self-gravitating Bose-Einstein condensates with short-range interactions: II.
  Numerical results}},
  \href{https://doi.org/10.1103/PhysRevD.84.043532}{\emph{Phys. Rev. D}
  {\bfseries 84} (2011) 043532}
  [\href{https://arxiv.org/abs/1103.2054}{{\ttfamily 1103.2054}}].

\bibitem{Chavanis:2011zi}
P.-H. Chavanis, \emph{{Mass-radius relation of Newtonian self-gravitating
  Bose-Einstein condensates with short-range interactions: I. Analytical
  results}}, \href{https://doi.org/10.1103/PhysRevD.84.043531}{\emph{Phys. Rev.
  D} {\bfseries 84} (2011) 043531}
  [\href{https://arxiv.org/abs/1103.2050}{{\ttfamily 1103.2050}}].

\bibitem{Robles:2018fur}
V.~H. Robles, J.~S. Bullock and M.~Boylan-Kolchin, \emph{{Scalar Field Dark
  Matter: Helping or Hurting Small-Scale Problems in Cosmology?}},
  \href{https://doi.org/10.1093/mnras/sty3190}{\emph{Mon. Not. Roy. Astron.
  Soc.} {\bfseries 483} (2019) 289}
  [\href{https://arxiv.org/abs/1807.06018}{{\ttfamily 1807.06018}}].

\bibitem{Desjacques:2019zhf}
V.~Desjacques and A.~Nusser, \emph{{Axion core\textendash{}halo mass and the
  black hole\textendash{}halo mass relation: constraints on a few parsec
  scales}}, \href{https://doi.org/10.1093/mnras/stz1978}{\emph{Mon. Not. Roy.
  Astron. Soc.} {\bfseries 488} (2019) 4497}
  [\href{https://arxiv.org/abs/1905.03450}{{\ttfamily 1905.03450}}].

\bibitem{Safarzadeh:2019sre}
M.~Safarzadeh and D.~N. Spergel, \emph{Ultra-light dark matter is incompatible
  with the milky way’s dwarf satellites},
  \href{https://doi.org/10.3847/1538-4357/ab7db2}{\emph{The Astrophysical
  Journal} {\bfseries 893} (2020) 21}
  [\href{https://arxiv.org/abs/1906.11848}{{\ttfamily 1906.11848}}].

\bibitem{Li:2018kyk}
X.~Li, L.~Hui and G.~L. Bryan, \emph{{Numerical and Perturbative Computations
  of the Fuzzy Dark Matter Model}},
  \href{https://doi.org/10.1103/PhysRevD.99.063509}{\emph{Phys. Rev. D}
  {\bfseries 99} (2019) 063509}
  [\href{https://arxiv.org/abs/1810.01915}{{\ttfamily 1810.01915}}].

\bibitem{Bechtol:2022koa}
K.~Bechtol et~al., \emph{{Snowmass2021 Cosmic Frontier White Paper: Dark Matter
  Physics from Halo Measurements}},  in \emph{{Snowmass 2021}}, 3, 2022,
  \href{https://arxiv.org/abs/2203.07354}{{\ttfamily 2203.07354}}.

\bibitem{DeFelice:2011bh}
A.~De~Felice and S.~Tsujikawa, \emph{{Conditions for the cosmological viability
  of the most general scalar-tensor theories and their applications to extended
  Galileon dark energy models}},
  \href{https://doi.org/10.1088/1475-7516/2012/02/007}{\emph{JCAP} {\bfseries
  02} (2012) 007} [\href{https://arxiv.org/abs/1110.3878}{{\ttfamily
  1110.3878}}].

\bibitem{LIGOScientific:2017zic}
{\scshape LIGO Scientific, Virgo, Fermi-GBM, INTEGRAL} collaboration,
  \emph{{Gravitational Waves and Gamma-rays from a Binary Neutron Star Merger:
  GW170817 and GRB 170817A}},
  \href{https://doi.org/10.3847/2041-8213/aa920c}{\emph{Astrophys. J. Lett.}
  {\bfseries 848} (2017) L13}
  [\href{https://arxiv.org/abs/1710.05834}{{\ttfamily 1710.05834}}].

\bibitem{Baker:2017hug}
T.~Baker, E.~Bellini, P.~G. Ferreira, M.~Lagos, J.~Noller and I.~Sawicki,
  \emph{{Strong constraints on cosmological gravity from GW170817 and GRB
  170817A}}, \href{https://doi.org/10.1103/PhysRevLett.119.251301}{\emph{Phys.
  Rev. Lett.} {\bfseries 119} (2017) 251301}
  [\href{https://arxiv.org/abs/1710.06394}{{\ttfamily 1710.06394}}].

\bibitem{Moore:2001bv}
G.~D. Moore and A.~E. Nelson, \emph{{Lower bound on the propagation speed of
  gravity from gravitational Cherenkov radiation}},
  \href{https://doi.org/10.1088/1126-6708/2001/09/023}{\emph{JHEP} {\bfseries
  09} (2001) 023} [\href{https://arxiv.org/abs/hep-ph/0106220}{{\ttfamily
  hep-ph/0106220}}].

\end{thebibliography}\endgroup
\end{document}